\documentclass[12pt,epsf]{article}

\renewcommand{\thefootnote}{\fnsymbol{footnote}}
\newlength{\pubnumber} 

\catcode`\@=11
\@addtoreset{equation}{section}

\def\section{\@startsection{section}{1}{\z@}{3.5ex plus 1ex minus .2ex}
 {2.3ex plus .2ex}{\large\bf}}
\def\subsection{\@startsection{subsection}{2}{\z@}{2.3ex plus .2ex}
 {2.3ex plus .2ex}{\bf}}

\usepackage{graphicx}% Include figure files
\usepackage{dcolumn}% Align table columns on decimal point
\usepackage{bm}% bold math

\usepackage[margin=1in]{geometry}

\usepackage{amsmath}
\usepackage{amsfonts}
\usepackage{multirow}
\usepackage{rotating}
\usepackage{palatino, url, multicol}
\usepackage{amssymb}
\usepackage{hyperref}

\def\beq{\begin{equation}}
\def\eeq{\end{equation}}
\def\beqn{\begin{eqnarray}}
\def\eeqn{\end{eqnarray}}

\def\nolabel{\nonumber }

\newcommand{\mmod}{\mbox{mod }}

\def\nolabel{\nonumber }
\def\Tr{\mathop{\rm Tr \,} \nolimits}

\begin{document}

\renewcommand{\thefootnote}{\arabic{footnote}}

\begin{titlepage}
\setcounter{page}{1}
\rightline{BU-HEPP-08-20}
\rightline{\tt }

\vspace{.06in}
\begin{center}
{\Large A Simple Introduction to Particle Physics \\
\rm \large Part I - Foundations and the Standard Model}\\
\vspace{.12in}

{\large
        Matthew B. Robinson, \footnote{m\_robinson@baylor.edu}
        Karen R. Bland, \footnote{karen\_bland@baylor.edu} \\
        Gerald B. Cleaver, \footnote{gerald\_cleaver@baylor.edu} and
        Jay R. Dittmann \footnote{jay\_dittmann@baylor.edu}}
\\
\vspace{.12in}
{\it        Department of Physics, One Bear Place \# 97316\\
            Baylor University\\
            Waco, TX 76798-7316\\}
\vspace{.06in}
\end{center}

\begin{abstract}

\addtolength{\parskip}{0.8\baselineskip}
\parindent 0pt

\noindent This is the first of a series of papers in which we present a brief
introduction to the relevant mathematical and physical ideas that form
the foundation of Particle Physics, including Group Theory,
Relativistic Quantum Mechanics, Quantum Field Theory and Interactions,
Abelian and Non-Abelian Gauge Theory, and the $SU(3)\otimes SU(2)
\otimes U(1)$ Gauge Theory that describes our universe apart from
gravity.  Our approach, at first, is an algebraic exposition of Gauge
Theory and how the physics of our universe comes out of Gauge Theory.  

With an algebraic understanding of Gauge Theory and the relevant
physics of the Standard Model from this paper, in a subsequent paper we
will ``back up" and reformulate Gauge Theory from a geometric
foundation, showing how it connects to the algebraic picture initially
built in these notes.  

Finally, we will introduce the basic ideas of String Theory, showing
both the geometric and algebraic correspondence with Gauge Theory as
outlined in the first two parts.  

These notes are not intended to be a comprehensive introduction to any
of the ideas contained in them.  Their purpose is to introduce the
``forest" rather than the ``trees".  The primary emphasis is on the
algebraic/geometric/mathematical underpinnings rather than the
calculational/phenomenological details.  Among the glaring omissions
are CPT theorems, evaluations of Feynman Diagrams, Renormalization, and
Anomalies.  The topics were chosen according to the authors'
preferences and agenda.  

These notes are intended for a student who has completed the standard
undergraduate physics and mathematics courses.	The material in the
first part is intended as a review and is therefore cursory. 
Furthermore, these notes should not and will not in any way take the
place of the related courses, but rather provide a primer for detailed
courses in QFT, Gauge Theory, String Theory, etc., which will fill in
the many gaps left by this paper.  
\end{abstract}

\end{titlepage}
\setcounter{footnote}{0}

%******************************************************************

\addtolength{\parskip}{0.8\baselineskip}
\parindent 0pt

\tableofcontents

\newpage

\section{Part I --- Preliminary Concepts}

\subsection{Review of Classical Physics}

\subsubsection{Hamilton's Principle}
\label{sec:hamiltonsprinciple}

Nearly all physics begins with what is called a \bf Lagrangian \rm for a
particle, which is initially defined as the kinetic energy minus the
potential energy,
\begin{eqnarray}
L \equiv T-V \nolabel 
\end{eqnarray}
where $T=T(q,\dot{q})$ and $V=V(q)$.  Then, the \bf Action \rm is defined as
the integral of the Lagrangian from an initial time to a final time,
\begin{eqnarray}
S \equiv \int_{t_i}^{t_f} dt L(q,\dot{q}) \nolabel 
\end{eqnarray}

It is important to realize that $S$ is a ``functional" of the
particle's world-line in $(q,\dot{q})$ space, not a function. 
This means that it depends on the entire path $(q,\dot{q})$, rather
than a given point on the path.  The only fixed points on the path are
$q(t_i)$, $q(t_f)$, $\dot{q}(t_i)$, and $\dot{q}(t_f)$.	The rest of the
path is generally unconstrained, and the value of $S$ depends on the
entire path.  

\bf Hamilton's Principle \rm says that nature extremizes the path a particle
will take in going from $q(t_i)$ at time $t_i$ to position $q(t_f)$ at
time $t_f$.  In other words, the path that extremizes the action will
be the path the particle will travel.  

But, because $S$ is a functional, depending on the entire path in
$(q,\dot{q})$ space rather than a point, it cannot be extremized in the
``Calculus I" sense of merely setting the derivative equal to 0. 
Instead, we must find the path for which the action is ``stationary". 
This means that the first-order term in the Taylor Expansion around
that path will vanish, or $\delta S = 0$ at that path.	

To find this, consider some arbitrary path $(q,\dot{q})$.  If it is a
path that minimizes the action, then we will have
\begin{eqnarray}
0 &=& \delta S = \delta \int_{t_i}^{t_f} dt L(q,\dot{q}) =
\int_{t_i}^{t_f} dt L(q+\delta q,\dot{q}+\delta \dot{q})-S \nolabel \\
&=& \int_{t_i}^{t_f} dtL(q,\dot{q}) + \int_{t_i}^{t_f} dt\bigg(\delta q
{\partial L \over \partial q} + \delta \dot{q} {\partial L \over
\partial \dot{q}} \bigg) - S \nolabel \\
&=& \int_{t_i}^{t_f} dt \bigg(\delta q {\partial L \over \partial q} +
{\partial L \over \partial \dot{q}} {d \over dt} \delta q\bigg)
\nolabel 
\end{eqnarray}
Integrating the second term by parts, and taking the variation of
$\delta q$ to be at 0 at $t_i$ and $t_f$, 
\begin{eqnarray}
\delta S = \int_{t_i}^{t_f} dt \bigg(\delta q{\partial L \over \partial
q} - \delta q {d \over dt} {\partial L \over \partial \dot{q}} \bigg) =
\int_{t_i}^{t_f} dt \delta q \bigg({\partial L \over \partial q} - {d
\over dt} {\partial L \over \partial \dot{q}} \bigg) = 0 \nolabel 
\end{eqnarray}
The only way to guarantee this for an arbitrary variation $\delta q$
from the path $(q,\dot{q})$ is to demand
\begin{eqnarray}
{d \over dt}{\partial L \over \partial \dot{q}} - {\partial L \over
\partial q} = 0 \nolabel
\end{eqnarray}
This equation is called the \bf Euler-Lagrange \rm equation, and it produces
the equations of motion of the particle.  

The generalization to multiple coordinates $q_i$ ($i=1,\ldots,n$) is
straightforward:
\begin{eqnarray}
{d \over dt} {\partial L \over \partial \dot{q}_i} - {\partial L \over
\partial q_i} = 0 \label{eq:eulerlagrange}
\end{eqnarray}

\subsubsection{Noether's Theorem}
\label{sec:noether}

Given a Lagrangian $L=L(q,\dot{q})$, consider making an infinitesimal
transformation
\begin{eqnarray}
q \rightarrow q+\epsilon \delta q \nolabel
\end{eqnarray}
where $\epsilon$ is some infinitesimal constant.  This transformation
will give 
\begin{eqnarray}
L(q,\dot{q}) \rightarrow L(q+\epsilon \delta q, \dot{q}+\epsilon \delta
\dot{q}) = L(q,\dot{q})+\epsilon \delta q{\partial L \over \partial
q}+\epsilon \delta \dot{q} {\partial L \over \partial \dot{q}} \nolabel 
\end{eqnarray}
If the Euler-Lagrange equations of motion are satisfied, so that
${\partial L \over \partial q} = {d \over dt} {\partial L \over
\partial \dot{q}}$, then under $q \rightarrow q + \epsilon \delta q$, 
\begin{eqnarray}
L \rightarrow L + \epsilon \delta q {\partial L \over \partial q} +
\epsilon \delta \dot{q} {\partial L \over \partial \dot{q}} = L +
\epsilon \delta q {d \over dt} {\partial L \over \partial \dot{q}} +
\epsilon {\partial L \over \partial \dot{q}}{d \over dt} \delta q  = L
+ {d \over dt} \bigg({\partial L \over \partial \dot{q}} \epsilon
\delta q\bigg) \nolabel 
\end{eqnarray}

So, under $q \rightarrow q+ \epsilon \delta q$, we have $\delta L = {d
\over dt} \big({\partial L \over \partial \dot{q}} \epsilon \delta q
\big)$.  We define the \bf Noether Current\rm, $j$, as 
\begin{eqnarray}
j \equiv {\partial L \over \partial \dot{q}} \delta q \nolabel 
\end{eqnarray}

Now, if we can find some transformation $\delta q$ that leaves the
action invariant, or in other words such that $\delta S = 0$, then ${d
j \over dt} = 0$, and therefore the current $j$ is a constant in time. 
In other words, $j$ is \it conserved\rm.  

As a familiar example, consider a projectile, described by the
Lagrangian
\begin{eqnarray}
L = {1\over 2} m (\dot{x}^2 + \dot{y}^2) - mgy \label{eq:projectile}
\end{eqnarray}
This will be unchanged under the transformation $x \rightarrow
x+\epsilon$, where $\epsilon$ is any constant (here, $\delta q=1$ in
the above notation), because $x \rightarrow x+\epsilon \Rightarrow
\dot{x} \rightarrow \dot{x}$.  So, $j = {\partial L \over \partial
\dot{q}} \delta q = m \dot{x}$ is conserved.  We recognize $m\dot{x}$
as the momentum in the $x$-direction, which we expect to be conserved
by conservation of momentum.  

So in summary, \bf Noether's Theorem \rm merely says that whenever there is a
continuous symmetry in the action, there is a corresponding conserved
quantity.   

\subsubsection{Conservation of Energy}

Consider the quantity 
\begin{eqnarray}
{d L \over dt} = {d \over dt} L (q,\dot{q}) = {\partial L \over
\partial q} {dq \over dt} + {\partial L \over \partial \dot{q}}
{d\dot{q} \over dt} + {\partial L \over \partial t} \nolabel 
\end{eqnarray}
Because $L$ does not depend explicitly on time, ${\partial L \over
\partial t} = 0$, and therefore
\begin{eqnarray}
{d L \over dt} = {\partial L \over \partial q}\dot{q} + {\partial L
\over \partial \dot{q}} \ddot{q} = \bigg({d\over dt} {\partial L \over
\partial \dot{q}} \bigg) \dot{q} + {\partial L \over \partial \dot{q}}
\ddot{q} = {d \over dt} \bigg({\partial L \over \partial \dot{q}}
\dot{q}\bigg) \nolabel 
\end{eqnarray}
where we have used the Euler-Lagrange equation to get the second
equality.  So, we have ${d L \over dt} = {d \over dt} \big({\partial L
\over \partial \dot{q}} \dot{q}\big)$, or
\begin{eqnarray}
{d \over dt} \bigg({\partial L \over \partial \dot{q}}\dot{q} - L
\bigg) = 0 \label{eq:conservationofenergy}
\end{eqnarray}

For a general non-relativistic system, $L=T-V$, so ${\partial L \over
\partial \dot{q}} = {\partial T \over \partial \dot{q}}$ because $V$ is
a function of $q$ only, and normally 
\begin{eqnarray}
T \propto \dot{q}^2 \quad \Rightarrow\quad  {\partial L \over \partial \dot{q}} \dot{q} = 2T \nolabel
\end{eqnarray}
So, ${\partial L \over \partial \dot{q}} \dot{q} - L = 2T
- (T-V) = T+V = E$, the total energy of the system, which is conserved
according to (\ref{eq:conservationofenergy}).  We identify $T+V \equiv
H$ as the \bf Hamiltonian\rm, or total energy function, of the system.   

Furthermore, we define ${\partial L \over \partial \dot{q}} \equiv p$ to be
the momentum of the system.  Then, the relationship between the
Lagrangian and the Hamiltonian is the Legendre transformation
\begin{eqnarray}
p \dot{q} - L = H \nolabel 
\end{eqnarray}

\subsubsection{Lorentz Transformations}

Consider some event that occurs at spatial position $(x, y, z)^T$, at
time $t$. (The superscript $T$ denotes the transpose, so this is a column vector.)  We
arrange this event in a column \mbox{4-vector} as $(ct,x,y,z)^T$, where $c$
is the speed of light (the units of $c$ give each element the
same units).  A more useful notation is to refer to this vector as
$a^{\mu} = (ct, x,y,z)^T$, where $\mu = 0,1,2,3$.  This 4-vector,
with the $\mu$ index raised, is called a ``vector", or a
``contravariant vector".  Then, we define the row vector $a_{\mu} =
(-ct,x,y,z)$.  This is called a ``covector", or a ``covariant vector". 
In general, the sign of the $0^{th}$ component (the component in the
first position) changes when going from vector to covector.  

There is something very deep going on here regarding the geometrical
picture between vectors and covectors, but we will not discuss it until
the next paper in this series.	

The dot product between two such vectors (a covector and vector) is
then defined as the product with one index raised and the other
lowered.  Whenever indices are contracted in such a way, it is
understood that they are to be summed over.\footnote{because we are
summing over components, we can write $a^{\mu}b_{\mu}$ or $a_{\mu}b^{\mu}$ --- 
they mean the same thing}
\begin{eqnarray}
a \cdot b = a^{\mu}b_{\mu} = a^0b_0+a^1b_1+a^2b_2+a^3b_3 =
-a^0b^0+a^1b^1+a^2b^2+a^3b^3 \nolabel
\end{eqnarray}
Or, plugging in the spacetime notation from above, where 
\begin{eqnarray}
a^{\mu} = (ct_1,x_1,y_1,z_1)^T \qquad \rm and \qquad \it b^{\mu} = (ct_2,x_2,y_2,z_2)^T \nolabel
\end{eqnarray}
we have
\begin{eqnarray}
a\cdot b = a_{\mu}b^{\mu} = -c^2t_1t_2+x_1x_2+y_1y_2+z_1z_2 \nolabel 
\end{eqnarray}

We can also discuss the differential version of this.  If $s^{\mu} =
(ct,x,y,z)$, then $ds^2 = -c^2dt^2+dx^2+dy^2+dz^2$.

In his theory of Special Relativity, Einstein postulated that all
inertial reference frames are equivalent, and that the speed of light
is the same in all frames.  To put this in more mathematical terms, if
observers in different inertial frames $1$ and $2$ each see an event,
they will see, respectively,
\begin{eqnarray}
ds_1^2 &=& -c^2dt_1^2+dx^2_1+dy^2_1+dz_1^2 \nolabel \\
ds_2^2 &=& -c^2 dt_2^2+dx^2_2+dy^2_2+dz^2_2 \nolabel
\end{eqnarray}
We then demand that $ds_1^2 = ds^2_2$.	To do this, we must find a
modification of the standard Galilean transformations that will leave
$ds^2$ unchanged.  The derivation for the correct transformations can
be found in any introductory or modern physics text, so we merely quote
the result.  If we assume that frame $2$ is moving only in the $z$-direction
with respect to frame $1$ (and that their $x$, $y$, and $z$ axes are aligned),
then we find that the transformations are 
\begin{eqnarray}
t_2 &=& \gamma(ct_1-\beta z_1) \nolabel \\
z_2 &=& \gamma(z_1 - \beta ct_1) \label{eq:lorentz}
\end{eqnarray}
where $\beta = {v\over c}$ and $\gamma = {1\over \sqrt{1-\beta^2}}$. 
These transformations, which preserve $ds^2$ when transforming one
frame to another, are called \bf Lorentz Transformations\rm.  

Discussions of the implications of these transformations, including
time dilation, length contraction, and the relationship between energy
and mass can be found in most introductory texts.  You are encouraged
to review the material if you are not familiar with it.  

\subsubsection{A More Detailed Look at Lorentz Transformations}
\label{sec:lorentzdetail}

As we have seen, we have a quantity $ds^2 = -c^2dt^2+dx^2+dy^2+dz^2$,
which does not change under transformations (\ref{eq:lorentz}). 
Thinking of physical ideas this way, in terms of ``what doesn't change
when something else changes", will prove to be an extraordinarily
powerful approach.  In order to understand Special Relativity in such a
way, we begin with a simpler example.  

Consider a spatial rotation around, say, the $z$-axis (or, equivalently,
mixing the $x$ and $y$ coordinates).  Such a transformation is called
an \bf Euler Transformation\rm, and takes the form
\begin{eqnarray}
t' &=& t \nolabel \\
x' &=& x  \cos \theta + y \sin \theta \nolabel \\
y' &=& -x  \sin \theta + y \cos \theta \nolabel \\
z' &=& z \label{eq:lorrot}
\end{eqnarray}
where $\theta$ is the angle of rotation, called the \bf Euler Angle\rm.
 We can simultaneously express a Lorentz transformation as a sort of
``rotation" that mixes a spatial dimension and a time dimension, as follows
(these transformations are equivalent to (\ref{eq:lorentz}):
\begin{eqnarray}
t' &=& t  \cosh\theta - x  \sinh \theta \nolabel \\
x' &=& -t  \sinh \theta + x  \cosh \theta \nolabel \\
y' &=& y \nolabel \\
z' &=& z \label{eq:lorboost}
\end{eqnarray}
where $\theta$ is defined by the relationship $\beta = \tan \theta$.  

We denote a transformation mixing two spatial dimensions simply a \bf
Rotation\rm, whereas a transformation mixing a spatial dimension and a
time dimension is a \bf Boost\rm.  Any two frames whose origins
coincide at $t=t'=0$ can be transformed into each other through some
combination of rotations and boosts.  

To rephrase this in more precise language, given a 4-vector
$x^{\mu}$, it will be related to the equivalent 4-vector in another
frame, $x'^{\mu}$, by some matrix $L$, according to $x'^{\mu} =
L^{\mu}_{\nu}x^{\nu}$ (where the summation convention discussed earlier
is in effect for the repeated index).  

We also introduce what is called the \bf Metric \rm matrix,
\begin{eqnarray}
\eta_{\mu \nu} = \eta^{\mu \nu} = 
\begin{pmatrix}
-1 & 0 & 0 & 0 \\
0 & 1 & 0 & 0 \\
0 & 0 & 1 & 0 \\
0 & 0 & 0 & 1
\end{pmatrix} \nolabel
\end{eqnarray}
In general, $\eta^{\mu \nu} \equiv (\eta_{\mu \nu})^{-1}$.

Using the metric, the dot product of any 4-vector $x^{\mu} = (ct,
x,y,z)^T$ can be easily written as $x^2 = x^{\mu}x_{\mu} = \eta_{\mu
\nu} x^{\mu} x^{\nu} = -c^2t^2 + x^2+y^2+z^2$.	In general, a Lorentz
transformation can be defined as a matrix $L^{\mu}_{\nu}$ (including
boosts and rotations) that leaves $\eta_{\mu \nu}x^{\mu}x^{\nu}$ 
unchanged.  

For example, a scalar, or an object with no uncontracted indices, like
$\phi$ or $x^{\mu}x_{\mu}$, is simply invariant under Lorentz
transformations ($\phi \rightarrow \phi$, $x^{\mu}x_{\mu} \rightarrow
x^{\mu}x_{\mu}$).  

A vector, or an object with only one uncontracted index, like $x^{\mu}$
or $a^{\mu}b_{\mu}^{\nu}$, transforms according to $x'^{\mu} =
L^{\mu}_{\nu}x^{\nu}$, or $(a^{\mu}b^{\nu}_{\mu})' =
L^{\nu}_{\alpha}(a^{\mu}b^{\alpha}_{\mu}$).  

Now, consider the dot product $x^2 = x^{\mu}x_{\mu} = \eta_{\mu
\nu}x^{\mu}x^{\nu}$.  If $x^2$ is invariant, then $x'^2 = x^2
\Rightarrow \eta_{\mu \nu}x'^{\mu}x'^{\nu} = \eta_{\mu
\nu}L^{\mu}_{\alpha}L^{\nu}_{\beta}x^{\alpha}x^{\beta}$ demands that
$\eta_{\mu \nu}L^{\mu}_{\alpha}L^{\nu}_{\beta} = \eta_{\alpha \beta}$. 
So, the constraint for Lorentz transformations is that they are the set
of all matrices such that 
\begin{eqnarray}
\eta_{\mu \nu}L^{\mu}_{\alpha}L^{\nu}_{\beta} = \eta_{\alpha \beta}
\nolabel 
\end{eqnarray}
We take this to be the \it defining \rm constraint for a Lorentz
transformation.  

\subsubsection{Classical Fields}
\label{sec:classical}

When deriving the Euler-Lagrange equations, we started with an action
$S$ which was an integral over time only ($S \equiv \int dt L$).  If we
are eventually interested in a relativistically acceptable theory, this
is obviously no good because it treats time and space differently (the
action is an integral over time but not over space).  

So, let's consider an action defined not in terms of the Lagrangian,
but of the ``Lagrangian per unit volume", or the \bf Lagrangian Density
\rm $\mathcal{L}$.  The Lagrangian will naturally be the integral of
$\mathcal{L}$ over all space, $L = \int d^nx \mathcal{L}$.  The integral
is in $n$-dimensions, so $d^nx$ means $dx^1dx^2dx^2\cdots dx^n$.  

Now, the action will be $S = \int dt L = \int dt d^nx \mathcal{L}$.  In
the normal $1+3$ dimensional Minkowski spacetime we live in, this will
be $S = \int dt d^3x\mathcal{L} = \int d^4x \mathcal{L}$.  

Before, $L$ depended not on $t$, but on the path $q(t)$, $\dot{q}(t)$. 
In a similar sense, $\mathcal{L}$ will not depend on $\bar x$ and $t$,
but on what we will refer to as \bf Fields\rm, $\phi(\bar x,t) =
\phi(x^{\mu})$, which exist in spacetime.  

Following a nearly identical argument as the one leading to
(\ref{eq:eulerlagrange}), we get the relativistic field generalization 
\begin{eqnarray}
\partial_{\mu} \bigg({\partial \mathcal{L} \over \partial
(\partial_{\mu} \phi_i)}\bigg) - {\partial \mathcal{L} \over \partial
\phi_i} = 0 \nolabel
\end{eqnarray}
for multiple fields $\phi_i$ ($i=1,\ldots,n$).  

Noether's Theorem says that, for $\phi \rightarrow \phi+\epsilon \delta
\phi$, we have a current $j^{\mu} \equiv {\partial \mathcal{L} \over
\partial (\partial_{\mu}\phi)} \delta \phi$, and if $\phi \rightarrow
\phi + \epsilon \delta \phi$ leaves $\delta \mathcal{L} = 0$, then
$\partial_{\mu}j^{\mu} = 0 \Rightarrow -{\partial j^0 \over \partial t}
+ \bar \nabla \cdot \bar j = 0 $, where $j^0$ is the \bf Charge
Density\rm, and $\bar j$ is the \bf Current Density\rm.  The total
charge will naturally be $Q \equiv \int_{all \; space} d^3xj^0$.  

Finally, we also have a \bf Hamiltonian Density \rm and momentum
\begin{eqnarray}
\mathcal{H} &\equiv& {\partial \mathcal{L} \over \partial
\dot{\phi}_{\mu}} \dot{\phi}_{\mu} - \mathcal{L}
\label{eq:hamiltoniandensity}\\
\Pi^{\mu} &\equiv& {\partial \mathcal{L} \over
\partial\dot{\phi}_{\mu}} \label{eq:momentumdensity}
\end{eqnarray}

One final comment for this section.  For the remainder of these notes,
we will ultimately be seeking a relativistic field theory, and
therefore we will never make use of Lagrangians.  We will always use
Lagrangian densities.  We will always use the notation $\mathcal{L}$
instead of $L$, but we will refer to the Lagrangian densities simply as
Lagrangians.  We drop the word ``densities'' for brevity, and because
there will never be ambiguity.	

\subsubsection{Classical Electrodynamics}

We choose our units so that $c=\mu_0=\epsilon_0=1$.  So, the magnitude
of the force between two charges $q_1$ and $q_2$ is $F={q_1q_2 \over
4\pi r^2}$.  In these units, Maxwell's equations are 
\begin{eqnarray}
\bar \nabla \cdot \bar E &=& \rho \label{eq:maxwell1} \\
\bar \nabla \times \bar B - {\partial \bar E \over \partial t} &=& \bar
J \label{eq:maxwell2} \\
\bar \nabla \cdot \bar B &=& 0 \label{eq:hommaxwell1} \\
\bar \nabla \times \bar E + {\partial \bar B \over \partial t} &=& 0
\label{eq:hommaxwell2}
\end{eqnarray}

If we define the \bf Potential \rm 4-vector $A^{\mu} = (\phi, \bar
A)$, then we can define $\bar B = \bar \nabla \times \bar A$ and $\bar
E = -\bar \nabla \phi - {\partial \bar A \over \partial t}$.  Writing
$\bar B$ and $\bar E$ this way will automatically solve the homogenous
Maxwell equations, (\ref{eq:hommaxwell1}) and (\ref{eq:hommaxwell2}).  

Then, we define the totally antisymmetric \bf Electromagnetic Field
Strength Tensor \rm $F^{\mu \nu}$ as
\begin{eqnarray}
F^{\mu \nu} \equiv \partial^{\mu}A^{\nu} - \partial^{\nu}A^{\mu} = 
\begin{pmatrix}
0 & -E_x & -E_y & -E_z \\
E_x & 0 & -B_z & B_y \\
E_y & B_z & 0 & -B_x \\
E_z & -B_y & B_x & 0
\end{pmatrix} \nolabel
\end{eqnarray}
We define the 4-vector current as $J^{\mu} = (\rho, \bar J)$. 
It is straightforward, though tedious, to show that
\begin{eqnarray}
\partial^{\lambda}F^{\mu \nu} + \partial^{\nu}F^{\lambda \mu} +
\partial^{\mu}F^{\nu \lambda} = 0 &\Rightarrow& \bar \nabla \cdot \bar
B = 0 \quad \mbox{and} \quad \bar \nabla \times \bar E + {\partial \bar B
\over \partial t} = 0 \nolabel \\
\partial_{\mu}F^{\mu \nu} = J^{\nu} &\Rightarrow& \bar \nabla \cdot
\bar E = \rho \quad \mbox{and} \quad \bar \nabla \times \bar B - {\partial
\bar E \over \partial t} = \bar J \nolabel
\end{eqnarray}

\subsubsection{Classical Electrodynamics Lagrangian}

Bringing together the ideas of the previous sections, we now want to construct a
Lagrangian density $\mathcal{L}$ which will, via Hamilton's Principle,
produce Maxwell's equations.  

First, we know that $\mathcal{L}$ must be a scalar (no uncontracted
indices).  From our intuition with ``Physics I" type Lagrangians, we
know that kinetic terms are quadratic in the derivatives of the
fundamental coordinates (i.e. ${1\over 2} m\dot{x}^2 = {1\over 2} m
({dx \over dt})\cdot ({dx \over dt})$).  The natural choice is to take
$A^{\mu}$ as the fundamental field.  It turns out that the correct
choice is 
\begin{eqnarray}
\mathcal{L}_{EM} = -{1\over 4} F_{\mu \nu}F^{\mu \nu} - J^{\mu}A_{\mu}
\label{eq:emlagrangian}
\end{eqnarray}
(note that the $F^2$ term is quadratic in $\partial^{\mu}A^{\nu}$). 
So, 
\begin{eqnarray}
S = \int d^4x \bigg[-{1\over 4}F_{\mu \nu}F^{\mu \nu} -
J^{\mu}A_{\mu}\bigg] \label{eq:emaction}
\end{eqnarray}

Taking the variation of (\ref{eq:emaction}) with respect to
$A^{\mu}$, 
\begin{eqnarray}
\delta S &=&\int d^4x \bigg[-{1\over 4}F_{\mu \nu} \delta F^{\mu \nu} -
{1\over 4} \delta F_{\mu \nu}F^{\mu \nu} - J^{\mu} \delta A_{\mu}
\bigg] \nolabel \\
&=& \int d^4x \bigg[-{1\over 2}F_{\mu \nu}\delta F^{\mu \nu} - J^{\mu}
\delta A_{\mu}\bigg] \nolabel \\
&=&\int d^4x \bigg[-{1\over 2}F_{\mu \nu}(\partial^{\mu}\delta A^{\nu}
- \partial^{\nu}\delta A^{\mu}) - J^{\mu}\delta A_{\mu}\bigg] \nolabel
\\
&=&\int d^4x \bigg[-F_{\mu \nu} \partial^{\mu} \delta A^{\nu} -
J^{\mu}\delta A_{\mu}\bigg] \nolabel 
\end{eqnarray}
Integrating the first term by parts, and choosing boundary
conditions so that $\delta A$ vanishes at the boundaries, 
\begin{eqnarray}
&=& \int d^4x \bigg[\partial_{\mu}F^{\mu \nu}\delta A_{\nu} -
J^{\nu}\delta A_{\nu}\bigg] \nolabel \\
&=& \int d^4x \bigg[\partial_{\mu}F^{\mu \nu} - J^{\nu}\bigg] \delta
A_{\nu} \nolabel
\end{eqnarray}
So, to have $\delta S = 0$, we must have $\partial_{\mu}F^{\mu \nu} =
J^{\nu}$, and if this is written out one component at a time, it will
give exactly the inhomogenous Maxwell equations (\ref{eq:maxwell1}) and
(\ref{eq:maxwell2}).  And as we already pointed out, the homogenous
Maxwell equations become identities when written in terms of $A^{\mu}$. 

As a brief note, the way we have chosen to write equation
(\ref{eq:emlagrangian}), in terms of a ``potential" $A_{\mu}$, and the
somewhat mysterious antisymmetric ``field strength" $F_{\mu \nu}$, is
indicative of an extremely deep and very general mathematical structure
that goes well beyond classical electrodynamics.  We will see this
structure unfold as we proceed through these notes.  We just want to
mention now that this is not merely a clever way of writing electric
and magnetic fields, but a specific example of a general theory.  

\subsubsection{Gauge Transformations}

\bf Gauge Transformations \rm are usually discussed toward the end of an
undergraduate course on E\&M.  Students are typically told that they
are extremely important, but the reason why is not obvious.  We will
briefly introduce them here, and while their significance may still not
be transparent, we will return to them several times throughout these
notes.	

Given some specific potential $A^{\mu}$, we can find the field strength
action as in (\ref{eq:emaction}).  However, $A^{\mu}$ does not uniquely
specify the action.  We can take any arbitrary function
$\chi(x^{\mu})$, and the action will be invariant under the
transformation
\begin{eqnarray}
A^{\mu} \rightarrow A'^{\mu} = A^{\mu} + \partial^{\mu}\chi
\label{eq:gaugetransformation}
\end{eqnarray}
or 
\begin{eqnarray}
A^{\mu} \rightarrow A'^{\mu} = (\phi - {\partial \chi \over \partial
t}, \bar A + \bar \nabla \chi) \nolabel
\end{eqnarray}
Under this transformation, we have 
\begin{eqnarray}
F'^{\mu \nu} &=& \partial^{\mu}A'^{\nu} - \partial^{\nu}A'^{\mu} =
\partial^{\mu}(A^{\nu}+\partial^{\nu}\chi) - \partial^{\nu}(A^{\mu}+
\partial^{\mu}\chi) \nolabel \\
&=& \partial^{\mu}A^{\nu} - \partial^{\nu}A^{\mu} + \partial^{\mu}
\partial^{\nu}\chi -\partial^{\mu}\partial^{\nu}\chi \nolabel \\
&=& F^{\mu \nu}
\label{eq:invarianceoffmn}
\end{eqnarray}
So, $F'^{\mu \nu} = F^{\mu \nu}$.  

Furthermore, $J^{\mu}A_{\mu} \rightarrow J^{\mu}A_{\mu} +
J^{\mu}\partial_{\mu}\chi$.  Integrating the second term by parts with
the usual boundary conditions, 
\begin{eqnarray}
\int d^4x J^{\mu}\partial_{\mu}\chi = -\int d^4x
(\partial_{\mu}J^{\mu})\chi \nolabel
\end{eqnarray}
But, according to Maxwell's equations, $\partial_{\mu}J^{\mu} =
\partial_{\mu}\partial_{\nu}F^{\mu \nu} \equiv 0$ because $F^{\mu \nu}$
is totally antisymmetric.  So, both $F^{\mu \nu}$ and
$J^{\mu}\partial_{\mu}\chi$ are invariant under
(\ref{eq:gaugetransformation}), and therefore the action of $S$ is
invariant under (\ref{eq:gaugetransformation}).  

While the importance of gauge transformations may not be obvious at
this point, it will become perhaps the most important idea in particle
physics.  As a note before moving on, recall previously when we
mentioned the idea of ``what doesn't change when something else
changes" when talking about Lorentz transformations.  A gauge
transformation is exactly this (in a different context):  the
fundamental fields are changed by $\chi$, but the equations which
govern the physics are unchanged.  

In the next section, we provide the mathematical tools to understand
why this idea is so important.	

\subsection{References and Further Reading}

The material in this section can be found in nearly any introductory
text on Classical Mechanics, Classical Electrodynamics, and Relativity.
 The primary sources for these notes are \cite{Cottingham},
\cite{Goldstein}, and \cite{Griffiths}.  

For further reading, we recommend \cite{Feynman}, \cite{Jackson},
\cite{Jose}, \cite{Naber2}, \cite{Schwinger}, and \cite{Woodhouse}.  

\newpage
\section{Part II --- Algebraic Foundations}

\subsection{Introduction to Group Theory}
\label{sec:groupintro}

There are several symbols in this section which may not be familiar. 
We therefore provide a summary of them for reference.
\begin{multicols}{3}
$
\mathbb{N} = \{0, 1, 2, 3, \ldots\} \\
\indent \mathbb{Z} = \{0, \pm 1, \pm 2, \pm 3, \ldots \} \\
\indent \mathbb{Q} = \rm Rational \; Numbers \it \\
\indent \mathbb{R} = \rm Real \; Numbers \it \\
\indent \mathbb{C} = \rm Complex \; Numbers \it \\
\indent \mathbb{Z}_n = \mathbb{Z} \; \mod \; n \\
\indent \Rightarrow \rm \; is \; read \; ``implies"\\
\indent \rm iff \; is \; read \; ``if\; and \; only \; if" \\
\indent \forall \rm \; is \; read \; ``for \; every" \\
\indent \exists \rm \; is \; read \; ``there \; exists" \\
\indent \in \; is \; read \; ``in" \\
\indent \ni \; is \; read \; ``such \; that" \\
\indent \dot{=} \; is \; ``represented \; by" \\
\indent \subset \; is \; ``subset\; of" \\
\indent \equiv \; is\;``defined\; as" \\
$
\end{multicols}

Now that we have reviewed the primary ideas of classical physics, we
are almost ready to start talking about particle physics.  However,
there is a bit of mathematical ``machinery" we will need first. 
Namely, \bf Group Theory\rm.  

Group theory is, in short, the mathematics of symmetry.  We are going
to begin talking about what will seem to be extremely abstract ideas,
but eventually we will explain how those ideas relate to physics.  As a
preface of what is to come, the most foundational idea here is, as we
said before, ``what doesn't change when something else changes".  A
group is a precise and well-defined way of specifying the thing or
things that change.  

\subsubsection{What is a Group?}

To begin with, we define the notion of a \bf Group\rm.	This definition
may seem cryptic, but it will be explained in the paragraphs that
follow.  

\it A group, denoted $(G,\star)$, is a set of objects, denoted $G$, and
some operation on those objects, denoted~$\star$, subject to the
following:
\begin{enumerate}
\parskip -4pt
\item $\forall \; g_1,\; g_2 \in G$, $g_1\star g_2 \in G$ also.
(closure)
\item $\forall \; g_1,g_2,g_3\in G$, it must be true that
$(g_1\star g_2)\star g_3 = g_1 \star (g_2 \star g_3)$. (associativity)
\item $\exists g \in G$, denoted $e$, $\ni \forall g_i\in G,\;
e\star g_i = g_i \star e = g_i$. (identity)
\item $\forall g\in G, \exists h \in G \ni h \star g = g \star h =
e$, (so $h = g^{-1}$). (inverse)
\end{enumerate}

\rm Now we explain what this means.  By ``objects" we literally mean
anything.  We could be talking about $\mathbb{Z}$ or $\mathbb{R}$, or
we could be talking about a set of Easter eggs all painted different
colors.  

The meaning of ``some operation", which we are calling $\star$, can
literally be anything you can do to those objects.  A formal definition
of what $\star$ means could be given, but it will be easier to
understand with examples.  

\bf Note: \rm The definition of a group doesn't demand that
$g_i \star g_j = g_j \star g_i$.  This is a very important point, but
we will discuss it in more detail later.  We mention it now so it is
not new later.	

\bf Example 1: \rm \quad $(G,\star) = (\mathbb{Z},+)$
\vspace*{-2ex}
\begin{quote}
Consider the set $G$ to be $\mathbb{Z}$, and the
operation to be $\star = +$, or simply addition.  

We first check closure.  If you take any two elements of $\mathbb{Z}$
and add them together, is the result in $\mathbb{Z}$?  In other words,
if $a,b \in \mathbb{Z}$, is $a+b\in \mathbb{Z}$?  Obviously the answer
is yes; the sum of two integers is an integer, so closure is met.  

Now we check associativity.  If $a,b,c \in \mathbb{Z}$, it is trivially
true that $a+(b+c) = (a+b)+c$.	So, associativity is met.  

Now we check identity.	Is there an element $e\in \mathbb{Z}$ such that
when you add $e$ to any other integer, you get that same integer? 
Clearly the integer 0 satisfies this.  So, identity is met.  

Finally, is there an inverse?  For any integer $a\in \mathbb{Z}$, will
there be another integer $b\in \mathbb{Z}$ such that $a+b = e = 0$? 
Again, this is obvious, $a^{-1} = -a$ in this case.  So, inverse is
met.  

So, $(G,\star) = (\mathbb{Z},+)$ is a group.  
\end{quote}

\bf Example 2: \rm \quad $(G,\star) = (\mathbb{R},+)$
\vspace*{-2ex}
\begin{quote}

Obviously, any two real numbers added together is also a real number.  

Associativity will hold (of course).  

The identity is again 0.  

And finally, once again, $-a$ will be the inverse of any $a \in \mathbb{R}$.  
\end{quote}

\bf Example 3: \rm  \quad $(G,\star) = (\mathbb{R},\cdot)$ (multiplication)
\vspace*{-2ex}
\begin{quote}
Closer is met; two real numbers multiplied together give a real number. 

Associativity obviously holds.	

Identity also holds.  Any real number $a\in \mathbb{R}$, when multipled
by 1 is $a$.	

Inverse, on the other hand, is trickier.  For any real number, is there
another real number you can multiply by it to get 1?	The instinctive
choice is $a^{-1} = {1\over a}$.  But, this doesn't quite work because
of $a=0$.  This is the \it only \rm exception, but because there's an
exception, $(\mathbb{R},\cdot)$ is not a group.  

\bf Note: \rm If we take the set $\mathbb{R} - \{0\}$ instead of
$\mathbb{R}$, then $(\mathbb{R}-\{0\},\cdot)$ is a group.  
\end{quote}

\bf Example 4: \rm \quad $(G,\star) = (\{1\},\cdot)$
\vspace*{-2ex}
\begin{quote}
This is the set with only the element $1$, and the operation is normal
multiplication.  This is indeed a group, but it is extremely
uninteresting, and is called the \bf Trivial Group\rm.	
\end{quote}

\bf Example 5: \rm \quad $(G,\star) = (\mathbb{Z}_3,+)$
\vspace*{-2ex}
\begin{quote}
This is the set of integers mod 3, containing only the elements $0$, $1$,
and $2$ (3~mod~3 is 0, 4 mod 3 is 1, 5 mod 3 is 2, etc.)

You can check yourself that this is a group.  
\end{quote}

\subsubsection{Finite Discrete Groups and Their Organization}

From the examples above, several things should be apparent about
groups.  One is that there can be any number of objects in a group.  We
have a special name for the number of objects in the group's set.  \it
The \bf Order \it of a group is the number of elements in it.  \rm

The order of $(\mathbb{Z},+)$ is infinite (there are an infinite number
of integers), as is the order of $(\mathbb{R},+)$ and
$(\mathbb{R}-\{0\},\cdot)$.  But, the order of $(\{1\},\cdot)$ is 1,
and the order of $(\mathbb{Z}_3,+)$ is 3.  

If the order of a group is finite, the group is said to be \bf
Finite\rm.  Otherwise it is \bf Infinite\rm.  

It is also clear that the elements of groups may be \bf Discrete\rm, or
they may be \bf Continuous\rm.	For example, $(\mathbb{Z},+)$,
$(\{1\},\cdot)$, and $(\mathbb{Z}_3,+)$ are all discrete, while
$(\mathbb{R},+)$ and $(\mathbb{R}-\{0\},\cdot)$ are both continuous.  

Now that we understand what a discrete finite group is, we can talk
about how to organize one.  Namely, we use what is called a \bf
Multiplication Table\rm.  A multiplication table is a way of organizing
the elements of a group as follows:
\newline
\begin{table} [h]
\centering
\begin{tabular}{|c||c|c|c|c|}
\hline
$(G,\star)$ & $e$ & $g_1$ & $g_2$ & $\cdots$ \\
\hline 
\hline
$e$ & $e\star e$ & $e\star g_1$ & $e \star g_2$ & $\cdots$ \\
\hline
$g_1$ & $g_1\star e$ & $ g_1 \star g_1 $ & $g_1 \star g_2$ & $\cdots$
\\
\hline
$g_2$ & $g_2\star e$ & $ g_2 \star g_1 $ & $g_2 \star g_2$ & $\cdots$
\\
\hline
$\vdots$ & $\vdots$ & $\vdots$ & $\vdots$ & $\ddots$ \\
\hline
\end{tabular}
\end{table}
\newline
We state the following property of multiplication tables without proof.
 \it A multiplication table must contain every element of the group
exactly one time in every row and every column\rm.  A few minutes
thought should convince you that this is necessary to ensure that the
definition of a group is satisfied.  

As an example, we will draw a multiplication table for the group of
order 2.  We won't look at specific numbers, but rather call the
elements $g_1$ and $g_2$.  We begin as follows:
\newline
\begin{table}[h]
\centering
\begin{tabular}{|c||c|c|}
\hline
$(G,\star)$ & $e$ & $g_1$ \\
\hline
\hline
$e$ & ? & ? \\
\hline
$g_1$ & ? & ? \\
\hline
\end{tabular}
\end{table}
\newline
Three of these are easy to fill in from the identity:
\newline
\begin{table}[h]
\centering
\begin{tabular}{|c||c|c|} 
\hline
$(G,\star)$ & $e$ & $g_1$ \\
\hline
\hline
$e$ & $e$ & $g_1$ \\
\hline
$g_1$ & $g_1$ & ? \\
\hline
\end{tabular}\label{order2}
\end{table}
\newline
And because we know that every element must appear exactly once, the
final question mark must be $e$.  So, there is only one possible group
of order 2.  

We will consider a few more examples, but we stress at this point that
the temptation to plug in numbers should be avoided.  Groups are
abstract things, and you should try to think of them in terms of the
abstract properties, not in terms of actual numbers.  

We can proceed with the multiplication table for the group of order 3. 
You will find that, once again, there is only one option. (Doing this is
instructive and it would be helpful to work this out yourself.)
\newline
\begin{table}[h]
\centering
\begin{tabular}{|c||c|c|c|}
\hline
$(G,\star)$ & $e$ & $g_1$ & $g_2$ \\
\hline
\hline
$e$ & $e$ & $g_1$ & $g_2$ \\
\hline
$g_1$ & $g_1$ & $g_2$ & $e$ \\
\hline
$g_2$ & $g_2$ & $e$ & $g_1$ \\
\hline
\end{tabular}
\end{table}
\newline

You are encouraged to work out the possibilities for groups of order 4.
\it (Hint: there are 4 possibilities.)\rm  

\subsubsection{Group Actions}

So far we have only considered elements of groups and how they relate
to each other.	The point has been that a particular group represents
nothing more than a structure.	There are a set of things, and they
relate to each other in a particular way.  Now, however, we want to
consider an extremely simple version of how this relates to nature.  

\newpage
\bf Example 6 \rm
\vspace*{-2ex}
\begin{quote}
Consider three Easter eggs, all painted different
colors (red, orange, and yellow), which we denote R, O, and Y.	Now,
assume they have been put into a row in the order (ROY).  If we want to
keep them lined up, not take any eggs away, and not add any eggs, what
we can we do to them?  We can do any of the following:
\begin{enumerate}
\item Let $e$ be doing nothing to the set, so $e(ROY) = (ROY)$. 
\item Let $g_1$ be a cyclic permutation of the three, $g_1(ROY) =
(OYR)$
\item Let $g_2$ be a cyclic permutation in the other direction,
$g_2(ROY)=(YRO)$
\item Let $g_3$ be swapping the first and second, $g_3(ROY) =
(ORY)$
\item Let $g_4$ be swapping the first and third, $g_4(ROY) =
(YOR)$
\item Let $g_5$ be swapping the second and third, $g_5(ROY) =
(RYO)$
\end{enumerate}

You will find that these 6 elements are closed, there is an identity,
and each has an inverse.\footnote{We should be very careful to draw a distinction
between the \it elements \rm of the group and the \it objects \rm the group acts on.  
The objects in this example are the eggs, and the permutations are the results of 
the group action.  Neither the eggs nor the permutations of the eggs are the elements
of the group.  The elements of the group are abstract objects which we are assigning 
to some operation on the eggs, resulting in a new permutation}
So, we can draw a multiplication table (you
are strongly encouraged to write at least part of this out on your
own):
\begin{table}[h]
\centering
\begin{tabular}{|c||c|c|c|c|c|c|}
\hline
$(G,\star)$ & $e$ & $g_1$ & $g_2$ & $g_3$ & $g_4$ & $g_5$ \\
\hline
$e$ & $e$ & $g_1$ & $g_2$ & $g_3$ & $g_4$ & $g_5$ \\
\hline
$g_1$ & $g_1$ & $g_2$ & $e$ & $g_5$ & $g_3$ & $g_4$ \\
\hline
$g_2$ & $g_2$ & $e$ & $g_1$ & $g_4$ & $g_5$ & $g_3$ \\
\hline
$g_3$ & $g_3$ & $g_4$ & $g_5$ & $e$ & $g_1$ & $g_2$ \\
\hline
$g_4$ & $g_4$ & $g_5$ & $g_3$ & $g_2$ & $e$ & $g_1$ \\
\hline
$g_5$ & $g_5$ & $g_3$ & $g_4$ & $g_1$ & $g_2$ & $e$ \\
\hline
\end{tabular} \label{s3}
\end{table}

\end{quote}

There is something interesting about this group.  Notice that $g_3
\star g_1 = g_4$, whereas $g_1 \star g_3 = g_5$.  So, we have the
surprising result that in this group it is not necessarily true that
$g_i\star g_j = g_j \star g_i$.  

This leads to a new way of classifying groups.	\it We say a group is
\bf Abelian \it if $g_i \star g_j = g_j \star g_i$ $\forall g_i,g_j \in
G$\rm.	If a group is not Abelian, it is \bf Non-Abelian\rm.  

Another term commonly used is \bf Commute\rm.  If $g_i\star g_j = g_j
\star g_i$, then we say that $g_i$ and $g_j$ commute.  So, an Abelian
group is \bf Commutative\rm, whereas a Non-Abelian group is \bf
Non-Commutative\rm.  

The Easter egg group of order 6 above is an example of a very important
type of group.	It is denoted $S_3$, and is called the \bf Symmetric
Group\rm.  It is the group that takes three objects to all permutations of
those three objects.  

The more general group of this type is $S_n$, the group that takes $n$ objects
to all permutations of those objects.  You can convince yourself that $S_n$ 
will always have order $n!$ ($n$ factorial).  

The idea above with the 3 eggs is that $S_3$ is the \it group\rm, while
the eggs are the objects that the group \it acts on\rm.  The particular
way an element of $S_3$ changes the eggs around is called the \bf Group
Action \rm of that element.  And each element of $S_3$ will move the
eggs around while leaving them lined up.  This ties in to our
overarching concept of ``what doesn't change when something else
changes".  The fact that there are 3 eggs with 3 particular colors
lined up doesn't change.  The order they appear in does.  

\subsubsection{Representations}

We suggested above that you think of groups as purely abstract things
rather than trying to plug in actual numbers.  Now, however, we want to
talk about how to see groups, or the elements of groups, in terms of
specific numbers.  But, we will do this in a very systematic way.  The
name for a specific set of numbers or objects that form a group is a
\bf Representation\rm.	The remainder of this section (and the next)
will primarily be about group representations.	

We already discussed a few simple representations when we discussed
$(\mathbb{Z},+)$, $(\mathbb{R}-\{0\},\cdot)$, and $(\mathbb{Z}_3,+)$. 
Let's focus on $(\mathbb{Z}_3,+)$ for a moment (the integers mod 3,
where $e=0$, $g_1=1$, $g_2=2$, with addition).	Notice that we could
alternatively define $e=1$, $g_1 = e^{{2\pi i \over 3}}$, and
$g_2=e^{{4\pi i \over 3}}$, and let $\star$ be multiplication.	So, in
the ``representation" with $(0,1,2)$ and addition, we had for example
\begin{eqnarray}
g_1\star g_2 = (1+2)~\mmod 3 = 3~\mmod 3 = 0 = e \nolabel 
\end{eqnarray}
whereas now with the multiplicative representation we have
\begin{eqnarray}
g_1 \star g_2 = e^{{2\pi i \over 3}} \cdot e^{{4 \pi i \over 3}} =
e^{2\pi i} = e^0 = 1 = e \nolabel
\end{eqnarray}
So the structure of the group is preserved in both representations.  

We have two completely different representations of the same group. 
This idea of different ways of expressing the same group is of extreme
importance, and we will be using it throughout the remainder of these
notes.	

We now see a more rigorous way of coming up with representations of a
particular group.  We begin by introducing some notation.  For a group
$(G,\star)$ with elements $g_1,g_2,\ldots$, we call the
Representation of that group $D(G)$, so that the elements of $G$
are $D(e)$, $D(g_1)$, $D(g_2)$ (where each $D(g_i)$ is a matrix of some
dimension).  We then choose $\star$ to be matrix multiplication.  So,
$D(g_i) \cdot D(g_j) = D(g_i \star g_j)$.  

It may not seem that we have done anything profound at this point, but
we most definitely have.  Remember above that we encouraged seeing
groups as abstract things, rather than in terms of specific numbers. 
This is because a group is fundamentally an \it abstract \rm object.  A
group is not a specific set of numbers, but rather a set of abstract
objects with a well-defined structure telling you how those elements
relate to each other.  

And the beauty of a representation $D$ is that, via normal matrix
multiplication, we have a sort of ``lens", made of familiar things
(like numbers, matrices, or Easter eggs), through which we can see into
this abstract world.  And because $D(g_i)\cdot D(g_j) = D(g_i \star
g_j)$, we aren't losing any of the structure of the abstract group by
using a representation.  

So now that we have some notation, we can develop a formalism to figure
out exactly what $D$ is for an arbitrary group.  

We will use Dirac vector notation, where the column vector
\begin{eqnarray}
\bar v = 
\begin{pmatrix}
v_1 \\ v_2 \\ v_3 \\ \vdots
\end{pmatrix}
 = |v \rangle \nolabel
\end{eqnarray}
and the row vector
\begin{eqnarray}
\bar v^T = 
\begin{pmatrix}
v_1 & v_2 & v_3 \; \cdots
\end{pmatrix}
 = \langle v | \nolabel 
\end{eqnarray}
So, the dot product between two vectors is
\begin{eqnarray}
\bar v \cdot \bar u = 
\begin{pmatrix}
v_1 & v_2 & v_3 \cdots
\end{pmatrix}
\begin{pmatrix}
u_1 \\ u_2 \\ u_3 \\ \vdots
\end{pmatrix} = v_1u_1+ v_2u_2+v_3u_3+\cdots \equiv \langle v | u
\rangle \nolabel 
\end{eqnarray}

Now, we proceed by relating each element of a finite discrete group to
one of the standard orthonormal unit vectors:
\begin{eqnarray}
e \rightarrow |e\rangle = |\hat e_1\rangle \qquad g_1 \rightarrow
|g_1\rangle = |\hat e_2\rangle \qquad g_2 \rightarrow |g_2\rangle  =
|\hat e_3\rangle \nolabel 
\end{eqnarray}
And we define the way an element in a representation $D(G)$ acts on
these vectors to be
\begin{eqnarray}
D(g_i)|g_j\rangle = |g_i \star g_j\rangle \nolabel 
\end{eqnarray}

Now, we can build our representation.  We will (from now on unless
otherwise stated) represent the elements of a group $G$ using matrices
of various sizes, and the group operation $\star$ will be standard
matrix multiplication.	The specific matrices that represent a given
element $g_k$ of our group will be given by 
\begin{eqnarray}
[D(g_k)]_{ij} = \langle g_i |D(g_k)|g_j\rangle \label{eq:regrep}
\end{eqnarray}
As an example, consider again the group of order 2 (we wrote out the
multiplication table above on page \pageref{order2}).  First, we find
the matrix representation of the identity, $[D(e)]_{ij}$, 
\begin{eqnarray}
\;[D(e)]_{11} = \langle e|D(e)|e\rangle &=& \langle e|e\star e\rangle =
\langle e| e\rangle = 1 \nolabel \\
\;[D(e)]_{12} = \langle e|D(e)|g_1\rangle &=& \langle e|e\star
g_1\rangle = \langle e| g_1\rangle = 0\nolabel	\\
\; [D(e)]_{21} = \langle g_1|D(e)|e\rangle &=& \langle g_1|e\star
e\rangle = \langle g_1|e\rangle = 0 \nolabel \\
\; [D(e)]_{22} = \langle g_1|D(e)|g_1\rangle &=& \langle g_1|e\star
g_1\rangle = \langle g_1|g_1\rangle = 1\nolabel 
\end{eqnarray}
So, the matrix representation of the identity is $D(e)~\dot{=}
\begin{pmatrix}
1 & 0 \\ 0 & 1
\end{pmatrix}$.  It shouldn't be surprising that the identity element
is represented by the identity	matrix.  

Next we find the representation of $D(g_1)$:
\begin{eqnarray}
\;[D(g_1)]_{11} = \langle e|D(g_1)|e\rangle &=& \langle e|g_1\star
e\rangle = \langle e| g_1\rangle = 0 \nolabel \\
\;[D(g_1)]_{12} = \langle e|D(g_1)|g_1\rangle &=& \langle e|g_1\star
g_1\rangle = \langle e| e\rangle = 1 \nolabel \\
\;[D(g_1)]_{21} = \langle g_1|D(g_1)|e\rangle &=& \langle g_1|g_1\star
e\rangle = \langle g_1|g_1\rangle = 1 \nolabel \\
\;[D(g_1)]_{22} = \langle g_1|D(g_1)|g_1\rangle &=& \langle
g_1|g_1\star g_1\rangle = \langle g_1|e\rangle = 0\nolabel 
\end{eqnarray}
So, the matrix representation of $g_1$ is $D(g_1)~\dot{=}
\begin{pmatrix}
0 & 1 \\ 1 & 0
\end{pmatrix}$.  It is straightforward to check that this is a true
representation,
\begin{eqnarray}
& & e\star e = 
\begin{pmatrix}
1 & 0 \\ 0 & 1
\end{pmatrix}
\begin{pmatrix}
1 & 0 \\ 0 & 1
\end{pmatrix} = 
\begin{pmatrix}
1 & 0 \\ 0 & 1
\end{pmatrix} = e \qquad \checkmark \nolabel \\
& & e\star g_1 = 
\begin{pmatrix}
1 & 0 \\ 0 & 1
\end{pmatrix}
\begin{pmatrix}
0 & 1 \\ 1 & 0
\end{pmatrix} = 
\begin{pmatrix}
0 & 1 \\ 1 & 0
\end{pmatrix} = g_1 \qquad \checkmark \nolabel \\
& & g_1\star e = 
\begin{pmatrix}
0 & 1 \\ 1 & 0
\end{pmatrix}
\begin{pmatrix}
1 & 0 \\ 0 & 1
\end{pmatrix} = 
\begin{pmatrix}
0 & 1 \\ 1 & 0
\end{pmatrix} = g_1 \qquad \checkmark\nolabel  \\
& & g_1\star g_1 = 
\begin{pmatrix}
0 & 1 \\ 1 & 0
\end{pmatrix}
\begin{pmatrix}
0 & 1 \\ 1 & 0
\end{pmatrix} = 
\begin{pmatrix}
1 & 0 \\ 0 & 1
\end{pmatrix} = e \qquad \checkmark \nolabel 
\end{eqnarray}

Instead of considering the next obvious example, the group of order
3, consider the group $S_3$ from above (the multiplication table is
on page \pageref{s3}).	The identity representation $D(e)$ is easy --- it
is just the $6\times 6$ identity matrix.  We encourage you to work out
the representation of $D(g_1)$ on your own, and check to see that it is
\begin{eqnarray}
D(g_1)~\dot{=}
\begin{pmatrix}
0 & 0 & 1 & 0 & 0 & 0 \\
1 & 0 & 0 & 0 & 0 & 0 \\
0 & 1 & 0 & 0 & 0 & 0 \\
0 & 0 & 0 & 0 & 1 & 0 \\
0 & 0 & 0 & 0 & 0 & 1 \\
0 & 0 & 0 & 1 & 0 & 0 \\
\end{pmatrix} \label{eq:s3g1}
\end{eqnarray}

All 6 matrices can be found this way, and multiplying them out will
confirm that they do indeed satisfy the group structure of $S_3$.  

\subsubsection{Reducibility and Irreducibility --- A Preview}
\label{sec:redprev}

You have probably noticed that equation (\ref{eq:regrep}) will always
produce a set of $n\times n$ matrices, where $n$ is the order of the
group.	There is actually a name for this particular representation. 
\it The $n\times n$ matrix representation of a group of order $n$ is
called the \bf Regular Representation\rm.  More generally, \it the
$m\times  m$ matrix representation of a group (of any order) is called
the \bf {\boldmath$m$}-dimensional representation.  \rm

But, as we have seen, there is more than one representation for a given
group (in fact, there are an infinite number of representations).  

One thing we can immediately see is that any group that is Non-Abelian
cannot have a $1\times 1$ matrix representation.  This is because
scalars ($1\times 1$ matrices) always commute, whereas matrices in
general do not.  

We saw above in equation (\ref{eq:s3g1}) that we can represent the
group $S_n$ by $n!\times n!$ matrices.	Or, more generally, we can
represent any group using $m \times m$ matrices, were $m$ equals order($G$).  This is
the regular representation.  But it turns out that it is usually
possible to find representations that are ``smaller'' than the regular
representation.  

To pursue how this might be done, note that we are working with matrix
representations of groups.  In other words, we are representing groups
in \it linear spaces\rm.  We will therefore be using a great deal of
linear algebra to find smaller representations.  This process, of
finding a smaller representation, is called \bf Reducing \rm a
representation.  Given an arbitrary representation of some group, the
first question that must be asked is ``is there a smaller
representation?"  If the answer is yes, then the representation is said
to be \bf Reducible\rm.  If the answer is no, then it is \bf
Irreducible\rm.  

Before we dive into the more rigorous approach to reducibility and
irreducibility, let's consider a more intuitive example, using $S_3$. 
In fact, we'll stick with our three painted Easter eggs, $R$, $O$, and
$Y$:  
\begin{enumerate}
\parskip -4pt
\item $e(ROY) = (ROY)$
\item $g_1(ROY) = (OYR)$
\item $g_2(ROY) = (YRO) $
\item $g_3(ROY) = (ORY) $
\item $g_4(ROY) = (YOR) $
\item $g_5(ROY) = (RYO) $
\end{enumerate}
We will represent the set of eggs by a column vector $|E\rangle = 
\begin{pmatrix}
R \\ O \\ Y
\end{pmatrix}$.  

Now, by inspection, what matrix would do to $|E\rangle$ what $g_1$ does
to $(ROY)$?  In other words, how can we fill in the ?'s in 
\begin{eqnarray}
\begin{pmatrix}
? & ? & ? \\
? & ? & ? \\
? & ? & ? \\
\end{pmatrix}
\begin{pmatrix}
R \\ O \\ Y
\end{pmatrix} = 
\begin{pmatrix}
O \\ Y \\ R
\end{pmatrix}  \nolabel
\end{eqnarray}
to make the equality hold?  A few moments thought will show that the
appropriate matrix is
\begin{eqnarray}
\begin{pmatrix}
0 & 1 & 0 \\
0 & 0 & 1 \\
1 & 0 & 0 \\
\end{pmatrix}
\begin{pmatrix}
R \\ O \\ Y
\end{pmatrix} = 
\begin{pmatrix}
O \\ Y \\ R
\end{pmatrix} \nolabel
\end{eqnarray}
Continuing this reasoning, we can see that the rest of the matrices are
\begin{eqnarray}
D(e)~\dot{=} 
\begin{pmatrix}
1 & 0 & 0 \\ 0 & 1 & 0 \\ 0 & 0 & 1
\end{pmatrix},\qquad
D(g_1)~\dot{=} 
\begin{pmatrix}
0 & 1 & 0 \\ 0 & 0 & 1 \\ 1 & 0 & 0
\end{pmatrix},\qquad
D(g_2)~\dot{=} 
\begin{pmatrix}
0 & 0 & 1 \\ 1 & 0 & 0 \\ 0 & 1 & 0
\end{pmatrix}\qquad \nolabel
\end{eqnarray}
\begin{eqnarray}
D(g_3)~\dot{=} 
\begin{pmatrix}
0 & 1 & 0 \\ 1 & 0 & 0 \\ 0 & 0 & 1
\end{pmatrix},\qquad
D(g_4)~\dot{=} 
\begin{pmatrix}
0 & 0 & 1 \\ 0 & 1 & 0 \\ 1 & 0 & 0
\end{pmatrix},\qquad
D(g_5)~\dot{=} 
\begin{pmatrix}
1 & 0 & 0 \\ 0 & 0 & 1 \\ 0 & 1 & 0
\end{pmatrix} \qquad \nolabel
\end{eqnarray}
You can do the matrix multiplication to convince yourself that this is
in fact a representation of $S_3$.  

So, in equation (\ref{eq:s3g1}), we had a $6\times 6$
matrix representation.	Here, we have a new representation of consisting of $3\times
3$ matrices.  We have therefore ``reduced" the representation.	In the
next section, we will look at more mathematically rigorous ways of
reducing representations.  

\subsubsection{Algebraic Definitions}

Before moving on, we must spend this section learning the definitions
of several terms which are used in group theory.  

\it If $H$ is a subset of $G$, denoted $H \subset G$, such that the
elements of $H$ form a group, then we say that $H$ forms a \bf Subgroup
\it of $G$\rm.	
We make this more precise with examples.  \\

\newpage
\bf Example 7 \rm
\vspace*{-2ex}
\begin{quote}
Consider (as usual) the group $S_3$, with the
elements labeled as before:
\begin{enumerate}
\parskip 0pt
\item $g_0(ROY) = (ROY)$
\item $g_1(ROY) = (OYR)$
\item $g_2(ROY) = (YRO)$
\item $g_3(ROY) = (ORY)$
\item $g_4(ROY) = (YOR)$
\item $g_5(ROY) = (RYO)$
\end{enumerate}
(where we are relabeling $g_0 \equiv e$ for later convenience).  The
multiplication table is given on page \pageref{s3}.  

Notice that $\{g_0, g_1, g_2\}$ form a subgroup.	You can see this by
noticing that the upper left 9 boxes in the multiplication table (the
$g_0, g_1, g_2$ rows and columns) all have only $g_0$'s, $g_1$'s, and
$g_2$'s.  So, here is a group of order 3 contained in $S_3$.	
\end{quote}

\bf Example 8 \rm
\vspace*{-2ex}
\begin{quote}
Consider the subset of $S_3$ consisting of $g_0$ and
$g_3$ only.  Both $g_0$ and $g_3$ are their own inverses, so the
identity exists, and the group is closed.  Therefore, we can say that
$\{g_0,g_3\}\subset S_3$ is a subgroup of $S_3$.  

In fact, if you write out the multiplication table for $g_0$ and $g_3$
only, you will see that it is exactly equivalent to the group of order
2 considered above.  This means that we can say that $S_3$ \it
contains \rm the group of order 2 (and we know from last time that
there is only one such group, though there are an infinite number of
representations of it).  The way we understand this is that the
abstract entity $S_3$, of which there is only one, contains the group
of order 2, of which there is only one.  However, the \it
representations \rm of $S_3$, of which there are an infinite number,
will \it each \rm contain the group of order 2 (of which there are
also an infinite number of representations).  
\end{quote}

\bf Example 9 \rm
\vspace*{-2ex}
\begin{quote}
Notice that the sets $\{g_0,g_3\}$, $\{g_0,g_4\}$,
and $\{g_0,g_5\}$ (all $\subset S_3$), are all the same as the group of
order 2.  This means that $S_3$ actually contains exactly three
copies of the group of order 2 in addition to the single copy of the
group of order 3.  

Again, this is speaking in terms of the abstract entity $S_3$.	We can
see this through the ``lens" of representation by the fact that \it any
\rm representation of $S_3$ will contain three different copies of the
group of order 2.  
\end{quote}

\newpage
\bf Example 10 \rm
\vspace*{-2ex}
\begin{quote}
 As a final example of subgroups, there are two subgroups of \it
any \rm group, no matter what the group.  One is the subgroup
consisting of only the identity, $\{g_0\} \subset G$.  All groups
contain this, but it is never very interesting.  

Secondly, $\forall G$, $G \subset G$, and therefore $G$ is always a
subgroup of itself.  We call these subgroups the ``trivial" subgroups.  
\end{quote}

We now introduce another important definition.

\it If $G$ is a group,
and $H$ is a subgroup of $G$ ($H\subset G$), then
\begin{itemize}
\parskip 0pt
\item The set $gH = \{g\star h |h\in H\}$ is called the \bf Left
Coset \it of $H$ in $G$ \\
\item The set $Hg = \{h\star g |h\in H\}$ is called the \bf Right
Coset \it of $H$ in $G$
\end{itemize}

\rm There is a right (or left) coset for each element $g \in G$, though
they are not necessarily all unique.  This definition should be
understood as follows; a coset is a \it set \rm consisting of the
elements of $H$ all multiplied on the right (or left) by some element
of $G$.  

\bf Example 11 \rm
\vspace*{-2ex}
\begin{quote}
For the subgroup $H = \{g_0,g_1\} \subset S_3$
discussed above, the left cosets are
\begin{itemize}
\parskip 0pt
\item[] $g_0\{g_0,g_1\} = \{g_0 \star g_0, g_0 \star g_1\} =
\{g_0,g_1\}$
\item[] $g_1\{g_0,g_1\} = \{g_1 \star g_0, g_1 \star g_1\} =
\{g_1,g_2\} $
\item[] $g_2\{g_0,g_1\} = \{g_2 \star g_0, g_2 \star g_1\} =
\{g_2,g_0\} $
\item[] $g_3\{g_0,g_1\} = \{g_3 \star g_0, g_3 \star g_1\} =
\{g_3,g_4\} $
\item[] $g_4\{g_0,g_1\} = \{g_4 \star g_0, g_4 \star g_1\} =
\{g_4,g_5\} $
\item[] $g_5\{g_0,g_1\} = \{g_5 \star g_0, g_5 \star g_1\} =
\{g_5,g_3\}$
\end{itemize}
So, the left cosets of $\{g_0,g_1\}$ in $S_3$ are $\{g_0,g_1\}$,
$\{g_1,g_2\}$, $\{g_2,g_0\}$, $\{g_3,g_4\}$, $\{g_4,g_5\}$, and
$\{g_5,g_3\}$.
\end{quote}

We can now understand the following definition.  \it $H$ is a \bf
Normal Subgroup \rm of $G$ if $\forall h \in H$, $g^{-1} \star h \star
g \in H$.  Or, in other words, if $H$ denotes the subgroup, it is a
normal subgroup if $gH = Hg$, which says that the left and right
cosets are all equal.	\rm 

As a comment, saying $gH$ and $Hg$ are equal doesn't mean that each
individual element in the coset $gH$ is equal to the corresponding
element in $Hg$, but rather that the two cosets \it contain \rm the
same elements, regardless of their order.  For example, if we had the
cosets $\{g_i, g_j, g_k\}$ and $\{g_j, g_k, g_i\}$, they would be equal
because they contain the same three elements.  

This definition means that if you take a subgroup $H$ of a group $G$,
and you multiply the \it entire set \rm on the left by some element of
$g\in G$, the resulting set will contain the exact same elements it
would if you had multiplied \it on the right \rm by the same element $g
\in G$.  Here is an example to illustrate.  

\bf Example 12 \rm
\vspace*{-2ex}
\begin{quote}
Consider the order 2 subgroup $\{g_0,g_3\}\subset
S_3$.  Multiplying on the left by, say, $g_4$, gives 
\begin{eqnarray}
g_4 \star \{g_0,g_3\} = \{g_4 \star g_0,g_4\star g_3\} = \{g_4,g_2\}
\nolabel 
\end{eqnarray}
And multiplying on the right by $g_4$ givs
\begin{eqnarray}
\{g_0,g_3\} \star g_4 = \{g_0 \star g_4, g_3 \star g_4\} = \{g_4,g_1\}
\nolabel 
\end{eqnarray}
So, because the final sets do not contain the same elements, $\{g_4,
g_2\} \neq \{g_4,g_1\}$, we conclude that the subgroup $\{g_0,g_3\}$ is
\it not \rm a normal subgroup of $S_3$.  
\end{quote}

\bf Example 13 \rm
\vspace*{-2ex}
\begin{quote}
Above, we found that $\{g_0,g_1,g_2\}\subset S_3$ is
a subgroup of order 3 in $S_3$.  To use a familiar label, remember that
we previously called the group of order 3 $(\mathbb{Z}_3,+)$.  So,
dropping the $`+'$, we refer to the group of order 3 as
$\mathbb{Z}_3$.  Is this subgroup normal?  We leave it to you to show
that it is.	
\end{quote}
%\newpage
\bf Example 14 \rm
\vspace*{-2ex}
\begin{quote}
Consider the group of integers under addition,
$(\mathbb{Z},+)$.  And, consider the subgroup $\mathbb{Z}_{even}
\subset \mathbb{Z}$, the even integers under addition (we leave it to
you to show that this is indeed a group).  

Now, take some odd integer $n_{odd}$ and act on the left:
\begin{eqnarray}
n_{odd} + \mathbb{Z}_{even} = \{n_{odd} + 0, n_{odd}\pm 2,n_{odd}\pm 4,
\ldots\} \nolabel 
\end{eqnarray}
and then on the right:
\begin{eqnarray}
\mathbb{Z}_{even} + n_{odd} = \{0 + n_{odd}, \pm 2 + n_{odd}, \pm 4 +
n_{odd},\ldots\} \nolabel 
\end{eqnarray}
Notice that the final sets are the same (because addition is
commutative).  So, $\mathbb{Z}_{even} \subset \mathbb{Z}$ is a normal
subgroup.
\end{quote} 

With a little thought, you can convince yourself that \it all \rm
subgroups of Abelian groups are normal.  

\it If $G$ is a group and $H \subset G$ is normal, then the \bf Factor
Group \it of $H$ in $G$, denoted $G/H$ (read ``G mod H"), is the group
with elements in the set $G/H \equiv \{gH  | g \in G\}$.  The group
operation $\star$ is understood to be 
\begin{eqnarray}
(g_i H ) \star (g_jH ) = (g_i \star g_j) H \nolabel 
\end{eqnarray}
\newpage

\bf Example 15 \rm
\vspace*{-2ex}
\begin{quote}
Consider again $\mathbb{Z}_{even}$.  Notice that we
can call $\mathbb{Z}_{even} = 2\mathbb{Z}$ because $2\mathbb{Z} = 2\{0,
\pm 1, \pm 2 \pm 3, \ldots\} = \{0, \pm 2, \pm 4, \ldots\} =
\mathbb{Z}_{even}$.  We know that $2\mathbb{Z} \subset \mathbb{Z}$
is normal, so we can build the factor group $\mathbb{Z}/2\mathbb{Z}$ as 
\begin{eqnarray}
\mathbb{Z}/2\mathbb{Z} = \{0+ 2\mathbb{Z}, \pm 1 + 2\mathbb{Z}, \pm 2 +
2\mathbb{Z}, \ldots\} \nolabel 
\end{eqnarray}

But, notice that 
\begin{eqnarray}
n_{even} + 2\mathbb{Z} &=& \mathbb{Z}_{even} \nolabel \\
n_{odd} + 2\mathbb{Z} &=& \mathbb{Z}_{odd} \nolabel 
\end{eqnarray}
So, the group $\mathbb{Z}/2\mathbb{Z}$ only has 2 elements; the set
of all even integers, and the set of all odd integers.	And we know
from before that there is only one group of order 2, which we denote
$\mathbb{Z}_2$.  So, we have found that $\mathbb{Z}/2\mathbb{Z} =
\mathbb{Z}_2$.	

You can also convince yourself of the more general result
\begin{eqnarray}
\mathbb{Z}/n\mathbb{Z} = \mathbb{Z}_n \nolabel 
\end{eqnarray}
\end{quote}
%\newpage
\bf Example 16 \rm
\vspace*{-2ex}
\begin{quote}
Finally, we consider the factor groups
$G/G$ and $G/e$.

\begin{itemize}
\item $G/G$ --- The set $G = \{g_0,g_1,g_2,\ldots\}$ will be the same coset for
any element of $G$ multiplied by it.  Therefore this factor group
consists of only one element, and therefore $G/G = e$, the trivial
group.	

\item $G/e$ --- The set $\{e\}$ will be a unique coset for any element of $G$,
and therefore $G/e = G$.  
\end{itemize}
\end{quote}

Something that might help you understand factor groups better is this:
the factor group $G/H$ is the group that is left over when everything
in $H$ is ``collapsed" to the identity element.  Think about the above
examples in terms of this picture.  

\it If $G$ and $H$ are both groups (not necessarily related in any
way), then we can form the \bf Product Group\it, denoted $K \equiv G
\otimes H$, where an arbitrary element of $K$ is $(g_i,h_j)$.  If the
group operation of $G$ is $\star_G$, and the group operation of $H$ is
$\star_H$, then two elements of $K$ are multiplied according to the
rule \begin{eqnarray}
(g_i,h_j)\star_K(g_k,h_l) \equiv (g_i \star_G g_k, h_j \star_H h_l)
\nolabel 
\end{eqnarray} \rm

\subsubsection{Reducibility Revisited}

Now that we understand subgroups, cosets, normal subgroups, and factor
groups, we can begin a more formal discussion of reducing
representations.  Recall that in deriving equation (\ref{eq:regrep}),
we made the designation 
\begin{eqnarray}
g_0 \rightarrow |\hat e_1 \rangle \qquad g_1 \rightarrow |\hat e_2
\rangle \qquad g_2 \rightarrow |\hat e_3\rangle \qquad \mbox{etc.} \nolabel 
\end{eqnarray}
This was used to create an order($G$)-dimensional Euclidian space
which, while not having any ``physical" meaning, and while obviously not
possessing any structure similar to the group, was and will continue to
be of great use to us.	

We have an $n$-dimensional space spanned by the orthonormal vectors
$|g_0\rangle, |g_1\rangle, \ldots, |g_{n-1}\rangle$, where $g_0$ is
understood to always refer to the identity element.  This brings us to
the first definition of this section.  \it For a group $G =
\{g_0,g_1,g_2,\ldots\}$, we call the \bf Algebra \it of $G$ the set 
\begin{eqnarray}
\mathbb{C}[\bf G\rm] \equiv \bigg\{ \sum_{i=0}^{n-1} c_i|g_i\rangle
\bigg| c_i \in \mathbb{C} \; \forall i \bigg\} \nolabel 
\end{eqnarray} \rm
In other words, $\mathbb{C}[\bf G\rm]$ is the set of all possible
linear combinations of the vectors $|g_i\rangle$ with complex
coefficients.  

We could have defined the algebra over $\mathbb{Z}$ or $\mathbb{R}$,
but we used $\mathbb{C}$ for generality at this point.	

Addition of two elements of $\mathbb{C}[\bf G\rm]$ is merely normal
addition of linear combinations, 
\begin{eqnarray}
\sum_{i=0}^{n-1} c_i |g_i\rangle + \sum_{i=0}^{n-1}d_i|g_i\rangle =
\sum_{i=0}^{n-1}(c_i+d_i)|g_i\rangle \nolabel 
\end{eqnarray}

This definition amounts to saying that, in the $n$-dimensional
Euclidian space we have created, with $n$ = order($G$), you can choose any point in the space
with complex coefficients, and this will correspond to a particular
linear combination of elements of $G$.	

Now that we have defined an algebra, we can talk about group actions. 
Recall that the $g_i$'s don't act on the $|g_j\rangle$'s, but rather
the representation $D(g_i)$ does.  We define the action $D(g_i)$ on an
element of $\mathbb{C}[\bf G\rm]$ as follows:
\begin{eqnarray}
D(g_i) \cdot \sum_{j=0}^{n-1} c_j |g_j\rangle = D(g_i) \cdot
(c_0|g_0\rangle + c_1|g_1\rangle+ \cdots + c_{n-1}|g_{n-1}\rangle)
\nolabel \\
= c_0|g_i\star g_0\rangle + c_1|g_i \star g_1\rangle + \cdots +
c_{n-1}|g_i \star g_{n-1}\rangle = \sum_{j=0}^{n-1} c_j|g_i \star
g_j\rangle\nolabel 
\end{eqnarray}

Previously, we discussed how elements of a group act on each other, and
we also talked about how elements of a group act on some other object
or set of objects (like three painted eggs).  We now generalize this
notion to a set of $q$ abstract objects a group can act on, denoted $M
= \{m_0,m_1,m_2,\ldots,m_{q-1}\}$.  Just as before, we build a vector
space, similar to the one above used in building an algebra.  The
orthonormal vectors here will be 
\begin{eqnarray}
m_0 \rightarrow |m_0\rangle, \qquad m_1 \rightarrow |m_1\rangle, \qquad
\ldots \qquad m_{q-1}  \rightarrow |m_{q-1}\rangle \nolabel 
\end{eqnarray}
This allows us to understand the following definition.

\it The set 
\begin{eqnarray}
\mathbb{C}\bf M\rm \equiv \bigg\{\sum_{i=0}^{q-1}c_i|m_i\rangle \bigg|
c_i \in \mathbb{C} \; \forall i\bigg\}\nolabel 
\end{eqnarray}
is called the \bf Module \it of $M$\rm.  
(We don't use the square brackets here to distinguish modules from
algebras).  In other words, the space spanned by the $|m_i\rangle$ is
the module.  

\bf Example 17 \rm
\vspace*{-2ex}
\begin{quote}
Consider, once again, $S_3$.	However, we generalize
from three eggs to three ``objects" $m_0,m_1$, and $m_2$.  So,
$\mathbb{C}\bf M\rm$ is all points in the 3-dimensional space of the
form $c_0|m_0\rangle + c_1|m_1\rangle + c_2 |m_2 \rangle$ with $c_i \in
\mathbb{C} \; \forall i$.  

Then, operating on a given point with, say, $g_1$ gives
\begin{eqnarray}
g_1 (c_0|m_0\rangle + c_1 |m_1\rangle + c_2 |m_2\rangle ) =
(c_0|g_1m_0\rangle + c_1|g_1m_1\rangle + c_2|g_1m_2\rangle)\nolabel 
\end{eqnarray}
and from the multiplication table on page \pageref{s3}, we know
\begin{eqnarray}
g_1m_0 = m_1, \qquad g_1m_1 = m_0, \qquad g_1m_2 = m_2\nolabel 
\end{eqnarray}
So, 
\begin{eqnarray}
(c_0|g_1m_0\rangle + c_1|g_1m_1\rangle + c_2|g_1m_2\rangle) &=&
(c_0|m_1\rangle + c_1|m_0\rangle + c_2|m_2\rangle) \nolabel \\
&=& c_1|m_0\rangle + c_0|m_1\rangle + c_2|m_2\rangle\nolabel 
\end{eqnarray}
So, the effect of $g_1$ was to swap $c_1$ and $c_0$.  This can be
visualized geometrically as a reflection in the $c_0 = c_1$ plane in
the 3-dimensional module space.  We can visualize every element of
$G$ in this way.  They each move points around the module space in a
well-defined way.  

This allows us to give the following definition.  \it If $\mathbb{C}\bf
V\it$ is a module, and $\mathbb{C}\bf W\it$ is a subspace of
$\mathbb{C}\bf V\it$ that is closed under the action of $G$, then
$\mathbb{C}\bf W\it$ is an \bf Invariant Subspace \it of $\mathbb{C}\bf
V\it$. \rm
\end{quote}

\bf Example 18 \rm
\vspace*{-2ex}
\begin{quote}
Working with $S_3$, we know that $S_3$ acts on a
3-dimensional space spanned by 
\begin{eqnarray}
|m_0\rangle = (1,0,0)^T, \qquad |m_1\rangle = (0, 1, 0 )^T, \qquad \mbox{and}
\qquad |m_2\rangle = (0, 0, 1)^T \nolabel 
\end{eqnarray}

Now, consider the subspace spanned by 
\begin{eqnarray}
c(|m_0\rangle + |m_1\rangle + |m_2\rangle) \label{eq:invsub}
\end{eqnarray}
where $c\in \mathbb{C}$, and $c$ ranges over all possible complex
numbers.  If we restrict $c$ to $\mathbb{R}$, we can visualize this
more easily as the set of all points in the line through the origin
defined by $\lambda(\hat i + \hat j + \hat k)$ (where $\lambda \in
\mathbb{R}$).  You can write out the action of any element of $S_3$ on
any point in this subspace, and you will see that they are unaffected. 
This means that the space spanned by (\ref{eq:invsub}) is an invariant
subspace.  
\end{quote}

As a note, all modules $\mathbb{C}\bf V \rm$ have two trivial invariant
subspaces.
\begin{itemize}
\parskip 0pt
\item $\mathbb{C}\bf V\rm$ is a trivial invariant subspace of
$\mathbb{C}\bf V\rm$
\item $\mathbb{C}\bf e\rm$ is a trivial invariant subspace of
$\mathbb{C}\bf V\rm$
\end{itemize}  

Finally, we can give a more formal definition of reducibility.	\it If
a representation $D$ of a group $G$ acts on the space of a module
$\mathbb{C}\bf M\it$, then the representation $D$ is said to be \bf
Reducible \it if $\mathbb{C}\bf M\it$ contains a non-trivial invariant
subspace.  If a representation is not reducible, it is \bf
Irreducible\rm.  

We encouraged you to write out the entire regular representation of
$S_3$ above.  If you have done so, you may have noticed that every
$6\times 6$ matrix appeared with non-zero elements only in the upper
left $3\times 3$ elements, and the lower right $3\times 3$ elements. 
The upper right and lower left are all 0.  This means that, for every
element of $S_3$, there will never be any mixing of the first 3
dimensions with the last 3.  So, there are two 3-dimensional
invariant subspaces in the module for this particular representation of
$S_3$ (the regular representation).  

We can now begin to take advantage of the fact that representations
live in linear spaces with the following definition.  

\it If $V$ is any $n$-dimensional space spanned by $n$ linearly
independent basis vectors, and $U$ and $W$ are both subspaces of $V$,
then we say that $V$ is the \bf Direct Sum \it of $U$ and $W$ if every
vector $\bar v \in V$ can be written as the sum $\bar v = \bar u + \bar
w$, where $\bar u \in U$ and $\bar w \in W$, and every operator $X$
acting on elements of $V$ can be separated into parts acting
individually on $U$ and $W$.  The notation for this is $V = U \oplus
W$.  \rm

In order to make this clearer, \it if $X_n$ is an $n\times n$ matrix,
it is the direct sum of $m\times m$ matrix $A_m$ and $k\times k$ matrix
$B_k$, denoted $X_n = A_m\oplus B_k$, iff $X$ is in \bf Block Diagonal
\it form,
\begin{eqnarray}
X_n = 
\begin{pmatrix}
A_m & 0 \\ 0 & B_k
\end{pmatrix}\nolabel 
\end{eqnarray}
where $n=m+k$, and $A_m$, $B_k$, and the $0$'s are understood as
matrices of appropriate dimension.  \rm

We can generalize the previous definition as follows, 
\begin{eqnarray}
X_n = A_{n_1}\oplus B_{n_2} \oplus \cdots \oplus C_{n_k} = 
\begin{pmatrix}
A_{n_1} & 0 & \cdots & 0 \\
0 & B_{n_2} & \cdots & 0 \\
\vdots & \vdots & \ddots & \cdots \\
0 & 0 & \vdots & C_{n_k} 
\end{pmatrix}\nolabel 
\end{eqnarray}
where $n=n_1+n_2+\cdots + n_k$.  

\bf Example 19 \rm
\vspace*{-2ex}
\begin{quote}
Let $A_3 = 
\begin{pmatrix}
1 & 1 & -2 \\ -1 & 5 & \pi \\ -17 & 4 & 11
\end{pmatrix}$, and let $B_2 = 
\begin{pmatrix}
1 & 2 \\ 3 & 4
\end{pmatrix}$.  Then, 
\begin{eqnarray}
B_2 \oplus A_3 = 
\begin{pmatrix}
1 & 2 & 0 & 0 & 0 \\
3 & 4 & 0 & 0 & 0 \\
0 & 0 & 1 & 1 & -2 \\
0 & 0 & -1 & 5 & \pi \\
0 & 0 & -17 & 4 & 11 \\
\end{pmatrix}\nolabel 
\end{eqnarray}
\end{quote}

To take stock of what we have done so far, we have talked about
algebras, which are the vector spaces spanned by the elements of a
group, and about modules, which are the vector spaces that representations
of groups act on.  We have also defined invariant subspaces as
follows:  Given some space and some group that acts on that space, moving the 
points around in a well-defined way, an invariant subspace is a subspace which 
always contains the same points.  The group doesn't remove any points from that
subspace, and it doesn't add any points to it.  It merely moves the points around
\it inside \rm that subspace.  
Then, we defined a representation as reducible if there are any
non-trivial invariant subspaces in the space that the group acts on.  

And what this amounts to is the following: a representation of any
group is reducible if it can be written in block diagonal form.  

But this leaves the question of what we mean when we say ``can be
written".  How can you ``rewrite" a representation?  This leads us to
the following definition.  \it Given a matrix $D$ and a non-singular
matrix S, the linear transformation 
\begin{eqnarray}
D \rightarrow D' = S^{-1}DS \nolabel 
\end{eqnarray}
is called a \bf Similarity Transformation\rm.  

Then, we can give the following definition.  \it Two matrices related
by a similarity transformation are said to be \bf Equivalent\rm.  

Because similarity transformations are linear transformations, if
$D(G)$ is a representation of $G$, then so is $S^{-1}DS$ for literally
\it any \rm non-singular matrix $S$.  To see this, if $g_i\star g_j =
g_k$, then $D(g_i)D(g_j) = D(g_k)$, and therefore
\begin{eqnarray}
S^{-1}D(g_i)S\cdot S^{-1}D(g_j)S = S^{-1}D(g_i)D(g_j)S =
S^{-1}D(g_k)S\nolabel 
\end{eqnarray}

So, if we have a representation that isn't in block diagonal form, how
can we figure out if it is reducible?  We must look for a matrix $S$
that will transform it into block diagonal form.  

You likely realize immediately that this is not a particularly easy
thing to do by inspection.  It turns out that there is a very
straightforward and systematic way of taking a given representation and
determining whether or not it is reducible, and if so, what the
irreducible representations are.  

However, the details of how this can be done, while very interesting,
are not necessary for the agenda of these notes.  Therefore, for the
sake of brevity, we will not pursue them.  What is important is that
you understand not only the details of general group theory and
representation theory (which we outlined above), but also the concept
of what it means for a group to be reducible or irreducible.  

\subsection{Introduction to Lie Groups}
\label{sec:lieintro}

In section \ref{sec:groupintro}, we considered groups which are of
finite order and discrete, which allowed us to write out a
multiplication table.  

Here, however, we examine a different type of group.  Consider the unit
circle, where each point on the circle is specified by an angle
$\theta$, measured from the positive $x$-axis.	
\newline
\begin{center}
\includegraphics[scale = .4]{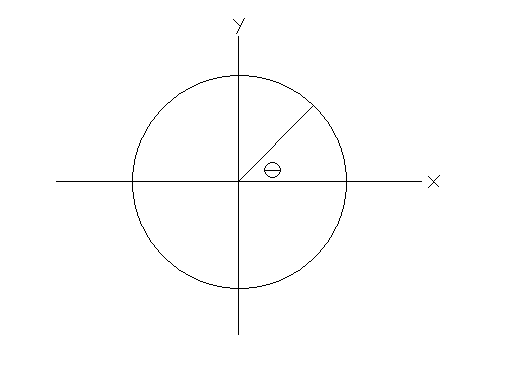}
\end{center}

We will refer to the point at $\theta=0$ as the ``starting point" (like
$ROY$ was for the Easter eggs).  Now, just as we considered all
possible orientations of $(ROY)$ that left the eggs lined up, we consider
all possible rotations the wheel can undergo.  With the eggs there
were only 6 possibilities.  Now however, for the wheel there are an
infinite number of possibilities for $\theta$ (any real number $\in
[0,2\pi)$).  

And note that if we denote the set of all angles as $G$, then all the
rotations obey closure $(\theta_1+\theta_2 = \theta_3 \in G, \; \forall
\theta_1,\theta_2 \in G)$, associativity (as usual), identity $(0 +
\theta = \theta+0 = \theta)$, and inverse (the inverse of $\theta$ is
$-\theta$).  

So, we have a group that is \it parameterized \rm by a continuous
variable $\theta$.  So, we are no longer talking about $g_i$'s, but
about $g(\theta)$.  

Notice that this particular group (the circle) is Abelian, which is why
we can (temporarily) use addition to represent it.  Also, note that we
obviously cannot make a multiplication table because the order of this
group is $\infty$.  

One simple representation is the one we used above: taking
$\theta$ and using addition.  A more familiar (and useful)
representation is the Euler matrix $g(\theta)~\dot{=} 
\begin{pmatrix}
\cos\theta & \sin \theta \\ -\sin \theta & \cos \theta
\end{pmatrix}$ with the usual matrix multiplication: 

\begin{eqnarray}
& & \begin{pmatrix}
\cos \theta_1 & \sin \theta_1 \\ -\sin \theta_1 & \cos \theta_1
\end{pmatrix}
\begin{pmatrix}
\cos \theta_2 & \sin \theta_2 \\ -\sin \theta_2 & \cos \theta_2
\end{pmatrix} \\
&=& 
\begin{pmatrix}
\cos\theta_1 \cos\theta_2-\sin\theta_1 \sin\theta_2 &
\cos\theta_1 \sin\theta_2+\sin\theta_1 \cos\theta_2 \\
-\sin\theta_1 \cos\theta_2-\cos\theta_1 \sin\theta_2 &
-\sin\theta_1 \sin\theta_2+\cos\theta_1 \cos\theta_2 
\end{pmatrix} \\
&=&
\begin{pmatrix}
\cos(\theta_1+\theta_2) & \sin(\theta_1+\theta_2) \\
-\sin(\theta_1+\theta_2) & \cos(\theta_1+\theta_2)
\end{pmatrix}
\end{eqnarray}
This will prove to be a much more useful representation than $\theta$
with addition.	

Groups that are parameterized by one or more continuous variables like
this are called \bf Lie Groups\rm.  Of course, the true definition of a
Lie group is much more rigorous (and complicated), and that definition
should eventually be understood.  However, the definition we have given
will suffice for the purposes of these notes.  

\subsubsection{Classification of Lie Groups}
\label{sec:lieclass}

The usefulness of group theory is that groups represent a mathematical
way to make changes to a system while leaving \it something \rm about
the system unchanged.  For example, we moved $(ROY)$ around, but the
structure ``3 eggs with different colors lined up" was preserved.  With
the circle, we rotated it, but it still maintained its basic structure
as a circle.  It is in this sense that group theory is a study of \bf
Symmetry\rm.  No matter which of ``these" transformations you do to the
system, ``this" stays the same---this is symmetry.  

To see the usefulness of this in physics, recall Noether's Theorem
(section \ref{sec:noether}).  When you do a \it symmetry \rm
transformation to a Lagrangian, you get a conserved quantity.  Think
back to the Lagrangian for the projectile (\ref{eq:projectile}).  The
transformation $x \rightarrow x+\epsilon$ was a symmetry because
$\epsilon$ could take any value, and the Lagrangian was unchanged (note
that $\epsilon$ forms the Abelian group $(\mathbb{R},+)$).  

So, given a Lagrangian, which represents the structure of a physical
system, a symmetry represents a way of changing the Lagrangian while
preserving that structure.  The particular preserved part of the system
is the conserved quantity $j$ we discussed in sections
\ref{sec:noether} and \ref{sec:classical}.  And as you have no doubt
noticed, nearly all physical processes are governed by \bf Conservation
Laws\rm:  conservation of momentum, energy, charge, spin, etc.  

So, group theory, and in particular Lie group theory, gives us an
extremely powerful way of understanding and classifying symmetries, and
therefore conserved charges.  And because it allows us to understand
conserved charges, group theory can be used to understand the entirety
of the physics in our universe.  

We now begin to classify the major types of Lie groups we will be working
with in these notes.  To start, we consider the most general
possible Lie group in an arbitrary number of dimensions, $n$.  This
will be the group that, for any point $p$ in the $n$-dimensional space,
can continuously take it \it anywhere \rm else in the space.  All that
is preserved is that the points in the space stay in the space.
 This means that we can have literally any $n\times n$ matrix, or \it
linear \rm transformation, so long as the matrix is invertible
(non-singular).  Thus, in $n$ dimensions the largest and most general
Lie group is the group of all $n\times n$ non-singular matrices.  We
call this group $GL(n)$, or the \bf General Linear \rm group.  The most
general field of numbers to take the elements of $GL(n)$ from is
$\mathbb{C}$, so we begin with $GL(n,\mathbb{C})$.  This is the group
of all $n\times n$ non-singular matrices with complex elements.  The
preserved quantity is that all points in $\mathbb{C}^n$ stay in
$\mathbb{C}^n$.

The most obvious subgroup of $GL(n,\mathbb{C})$ is $GL(n,\mathbb{R})$,
or the set of all $n\times n$ invertible matrices with real elements. 
This leaves all points in $\mathbb{R}^n$ in $\mathbb{R}^n$.  

To find a further subgroup, recall from linear algebra and vector
calculus that in $n$ dimensions, you can take $n$ vectors at the origin
such that for a parallelepiped, we could obtain
\begin{center}
\includegraphics[scale=.3]{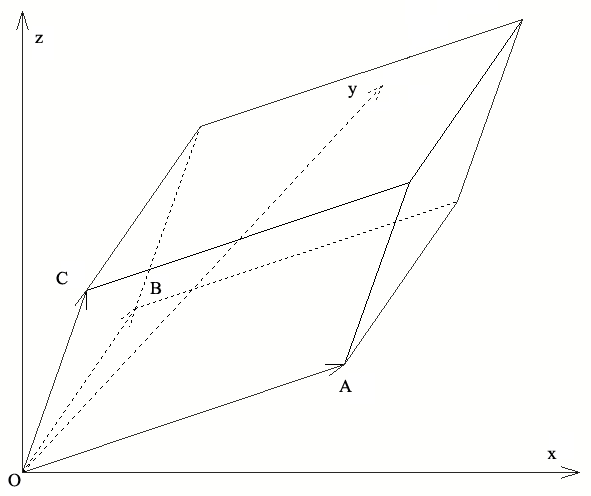}
\end{center}

Then, if you arrange the components of the $n$ vectors into the rows
(or columns) of a matrix, the determinant of that matrix will be the
volume of the parallelepiped.  

So, consider now the set of all General Linear transformations that
transform all vectors from the origin (or in other words, points in the
space) in such a way that the volume of the corresponding
parallelepiped is preserved.  This will demand that we only consider
General Linear matrices with determinant 1.  Also, the set of all
General Linear matrices with unit determinant will form a group because
of the general rule $\det|A\cdot B| = \det|A|\cdot \det|B|$.  So, if
$\det|A|=1$ and $\det|B|=1$, then $\det|A\cdot B| = 1$.  We call this
subgroup of $GL(n,\mathbb{C})$ the \bf Special Linear \rm group, or
$SL(n,\mathbb{C})$.  The natural subgroup of this is
$SL(n,\mathbb{R})$.  This group preserves not only the points in the
space (as $GL$ did), but also the volume, as described above.  

Now, consider the familiar transformations on vectors in
$n$-dimensional space of generalized Euler angles.  These are
transformations that rotate all points around the origin.  These
rotation transformations leave the radius squared $(r^2)$ invariant. 
And, because $\bar r^2 = \bar r^T \cdot \bar r$, if we transform with a
rotation matrix $R$, then $\bar r \rightarrow \bar r' = R\bar r$, and
$\bar r^T \rightarrow \bar r'^T = \bar r^T R^T$, so $\bar r'^T \cdot
\bar r' = \bar r^TR^T \cdot R \bar r$.	But, as we said, we are
demanding that the radius squared be invariant under the action of $R$,
and so we demand $\bar r^TR^T\cdot R\bar r = \bar r^T \cdot \bar r$. 
So, the constraint we are imposing is $R^T\cdot R = \mathbb{I}$, which
implies $R^T = R^{-1}$.  This tells us that the rows and columns of $R$
are orthogonal.  Therefore, we call the group of generalized rotations,
or generalized Euler angles in $n$ dimensions, $O(n)$, or the \bf
Orthogonal \rm group.  We don't specify $\mathbb{C}$ or $\mathbb{R}$
here because it will be understood that we are always talking about
$\mathbb{R}$.  

Also, note that because $\det|R^T\cdot R| = \det|\mathbb{I}| \Rightarrow
(\det|R|)^2 = 1 \Rightarrow \det|R| = \pm 1$.  We again denote the
subgroup with $\det|R| = +1$ the \bf Special Orthogonal \rm group, or
$SO(n)$.  To understand what this means, consider an orthogonal matrix
with determinant $-1$, such as 
\begin{eqnarray}
M = 
\begin{pmatrix}
1 & 0 & 0 \\ 0 & 1 & 0 \\ 0 & 0 & -1
\end{pmatrix} \nolabel 
\end{eqnarray}
This matrix is orthogonal, and therefore is an element of the group
$O(3)$, but the determinant is $-1$.  This matrix will take the point
$(x,y,z)^T$ to the point $(x,y,-z)^T$.	This changes the handedness of
the system (the right hand rule will no longer work).  So, if we limit
ourselves to $SO(n)$, we are preserving the space, the radius, the
volume, and the handedness of the space.  

For vectors in $\mathbb{C}$ space, we do not define orthogonal matrices
(although we could).  Instead, we discuss the complex version of the
radius, where instead of $\bar r^2 = \bar r^T \cdot \bar r$, we have
$\bar r^2 = \bar r^{\dagger} \cdot \bar r$, where the dagger denotes
the Hermitian conjugate, $\bar r^{\dagger} = (\bar r^{\star})^T$, where
$\star$ denotes complex conjugate.  

So, with the elements in $R$ being in $\mathbb{C}$, we have $\bar r
\rightarrow R\bar r$, and $\bar r^{\dagger} \rightarrow \bar
r^{\dagger} R^{\dagger}$.  So, $\bar r^{\dagger}\cdot \bar r
\rightarrow \bar r^{\dagger} R^{\dagger} \cdot R \bar r$, and by the
same argument as above with the orthogonal matrices, this demands that
$R^{\dagger} \cdot R = \mathbb{I}$, or $R^{\dagger} = R^{-1}$.	We
denote such matrices \bf Unitary\rm, and the set of all such $n\times
n$ invertible matrices form the group $U(n)$.  Again, we understand the
unitary groups to have elements in $\mathbb{C}$, so we don't specify that. 
And, we will still have a subset of unitary matrices $R$ with $\det|R| =
1$ called $SU(n)$, the \bf Special Unitary \rm groups.	

We can summarize the hierarchy we have just described in the following
diagram:
\begin{center}
\includegraphics{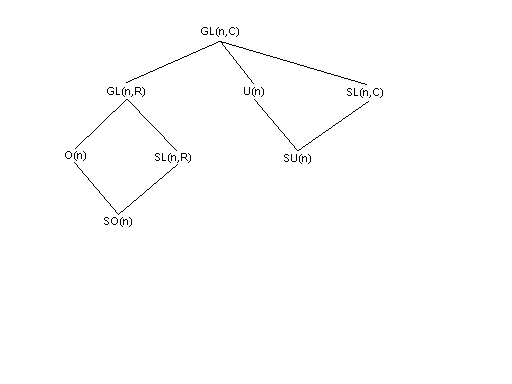}
\end{center}

We will now describe one more category of Lie groups before
moving on.  We saw above that the group $SO(n)$ preserves the radius
squared in real space.	In coordinates, this means that $\bar r^2 =
x_1^2+x_2^2+\cdots+x_n^2$, or more generally the dot product $\bar x
\cdot \bar y = x_1y_1+x_2y_2+\cdots + x_ny_n$ is preserved.  

However, we can generalize this to form a group action that preserves
not the radius squared, but the value (switching to indicial notation
for the dot product) $x^ay_a = -x_1y_1 - x_2y_2 - \cdots - x_my_m +
x_{m+1}y_{m+1} + \cdots + x_{m+n}y_{m+n}$.  We call the group that
preserves this quantity $SO(m,n)$.  The space we are working in is
still $\mathbb{R}^{m+n}$, but we are making transformations that
preserve something different than the radius.  

Note that $SO(m,n)$ will have an $SO(m)$ subgroup and an $SO(n)$
subgroup, consisting of rotations in the first $m$ and last $n$
components separately.	

Finally, notice that the specific group of this type, $SO(1,3)$, is the
group that preserves the value $s^2 = -x_1y_1+x_2y_2+x_3y_3+x_4y_4$,
or written more suggestively, $s^2 = -c^2t^2 + x^2+y^2+z^2$. 
Therefore, the group $SO(1,3)$ is the \bf Lorentz Group\rm.  Any action
that is invariant under $SO(1,3)$ is said to be a Lorentz Invariant
theory (as all theories should be).  We will find that thinking of
Special Relativity in these terms, rather than in the terms of Part I,
will be much more useful.  

It should be noted that there are many other types of Lie groups.  We
have limited ourselves to the ones we will be working with in these
notes.	

\subsubsection{Generators}
\label{sec:generatorssection}

Now that we have a good ``birds eye view" of Lie groups, we can begin
to pick apart the details of how they work.  

As we said before, a Lie group is a group that is parameterized by a
set of continuous parameters, which we call $\alpha_i$ for
$i=1,\ldots,n$, where $n$ is the number of parameters the group depends
on.  The elements of the group will then be denoted $g(\alpha_i)$.  

Because all groups include an identity element, we will choose to
parameterize them in such a way that $g(\alpha_i)\big|_{\alpha_i=0}=e$, the
identity element.  So, if we are going to talk about representations,
$D_n(g(\alpha_i))\big|_{\alpha_i = 0} = \mathbb{I}$, where $\mathbb{I}$
is the $n\times n$ identity matrix for whatever dimension ($n$)
representation we want.  

Now, take $\alpha_i$ to be very small with $\delta \alpha_i <<1$.  So,
$D_n(g(0+\delta \alpha_i))$ can be Taylor expanded:
\begin{eqnarray}
D_n(g(\delta \alpha_i)) = \mathbb{I}+\delta \alpha_i {\partial
D_n(g(\alpha_i)) \over \partial \alpha_i} \big|_{\alpha_i = 0} + \cdots
\nolabel 
\end{eqnarray}
The terms ${\partial D_n \over \partial \alpha_i} \big|_{\alpha_i = 0}$
are extremely important, and we give them their own expression:
\begin{eqnarray}
X_i \equiv -i {\partial D_n \over \partial \alpha_i} \bigg|_{\alpha_i
=0} \label{eq:generators}
\end{eqnarray}
(we have included the $-i$ in order to make $X_i$ Hermitian, which will
be necessary later).  

So, the representation for infinitesimal $\delta \alpha_i$ is then
\begin{eqnarray}
D_n(\delta \alpha_i) = \mathbb{I}+ i \delta \alpha_i X_i +
\cdots\nolabel 
\end{eqnarray}
(where we have switched our notation from $D_n(g(\alpha))$ to
$D_n(\alpha)$ for brevity).  

The $X_i$'s are constant matrices which we will determine later.  

Now, let's say that we want to see what the representation will look
like for a finite value of $\alpha_i$ rather than an infinitesimal
value.	A finite transformation will be the result of an infinite
number of infinitesimal transformations.  Or in other words, $\alpha_i
= N \delta \alpha_i$ as $N \rightarrow \infty$.  So, $\delta \alpha_i =
{\alpha_i \over N}$, and an infinite number of infinitesimal
transformations is 
\begin{eqnarray}
\lim_{N\rightarrow \infty} (1+i\delta\alpha_i X_i)^N =
\lim_{N\rightarrow \infty}\big(1+i{\alpha_i \over N} X_i\big)^N\nolabel 
\end{eqnarray}
If you expand this out for several values of $N$, you will see that it
is exactly
\begin{eqnarray}
\lim_{N\rightarrow \infty}\big(1+i{\alpha_i \over N}X_i\big)^N =
e^{i\alpha_i X_i}\nolabel 
\end{eqnarray}

We call the $X_i$'s the \bf Generators \rm of the group, and there is
one for each parameter required to specify a particular element of the
group.	For example, consider $SO(3)$, the group of rotations in
3 dimensions.  We know from vector calculus that an element of
$SO(3)$ requires 3 angles, usually denoted $\theta, \phi$, and
$\psi$.  Therefore, $SO(3)$ will require 3 generators, which will be
denoted $X_{\theta}, X_{\phi}$, and $X_{\psi}$.  We will discuss how the
generators can be found soon.	

In general, there will be several (in fact, infinite) different sets of
$X_i$'s that define a given group (just as there are an infinite
number of representations of any finite group).  What we will find is
that up to a similarity transformation, a particular set of generators
defines a particular representation of a group.  

So, $D_n(\alpha_i) = e^{i\alpha_iX_i}$ for any group (the $i$ index in
the exponent is understood to be summed over all parameters and
generators).  The best way to think of the parameter space for the
group is as a vector space, where the generators describe the behavior
near the identity, but form a basis for the entire vector space.  By
analogy, think of the unit vectors $\hat i$, $\hat j$, and $\hat k$ in
$\mathbb{R}^3$.  They are defined at the origin, but they can be
combined with real numbers/parameters to specify any arbitrary point in
$\mathbb{R}^3$.  In the same way, the generators are the ``unit
vectors" of the parameter space (which in general is a much more
complicated space than Euclidian space), and the parameters (like
$\theta, \phi$, and $\psi$) specify where in the parameter space you
are in terms of the generators.  That point in the parameter space will
then correspond to a particular element of the group.  

We call the number of generators of a group (or equivalently the number
of parameters necessary to specify an element), the \bf Dimension \rm
of the group.  For example, the dimension of $SO(3)$ is 3.  The
dimension of $SO(2)$ (rotations in the plane) however is only 1 (only
$\theta$ is needed), so there will be only one generator.  

\subsubsection{Lie Algebras}

In section \ref{sec:groupintro} we discussed algebras.	An algebra is a
space spanned by elements of the group with $\mathbb{C}$ coefficients
parameterizing the Euclidian space we defined.	Obviously we can't
define an algebra in the same way for Lie groups, because the elements
are continuous.  But, as discussed in the last section, a particular
element of a Lie group is defined by the values of the parameters in
the parameter space spanned by the generators.	We will see that the
generators will form the algebras for Lie groups.  

Consider two elements of the same group with generators $X_i$, one with
parameter values $\alpha_i$ and the other with parameter values
$\beta_i$.  The product of the 2 elements will then be
$e^{i\alpha_iX_i}e^{i\beta_jX_j}$.  Because we are assuming this is a
group, we know that the product must be an element of the group (due to
closure), and therefore the product must be specified by some set of
parameters $\delta_k$, so $e^{i\alpha_iX_i}e^{i\beta_jX_j} = e^{i
\delta_k X_k}$.  Note that the product won't necessarily simply be
$e^{i\alpha_iX_i}e^{i\beta_jX_j} = e^{i(\alpha_iX_i + \beta_jX_j)}$
because the generators are matrices and therefore don't in general
commute.  

So, we want to figure out what $\delta_i$ will be in terms of
$\alpha_i$ and $\beta_i$.  We do this as follows.  
\begin{eqnarray}
i \delta_i X_k = \ln(e^{i \delta_k X_k}) =
\ln(e^{i\alpha_iX_i}e^{\beta_jX_j}) = \ln(1+
e^{i\alpha_iX_i}e^{\beta_jX_j} -1) \equiv \ln(1+x) \nolabel 
\end{eqnarray}
where we have defined $x \equiv e^{i\alpha_iX_i}e^{ \beta_jX_j} - 1$. 
We will proceed by expanding only to second order in $\alpha_i$ and
$\beta_j$, though the result we will obtain will hold at arbitrary
order.	By Taylor expanding the exponential terms,
\begin{eqnarray}
e^{i\alpha_iX_i}e^{\beta_jX_j} - 1 &=& (1+i\alpha_iX_i+{1\over
2}(i\alpha_iX_i)^2+ \cdots)(1+i\beta_jX_j+{1\over
2}(i\beta_jX_j)^2+\cdots)-1 \nolabel \\
&=& 1+i\beta_jX_j - {1\over 2}(\beta_jX_j)^2+i\alpha_iX_i -
\alpha_iX_i\beta_jX_j-{1\over 2}(\alpha_iX_i)^2 - 1 \nolabel \\
&=& i(\alpha_iX_i + \beta_jX_j) - \alpha_iX_i\beta_jX_j - {1\over
2}\big((\alpha_iX_i)^2+(\beta_jX_j)^2\big) \nolabel
\end{eqnarray}

Then, using the general Taylor expansion $\ln(1+x) = x-{x^2\over
2}+{x^3\over 3}-{x^4 \over 4} + \cdots$, and again keeping terms only to
second order in $\alpha$ and $\beta$, we have 
\begin{eqnarray}
x-{x^2\over 2} &=&
\bigg[i(\alpha_iX_i+\beta_jX_j)-\alpha_iX_i\beta_jX_j -
{1\over2}[(\alpha_iX_i)^2+(\beta_jX_j)^2]\bigg] \nolabel \\
\;& & - {1\over 2}\bigg[i(\alpha_iX_i+\beta_jX_j)-\alpha_iX_i\beta_jX_j
- {1\over 2}[(\alpha_iX_i)^2+(\beta_jX_j)^2]\bigg]^2 \nolabel \\
&=& i (\alpha_iX_i+\beta_jX_j)-\alpha_iX_i\beta_jX_j - {1\over
2}\bigg[(\alpha_iX_i)^2+(\beta_jX_j)^2\bigg] \nolabel \\
\; & & - {1\over
2}\bigg[-(\alpha_iX_i+\beta_jX_j)(\alpha_iX_i+\beta_jX_j)\bigg]
\nolabel \\
&=& i(\alpha_iX_i+\beta_jX_j)-\alpha_iX_i\beta_jX_j - {1\over
2}\bigg[(\alpha_iX_i)^2+(\beta_jX_j)^2\bigg] \nolabel \\
\; & & + {1\over 2}\bigg[(\alpha_iX_i)^2+(\beta_jX_j)^2+
\alpha_i\beta_j(X_iX_j+X_jX_i)\bigg] \nolabel \\
&=& i(\alpha_iX_i+\beta_jX_j)+{1\over 2}\alpha_i\beta_j(X_jX_i-X_iX_j)
\nolabel \\
&=& i(\alpha_iX_i+\beta_jX_j)-{1\over2} \alpha_i\beta_j[X_i,X_j]
\nolabel \\
&=& i(\alpha_iX_i + \beta_jX_j)-{1\over 2}[\alpha_iX_i,\beta_jX_j]
\nolabel
\end{eqnarray}
So finally we can see 
\begin{eqnarray}
i\delta_kX_k = i(\alpha_iX_i+\beta_jX_j) - {1\over
2}[\alpha_iX_i,\beta_j,X_j] \nolabel 
\end{eqnarray}
or 
\begin{eqnarray}
e^{i\alpha_iX_i}e^{i\beta_jX_j} = e^{i(\alpha_iX_i+\beta_jX_j) -
{1\over 2}[\alpha_iX_i, \, \beta_j X_j]} \label{eq:BCH}
\end{eqnarray}

Equation (\ref{eq:BCH}) is called the \bf Baker-Campbell-Hausdorff \rm
formula, and it is one of the most important relations in group theory
and in physics.  Notice that, if the generators commute, this reduces
to the normal equation for multiplying exponentials.  You can think of
equation (\ref{eq:BCH}) as the generalization of the normal exponential
multiplication rule.  

Now, it is clear that the commutator $[X_i,X_j]$ must be proportional
to some linear combination of the generators of the group (because of
closure).  So, it must be the case that 
\begin{eqnarray}
[X_i,X_j] = if_{ijk}X_k \label{eq:structureconstants}
\end{eqnarray}
for some set of constants $f_{ijk}$.  These constants are called the
\bf Structure Constants \rm of the group, and if they are completely
known, the commutation relations between all the generators are known,
and so the entire group can be determined in any representation you
want.  

The generators, under the specific commutation relations defined by the
structure constants, form the \bf Lie Algebra \rm of the group, and it
is this commutation structure which forms the structure of the Lie
group.	

\subsubsection{The Adjoint Representation}
\label{sec:adjoint}

We will talk about several representations for each group we discuss,
but we will mention a very important one now.  We mentioned before that
the structure constants $f_{ijk}$ completely determine the entire
structure of the group.  

We begin by using the Jacobi identity,
\begin{eqnarray}
\big[X_i,[X_j,X_k]\big]+\big[X_j,[X_k,X_i]\big]+\big[X_k,[X_i,X_j]\big]
= 0 \label{eq:jacobi}
\end{eqnarray}
(if you aren't familiar with this identity, try multiplying it out.
You will find that it is identically true --- all the terms cancel
exactly).  But, from equation (\ref{eq:structureconstants}), we can
write
\begin{eqnarray}
\big[X_i,[X_j,X_k]\big] = if_{jka}[X_i,X_a] = if_{jka}f_{iab}X_b
\nolabel 
\end{eqnarray}
Plugging this into (\ref{eq:jacobi}) we get 
\begin{eqnarray}
& &if_{jka}f_{iab}X_b+if_{kia}f_{jab}X_b + if_{ija}f_{kab}X_b = 0
\nolabel \\
&\Rightarrow& (f_{jka}f_{iab}+f_{kia}f_{jab}+f_{ija}f_{kab})iX_b = 0
\nolabel \\
&\Rightarrow& f_{jka}f_{iab}+f_{kia}f_{jab}+f_{ija}f_{kab} = 0
\label{eq:adjointjacobi}
\end{eqnarray}

So, if we define the matrices 
\begin{eqnarray}
[T^a]_{bc} \equiv -if_{abc} \label{eq:adjointmatrices}
\end{eqnarray}
then it is easy to show that (\ref{eq:adjointjacobi}) leads to 
\begin{eqnarray}
[T^a,T^b] = if_{abc}T^c \nolabel 
\end{eqnarray}

So, the structure constants themselves form a representation of the
group (as defined by (\ref{eq:adjointmatrices}).  We call this
representation the \bf Adjoint Representation\rm, and it will prove to
be extremely important.  

Notice that the indices labeling the rows and columns in
(\ref{eq:adjointmatrices}) each run over the same values as the indices
labeling the $T$ matrices.  This tells us that the adjoint
representation is made of $n\times n$ matrices, where $n$ is the
dimension of the group, or the number of parameters in the group.  For
example, $SO(3)$ requires 3 parameters to specify an element
$(\theta,\phi,\psi)$, so the adjoint representation of $SO(3)$ will
consist of $3\times3$ matrices.  $SO(2)$ on the other hand is Abelian,
and therefore all of the structure constants vanish.  Therefore there
is no adjoint representation of $SO(2)$.  

We now go on to consider several specific groups in detail.  

\subsubsection{$SO(2)$}

We start by looking at an extremely simple group, $SO(2)$.  This is the
group of rotations in the plane that leaves $\bar r^2 = x^2+y^2 = 
\begin{pmatrix}
x & y
\end{pmatrix}\cdot 
\begin{pmatrix}
x \\ y
\end{pmatrix} = \bar v^T \cdot \bar v$ invariant.  So for some
generator $X$ (which we will now find) of $SO(2)$, $\bar v \rightarrow
R(\theta)\bar v = e^{i\theta X}\bar v$, and $\bar v^T \rightarrow \bar
v^Te^{i\theta X^T}$.  So, expanding to first order only, $\bar v^T
e^{i\theta X^T}e^{i\theta X}\bar v = \bar v^T(1+i\theta X^T + i\theta
X)\bar v = \bar v^T \cdot \bar v + \bar v^T i\theta (X+X^T) \bar v$. 
And because we demand that $r^2$ be invariant, we demand that $X+X^T =
0 \Rightarrow X=-X^T$.	So, $X$ must be antisymmetric.	Therefore we
take
\begin{eqnarray}
X \equiv {1\over i}
\begin{pmatrix}
0 & 1 \\ -1 & 0
\end{pmatrix}\nolabel 
\end{eqnarray}
(the ${1\over i}$ is included to balance the $i$ we inserted in
equation (\ref{eq:generators}) to ensure that $X$ is Hermitian).  

So, an arbitrary element of $SO(2)$ will be
\begin{eqnarray}
e^{i\theta X} &=& e^{i \theta {1\over i} 
\begin{pmatrix}
0 & 1 \\ -1 & 0
\end{pmatrix}} = e^{\theta
\begin{pmatrix}
0 & 1 \\ -1 & 0
\end{pmatrix}} \nolabel \\
&=& 
\begin{pmatrix}
0 & 1 \\ -1 & 0
\end{pmatrix}^0 + \theta
\begin{pmatrix}
0 & 1 \\ -1 & 0
\end{pmatrix}^1 + {1\over 2}\theta^2
\begin{pmatrix}
0 & 1 \\ -1 & 0
\end{pmatrix}^2 + \cdots \nolabel \\
&=& 
\begin{pmatrix}
1 & 0 \\ 0 & 1
\end{pmatrix} + \theta
\begin{pmatrix}
0 & 1 \\ -1 & 0 
\end{pmatrix} - {1\over 2}\theta^2
\begin{pmatrix}
1 & 0 \\ 0 & 1
\end{pmatrix} - {1\over 3!} \theta^3 
\begin{pmatrix}
0 & 1 \\ -1 & 0
\end{pmatrix} + \cdots \nolabel \\
&=&
\begin{pmatrix}
1 - {1\over 2}\theta^2 + \cdots & \theta - {1\over 3!}\theta^3 + \cdots
\\
-(\theta - {1\over 3!}\theta^3 + \cdots) & 1 - {1\over 2}\theta^2 +
\cdots \nolabel 
\end{pmatrix} \\
&=&
\begin{pmatrix}
\cos\theta & \sin \theta \\ -\sin \theta & \cos \theta
\end{pmatrix}\nolabel 
\end{eqnarray}
which is exactly what we would expect for a matrix describing rotations
in the plane.  

Also, notice that because $SO(2)$ is Abelian, the commutation relations
trivially vanish ($[X,X]\equiv 0$), and so all of the structure
constants are zero.  

Now that we have found an explicit example of a generator, and seen an
example of how generators relate to group elements, we move on to
slightly more complicated examples.  

\subsubsection{$SO(3)$}

We could easily generalize the argument from the proceeding section and
find the generators of $SO(3)$ in the same way, but in order to
illustrate more clearly how generators work, we will approach $SO(3)$
differently by working backwards.  Above, we found the generators and
used them to calculate the group elements.  Here, we begin with the
known group elements of $SO(3)$, which are just the standard Euler
matrices for rotations in 3-dimensional space:
\begin{eqnarray}
& & R_x(\phi) = 
\begin{pmatrix}
1 & 0 & 0 \\ 0 & \cos\phi & \sin\phi \\ 0 & -\sin\phi & \cos\phi
\end{pmatrix} \label{eq:rotx} \\
& & R_y(\psi) = 
\begin{pmatrix}
\cos \psi & 0 & -\sin\psi \\ 0 & 1 & 0 \\ \sin\psi & 0 & \cos\psi
\end{pmatrix} \label{eq:roty} \\
& & R_z(\theta) = 
\begin{pmatrix}
\cos \theta & \sin \theta & 0 \\-\sin\theta & \cos\theta & 0 \\ 0 & 0 & 1
\end{pmatrix} \label{eq:rotz}
\end{eqnarray}

Now, recall the definition of the generators, equation
(\ref{eq:generators}).	We can use it to find the generators of
$SO(3)$, which we will denote $J_x$, $J_y$, and $J_z$.	
\begin{eqnarray}
J_x = {1\over i} {dR_x(\phi) \over d\phi}\bigg|_{\phi=0} = {1\over i}
\begin{pmatrix}
0 & 0 & 0 \\ 0 & -\sin \phi & \cos \phi \\ 0 & -\cos \phi & \sin \phi
\end{pmatrix} \bigg|_{\phi=0}  = {1\over i} 
\begin{pmatrix}
0 & 0 & 0 \\ 0 & 0 & 1 \\ 0 & -1 & 0
\end{pmatrix} \nolabel 
\end{eqnarray}
And similarly 
\begin{eqnarray}
J_y = {1\over i} 
\begin{pmatrix}
0 & 0 & -1 \\ 0 & 0 & 0 \\ 1 & 0 & 0
\end{pmatrix}, \qquad
J_z = {1\over i}
\begin{pmatrix}
0 & 1 & 0 \\ -1 & 0 & 0 \\ 0 & 0 & 0
\end{pmatrix}\nolabel 
\end{eqnarray}

You can plug these into the exponentials with the appropriate
parameters ($\phi, \psi$, or $\theta$) and find that $e^{i\phi
J_x}$, $e^{i\psi J_y}$, and $e^{i\theta J_z}$ reproduce (\ref{eq:rotx}),
(\ref{eq:roty}), and (\ref{eq:rotz}), respectively.  

Furthermore, you can multiply out the commutators to find
\begin{eqnarray}
[J_x,J_y] = iJ_z, \qquad [J_y,J_z] = iJ_x, \qquad [J_z,J_x] = iJ_y
\nolabel 
\end{eqnarray}
or
\begin{eqnarray}
[J_i,J_j] = i\epsilon_{ijk}J_k \nolabel 
\end{eqnarray}
which tells us that the structure constants for $SO(3)$ are 
\begin{eqnarray}
f_{ijk} = \epsilon_{ijk} \label{eq:so3structure}
\end{eqnarray}
where $\epsilon_{ijk}$ is the totally antisymmetric tensor.  The
structure constants being non-zero is consistent with $SO(3)$ being a
non-Abelian group.  

\subsubsection{$SU(2)$}

We will approach $SU(2)$ yet another way: by starting with the
structure constants.  It turns out they are the same as the structure
constants for $SO(3)$:
\begin{eqnarray}
f_{ijk} = \epsilon_{ijk} \label{eq:su2structurecon}
\end{eqnarray}
To see why, recall that $SU(2)$ are rotations in two complex
dimensions.  The most general form of such a matrix $U \in SU(2)$ is $U
= 
\begin{pmatrix}
a & b \\ c & d
\end{pmatrix}$.  The ``Special" part of $SU(2)$ demands that the
determinant be equal to 1, or 
\begin{eqnarray}
ad - bc = 1 \nolabel 
\end{eqnarray}
and the ``Unitary" part demands that $U^{-1} = U^{\dagger}$.  So, 
\begin{eqnarray}
U^{-1} = 
\begin{pmatrix}
d & -b \\ -c & a
\end{pmatrix} = U^{\dagger} = 
\begin{pmatrix}
a^{\star} & c^{\star} \\ b^{\star} & d^{\star}
\end{pmatrix}\nolabel 
\end{eqnarray}
or in other words, 
\begin{eqnarray}
U = 
\begin{pmatrix}
a & b \\ -b^{\star} & a^{\star}
\end{pmatrix}\nolabel 
\end{eqnarray}
where we demand $|a|^2+|b|^2 = 1$.  

Both $a$ and $b$ are in $\mathbb{C}$, and therefore have 2 real
components each, so $U$ has 4 real parameters.  The constraint
$|a|^2+|b|^2 = 1$ fixes one of them, leaving 3 real parameters, just
like in $SO(3)$.  This is a loose explanation of why $SU(2)$ and
$SO(3)$ have the same structure constants.  They are both rotational
groups with 3 real parameters.  

This also tells us that $SU(2)$ will have 3 generators.  

\subsubsection{$SU(2)$ and Physical States}

The elements of any Lie group (in a $d$-dimensional representation
consisting of $d\times d$ matrices) will act on vectors, just like the
$3\times 3$ matrices representing $S_3$ acted on $\begin{pmatrix} R & O
& Y \end{pmatrix}^T$ in section \ref{sec:redprev}.  The most natural
way to understand the space a Lie group acts on is to study the
eigenvectors and eigenvalues of the generators of the representation
you are using (the reason for this is beyond the scope of these notes
at this point, but will become more clear as we proceed).  These
eigenvectors will obviously form a basis of the eigenspace of the
physical space the group is acting on.	

Using similarity transformations, one or more of the generators of a
Lie group can be diagonalized.	For now, trust us that with $SU(2)$, it
is only possible to diagonalize one of the three generators at a time
(you may convince yourself of this by studying the commutation
relations).  We will call the generators of $SU(2)$ $J^1, J^2$, and
$J^3$, and by convention we take $J^3$ to be the diagonal one.	So,
consequently, the eigenvectors of $J^3$ will be the basis vectors of
the physical vector space upon which $SU(2)$ acts.  

Now, we know that $J^3$ (whatever it is ... we don't know at this
point) will in general have more than one eigenvalue.  Let's call the
greatest eigenvalue of $J^3$ (whatever it is) $j$, and the eigenvectors
of $J^3$ will be denoted $|j;m\rangle$ (the first $j$ is merely a label
--- the second value describes the vector), where $m$ is the eigenvalue
of the eigenvector.  The eigenvector corresponding to the greatest
eigenvalue $j$ will obviously then be $|j;j\rangle$.  So,
$J^3|j;j\rangle = j|j;j\rangle$, or more generally $J^3|j;m\rangle =
m|j;m\rangle$.	

Now let's assume that we know $|j;j\rangle$.  There is a trick we can
employ to find the rest of the states.	Define the following linear
combinations of the generators:
\begin{eqnarray}
J^{\pm} \equiv {1\over \sqrt{2}}(J^1 \pm i J^2) \label{eq:jpjm}
\end{eqnarray}
Now, using the fact that the $SU(2)$ generators obey the commutation
relations in equation (\ref{eq:so3structure}), it is easy to show the
following relations, 
\begin{eqnarray}
[J^2,J^{\pm}] = \pm J^{\pm} \qquad \mbox{and} \qquad [J^+,J^-] = J^3
\label{eq:pmcom}
\end{eqnarray}
Notice that, because by definition $J^i$ are all Hermitian, we have 
\begin{eqnarray}
(J^-)^{\dagger} = J^+ \label{eq:daggerplus}
\end{eqnarray}

Consider some arbitrary eigenvector $|j;m\rangle$.  We know the
eigenvalue of this will be $m$, so $J^3|j;m\rangle = m|j;m\rangle$. 
But now let's create some new state by acting  on $|j;m\rangle$ with
either of the operators (\ref{eq:jpjm}).  The new state will be
$J^{\pm}|j;m\rangle$, but what will the $J^3$ eigenvalue be?  Using the
commutation relations in (\ref{eq:pmcom}), 
\begin{eqnarray}
J^3J^{\pm}|j;m\rangle = (\pm J^{\pm}+J^{\pm}J^3)|j;m\rangle =
(m\pm1)J^{\pm}|j;m\rangle \nolabel 
\end{eqnarray}
So, the vector $J^+|j;m\rangle$ is the eigenvector with eigenvalue
$m+1$, and the vector $J^-|j;m\rangle$ is the eigenvector with
eigenvalue $m-1$.  

If we have some arbitrary eigenvector $|j;m\rangle$, we can use
$J^{\pm}$ to move up or down to the eigenvector with the next highest
or lowest eigenvalue.  For this reason, $J^{\pm}$ are called the \bf
Raising \rm and \bf Lowering \rm operators.  They raise and lower the
eigenvalue of the state by one.  

Clearly, the eigenvector with the greatest eigenvalue $j$, with
eigenvector $|j;j\rangle$, cannot be raised any higher, so we define
$J^+|j;j\rangle \equiv 0$.  We will see that there is also a lowest
eigenvalue $j'$, so we similarly define $J^-|j;j'\rangle \equiv 0$.  

Now, considering once again $|j;j\rangle$.  We know that if we operate
on this state with $J^-$, we will get the eigenvector with the
eigenvalue $j-1$.  But, we don't know exactly what this state will be
(knowing the eigenvalue doesn't mean we know the actual state).  But,
we know it will be proportional to $|j;j-1\rangle$.  So, we set
$J^-|j;j\rangle = N_j|j;j-1\rangle$, where $N_j$ is the proportionality
constant.  To find $N_j$, we take the inner product (and using
(\ref{eq:daggerplus})):
\begin{eqnarray}
\langle j;j|J^+J^-|j;j\rangle = |N_j|^2\langle j;j-1|j;j-1\rangle
\nolabel 
\end{eqnarray}
But we can also write
\begin{eqnarray}
\langle j;j|J^+J^-|j;j\rangle &=& \langle j;j|(J^+J^- -
J^-J^+)|j;j\rangle = \langle j;j|\big[J^+,J^-\big]|j;j\rangle \nolabel
\\
&=& \langle j;j|J^3|j;j\rangle = j\langle j;j|j;j\rangle = j
\label{eq:littlej}
\end{eqnarray}
where we used the fact that $J^+|j;j\rangle = 0 $ to get the first
equality, and (\ref{eq:pmcom}) to get the third equality.  We also
assumed that $|j;j\rangle$ is normalized.  

So, (\ref{eq:littlej}) tells us 
\begin{eqnarray}
\langle j;j-1|j;j-1\rangle = 1 \iff N_j \equiv \sqrt{j} \label{eq:n1}
\end{eqnarray}
And our normalized state is therefore 
\begin{eqnarray}
{J^- \over N_j} |j;j\rangle = {J^-\over \sqrt{j}}|j;j\rangle =
|j;j-1\rangle \nolabel 
\end{eqnarray}

Repeating this to find $N_{j-1}$, we have
\begin{eqnarray}
|N_{j-1}|^2\langle j;j-2|j;j-2\rangle &=& \langle
j;j-1|J^+J^-|j;j-1\rangle \nolabel \\
&=& \langle j;j|{J^+ \over \sqrt{j}}J^+J^-{J^- \over
\sqrt{j}}|j;j\rangle \nolabel \\
&=& {1\over j}\langle j;j|J^+J^+J^-J^-|j;j\rangle \nolabel \\
&=& {1\over j}\langle j;j|J^+(J^3+J^-J^+)J^-|j;j\rangle \nolabel \\
&=& {1\over j}\langle j;j|(J^+J^3J^-+J^+J^-J^+J^-)|j;j\rangle \nolabel
\\
&=& {1\over j}\langle
j;j|(J^+(-J^-+J^-J^3)+J^+J^-(J^3+J^-J^+))|j;j\rangle \nolabel \\
&=& {1\over j}[\langle j;j|(-J^+J^- + jJ^+J^-+jJ^+J^-)|j;j\rangle]
\nolabel \\
&=& {1\over j}\langle j;j|(-[J^+,J^-]+2j[J^+,J^-])|j;j\rangle \nolabel
\\
&=& {1\over j}\langle j;j|(-J^3+2jJ^3)|j;j\rangle \nolabel \\
&=& {1\over j} (2j^2 - j) = 2j-1 \label{eq:longder}
\end{eqnarray}
So, $|N_{j-1}|^2 = 2j-1$, or $N_{j-1} = \sqrt{2j-1}$.  

We can continue this process, and we will find that the general result
is 
\begin{eqnarray}
N_{j-k} = {1 \over \sqrt{2}}\sqrt{(2j-k)(k+1)} \label{eq:su3propconst}
\end{eqnarray}
and the general states are defined by 
\begin{eqnarray}
|j;j-k\rangle = {1 \over N_{j-k}}(J^-)^k|j;j\rangle \nolabel 
\end{eqnarray}

Notice that these expressions recover (\ref{eq:n1}) and
(\ref{eq:longder}) for $k=0$ and $k=1$, respectively.  

Furthermore, notice that when $k=2j$,
\begin{eqnarray}
N_{j-2j} = {1\over \sqrt{2}}\sqrt{(2j-2j)(2j+1)} \equiv 0 \nolabel 
\end{eqnarray}
So, the state $|j;j-k\rangle \big|_{k=2j} = |j;-j\rangle$ is the state
with the lowest eigenvalue, and by definition $J^-|j;-j\rangle \equiv
0$.  

So, in a general representation of $SU(2)$, we have $2j+1$ states: 
\begin{eqnarray}
\{j,j-1,j-2,\ldots, -j+2,-j+1,-j\} \nolabel 
\end{eqnarray}
This therefore demands that $j={n\over 2}$ for some integer $n$.  In
other words, the highest eigenvalue of an $SU(2)$ eigenvector can be
$0,{1\over 2},1,{3\over 2},2$, etc.  

Furthermore, using these states, it is easy to show
\begin{eqnarray}
\langle j;m'|J^3|j;m\rangle &=& m\delta_{m',m} \nolabel \\
\langle j;m'|J^+|j;m\rangle &=& {1\over \sqrt{2}} \sqrt{(j+m+1)(j-m)}
\delta_{m',m+1} \nolabel \\
\langle j;m'|J^-|j;m\rangle &=& {1\over \sqrt{2}} \sqrt{(j+m)(j-m+1)}
\delta_{m',m-1} \label{eq:expvals}
\end{eqnarray}

\subsubsection{$SU(2)$ for $j={1\over 2}$}

We will skip the $j=0$ case because it is trivial (though we will
discuss it later when we return to physics).  

For $j={1\over 2}$, the two eigenvalues of $J^3$ will be ${1\over 2}$
and ${1\over 2}-1$ = $-{1\over 2}$.  So, denoting the $J^3$ generator of
$SU(2)$ when $j={1\over 2}$ as $J^3_{1/2}$, we have 
\begin{eqnarray}
J^3_{1/2} = 
\begin{pmatrix}
1/2 & 0 \\ 0 & 1/2
\end{pmatrix}\nolabel 
\end{eqnarray}

Now, inverting (\ref{eq:jpjm}) to get 
\begin{eqnarray}
J^1 = {1\over \sqrt{2}}(J^-+J^+) \qquad \mbox{and} \qquad J^2 = {i \over
\sqrt{2}}(J^--J^+) \nolabel
\end{eqnarray}
and using the standard matrix equation $[J^a_j]_{m',m} = \langle
j,m'|J^a|j,m\rangle$, and the explicit products in (\ref{eq:expvals}),
we can find (for example)
\begin{eqnarray}
\bigg\langle {1\over 2};-{1\over 2}\bigg|J^1\bigg|{1\over 2};-{1\over
2}\bigg\rangle = \bigg\langle {1\over 2};-{1\over 2}\bigg|{1\over
\sqrt{2}}(J^-+J^+)\bigg|{1\over 2};-{1\over 2}\bigg\rangle = \cdots =0
\nolabel
\end{eqnarray}
So $[J^1]_{11} = 0$.  
Then, 
\begin{eqnarray}
\bigg\langle {1\over 2};-{1\over 2}\bigg|J^1\bigg|{1\over 2};{1\over
2}\bigg\rangle = \bigg\langle {1\over 2};-{1\over 2}\bigg|{1\over
\sqrt{2}}(J^-+J^+)\bigg|{1\over 2};{1\over 2}\bigg\rangle = \cdots
={1\over 2}\nolabel
\end{eqnarray}
So $[J^1]_{12} = {1\over 2}$.  

We can continue this to find all the elements for each generator for
$j=1/2$.  The final result will be
\begin{eqnarray}
J^1_{1/2} = {1\over 2}
\begin{pmatrix}
0 & 1 \\ 1 & 0
\end{pmatrix} = {\sigma^1 \over 2}, \quad 
J^2_{1/2} = {1\over 2}
\begin{pmatrix}
0 & -i \\ i & 0
\end{pmatrix} = {\sigma^2 \over 2}, \quad
J^3_{1/2} = {1\over 2}
\begin{pmatrix}
1 & 0 \\ 0 & -1
\end{pmatrix} = {\sigma^3 \over 2} \label{eq:paulimat}
\end{eqnarray}
where the $\sigma^i$ matrices are the \bf Pauli Spin Matrices\rm.  This is no
accident!  We will discuss this in much, much more detail later, but
for now recall that we said that $SU(2)$ is the group of
transformations in 2-dimensional complex space (with one of the real
parameters fixed, leaving 3 real parameters).  We are going to see
that $SU(2)$ is the group which represents quantum mechanical spin,
where $j$ is the value of the spin of the particle.  In other words,
particles with spin $1/2$ are described by the $j=1/2$ representation
(the $2\times 2$ representation in (\ref{eq:paulimat})), and particles
with spin 1 are described by the $j=1$ representation, and so on.  In
other words, $SU(2)$ describes quantum mechanical spin in 3
dimensions in the same way that $SO(3)$ describes normal ``spin" in 3
dimensions.  We will talk about the physical implications, reasons, and
meaning of this later.	

However, as a warning, be careful at this point not to think too much
in terms of physics.  You have likely covered $SU(2)$ in great detail
in a quantum mechanics course (though you may not have known it was
called ``$SU(2)$"), but the approach we are taking here has a different
goal than what you have likely seen before.  The properties of $SU(2)$
we are seeing here are actually very, very specific and simplified
illustrations of much deeper concepts in Lie groups, and in order to
understand particle physics we must understand Lie groups in this way. 
So for now, try to fight the temptation to merely understand everything
we are doing in terms of the physics you have seen before and learn
this as we are presenting it: pure mathematics.  We will focus on how
it applies to physics later, in its fuller and more fundamental way
than introductory quantum mechanics makes apparent.  

\subsubsection{$SU(2)$ for $j=1$}

You can follow the same procedure we used above to find
\begin{eqnarray}
J^1_1 = {1\over \sqrt{2}}
\begin{pmatrix}
0 & 1 & 0 \\ 1 & 0 & 1 \\ 0 & 1 & 0
\end{pmatrix}, \quad
J^2_1 = {1\over \sqrt{2}}
\begin{pmatrix}
0 & -i & 0 \\ i & 0 & -i \\ 0 & i & 0
\end{pmatrix}, \quad 
J^3_1 = 
\begin{pmatrix}
1 & 0 & 0 \\ 0 & 0 & 0 \\ 0 & 0 & -1
\end{pmatrix} \label{eq:su2adjoint}
\end{eqnarray}

Notice that only $J^3_1$ is diagonal (as before), and that the
eigenvalues are $\{1,0,-1\}$, or $\{j,j-1,j-2=-j\}$ as we'd expect.  

\subsubsection{$SU(2)$ for Arbitrary $j$}
\label{sec:arbitraryj}

For any given $j$, we have 3 generators $J^1_j,J^2_j$, and $J^3_j$,
and for whatever dimension ($d=2j+1$) the physical space we are working
in, we have $d$ eigenvectors
\begin{eqnarray}
|j;j\rangle = 
\begin{pmatrix}
1 \\ 0 \\ 0 \\ \vdots \\ 0
\end{pmatrix}, \quad |j;j-1\rangle = 
\begin{pmatrix}
0 \\ 1 \\ 0 \\ \vdots \\ 0
\end{pmatrix}, \quad |j;j-2\rangle = 
\begin{pmatrix}
0 \\ 0 \\ 1 \\ \vdots \\ 0
\end{pmatrix}, \quad \cdots \quad |j;-j\rangle = 
\begin{pmatrix}
0 \\ 0 \\ 0 \\ \vdots \\ 1
\end{pmatrix}\nolabel
\end{eqnarray}
with eigenvalues $\{j,j-1,j-2,\cdots, -j\}$, respectively.  

Then, for any $j$, we can form the linear combinations $J^{\pm}_j
\equiv {1\over \sqrt{2}}(J^1_j \pm i J^2_j)$.  For example, for $j=1/2$
these are
\begin{eqnarray}
J^+_{1/2} = {1\over \sqrt{2}} \bigg[{1\over 2}
\begin{pmatrix}
0 & 1 \\ 1 & 0
\end{pmatrix} + {i \over 2}
\begin{pmatrix}
0 & -i \\ i & 0
\end{pmatrix}\bigg] = {1\over 2\sqrt{2}}
\begin{pmatrix}
0 & 2 \\ 0 & 0
\end{pmatrix} = {1\over \sqrt{2}}
\begin{pmatrix}
0 & 1 \\ 0 & 0 
\end{pmatrix}\nolabel
\end{eqnarray}
and similarly
\begin{eqnarray}
J^-_{1/2} = \cdots = {1\over \sqrt{2}}
\begin{pmatrix}
0 & 0 \\ 1 & 0
\end{pmatrix}\nolabel
\end{eqnarray}

So, the two $j=1/2$ eigenvectors will be $\big|{1\over 2};{1\over
2}\rangle = \begin{pmatrix} 1 \\ 0 \end{pmatrix}$ and $\big|{1\over
2};-{1\over 2}\rangle = \begin{pmatrix} 0 \\ 1 \end{pmatrix}$.	So, 
\begin{eqnarray}
J^+_{1/2}\bigg|{1\over 2};{1\over 2}\bigg\rangle = {1\over \sqrt{2}} 
\begin{pmatrix}
0 & 1 \\ 0 & 0
\end{pmatrix}
\begin{pmatrix}
1 \\ 0
\end{pmatrix} = 0 \nolabel\\
J^-_{1/2}\bigg|{1\over 2};-{1\over 2}\bigg\rangle = {1\over \sqrt{2}}
\begin{pmatrix}
0 & 0 \\ 1& 0
\end{pmatrix}
\begin{pmatrix}
0 \\ 1
\end{pmatrix} = 0 \nolabel
\end{eqnarray}
and similarly
\begin{eqnarray} 
J^-_{1/2}\bigg|{1\over 2};{1\over 2}\bigg\rangle = \bigg|{1\over
2};-{1\over 2}\bigg\rangle \nolabel\\
J^+_{1/2}\bigg|{1\over 2};-{1\over 2}\bigg\rangle = \bigg|{1\over
2};{1\over 2}\bigg\rangle \nolabel
\end{eqnarray}
which is exactly what we would expect.	

The same calculation can be done for the $j=1$ case and we will find
the same results, except that the $j=1$ state (the first eigenvector)
can be lowered \it twice\rm.  The first time $J^-_{1/2}$ acts it takes
it to the state with eigenvalue 0, and the second time it acts it
takes it to the state with eigenvalue $-1$.  Acting a third time will
destroy the state (take it to 0).  Analogously, the lowest state,
with eigenvalue $j=-1$ can be raised twice.  

We can do the same analysis for any $j=$ integer or half integer.  

As we said before, we interpret $j$ as the quantum mechanical spin of a
particle, and the group $SU(2)$ describes that rotation.  It is
important to recognize that quantum spin is not a rotation through
spacetime (it would be described by $SO(3)$ if it was), but rather
through the mathematically constructed spinor space.  We will talk more
about this space later.  

So for a given particle with spin, we can talk about both its rotation
through physical spacetime using $SO(3)$, as well as its rotation
through complex spinor space using $SU(2)$.   Both values will be
physically measurable and will be conserved quantities.  The total
angular momentum of the particle will be the combination of both spin
and spacetime angular momentum.  Again, we will talk much more about
the spin of physical particles when we return to a discussion of
physics.  We only mention this now to give a preview of where this is
going.	However, spin is not the only thing $SU(2)$ describes.	We will
also find that it is the group which governs the weak nuclear force
(whereas $U(1)$ describes the electromagnetic force, and $SU(3)$
describes the strong force ... much, much more on this later).	

\subsubsection{Root Space}
\label{sec:rootspace}

As a comment before beginning this section, it is likely that you will
find this to be the most difficult section of these notes.  The
material here is both extremely difficult (especially the first time it
is encountered), and extremely important to the development of particle
physics.  In fact, this section is the most central to what will come
later in these notes.  If the contents are not clear you are encouraged
to read this section multiple times until it becomes clear.  It may
also be helpful to study this section while looking closely at the
examples in the sections forming the remainder of this part of these
notes.	They illustrate the point of where we are going with all of
this.  

We saw in the previous section that we can view the physical space that a
group is acting on by using the eigenvectors of the diagonal generators
as a basis.  These eigenvectors can be arranged in order of decreasing
eigenvalue.  Then, the non-diagonal generators can be used to form
linear combinations that act as raising and lowering operators, which
transform one eigenvector to another, changing the eigenvalue by an
amount defined by the commutation relations of the generators.	

We now see that this generalizes very nicely.  

An arbitrary Lie group is defined in terms of its generators.  As we
said at the end of section \ref{sec:generatorssection}, it is best to
think of the generators as being analogous to the basis vectors
spanning some space.  Of course, the space the generators span is much
more complicated than $\mathbb{R}^n$ in general, but the generators
span the space the same way.  In this sense, the generators form a
linear vector space.  So, we must define an inner product for them. 
For reasons that are beyond the scope of these notes, we will choose
the generators and inner product so that, for generators $T^a$ and
$T^b$,
\begin{eqnarray}
\langle T^a,T^b\rangle \equiv {1\over \kappa} \Tr(T^aT^b) = \delta^{ab}
\label{eq:killing}
\end{eqnarray}
where $\kappa$ is some normalization constant.	

Also, in the set of generators of a Lie group, there will be a closed
subalgebra of generators which all commute with each other, but not
with generators outside of this subalgebra.  In other words, this is
the set of generators which can be simultaneously diagonalized through
some similarity transformation.  For $SU(2)$, we saw that there was only
one generator in this subalgebra which we chose to be $J^3_j$ (recall
that a matrix will only commute with all other matrices if it is equal
to the identity matrix times a constant, whereas two diagonal matrices
will always commute regardless of what their diagonal elements are).  

Let's say that a particular Lie group has $N$ generators total, or is
an $N$-dimensional group.  Then, let's say that there are $M<N$
generators in the mutually commuting subalgebra.  We call those $M$
generators the \bf Cartan Subalgebra\rm, and the generators in it are
called \bf Cartan Generators\rm.  We define the number $M$ as the \bf
Rank \rm of the group.	

By convention we will label the Cartan generators $H^i$
($i=1,\ldots,M$) and the non-Cartan generators $E^i$ ($i=1,\ldots,N-M$).  

For example, with $SU(2)$ we had $H^1 = J^3_j$, and $E^1 = J^1_j$, $E^2
= J^2_j$.  

Before moving on, we point out that this should seem familiar.	If you
think back to an introductory class in quantum mechanics, recall that
we always choose some set of variables that all commute with each
other (usually we choose \it either \rm position or momentum because
$[x,p]\neq 0$).  Then, we expand the physical states in terms of the
position \it or \rm momentum eigenvectors.  Here, we are doing the
exact same thing, only in a much more general context.	

Now, the $H^i$'s are simultaneously diagonalized, so we will write the
physical states in terms of their eigenvalues.	In an $n$-dimensional
representation $D_n$, the generators are $n\times n$ matrices, so the
eigenvectors are $n$-dimensional.  So, there will be a total of $n$
eigenvectors, and each will have one eigenvalue with each of the $M$
Cartan generators $H^i$.  So, for each of these eigenvectors, which we
temporarily denote $|j\rangle$, for $j=1,\ldots,n$, we have the $M$
eigenvalues with $M$ Cartan generators, which we call $t^i_j$ (where
$j=1,\ldots,n$ labels the eigenvectors, and $i=1,\dots,M$ labels the
eigenvalues), and we form what is called a \bf Weight Vector\rm
\begin{eqnarray}
\bar t_j \equiv 
\begin{pmatrix}
t^1_j \\ t^2_j \\ t^3_j \\ \vdots \\t^M_j
\end{pmatrix} \label{eq:weightvectors}
\end{eqnarray}
where $j=1,\ldots,n$.  The individual components of these vectors, the
$t^i_j$'s, are called the \bf Weights\rm.  

So for a given representation $D_n$, we now denote the state $|D_n;\bar
t_j\rangle$ (instead of $|j\rangle$).  So, our eigenvalues will be 
\begin{eqnarray}
H^i|D_n;\bar t_j\rangle = t^i_j|D_n;\bar t_j\rangle \label{eq:honstate}
\end{eqnarray}

As we mentioned before, the adjoint representation is a particularly
important representation.  If you do not remember the details of the
adjoint representation, go reread section \ref{sec:adjoint}.  Here, the
generators are defined by equation (\ref{eq:adjointmatrices}),
$[T^a]_{bc} \equiv -i f_{abc}$.  Recall that each index runs from 1
to $N$, so that the generators in the adjoint representation are
$N\times N$ matrices, and the eigenvectors are $N$-dimensional.  

Also, as a point of nomenclature, weights in the adjoint representation
are called \bf Roots\rm, and the corresponding vectors (as in
(\ref{eq:weightvectors})) are called \bf Root Vectors\rm.  

This means that there is exactly one eigenvector for each generator,
and therefore one root vector for each generator.  So, in equation
(\ref{eq:weightvectors}), $j=1,\ldots,N$.  We make this more obvious by
explicitly assigning each eigenvector to a generator as follows. 
First, because we now have the same number of generators, eigenvectors,
and root vectors, we label the generators by the root vectors $T^{\bar
t_j}$ instead of $T^j$.  Also, we now refer to general eigenstates as
$|Adj;T^{\bar t_j}\rangle$, where $j=1,\ldots,N$ and $\bar t_j$ is the
$M$-dimensional root vector corresponding to $T^{\bar t_j}$.  And, we
also divide the states $|Adj;T^{\bar t_j}\rangle$ into two groups:
those corresponding to the $M$ Cartan generators $|Adj;H^{\bar
h_j}\rangle$ (where $j=1,\ldots,M$ and $\bar h_j$ is the $M$-dimensional
root vector corresponding to $H^{\bar h_j}$), and those
corresponding to the $N-M$ non-Cartan generators $|Adj;E^{\bar
e_j}\rangle$ (where $j=1,\ldots N-M$ and $\bar e_j$ is the $M$-dimensional
root vector corresponding to $E^{\bar e_j}$).  

Don't be alarmed by the superscripts being vectors.  We are using this
notation for later convenience, and $T^{\bar t_i}$ here means the same
thing $T^j$ did before (the $j^{th}$ generator).  This notation, which
we use only for the adjoint representation, is simply taking advantage
of the fact that in the adjoint representation, the total number of
generators, the number of eigenvectors of the Cartan generators, the
dimension of the representation, and the number of weight/root vectors
is the same.  

Also, with the adjoint representation states $|Adj;T^{\bar
t_j}\rangle$, we can use equation (\ref{eq:killing}) to define the
inner product between states as 
\begin{eqnarray}
\langle Adj; T^{\bar t_j}|Adj;T^{\bar t_k}\rangle = {1\over
\kappa}\Tr(T^{\bar t_j}T^{\bar t_k}) = \delta^{jk} \label{eq:modkilling}
\end{eqnarray}
We will make use of this equation soon.  

The matrix elements of a given generator will then be given by the
familiar equation 
\begin{eqnarray}
-if_{abc} = [T^{\bar t_a}]_{bc} \equiv \langle Adj;T^{\bar t_b}|T^{\bar
t_a}|Adj;T^{\bar t_c}\rangle\nolabel
\end{eqnarray}
We want to know what an arbitrary generator $T^{\bar t_a}$ will do to
an arbitrary state $|Adj;T^{\bar t_b}\rangle$ in the adjoint
representation.  So,
\begin{eqnarray}
T^{\bar t_a}|Adj;T^{\bar t_b}\rangle &=& \sum_c |Adj;T^{\bar
t_c}\rangle \langle Adj; T^{\bar t_c}|T^{\bar t_a}|Adj;T^{\bar
t_b}\rangle = \sum_c|Adj;T^{\bar t_c}\rangle [T^{\bar t_a}]_{cb}
\nolabel \\
&=& \sum_c|Adj;T^{\bar t_c}\rangle (-if_{acb}) = \sum_c
if_{abc}|Adj;T^{\bar t_c}\rangle\nolabel
\end{eqnarray}
And, because there is exactly one eigenvector for each generator, the
state $|Adj;T^{\bar t_c}\rangle$ corresponds to the generator $T^c$. 
And because we know that 
\begin{eqnarray}
if_{abc}T^{\bar t_c} = [T^{\bar t_a},T^{\bar t_b}]\nolabel
\end{eqnarray}
(where $c$ is understood to be summed) by definition of the structure
constants, we can infer that
\begin{eqnarray}
T^{\bar t_a}|Adj;T^{\bar t_b}\rangle = \sum_c if_{abc}|Adj;T^{\bar
t_c}\rangle = |Adj;[T^{\bar t_a},T^{\bar t_b}]\rangle
\label{eq:adjointcomm}
\end{eqnarray}
where $[T^{\bar t_a},T^{\bar t_b}]$ is simply the commutator.  

The derivation of equation (\ref{eq:adjointcomm}) is extremely
important, and it is vital that you understand it.  However, it is also
one of the more difficult results of this already difficult section. 
You are therefore encouraged (again) to read through this section,
comparing it with examples several times until it becomes clear.  

So, let's apply this to combinations of the two types of generators we
have, $H^{\bar h_a}$'s and $E^{\bar e_a}$'s.  If we have a Cartan
generator acting on a state corresponding to a Cartan generator, we
have (from equation (\ref{eq:honstate}))
\begin{eqnarray}
H^{\bar h_a}|Adj;H^{\bar h_b}\rangle = h^a_b|Adj;H^{\bar h_b}\rangle
\nolabel
\end{eqnarray}
But from (\ref{eq:adjointcomm}) we have
\begin{eqnarray}
H^{\bar h_a}|Adj;H^{\bar h_b}\rangle = |Adj;[H^{\bar h_a},H^{\bar
h_b}]\rangle \nolabel
\end{eqnarray}
By definition, the Cartan generators commute, so $[H^{\bar
t_a},H^{\bar t_b}] \equiv 0$, and therefore
\begin{eqnarray}
\bar h_b \equiv 0 \label{eq:heqzero}
\end{eqnarray}
So we can drop them from our notation, leaving the eigentstates
corresponding to non-Cartan generators denoted $|Adj;H^j\rangle$.  

On the other hand, if we have a Cartan generator acting on an
eigenstate corresponding to a non-Cartan generator, equation
(\ref{eq:honstate}) gives
\begin{eqnarray}
H^a|Adj;E^{\bar e_b}\rangle = e^a_b|Adj;[H^a,E^{\bar e_b}]\rangle
\label{eq:comp1}
\end{eqnarray}
And equation (\ref{eq:adjointcomm}) gives 
\begin{eqnarray}
H^a|Adj;E^{\bar e_b}\rangle = |Adj;[H^a,E^{\bar e_b}]\rangle
\label{eq:comp2}
\end{eqnarray}

Now, we don't know \it a priori \rm what $[H^a,E^{\bar e_b}]$ is, but
comparing (\ref{eq:comp1}) and (\ref{eq:comp2}), we see
\begin{eqnarray}
|Adj;e^a_bE^{\bar e_b}\rangle = |Adj;[H^a,E^{\bar e_b}]\rangle\nolabel
\end{eqnarray}
And because we know that each of these vectors corresponds directly to
the generators, we have the final result 
\begin{eqnarray}
[H^a,E^{\bar e_b}] = e^a_bE^{\bar e_b} \label{eq:cartannoncartancomrel}
\end{eqnarray}

Now we want to know what a non-Cartan generator does to a given
eigentstate.  Consider an arbitrary state $|Adj;T^{\bar t_b}\rangle$
with $H^c$ eigenvalue $t^c_b$.	We can act on this with $E^{\bar e_a}$
to create the new state $E^{\bar e_a}|Adj;T^{\bar t_b}\rangle$.  So
what will the $H^c$ eigenvalue of this new state be?  Using
(\ref{eq:cartannoncartancomrel}),
\begin{eqnarray}
H^cE^{\bar e_a}|Adj;T^{\bar t_b}\rangle &=& (H^cE^{\bar e_a} - E^{\bar
e_a}H^c+E^{\bar e_a}H^c)|Adj;T^{\bar t_b}\rangle =  ([H^c,E^{\bar
e_a}]+E^{\bar e_a}H^c)|Adj;T^{\bar t_b}\rangle \nolabel \\
&=& (e^c_aE^{\bar e_a}+E^{\bar e_a}t^c_b)|Adj;T^{\bar t_b}\rangle =
(t^c_b+e^c_a)E^{\bar e_a}|Adj;T^{\bar t_b}\rangle \nolabel \\
&=& (\bar t_b+\bar e_a)^cE^{\bar e_a}|Adj;T^{\bar t_b}\rangle
\label{eq:addrootvectors}
\end{eqnarray}
So, by acting on the one of the eigenstates with a non-Cartan generator
$E^{\bar e_a}$, we have shifted the $H^c$ eigenvalue by one of the
coordinates of the root vector.  What this means is that the non-Cartan
generators play a role analogous to the raising and lowering operators we
saw in $SU(2)$, except instead of merely shifting the state ``up" and
``down", it moves the states around through some $M$-dimensional space. 

From this, we can also see that if
there is an operator that can transform from one state to another,
there must be a corresponding operator that will make the opposite
transformation.  Therefore, for every operator $E^{\bar e_a}$, we
expect to have the operator $E^{-\bar e_a}$, and corresponding
eigenstate $|Adj;E^{-\bar e_a}\rangle$.  

Finally, consider the state $E^{\bar e_a}|Adj;E^{-\bar e_a}\rangle$. 
We know from (\ref{eq:adjointcomm}) that $E^{\bar e_a}|Adj;E^{-\bar
e_a}\rangle = |Adj;[E^{\bar e_a},E^{-\bar e_a}]\rangle$.  The
eigenvalue of this state can be found using equation
(\ref{eq:addrootvectors}):
\begin{eqnarray}
H^bE^{\bar e_a}|Adj;E^{-\bar e_a}\rangle = (-\bar e_a+\bar
e_a)^bE^{\bar e_a}|Adj;E^{-\bar e_a}\rangle \equiv 0\nolabel
\end{eqnarray}
But according to equation (\ref{eq:heqzero}), states with 0
eigenvalue are states corresponding to Cartan generators.  Therefore we
conclude that the state $E^{\bar e_a}|Adj;E^{-\bar e_a}\rangle$ is
proportional to some linear combination of the Cartan states, 
\begin{eqnarray}
E^{\bar e_a}|Adj;E^{-\bar e_a}\rangle = \sum_bN_b|Adj;H^b\rangle
\label{eq:cartanlincomb}
\end{eqnarray}
where the $N_b$'s are the constants of proportionality.  To find the
constants $N_b$, we follow an approach similar to the one we used in
deriving (\ref{eq:su3propconst}).  Taking the inner product and using
(\ref{eq:adjointcomm}), 
\begin{eqnarray}
\langle Adj;H^c|E^{\bar e_a}|Adj;E^{-\bar e_a}\rangle &=& \sum_b
N_b\langle Adj;H^c|Adj;H^b\rangle = \sum_bN_b\delta^{cb} = N_c \\
&\Rightarrow& \langle Adj;H^c|Adj;[E^{\bar e_a},E^{-\bar e_a}]\rangle =
N_c\nolabel
\end{eqnarray}
Then, using (\ref{eq:modkilling})
\begin{eqnarray}
\langle Adj;H^c|Adj;[E^{\bar e_a},E^{-\bar e_a}]\rangle &=& {1\over
\kappa}\Tr(H^c[E^{\bar e_a},E^{-\bar e_a}]) \nolabel \\
&=& {1\over \kappa}\Tr(E^{-\bar e_a}[H^c,E^{\bar e_a}]) \nolabel \\
&=& {1\over \kappa}e^c_a\Tr(E^{-\bar e_a}E^{\bar e_a}) \nolabel \\
&=& e^c_a \delta^{aa} \nolabel \\
&=& e^c_a\nolabel
\end{eqnarray}
So, 
\begin{eqnarray}
N_c = e^c_a\nolabel
\end{eqnarray}
And therefore equation (\ref{eq:cartanlincomb}) is now
\begin{eqnarray}
E^{\bar e_a}|Adj;E^{-\bar e_a}\rangle = |Adj;[E^{\bar e_a},E^{-\bar
e_a}]\rangle = e^b_a|Adj;H^b\rangle\nolabel
\end{eqnarray}
where the sum over $b$ is understood.  This leads to our final result,
\begin{eqnarray}
[E^{\bar e_a},E^{-\bar e_a}] = e^b_a H^b
\label{eq:finalresultrootspace}
\end{eqnarray}

Though we did all of this using the adjoint representation we have
seen before, this structure is the same in any representation, and
therefore everything we have said is valid in any $D_n$.  We worked in
the adjoint simply because that makes the results easiest to obtain. 
The extensive use we made of labeling the eigenvectors with the
generators can only be done in the adjoint representation because only
in the adjoint does the number of eigenvectors equal the number of
eigenstates. However, this will not be a problem.  The important
results from this section are (\ref{eq:cartannoncartancomrel}) and
(\ref{eq:finalresultrootspace}), which are true in any representation. 
Part of what we will do later is find these structures in other
representations.  

The importance of the ideas in this section cannot be stressed enough. 
However, the material is somewhat abstract.  So, we consider a few
examples of how all this works.  

\subsubsection{Adjoint Representation of $SU(2)$}
\label{sec:adjointsu2}

We now illustrate what we did in section \ref{sec:rootspace} with
$SU(2)$.  We will work in the adjoint representation to make the
correspondence with section \ref{sec:rootspace} as transparent as
possible.  

$SU(2)$ has 3 generators, and therefore the adjoint representation
will consist of $3\times 3$ matrices.  This is simply the $j=1$
representation, which we wrote out in equation (\ref{eq:su2adjoint}).

First, it is easy to verify that (\ref{eq:killing}) and
(\ref{eq:modkilling}) hold for $\kappa = 2$.  

Next we look at the eigenstates.  We know they will be the normal
vectors
\begin{eqnarray}
v_1 = \begin{pmatrix} 1 \\ 0 \\ 0 \end{pmatrix}, \qquad v_2 =
\begin{pmatrix} 0 \\ 1 \\ 0 \end{pmatrix}, \qquad v_3 = \begin{pmatrix}
0 \\ 0 \\ 1 \end{pmatrix}\nolabel
\end{eqnarray}
(we will relabel them to be consistent with section \ref{sec:rootspace}
shortly).  

Obviously only $J^3_1$ is diagonal, so $SU(2)$ has rank $M =1$.  We
define

\begin{eqnarray}
H^1 = J^3_1 &=& \begin{pmatrix}
1 & 0 & 0 \\ 0 & 0 & 0 \\ 0 & 0 & -1 \end{pmatrix}\nolabel \\
E^1 = J^1_1 = {1\over 2}
\begin{pmatrix}
0 & 1 & 0 \\ 1 & 0 & 1 \\ 0 & 1 & 0 
\end{pmatrix}, &\quad& E^2 = J^2_1 = {1\over2}
\begin{pmatrix}
0 & -i & 0 \\ i & 0 & -i \\ 0 & i & 0
\end{pmatrix}\nolabel
\end{eqnarray}

Because the rank is 1, the root vectors will be 1-dimensional
vectors, or scalars.  We find them easily by finding the eigenvalues of
each eigenvector with $H^1$:
\begin{eqnarray}
H^1v_1 = (+1)v_1, \quad H^1v_2 = (0)v_2,\quad H^1v_3 = (-1)v_3\nolabel
\end{eqnarray}
So the root vectors are
\begin{eqnarray}
\bar t_1 = t_1 = +1 \qquad \bar t_2 = t_2 = 0 \qquad \bar t_3 = t_3 =
-1 \label{eq:rootvectorsare}
\end{eqnarray}
We can graph these on the real line as shown below,
\begin{center}
\includegraphics[scale = .7]{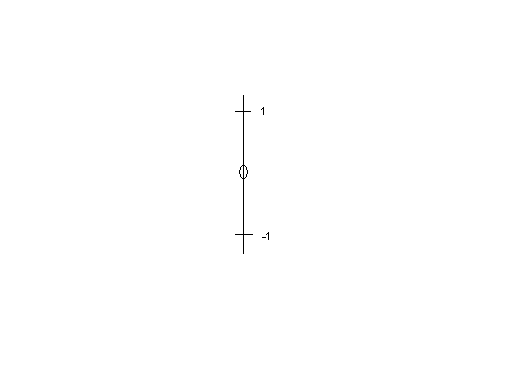} \label{su2picture}
\end{center}

Now our initial guess will be to associate $v_3$ with $J^3_1 = H^1$,
and then $v_1 = E^1$ and $v_2 = E^2$.  But we want to exploit what we
learned in section \ref{sec:rootspace}, and therefore we must make sure
that (\ref{eq:cartannoncartancomrel}) and
(\ref{eq:finalresultrootspace}) hold.  

Starting with (\ref{eq:cartannoncartancomrel}), we check (leaving the
tedious matrix multiplication up to you)
\begin{eqnarray}
& &\;[H^1,E^1] = \cdots = {1\over 2}
\begin{pmatrix}
0 & 1 & 0 \\ -1 & 0 & 1 \\ 0 & -1 & 0
\end{pmatrix} \label{eq:h1e1} \\
& &\;[H^1,E^2] = \cdots = -{i\over 2}
\begin{pmatrix}
0 & 1 & 0 \\ 1 & 0 & 1 \\ 0 & 1 & 0
\end{pmatrix} \label{eq:h1e2}
\end{eqnarray}

But we have a problem.	According to (\ref{eq:cartannoncartancomrel}),
$[H^1,E^i]$ should be proportional to $E^i$, but this is not the case
here.  However notice that in (\ref{eq:h1e1}),
\begin{eqnarray}
{1\over 2}\begin{pmatrix}
0 & 1 & 0 \\ -1 & 0 & 1 \\ 0 & -1 & 0
\end{pmatrix} = iE^2\nolabel
\end{eqnarray}
and in (\ref{eq:h1e2}),
\begin{eqnarray}
-{i\over 2}
\begin{pmatrix}
0 & 1 & 0 \\ 1 & 0 & 1 \\ 0 & 1 & 0
\end{pmatrix} = -i E^1\nolabel
\end{eqnarray}
Writing this more suggestively,
\begin{eqnarray}
[H^1,E^1] = iE^2, \quad [H^1,iE^2] = E^1 \label{eq:handie}
\end{eqnarray}
So, if we take the linear combinations of equations (\ref{eq:handie}),
we get $[H^1,\alpha E^1\pm \beta iE^2] = \beta E^1 \pm \alpha i E^2$,
which has the correct form of equation (\ref{eq:cartannoncartancomrel})
as long as $\alpha = \beta$.  Therefore we are now working with the
operators $E^{\pm} \equiv \alpha (E^1 \pm iE^2)$.  

Now we seek to impose (\ref{eq:finalresultrootspace}).	We start by
evaluating
\begin{eqnarray}
[E^+,E^-] &=& \alpha^2[E^1+iE^2,E^1-iE^2] \nolabel \\
&=& \alpha^2\big([E^1,E^1]-i[E^1,E^2]+i[E^2,E^1]+[E^2,E^2]\big)
\nolabel \\
&=& -2i\alpha^2[E^1,E^2] = \cdots \nolabel \\
&=& -2i\alpha^2iH^1 = 2\alpha^2H^1\nolabel
\end{eqnarray}
Then, from equations (\ref{eq:rootvectorsare}) and the definition of
$E^{\pm}$, we see that $\pm e^1_1 = \pm (t_1-t_2) = \pm(1-0) = \pm 1$. 
So we therefore set $\alpha^2 = {1\over 2} \Rightarrow \alpha = {1\over
\sqrt{2}}$, and we find that the appropriate non-Cartan generators
(including the 1 to be consistent with the notation in section
\ref{sec:rootspace}) are
\begin{eqnarray}
E^{\pm 1} = {1\over \sqrt{2}} (E^1\pm iE^2) \label{eq:linearcombose}
\end{eqnarray}
which is exactly what we had in equation (\ref{eq:jpjm}) above.  So, we
have derived the trick used to understand quantum mechanical spin in
introductory quantum courses!  

\subsubsection{$SU(2)$ for Arbitrary $j$ $\ldots$ Again}

Now that we have our operators in the adjoint representation, we can
consider any arbitrary representation. As we saw in section
\ref{sec:arbitraryj}, we can form the linear combinations in equation
(\ref{eq:linearcombose}) for any $j=$ integer or half integer.	The
weight vectors will always look like those in the diagram on page
\pageref{su2picture} (in other words, raising and lowering operators
always raise or lower their eigenvalue by 1).  
\newpage
The space of physical states, on the other hand, changes for each
representation.  For $j={1\over 2}$, we have 
\begin{center}
\includegraphics[scale = .7]{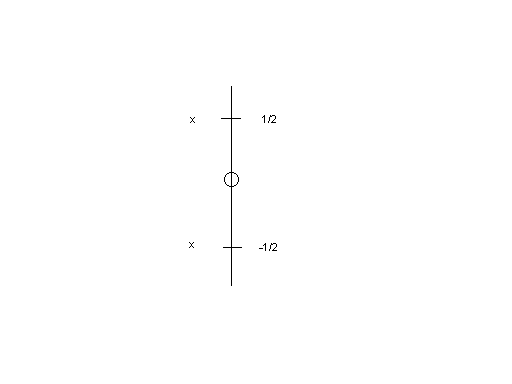}
\end{center}
For $j=1$,
\begin{center}
\includegraphics[scale = .7]{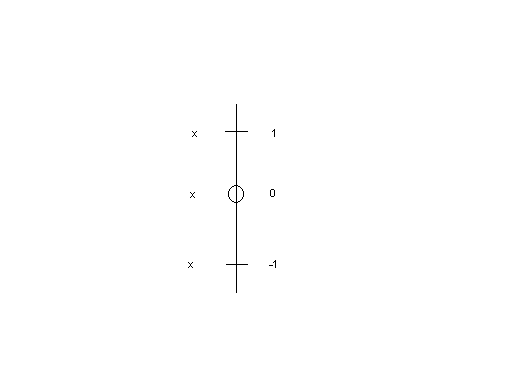}
\end{center}
\newpage
For $j={3/2}$,
\begin{center}
\includegraphics[scale = .7]{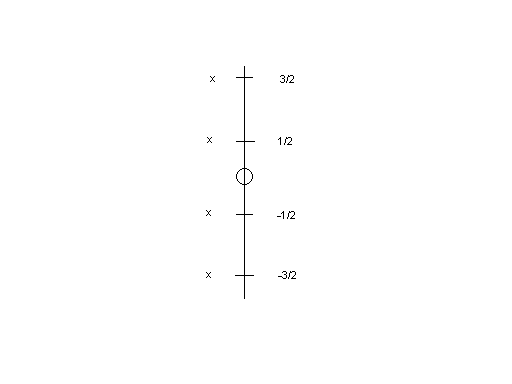}
\end{center}
and so on.  

Notice that the vectors graphed in the diagram on page
\pageref{su2picture} are the exact vectors required to move from point
to point in each of these graphs.  This is obviously not a coincidence. 

\subsubsection{$SU(3)$}
\label{sec:su3}

Now that we have said pretty much everything we can about $SU(2)$,
which is only Rank 1 (and therefore not all that interesting), we
move on to $SU(3)$.  However, we will expedite the process by stating
the structure constants up front. The non-zero structure constants are 
\begin{eqnarray}
f_{123}=1,\quad f_{147} = f_{165} = f_{246} = f_{257} = f_{345} =
f_{376} = {1\over 2}, \quad f_{458} = f_{678} = {\sqrt{3}\over
2}\nolabel
\end{eqnarray}

The most convenient representation is the \bf Fundamental Representation \rm
(consisting of $3\times 3$ matrices).  They are $T^a = {1\over 2}
\lambda^a$ for $a=1,\ldots,8$, where 
\begin{eqnarray}
\lambda^1 = 
\begin{pmatrix}
0 & 1 & 0 \\ 1 & 0 & 0 \\ 0 & 0 & 0
\end{pmatrix}, \quad \lambda^2 = 
\begin{pmatrix}
0 & -i & 0 \\ i & 0 & 0 \\ 0 & 0 & 0
\end{pmatrix}, \quad \lambda^3 = 
\begin{pmatrix}
1 & 0 & 0 \\ 0 & -1 & 0 \\ 0 & 0 & 0
\end{pmatrix}, \quad \lambda^4 = 
\begin{pmatrix}
0 & 0 & 1 \\ 0 & 0 & 0 \\ 1 & 0 & 0
\end{pmatrix} \nolabel	\\
\lambda^5 = 
\begin{pmatrix}
0 & 0 & -i \\ 0 & 0 & 0 \\ i & 0 & 0 
\end{pmatrix}, \quad \lambda^6 = 
\begin{pmatrix}
0 & 0 & 0 \\ 0 & 0 & 1 \\ 0 & 1 & 0
\end{pmatrix}, \quad \lambda^7 = 
\begin{pmatrix}
0 & 0 & 0 \\ 0 & 0 & -i \\ 0 & i & 0
\end{pmatrix}, \quad \lambda^8 = {1\over \sqrt{3}}
\begin{pmatrix}
1 & 0 & 0 \\ 0 & 1 & 0 \\ 0 & 0 & -2
\end{pmatrix} \nolabel \\
\label{eq:gellman}
\end{eqnarray}
Clearly, only two of these are diagonal, $\lambda^3$ and $\lambda^8$. 
So, $SU(3)$ is a rank 2 group.  

Before moving on, we summarize a few results (without proofs).	An
arbitrary $SU(n)$ group will always have $n^2-1$ generators, and will
be rank $n-1$.	An arbitrary $SO(n)$  group (for $n$ even) will always
have ${n(n-1)\over 2}$ generators.  We won't worry about the rank of the
orthogonal groups.  

Working in the adjoint representation of $SU(3)$ would involve $8
\times 8$ matrices, which would obviously be very tedious.  So, we
exploit the fact that the techniques we developed in section
\ref{sec:rootspace} are valid in any representation, and stick with the
Fundamental Representation defined by the generators in
(\ref{eq:gellman}).  

Proceeding as in section \ref{sec:adjointsu2}, we note that the
eigenvectors will again be 
\begin{eqnarray}
v_1 = 
\begin{pmatrix}
1 \\ 0 \\ 0 
\end{pmatrix}, \quad v_2 = 
\begin{pmatrix}
0 \\ 1 \\ 0
\end{pmatrix}, \quad v_3 = 
\begin{pmatrix}
0 \\ 0 \\ 1
\end{pmatrix}\nolabel
\end{eqnarray}
(we will relabel them to be consistent with section \ref{sec:rootspace}
shortly).  

Then, the Cartan generators are 
\begin{eqnarray}
H^1 = {1\over 2}
\begin{pmatrix}
1 & 0 & 0 \\ 0 & -1 & 0 \\ 0 & 0 & 0
\end{pmatrix}, \qquad H^2 = {1\over 2 \sqrt{3}}
\begin{pmatrix}
1 & 0 & 0 \\ 0 & 1 & 0 \\ 0 & 0 & -2
\end{pmatrix}\nolabel
\end{eqnarray}
and the non-Cartan Generators are simply 
\begin{eqnarray}
E^1 = T^1, \quad E^2 = T^2, \quad E^3 = T^4, \quad E^4 = T^5, \quad E^5
= T^6, \quad E^6 = T^7\nolabel
\end{eqnarray}

So we have 6 eigenvalues to find,
\begin{eqnarray}
H^1v_1 &=& \bigg({1\over 2}\bigg)v_1,\qquad H^1v_2 = \bigg(-{1\over
2}\bigg)v_2, \quad H^1v_3 = (0)v_3 \nolabel \\
H^2v_1 &=& \bigg({1\over 2\sqrt{3}}\bigg)v_1, \quad H^2v_2 =
\bigg({1\over 2\sqrt{3}}\bigg)v_2,\quad H^2 v_3 = \bigg(-{1\over
\sqrt{3}}\bigg)v_3\nolabel
\end{eqnarray}

So the weight vectors will be 2-dimensional (because the rank is
2).  They are
\begin{eqnarray}
\bar t^1 = 
\begin{pmatrix}
{1\over 2} & {1\over 2\sqrt{3}}
\end{pmatrix}^T, \quad \bar t^2 = 
\begin{pmatrix}
-{1\over 2} & {1\over 2\sqrt{3}}
\end{pmatrix}^T, \quad \bar t^3 = 
\begin{pmatrix}
0 & -{1\over \sqrt{3}}
\end{pmatrix}^T \label{eq:su3initial}
\end{eqnarray}
We can graph these in $\mathbb{R}^2$ as shown below,
\begin{center}
\includegraphics[scale = .7]{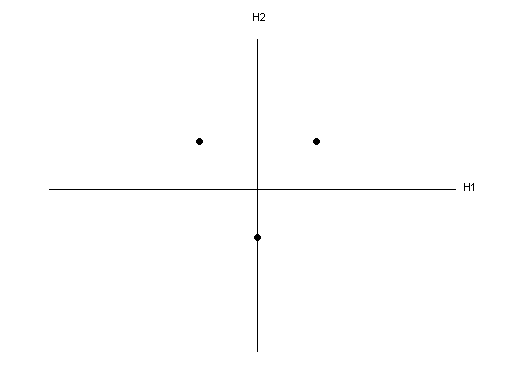} \label{su3fundy}
\end{center}

Now, repeating nearly the identical argument we started with equation
(\ref{eq:h1e1}) and repeating it for all 6 non-Cartan generators, we
find that in order to maintain (\ref{eq:cartannoncartancomrel}) and
(\ref{eq:finalresultrootspace}), we must work with the operators
\begin{eqnarray}
{1\over \sqrt{2}}(T^1\pm iT^2) = {1\over \sqrt{2}}(E^1\pm i E^2)
\nolabel \\
{1\over \sqrt{2}}(T^4 \pm i T^5) = {1\over \sqrt{2}}(E^3 \pm i E^4)
\nolabel \\
{1\over \sqrt{2}}(T^6\pm iT^7) = {1\over \sqrt{2}}(E^5 \pm i E^6)
\label{eq:gluons}
\end{eqnarray}
The weight vectors associated with these will be, respectively,
\begin{eqnarray}
\pm (\bar t_1 - \bar t_2) = \pm 
\begin{pmatrix}
1 \\ 0
\end{pmatrix}, \quad \pm (\bar t_1 - \bar t_3) = \pm 
\begin{pmatrix}
1/2 \\ \sqrt{3}/2
\end{pmatrix}, \quad \mbox{and} \quad \pm (\bar t_2 - \bar t_3) = \pm
\begin{pmatrix}
-1/2 \\ \sqrt{3}/2
\end{pmatrix} \nolabel\\ \label{eq:su3noncartan}
\end{eqnarray}
So, the non-Cartan generators are 
\begin{eqnarray}
E^{\pm
\begin{pmatrix}
1 \\ 0
\end{pmatrix}}, \qquad E^{\pm
\begin{pmatrix}
1/2 \\ \sqrt{3}/2
\end{pmatrix}}, \qquad E^{\pm
\begin{pmatrix}
-1/2 \\ \sqrt{3}/2
\end{pmatrix}}\nolabel
\end{eqnarray}

We are no longer in the adjoint representation, so we had to be more
deliberate about choosing these linear combinations than we could be in
section \ref{sec:adjointsu2}.  What we did here is more general; we
chose them to be the differences in the three weight vectors in
equation (\ref{eq:su3initial}), \it so that \rm these vectors would
naturally transform from one eigenvector to another (just as the
raising and lowering operators do, as we found for $SU(2)$ and more
generally in section \ref{sec:rootspace}).  The remarkable property of
Lie groups is that this is always possible in any representation.  

We can graph the 6 vectors in (\ref{eq:su3noncartan}), along with the
two Cartan weight vectors, which we know from (\ref{eq:heqzero}) are 0:
\begin{center}
\includegraphics[scale = .7]{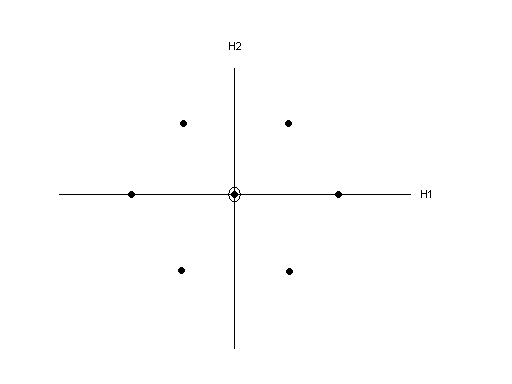}
\end{center}

And again, just as with $SU(2)$, notice that the 6 non-zero vectors
are the exact vectors that would be necessary to move from point to
point on the diagram on page \pageref{su3fundy}.  So once again, we see
that the non-Cartan generators act as raising and lowering operators
which transform between the eigenstates of the Cartan generators. 
Notice that there were 6 non-Cartan generators, and they formed
linear combinations to form 6 raising and lowering operators.  

\subsubsection{What is the Point of All of This?}
\label{sec:whatpoint}

Before finally getting back to physics, we give a spoiler of how Lie
theory is used in physics.  What we are going to find is that some
physical interaction (electromagnetism, weak force, strong force) will
ultimately be described by a Lie group in some particular
representation.  The particles that interact with that force will be
described by the eigenvectors of the Cartan generators of the group,
and the eigenvalues of those eigenvectors will be the physically
measurable charges.  Clearly, the number of charges associated with the
interaction is equal to the number of dimensions of the representation.
For example, you likely are aware that the strong force has 3
charges, called ``colors" (red, green, and blue).  So, the strong force
(we will see) will be in a 3-dimensional representation of the group
that describes it.  

We will find that all forces carrying particles (photons, gluons, $W$
and $Z$ bosons) will be described by the generators of their respective
Lie group.  The Cartan generators will be force-carrying particles
which can interact with any particle charged under that group by
transferring energy and momentum, but do not change the charge (photons
and $Z$ bosons).  This makes sense because Cartan generators are \it
not \rm raising or lowering operators.	On the other hand, the non-Cartan
generators will be force carrying particles which interact with
any particle charged under that group by not only transferring energy
and momentum, but also changing the charge ($W$ bosons and gluons).  

We won't be able to come back to discussing how this works until much
later, and until examples are worked out, this may not be clear.  We
merely wanted to give an idea of where we are going with this.	

\subsection{References and Further Reading}

The material in section \ref{sec:groupintro} came primarily from
\cite{Georgi} and \cite{Sagan}.  The material in \ref{sec:lieintro}
came from \cite{Georgi}, \cite{Gilmore}, and \cite{Hall}.  The sections
on $SU(2)$ also came from \cite{Sakurai}.  

For further reading, we recommend \cite{Cahn}, \cite{Fraleigh},
\cite{Humphreys}, \cite{Hungerford}, and \cite{Rotman}.  
\newpage
\section{Part III --- Quantum Field Theory}

\subsection{A Primer to Quantization}
\label{sec:primquant}

\subsubsection{Quantum Fields}
\label{sec:test}

Our ultimate goal in the exposition that follows is to formulate a \it
relativistic quantum mechanical \rm theory of \it interactions\rm.  So,
beginning with the fundamental equation of quantum mechanics,
Schroedinger's equation,
\begin{eqnarray}
H\Psi = i\hbar {\partial \Psi \over \partial t}\nolabel
\end{eqnarray}
we know that for a non-interacting, non-relativistic particle, $H =
{\bar p^2 \over 2m} = -{\hbar \over 2m}\bar \nabla^2$, so 
\begin{eqnarray}
-{\hbar \over 2m}\bar \nabla^2 \Psi = i\hbar {\partial \Psi \over
\partial t} \label{eq:nonrelschroed}
\end{eqnarray}

Of course, $\Psi$ is in this case a \bf Scalar Field \rm, and therefore only
has one state.	So, it describes a spin-0 particle (or, in the
language we have learned in the previous sections, it sits in a $j=0$
representation of $SU(2)$, which is the trivial representation).  And,
since $\Psi$ does not have any spacetime indices, it also transforms
trivially under the Lorentz group $SO(1,3)$.  

Notice, however, that we have a fundamental barrier in making a
relativistic theory - the spatial derivative in
(\ref{eq:nonrelschroed}) acts quadratically ($\bar \nabla^2$), whereas
the time derivative is linear.	Clearly, treating space and time
differently in this way is unacceptable for a relativistic theory. 
That is a hint of a much more fundamental problem with quantum
mechanics; space is always treated as an operator, but time is always
treated as a parameter.  This fundamental asymmetry is what ultimately
prevents a straightforward generalization to relativistic quantum
theory.  

To fix this problem, we have two choices: either promote time to an
operator along with space, or demote space back to a parameter and
quantize in a new way.	

The first option would result in the Hermitian operators $\hat X, \hat
Y, \hat Z$, and $\hat T$.  It turns out that this approach is very
difficult and less useful as far as building a relativistic quantum
theory.  So, we will take the second option.  

In demoting position to a parameter along with time, we obviously have
sacrificed the operators which we imposed commutation relations on to
get a ``quantum" theory in the first place.  And because we obviously
can't impose commutation relations on parameters (because they are
scalars), quantization appears impossible.  So, we are going to have to
make a fairly radical reinterpretation.  

Rather than letting the coordinates be Hermitian operators that act on
the state in the Hilbert space representing a particle, \it we now
interpret the particle as the Hermitian operator\rm, and \it this \rm
operator (or particle) will be parameterized by the spacetime
coordinates.  The physical state that the particle operators act on is
then the vacuum itself, $|0\rangle$.  So, whereas before you acted on
the ``electron" $|\Psi\rangle$ with the operator $\hat x$, now the
``electron" (parameterized by $x$) $\Psi(x^{\mu})$ acts on the vacuum
$|0\rangle$, creating the state $\big(\Psi(x^{\mu})|0\rangle\big)$.  In
other words, the operator representing an electron excites the vacuum
(empty space) resulting in an electron.  We will see that all quantum
fields contain appropriate raising and lowering operators to do just
this.  

This approach, where the quantum mechanical entities are no longer the
coordinates acting on the fields, but the fields themselves, is called
\bf Quantum Field Theory \rm (QFT).  

So, whereas before, quantization was defined by imposing commutation
relations on the coordinate operators $[x,p]\neq0$, we now quantize by
imposing commutation relations on the field operators, $[\Psi_1,\Psi_2]
\neq 0$.  

Because we must still write down the equations of motion which govern
the dynamics of the fields, we will need to spend the rest of this
section coming up with the classical equations governing the fields we
want to work with.  We will quantize them in the next section.

\subsubsection{Spin-0 Fields}
\label{sec:spin0fields}

As we said above, Schroedinger's equation (\ref{eq:nonrelschroed})
describes the time evolution of a spin-0 field, or a scalar field.
Generalizing to higher spins will come later.  Now, we see how
to make this description relativistic.	

The most obvious guess for a relativistic form is to simply plug in the
standard relativistic Hamiltonian
\begin{eqnarray}
H = \sqrt{\bar p^2c^2+m^2c^4} \label{eq:relativistichamiltonian}
\end{eqnarray}
Note that, using the standard Taylor expansion $\sqrt{1+x^2} \approx
1+{1\over 2}x$ for $x<<1$ gives $H \approx mc^2+{\bar p^2 \over 2m}$,
for $\bar p^2 <<c^2$, which is the standard non-relativistic form (plus
a constant) we'd expect from a low speed limit.  

Plugging (\ref{eq:relativistichamiltonian}) into
(\ref{eq:nonrelschroed}), we have
\begin{eqnarray}
i\hbar {\partial \phi \over \partial t} = \sqrt{-\hbar^2 c^2\bar
\nabla^2 + m^2c^4}\phi\nolabel
\end{eqnarray}
But there are two problems with this:
\begin{enumerate}
\item The space and time derivatives are still treated
differently, so this is inadequate as a relativistic equation, and
\item Taylor expanding the square root will give an infinite
number of derivatives acting on $\phi$, making this theory non-local.  
\end{enumerate}

One solution is to square the operator on both sides, giving
\begin{eqnarray}
-\hbar^2{\partial^2 \phi \over \partial t^2} &=& (-\hbar^2c^2\bar
\nabla^2 + m^2c^4)\phi \nolabel \\
&\Rightarrow& (-\partial^0\partial_0 + \bar \nabla^2 -{m^2c^2\over
\hbar^2})\phi = 0\nolabel
\end{eqnarray}
Or, if we choose the so called ``natural units" or ``God units", where
$c=\hbar = 1$, we have 
\begin{eqnarray}
(\partial^2  - m^2)\phi = 0 \label{eq:kleingordon}
\end{eqnarray}
Equation (\ref{eq:kleingordon}) is called the \bf Klein Gordon \rm
equation.  It is nothing more than an operator version of the standard
relativistic relation $E^2 = m^2c^4+\bar p^2c^2$.  

Note that because we will be quantizing fields and not coordinates,
there is absolutely nothing ``quantum" about the Klein Gordon equation.
 It is, at this point, merely a relativistic wave equation for a
classical, spinless, non-interacting field.  

Finally, we note one major problem with the Klein Gordon equation. 
When we squared the Hamiltonian $H=\sqrt{m^2c^4+\bar p^2c^2}$ to get
$H^2 = m^2c^4+\bar p^2c^2$, the energy eigenvalues became
$E=\pm\sqrt{m^2c^4+\bar p^2c^2}$.  It appears that we have a negative
energy	eigenvalue!  Obviously this is unacceptable in a physically
meaningful theory, because negative energy means that we don't have a
true vacuum, and therefore a particle can cascade down forever, giving
off an infinite amount of radiation.  

We will see that this problem plagues the spin-$1/2$ particles as well,
so we wait to talk about the solution until then.  

\subsubsection{Why $SU(2)$ for Spin?}

Because we are talking about particles ``spinning", a common question
is why don't we use $SO(3)$ instead of $SU(2)$?  The original answer to
the question is historical.  The experiments done in the early days of
quantum mechanics were not consistent with the particles having a
rotational degree of freedom in spacetime.  Rather, the data indicated
that, along any given axis, the spin could have only one of two
possible values, and $SO(3)$ does not explain this.  Here, however, we
consider a more mathematical explanation.  

First, recall that spin is a purely quantum mechanical phenomenon, with
no classical analogue.	Because the data demanded two possible spin
states, the field describing the particle had to have 2 spin
components, $\Psi = \begin{pmatrix} \Psi_1 \\ \Psi_2 \end{pmatrix}$. 
Now, if we seek a 2-dimensional representation of $SO(3)$, we find
that there is only one: $D_0\oplus D_0$, the trivial representation
consisting of all 1's.  This means 
\begin{eqnarray}
\begin{pmatrix}
\Psi'_1 \\ \Psi'_2
\end{pmatrix} = D_0\oplus D_0 
\begin{pmatrix}
\Psi_1 \\ \Psi_2
\end{pmatrix} = 
\begin{pmatrix}
1 & 0 \\ 0 & 1
\end{pmatrix}
\begin{pmatrix}
\Psi_1 \\ \Psi_2
\end{pmatrix} = 
\begin{pmatrix}
\Psi_1 \\ \Psi_2
\end{pmatrix}\nolabel
\end{eqnarray}
which is no transformation.  This is the only 2-dimensional
representation of $SO(3)$ that is possible.  

The solution to the problem is found in one of the many peculiarities
of quantum mechanics.  The only physically measurable quantity in
quantum theory is the probability amplitude, which is proportional to
the square of $\Psi$.  Therefore, the state $\begin{pmatrix} \Psi_1 \\
\Psi_2 \end{pmatrix}$ is physically identical to $\begin{pmatrix}
-\Psi_1 \\ - \Psi_2\end{pmatrix}$.  

Now consider a general element of $SO(3)$: $e^{i(\phi J_x+\psi J_y +
\theta J_z)}$.	On the other hand, a general element of $SU(2)$ will be
$e^{i(\phi {\sigma_x \over 2}+ \psi {\sigma_y \over 2} + \theta
{\sigma_z \over 2})}$. 

Now consider rotating the system by an angle of $2\pi$ around, say, the
$z$ axis.  The $SO(3)$ element corresponding to this rotation will be
$e^{i2\pi J_z}$, while the $SU(2)$ element will be $e^{i\pi \sigma_z}$.
 The factor of 1/2 difference means that the spinor space rotates
through only half the angle of the $SO(3)$ does.  So, in the $2\pi$
rotation, $U\in SU(2) \rightarrow -U$, whereas $R \in SO(3)\rightarrow
R$.  Therefore, both $U$ and $-U$ correspond to $R$.  There is a 2 to
1 correspondence between $SU(2)$ and $SO(3)$.  

And, as we said above, spin is a purely quantum mechanical effect and
experimentally only allows 2 values, but $SO(3)$ has no such
representation, whereas the $j=1/2$ representation of $SU(2)$ does.  We
therefore use $SU(2)$.	And, because $SU(2)$ is $2\rightarrow1$ with
$SO(3)$, but spin is quantum mechanical, both $U$ and $-U$ can
consistently correspond to the same $R\in SO(3)$.  The minus sign
difference is not subject to measurement; only $|\Psi|^2$ is physically
measurable.  

An important thing to understand is that ``spin'' is not a rotation
through spacetime in any meaningful way.  It is a rotation in ``spinor
space", which is an \it internal \rm degree of freedom.  Like many
things in quantum mechanics, spinor space is a mathematical structure. 
All we can say for certain is what we can measure, or know
($|\Psi|^2$), not what ``is".  

\subsubsection{Spin ${1\over 2}$ Particles}

Finding equation (\ref{eq:kleingordon}) was easy because scalar fields
have no spacetime indices and no spinor indices, and they therefore
transform trivially under $SU(2)$ and the Lorentz group.  

A particle of spin $1/2$ however, will have two complex components, one
for spin $+1/2$, and the other for spin $-1/2$.  So, we describe such a
particle as the two-component \bf Spinor \rm
\begin{eqnarray}
\psi = \begin{pmatrix}
\psi_1 \\ \psi_2
\end{pmatrix}\nolabel
\end{eqnarray}
where $\psi_1$ and $\psi_2$ are both $\in \mathbb{C}$.	So, we want
some differential operator in the form of $2\times 2$ matrices to act
on such a field to form the equation of motion.  

Following Dirac's approach, he reasoned that given such a $2\times 2$
operator, the equation of motion should somehow ``imply" the Klein
Gordon equation (which merely makes the theory relativistic).  So his
goal (and our goal) is to find an equation with a $2\times 2$ matrix
differential operator acting on $\psi$ that results in
(\ref{eq:kleingordon}).  

Dirac's approach was to find an operator of the form 
\begin{eqnarray}
\displaystyle{\not} D = \gamma^{\mu}\partial_{\mu} = \gamma^0\partial_0
+ \gamma^1\partial_1 + \gamma^2\partial_2 + \gamma^3 \partial_3\nolabel
\end{eqnarray}
where the $\gamma$'s are $2\times 2$ matrices, and the equation of
motion is then $\displaystyle{\not} D\psi = -im\psi$.  The challenge is
in finding the appropriate $2\times 2$ $\gamma$ matrices.  Dirac
reasoned that, in order to be properly relativistic, operating twice
with $\displaystyle{\not} D$ should give the Klein Gordon equation.  In
other words, 
\begin{eqnarray}
\displaystyle{\not}D = -im\psi &\Rightarrow& \displaystyle{\not} D
\displaystyle{\not} D \psi = -im \displaystyle{\not} D \psi \nolabel \\
&\Rightarrow& \gamma^{\mu}\partial_{\mu}\gamma^{\nu}\partial_{\nu}\psi
= -im(-im\psi) \nolabel \\
&\Rightarrow& \gamma^{\mu}\gamma^{\nu}\partial_{\mu}\partial_{\nu} \psi
= -m^2\psi \nolabel \\
&\Rightarrow&
\big(\gamma^{\mu}\gamma^{\nu}\partial_{\mu}\partial_{\nu}+m^2\big) \psi
= 0 \label{eq:sumwithsymmetry}
\end{eqnarray}

This will yield the Klein Gordon equation if $\gamma^{\mu}\gamma^{\nu}
= -\eta^{\mu \nu} \mathbb{I}$.	Or, using the symmetry of the sum in
(\ref{eq:sumwithsymmetry}), it will yield the Klein Gordon equation if we demand ${1\over
2}(\gamma^{\mu}\gamma^{\nu}+ \gamma^{\nu}\gamma^{\mu}) = -\eta^{\mu
\nu} \mathbb{I}$.  Consider
\begin{eqnarray}
\{\gamma^{\mu}, \gamma^{\nu}\} =
\gamma^{\mu}\gamma^{\nu}+\gamma^{\nu}\gamma^{\mu} = -2\eta^{\mu \nu}
\mathbb{I} \label{eq:clifford}
\end{eqnarray}
If the $\gamma$ matrices satisfy (\ref{eq:clifford}), then
(\ref{eq:sumwithsymmetry}) gives
\begin{eqnarray}
(\gamma^{\mu}\gamma^{\nu}\partial_{\mu}\partial_{\nu} + m^2)\psi = 0
\Rightarrow (-\eta^{\mu \nu} \partial_{\mu} \partial_{\nu} + m^2) \psi
= 0 \Rightarrow (\partial^2 - m^2)\psi = 0\nolabel
\end{eqnarray}
which is exactly the Klein Gordon equation (\ref{eq:kleingordon}).  

So, we have the Dirac equation
\begin{eqnarray}
\big(\displaystyle{\not} D + im \big) \psi = 0 \label{eq:dirac}
\end{eqnarray}
but we still have a problem.  Namely, there does not exist a set of
$2\times 2$ matrices that solve (\ref{eq:clifford}).  Nor does there
exist a set of $3\times 3$ matrices.  The smallest possible size where
this is possible is $4\times 4$.  Obviously, if we want to describe a
spin-$1/2$ particle with exactly 2 spin states, using 4 spin
components does not seem right.  But, we will accept the necessity of
$4\times 4$  Dirac matrices and move on.  

Instead of using $\psi = 
\begin{pmatrix}
\psi_1 \\ \psi_2
\end{pmatrix}$, we will define the two 2-dimensional spinors
\begin{eqnarray}
\psi_L \equiv 
\begin{pmatrix}
\psi_1 \\ \psi_2
\end{pmatrix} \qquad \mbox{and} \qquad \psi_R \equiv
\begin{pmatrix}
\psi_3 \\ \psi_4
\end{pmatrix} \label{eq:leftright}
\end{eqnarray} 
and the 4-component spinor
\begin{eqnarray}
\psi \equiv \begin{pmatrix} \psi_L \\ \psi_R \end{pmatrix}
\label{eq:diracspinor}
\end{eqnarray}

Now it is possible to solve (\ref{eq:clifford}).  Such a problem is
actually very familiar to algebraists, and we will not delve into the
details of how this is done.  Instead, we merely state one solution
(there are many, up to a similarity transformation).  We define the
4 $\times$ 4 matrices
\begin{eqnarray}
\gamma^i = 
\begin{pmatrix}
0 & -\sigma^i \\ \sigma^i & 0
\end{pmatrix} \qquad \mbox{and} \qquad \gamma^0 = 
\begin{pmatrix}
0 & \sigma^0 \\ \sigma^0 & 0
\end{pmatrix} \label{eq:gammamat}
\end{eqnarray}
where $\sigma^0$ is the 2 $\times$ 2 identity matrix, and $\sigma^i$ are
the Pauli spin matrices.  It should be no surprise that they show up in
attempting to describe spin-$1/2$ particles.  What is interesting is
that we did not assume them---we derived them using
(\ref{eq:clifford}).  

Before moving on, notice that we have initiated a convention that will
be used throughout the rest of these notes.  Whenever a greek index is
used, it runs over all spacetime indices.  Whenever a latin index is
used, it runs over only the spatial part.  So in (\ref{eq:gammamat}),
$i$ runs $1,2,3$.  

Now that we have an explicit form of the Dirac gamma matrices, we can
write out (\ref{eq:dirac}) explicitly:
\begin{eqnarray}
\begin{pmatrix}
0 & 0 & \partial_0 - \partial_3 & -\partial_1+i\partial_2 \\
0 & 0 & -\partial_1 - i\partial_2 & \partial_0 + \partial_3 \\
\partial_0 + \partial_3 & \partial_1 - i\partial_2 & 0 & 0 \\
\partial_1 + i\partial_2 & \partial_0 - \partial_3 & 0 & 0 
\end{pmatrix}
\begin{pmatrix}
\psi_1 \\ \psi_2 \\ \psi_3 \\ \psi_4
\end{pmatrix} = -im
\begin{pmatrix}
\psi_1 \\ \psi_2 \\ \psi_3 \\ \psi_4
\end{pmatrix}\nolabel
\end{eqnarray}
Or, in terms of $\psi_L$ and $\psi_R$, 
\begin{eqnarray}
i \bar \sigma^{\mu}\partial_{\mu} \psi_R= + m\psi_L \nolabel\\
i \sigma^{\mu}\partial_{\mu}\psi_L = + m\psi_R\nolabel
\end{eqnarray}
where we have defined the 4-vectors $\sigma^{\mu} =
(\sigma^0,\sigma^1\sigma^2,\sigma^3)$ and $\bar \sigma^{\mu} =
(\sigma^0, -\sigma^1, -\sigma^2,-\sigma^3)$.

\subsubsection{The Lorentz Group}

This section is intended to give a deeper understanding of why we were
unable to find a 2 $\times$ 2 matrix representation to solve
(\ref{eq:clifford}).  

Recall that the driving idea behind the derivation of the Dirac
equation (\ref{eq:dirac}) was to make it imply the Klein Gordon
equation, or in other words to be a relativistic theory.  Put another
way, it was to create a theory that was invariant under the Lorentz
group $SO(1,3)$.  So, let's take a closer look at the Lorentz group.  

We know from section \ref{sec:lorentzdetail} that the Lorentz group
consists of 3 rotations and 3 boosts.  We gave the general forms of
these transformations in equations (\ref{eq:lorrot}) and
(\ref{eq:lorboost}).  It is easy, using those general expressions in
addition to (\ref{eq:generators}), to find all 6 generators, and then
multiply them out to get the commutation relations.  We spare the (easy
but tedious) details and simply state the commutation relations.  If we
label the generators of rotation $J^i$ ($i=1,2,3$), and the generators
of boosts $K^i$ ($i=1,2,3$), then the commutation relations are 
\begin{eqnarray}
\;[J^i,J^j] &=& i\epsilon^{ijk}J^k \nolabel\\
\;[J^i,K^j] &=& i\epsilon^{ijk}K^k \nolabel\\
\;[K^i,K^j] &=& -i\epsilon^{ijk}J^k\nolabel
\end{eqnarray}

In order to make the actual structure of this group more obvious, we
define two new linear combinations of these generators:
\begin{eqnarray}
N^i = {1\over 2}(J^i - i K^i) \qquad N^{i\dagger} = {1\over
2}(J^i+iK^i)\nolabel
\end{eqnarray}
Writing out the commutation relations for $N^i$ and $N^{i\dagger}$, we
get
\begin{eqnarray}
\;[N^i,N^j] &=& i\epsilon^{ijk}N^k \nolabel\\
\;[N^{i\dagger},N^{j\dagger}] &=& i\epsilon^{ijk}N^{k\dagger}
\nolabel\\
\;[N^i,N^{j\dagger}] &=& 0\nolabel
\end{eqnarray}

So, both $N^i$ and $N^{i\dagger}$ separately form an $SU(2)$.  In more
mathematical terms, we say that $SO(1,3)$ is \bf Isomorphic \rm to
$SU(2)\otimes SU(2)$, which we denote $SO(1,3)\cong SU(2)\otimes
SU(2)$.  While the idea of an isomorphism is a very rich mathematical
idea, for now you can simply think of it as a way of saying that two
groups have the same group structure.  

So, because a given representation of $SU(2)$ is defined by the value
of $j$, we can see that a particular representation of the Lorentz
group $SO(1,3)\cong SU(2)\otimes SU(2)$ is defined by \it two \rm
values of $j$, or by the doublet $(j,j')$.  The smallest possible
representation then is $(j,j') = (0,0)$.  This has one state from $j=0$
and one state from $j'=0$, and therefore has 1~$\times$~1 = 1 state
total.	Therefore, this representation describes a scalar field.  

Then, there is the state $(0,1/2)$, which will have one state from
$j=0$, but two states from $j=1/2$, for a total of $1\times 2 = 2$
states.  Therefore, this describes a single spin-$1/2$ field.  We call
this field $\psi_L$, and the $(0,1/2)$ representation the \bf Left-Handed
Spinor Representation \rm of the Lorentz group.	

Clearly, we will also have the representation $(1/2,0)$, which also has
2 states, corresponding to the $\psi_R$ field.  This is called the \bf
Right-Handed Spinor Representation \rm of the Lorentz group.  This is
the reason for the notation used in (\ref{eq:leftright}).  The left-handed
$(0,1/2)$ representation acts on $\psi_L$ and the right-handed
$(1/2,0)$ representation acts on $\psi_R$.  

Next is the representation $(1/2,1/2)$, which has two states from
$j=1/2$ and two from the $j'=1/2$ for a total of $2\times 2 = 4$
states.  It turns out that this representation is the spacetime vector
representation we use to act on spacetime vectors for the standard
Lorentz transformations discussed in section \ref{sec:lorentzdetail}.  

Now, an $SU(2)$ representation specified by some $j$ is an irreducible
representation, and therefore the tensor products $SU(2)\otimes SU(2)$
specified by the doublet $(j,j')$ are irreducible.  This means that
there are no irreducible subspaces, and so given a representation
$(j,j')$, there is a particular transformation taking the state
$(j,j')$ to $(j',j)$.  For the $(0,0)$ and $(1/2,1/2)$ representations
this doesn't affect anything.  However, this fact means that the
$(0,1/2)$ and $(1/2,0)$ representations must always appear together. 
To put this in more mathematical language, our choices for
representations of the Lorentz group are 
\begin{eqnarray}
(0,0), \qquad (1/2,1/2) \qquad \mbox{and} \qquad (1/2,0)\oplus (0,1/2)\nolabel
\end{eqnarray}
which are 1, 4, and 4-dimensional representations, respectively. 
Furthermore, they are the representations which transform Klein Gordon
scalar/spinor-0 fields, spacetime 4-vectors, and  spin-$1/2$
spinors, respectively.	

The physical meaning of this fact is that relativity demands that if
you want a theory with spin-$1/2$ particles, you cannot have them
existing by themselves.  They must come in pairs, each transforming
under an $SU(2)$ representation of opposite handedness.  In the next
two sections we will discuss ways of interpreting this fact, starting
with Dirac's original approach which, while brilliant, didn't
ultimately work.  Then we will consider what appears to be the correct
view.  

\subsubsection{The Dirac Sea Interpretation of Antiparticles}

Initially, it may seem that the impossibility of finding a 2~$\times$~2
matrix solution to (\ref{eq:clifford}) means that we can't have fields
with 2 spinor states.  However, we saw in the last section that we
aren't limited to scalars and spacetime 4-component spinors.	We can
also have two fields, $\psi_L$ and $\psi_R$, which can be paired
together to form two spin-$1/2$ fields in a 4-component spinor $\psi
= \begin{pmatrix}
\psi_L \\ \psi_R
\end{pmatrix}$.  So, Dirac was faced with the challenge of both
interpreting this, while at the same time dealing with the negative
energy states mentioned in section \ref{sec:spin0fields}.  

Dirac's solution, though today abandoned, was brilliant enough to
mention.  He suggested that because spin-$1/2$ particles obey the \bf Pauli
Exclusion Principle\rm, there could be an infinite number of particles \it
already \rm in the negative energy levels, and so they are already
occupied, preventing any more particles from falling down and giving
off infinite energy.  Thus, the negative energy problem was solved.  

Furthermore, he said that it is possible for one of the particles in
this infinite negative sea to be excited and jump up into a positive
energy state, leaving behind a hole.  This would appear to us,
experimentally, as a particle with the same mass, but the opposite
charge.  He called such particles \bf Antiparticles\rm.  For example,
the antiparticle of the electron is the antielectron, or the positron
(same mass, opposite charge).  The positron is not a particle in the
same sense as the electron, but rather is a hole in an infinite sea of
electrons.  And where this negative charge is missing, all that is left
is a hole which appears as a positively charged particle.  

So, $\psi_L$ describes a particle, and due to the infinite sea of
negative particles, there can always be a hole, which will be described
by $\psi_L$.  Everything about this worked out mathematically, and when
antiparticles were detected about 5 years after Dirac's prediction
of them, it appeared that Dirac's suggestion was correct.  

However, there were two major problems with Dirac's idea, and they
ultimately proved fatal to the ``Dirac Sea" interpretation:
\begin{enumerate}
\item This theory, which was supposed to be a theory of single
particles, now requires an infinite number of them.
\item Particles like photons, pions, mesons, or Klein-Gordon
scalars don't obey the Pauli Exclusion Principle, but still have
negative energy states, and therefore Dirac's argument doesn't work.  
\end{enumerate}

However, his labeling them ``antiparticles" has stuck, and we therefore
still refer to the right-handed part of the spin-$1/2$ field as the
antiparticle, whereas the left-handed part is still the particle.  

For these reasons, we must have some other way of understanding the
existence of the antiparticles.  

\subsubsection{The QFT Interpretation of Antiparticles}

In presenting the problem of negative energy states, we have been
somewhat intentionally sloppy.	To take stock, we have two equations of
motion: the Klein Gordon (\ref{eq:kleingordon}) for scalar/spin-0
fields, and the Dirac equation (\ref{eq:dirac}) for spin-$1/2$
particles.  

And in our discussion of negative energy states, we were ``pretending"
that the $\psi$'s and $\phi$'s are ``states" with negative energy. 
But, as we said in section \ref{sec:test}, QFT offers a different
interpretation of the fields.  Namely, the fields are not states --- they
are operators.	And consequently they can't have energy.  A state is
\it made \rm by acting on the vacuum with either of the \it operators
\rm $\phi$ or $\psi$, and then the state $\phi|0\rangle$ or
$\psi|0\rangle$ has some energy.  

So, QFT allows us to see the antiparticle as a real, actual particle,
rather than the absence of a particle.	And, we do not need the
conceptually difficult idea of an infinite sea of negative energy
particles.  The vacuum $|0\rangle$, with no particles in it, is now our
state with the lowest possible energy level.  And, as we will see,
there are never negative energy states with these particles.  

How exactly $|0\rangle$ works will become clearer when we quantize. 
The point to be understood for now is that QFT solves the problem of
negative energy by reinterpreting what is a state and what is an
operator.  The fields $\phi$ and $\psi$ are operators, not states, and
therefore they do not have energy associated with them (any more than
the operator $\hat x$ or $\hat p_x$ did in non-relativistic quantum
mechanics).  So, without any problems of negative energy, we merely
accept that nature, due to relativity, demands that particles come in
particle/antiparticle pairs, and we move on.  

\subsubsection{Lagrangians for Scalars and Dirac Particles}

Now that we have the equations of motion (\ref{eq:kleingordon}) and
(\ref{eq:dirac}), we want to know the actions that lead to these
equations of motion.  In order to save time, we will merely write down
the answers and let you take the variations to see that they do indeed
lead to the Klein Gordon and Dirac equations of motion for $\phi$ and
$\psi$.  

They are 
\begin{eqnarray}
\mathcal{L}_{KG} &=& -{1\over 2}\partial^{\mu}\phi \partial_{\mu}\phi -
{1\over 2}m^2 \phi \label{eq:kleingordonlagrangian} \\
\mathcal{L}_D &=& i\psi_L^{\dagger}\bar
\sigma^{\mu}\partial_{\mu}\psi_L+i\psi^{\dagger}_R\sigma^{\mu}
\partial_{\mu}\psi_R - m(\psi^{\dagger}_L\psi_R+\psi^{\dagger}_R\psi_L)
\label{eq:diraclagrangian}
\end{eqnarray}
where the dagger represents the Hermitian conjugate, $\psi^{\dagger}_L
= (\psi_1^{\star}, \psi_2^{\star})$, $\psi^{\dagger}_R =
(\psi_3^{\star}, \psi_4^{\star})$, as usual.  You can actually take the
variation of $\mathcal{L}_D$ with respect to either $\psi_L^{\dagger}$
and $\psi_R^{\dagger}$ to get the equations of motion for $\psi_L$ and
$\psi_R$, or you can take the variations with respect to $\psi_L$ and
$\psi_R$ to get the equations for $\psi_L^{\dagger}$ and
$\psi_R^{\dagger}$.  The two sets of equations are simply the
conjugates of each other, and therefore represent a single set of
equations.  

In order to simplify (\ref{eq:diraclagrangian}), the convention is to
use the Dirac gamma matrices (\ref{eq:gammamat}) to define $\bar \psi
\equiv \psi^{\dagger}\gamma^0$ (where  $\psi$ here is the 4-component
spinor in equation (\ref{eq:diracspinor})).  Using this, all 4 terms
in (\ref{eq:diraclagrangian}) can be summarized as 
\begin{eqnarray}
\mathcal{L}_D = \bar \psi (i \gamma^{\mu}\partial_{\mu} - m)\psi
\label{eq:truediraclagrangian}
\end{eqnarray}

\subsubsection{Conserved Currents}
\label{sec:conservedcurrents}

In Part I we discussed how symmetries and conserved quantities are
related.  Let's consider a few examples of this using the Lagrangians
we have now defined.  

Consider a massless Klein Gordon scalar particle, described by
$\mathcal{L} = -{1\over 2} \partial^{\mu}\phi \partial_{\mu}\phi$. 
Following what we did starting with equation (\ref{eq:projectile}),
consider the transformation $\phi \rightarrow \phi+\epsilon$, where
$\epsilon$ is a constant.  Because $\partial^{\mu}\phi \rightarrow
\partial^{\mu}\phi + \partial^{\mu} \epsilon = \partial^{\mu} \phi$,
the Lagrangian is invariant.  So (using $\delta \phi$~=~1), our
conserved quantity is 
\begin{eqnarray}
j^{\mu} = {\partial \mathcal{L} \over \partial (\partial_{\mu} \phi)}
\delta \phi = -\partial^{\mu} \phi\nolabel
\end{eqnarray}

Or, consider the Klein Gordon Lagrangian with complex scalar fields
$\phi$ and $\phi^{\dagger}$, which we write as $\mathcal{L} =
-\partial^{\mu}\phi^{\dagger}\phi_{\mu}\phi - m^2 \phi^{\dagger}\phi$. 
We can make the transformation $\phi \rightarrow e^{i\alpha}\phi$ and
$\phi^{\dagger} \rightarrow \phi^{\dagger}e^{-i\alpha}$ (where $\alpha$
is an arbitrary real constant).  This type of transformation is called
a $U(1)$ transformation, because $e^{i\alpha}$ is an element of the
group of all 1~$\times$~1 unitary matrices, as discussed in section
\ref{sec:lieclass}.  

The conserved quantity associated with this $U(1)$ symmetry is 
\begin{eqnarray}
j^{\mu} = {\partial \mathcal{L} \over \partial (\partial_{\mu}
\phi)}\delta \phi + {\partial \mathcal{L} \over \partial
(\partial_{\mu}\phi^{\dagger})}\delta \phi^{\dagger} = i (\phi
\partial^{\mu}\phi^{\dagger} - \phi^{\dagger} \partial^{\mu}
\phi)\nolabel
\end{eqnarray}

Or consider the Dirac Lagrangian.  Notice that it is invariant under
the $U(1)$ transformation $\psi \rightarrow e^{i\alpha}$, with current 
\begin{eqnarray}
j^{\mu} = \bar \psi \gamma^{\mu} \psi \label{eq:u1current}
\end{eqnarray}

In both of the previous examples, notice that the $U(1)$ symmetry
changes the field at \it all \rm points in space at once, and all in
the same way.  In other words, it is a single overall constant phase
$e^{i\alpha}$.	We therefore call such a symmetry a \bf Global 
Symmetry\rm.  The implications of this are likely not clear at this point.
 We merely wish to call your attention to the fact that $e^{i\alpha}$
has no spacetime dependence.  

\subsubsection{The Dirac Equation with an Electromagnetic Field}

Previously we found the Lagrangian for an electromagnetic field
(\ref{eq:emlagrangian}).  Our goal now is to find a Lagrangian that
describes the electromagnetic field and a spin-$1/2$ particle that
couples to the electromagnetic field, and additionally the interaction
between them.  We start by writing down a Lagrangian without any
interaction.  This will simply be the sum of the two terms,
\begin{eqnarray}
\mathcal{L} = \mathcal{L}_D+\mathcal{L}_{EM} = \bar \psi (i
\gamma^{\mu}\partial_{\mu} - m)\psi - {1\over 4} F_{\mu \nu}F^{\mu \nu}
- J^{\mu}A_{\mu \nu} \label{eq:emintlag1}
\end{eqnarray}
But, because the Dirac part has no terms in common with the
electromagnetic part, the equations of motion and the conserved
quantities for both $\psi$ and $A^{\mu}$ will be exactly the same, as if
the other weren't present at all.  In other words, both fields go about
their way as if the other weren't there---there is no interaction in
this theory.  Because this makes for a boring universe (and horrible
phenomenology), we need to find some way of coupling the two fields
together to produce some sort of interaction.  

Interaction is added to a physical theory by adding another term to the
Lagrangian called the \bf Interaction Term\rm.	So, the final
Lagrangian will have the form $\mathcal{L} =
\mathcal{L}_D+\mathcal{L}_{EM}+\mathcal{L}_{int}$.  

Now, for reasons that will become clear in the next section (and even
more clear when we quantize), we do this by coupling the
electromagnetic field $A^{\mu}$ to the current resulting from the
$U(1)$ symmetry in $\mathcal{L}_D$, which we discussed in section
\ref{sec:conservedcurrents}, and wrote out in equation
(\ref{eq:u1current}).  In other words, our interaction term will be
proportional to $A^{\mu}j_{\mu}$.  

So, adding a constant of proportionality $q$ (which we will see has the
physical interpretation of a coupling constant, weighting the
probability of an interaction to take place, or equivalently the
physical interpretation of electric charge), our Lagrangian is now
\begin{eqnarray}
\mathcal{L} &=& \bar \psi(i\gamma^{\mu}\partial_{\mu} - m)\psi -
{1\over 4}F_{\mu \nu}F^{\mu \nu} - J^{\mu}A_{\mu} - qj^{\mu}A_{\mu}
\nolabel \\
&=& \bar \psi(i\gamma^{\mu}\partial_{\mu} - m)\psi - {1\over 4}F_{\mu
\nu}F^{\mu \nu} - (J^{\mu} + q\bar \psi \gamma^{\mu} \psi)A_{\mu}
\label{eq:emintlag2}
\end{eqnarray}
Notice that $\mathcal{L}$ is still invariant under the global $U(1)$
symmetry, and the $U(1)$ current is still $J^{\mu} = \bar \psi
\gamma^{\mu} \psi$.  

Also, notice that the Lagrangians in (\ref{eq:emintlag1}) and
(\ref{eq:emintlag2}) are the same except for a shift in the current
term, $J^{\mu} \rightarrow J^{\mu}+qj^{\mu}$.  Recall that physically,
$J^{\mu} = (\rho, \bar J)$ represents the charge and current creating
the field.  The fact that $J^{\mu}$ has shifted in (\ref{eq:emintlag2})
simply means that the spin-$1/2$ particle in this theory contributes to
the field, which is exactly what we would expect it to do.  

If we set $q=e$, the electric charge, this Lagrangian becomes upon
quantization the Lagrangian of Quantum Electrodynamics ($QED$), which
to date makes the most accurate experimental predictions ever.	

In the next section, we will re-derive this Lagrangian in a more
fundamental way.  

\subsubsection{Gauging the Symmetry}
\label{sec:gaugingthesymmetry}

Physically speaking, this section is among the most important in these
notes.	Read this section again and again until you understand every
step.  

Consider once again the Dirac Lagrangian (\ref{eq:dirac}).  As we said
in section \ref{sec:conservedcurrents}, it is invariant under the \it
global \rm $U(1)$ transformation $\psi \rightarrow e^{i\alpha}\psi$. 
It is global in that it acts on the field the exact same way at every
point in spacetime.  The idea behind this section is that we are going
to make this symmetry \bf Local\rm, so that $\alpha$ depends on
spacetime ($\alpha = \alpha(x^{\mu}))$, and then try to force the
Lagrangian to maintain its invariance under the \it local \rm $U(1)$
transformation.  Making a global symmetry local is referred to as \bf
Gauging \rm the symmetry.  

We start by making the local $U(1)$ transformation:
\begin{eqnarray}
\mathcal{L} = \bar \psi (i \gamma^{\mu} \partial_{\mu} - m)\psi
\rightarrow \bar \psi e^{-\alpha(x)}(i\gamma^{\mu}\partial_{\mu} -
m)e^{i\alpha(x)}\psi\nolabel
\end{eqnarray}
and because the differential operators will now act on $\alpha(x)$ as
well as $\psi$, we get extra terms:
\begin{eqnarray}
\mathcal{L}  \rightarrow \bar \psi
e^{-\alpha(x)}(i\gamma^{\mu}\partial_{\mu} - m)e^{i\alpha(x)}\psi &=&
\bar \psi (i\gamma^{\mu}\partial_{\mu} - m)\psi - \bar \psi
\gamma^{\mu}\psi \partial_{\mu}\alpha(x) \nolabel \\
&=& \bar \psi(i\gamma^{\mu}\partial_{\mu} - m -
\gamma^{\mu}\partial_{\mu}\alpha(x))\psi\nolabel
\end{eqnarray}
If we want to demand that $\mathcal{L}$ still be invariant under this
local $U(1)$ transformation, we must find a way of canceling the $\bar
\psi \gamma^{\mu} \psi \partial_{\mu}\alpha(x)$ term.  We do this in
the following way.  

Define some arbitrary field $A_{\mu}$ which under the $U(1)$
transformation $e^{i\alpha(x)}$ transforms according to
\begin{eqnarray}
A_{\mu} \rightarrow A_{\mu} - {1\over q}\partial_{\mu}\alpha(x)
\label{eq:emgaugefieldtrans}
\end{eqnarray}
We call $A_{\mu}$ the \bf Gauge Field \rm for reasons that will be
clear soon, and $q$ is a constant we have included for later
convenience.  

We introduce $A_{\mu}$ by replacing the standard derivative
$\partial_{\mu}$ with the \bf Covariant Derivative \rm
\begin{eqnarray}
D_{\mu} \equiv \partial_{\mu} + iqA_{\mu}
\label{eq:covariantderivative}
\end{eqnarray}
If you have studied general relativity or differential geometry at any
point, you are familiar with covariant derivatives.  There is an
incredibly rich geometric picture of all of this, but it is beyond the
scope of these notes.  We will deal with it later in this series,
however.  

As a comment regarding vocabulary, to say that a particle ``carries
charge" mathematically means that it has the corresponding term in its
covariant derivative.  So, if a particle's covariant derivative is
equal to the normal differential operator $\partial^{\mu}$, then the
particle has no charge, and it will not interact with anything.  But if
it carries charge, it will have a term corresponding to that charge in
its covariant derivative.  This will become clearer as we proceed.  

So, our Lagrangian is now
\begin{eqnarray}
\mathcal{L} = \bar \psi (i \gamma^{\mu}D_{\mu} - m)\psi = \bar \psi
(i\gamma^{\mu}[\partial_{\mu}+iqA_{\mu}] - m) \psi = \bar \psi
(i\gamma^{\mu}\partial_{\mu} - m - q\gamma^{\mu}A_{\mu})\psi\nolabel
\end{eqnarray}
And under the local $U(1)$ we have
\begin{eqnarray}
\mathcal{L} &\rightarrow & \bar \psi e^{-i\alpha(x)}(i \gamma^{\mu}
\partial_{\mu} - m - q \gamma^{\mu}[A_{\mu} - {1\over
q}\partial_{\mu}\alpha(x)])e^{i\alpha(x)}\psi \nolabel \\
&=& \bar \psi (i \gamma^{\mu}\partial_{\mu} - m -
\gamma^{\mu}\partial_{\mu}\alpha(x) - q\gamma^{\mu}
A_{\mu}+\gamma^{\mu}\partial_{\mu}\alpha(x))\psi \nolabel \\
&=& \bar \psi(i\gamma^{\mu} \partial_{\mu} - m -
q\gamma^{\mu}A_{\mu})\psi = \bar \psi (i \gamma^{\mu}D_{\mu} - m)\psi \nolabel \\
&=& \mathcal{L}\nolabel
\end{eqnarray}
So, the addition of the field $A_{\mu}$ has indeed restored the $U(1)$
symmetry.  Notice that now it is not only invariant under this local
$U(1)$, but also still under the global $U(1)$ we started with, with
the same conserved  $U(1)$ current $j^{\mu} = \bar \psi \gamma^{\mu}
\psi$.	This allows us to rewrite the Lagrangian as 
\begin{eqnarray}
\mathcal{L} = \bar \psi (i\gamma^{\mu}D_{\mu} - m) \psi = \bar \psi
(i\gamma^{\mu}\partial_{\mu} - m)\psi - qj^{\mu}A_{\mu}
\label{eq:nodynamics}
\end{eqnarray}

But we have a problem.	If we want to know what the dynamics of
$A_{\mu}$ will be, we naturally take the variation of the Lagrangian
with respect to $A_{\mu}$.  But because there are no derivatives of
$A_{\mu}$, the Euler-Lagrange equation is merely ${\partial \mathcal{L}
\over \partial A_{\mu}} = -q\bar \psi \gamma^{\mu} \psi = 0$.  But
$-q\bar \psi \gamma^{\mu}\psi = -qj^{\mu}$.  So the equation of motion
for $A_{\mu}$ says that the current vanishes, or that $j^{\mu} = 0$,
and so the Lagrangian is reduced back to
(\ref{eq:truediraclagrangian}), which was not invariant under the local
$U(1)$.  

We can state this problem in another way.  All physical fields have
some sort of dynamics.	If they don't then they are merely a constant
background field that never changes and does nothing.  As it is
written, equation (\ref{eq:nodynamics}) has a field $A_{\mu}$ but
$A_{\mu}$ has no kinetic term, and therefore no dynamics.  

So, to fix this problem we must include some sort of dynamics, or
kinetic terms, for $A_{\mu}$.  

The way to do this turns out to involve a considerable amount of
geometry which would be out of place in these notes.  We will cover the
necessary ideas in a later paper in this series and derive the
following expressions.	For now we merely give the results and ask you
for patience until we have the machinery to derive them.  

For an arbitrary field $A_{\mu}$, the appropriate gauge-invariant
kinetic term is 
\begin{eqnarray}
\mathcal{L}_{Kin,A} = -{1\over 4} F_{\mu \nu}F^{\mu \nu}\nolabel
\end{eqnarray} 
where 
\begin{eqnarray}
F^{\mu \nu} \equiv {i \over q}[D^{\mu},D^{\nu}] \label{eq:deffmn}
\end{eqnarray}
and $q$ is the constant of proportionality introduced in the
transformation of $A_{\mu}$ in equation (\ref{eq:emgaugefieldtrans}).
$D^{\mu}$ is the covariant derivative defined in
(\ref{eq:covariantderivative}).  

Writing out (\ref{eq:deffmn}) (and using an arbitrary test function
$f(x)$),
\begin{eqnarray}
F^{\mu \nu}f(x) &=& {i\over q} [D^{\mu},D^{\nu}]f(x) \nolabel \\
&=& {i\over q} \big[(\partial^{\mu} +
iqA^{\mu})(\partial^{\nu}+iqA^{\nu}) -
(\partial^{\nu}+iqA^{\nu})(\partial^{\mu}+iqA^{\mu}\big]f(x) \nolabel
\\
&=& {i \over q}\big[\partial^{\mu}\partial^{\nu}f(x) +
iq\partial^{\mu}(A^{\nu}f(x))+iqA^{\mu}\partial^{\nu}f(x) -
q^2A^{\mu}A^{\nu}f(x)  \nolabel \\
& & -\partial^{\nu}\partial^{\mu}f(x) +
iq\partial^{\nu}(A^{\mu}f(x))+iqA^{\nu}\partial^{\mu}f(x) -
q^2A^{\nu}A^{\mu}f(x)\big]  \nolabel \\
&=& {i\over q}\big[iqf(x)\partial^{\mu}A^{\nu} +
iqA^{\nu}\partial^{\mu}f(x) + iqA^{\mu}\partial^{\nu}f(x) -
q^2A^{\mu}A^{\nu}f(x) \nolabel \\
& & -iqf(x)\partial^{\nu}A^{\mu} - iqA^{\mu}\partial^{\nu}f(x) -
iqA^{\nu}\partial^{\mu}f(x) + q^2A^{\nu}A^{\mu}f(x)\big] \nolabel \\
&=& \big[\partial^{\mu}A^{\nu} - \partial^{\nu}A^{\mu} +
iq[A^{\mu},A^{\nu}]\big] f(x)\nolabel
\end{eqnarray}

But for each value of $\mu$, $A^{\mu}$ is a scalar function, so the
commutator term vanishes, leaving (dropping the test function $f(x)$)
\begin{eqnarray}
F^{\mu \nu} = {i\over q}[D^{\mu},D^{\nu}] = \partial^{\mu} A^{\nu} -
\partial^{\nu}A^{\mu} \label{eq:qwerqwer}
\end{eqnarray}

So, writing out the entire Lagrangian we have
\begin{eqnarray}
\mathcal{L} = \bar \psi (i\gamma^{\mu}D_{\mu} - m)\psi -{1\over
4}F_{\mu \nu}F^{\mu \nu} \nolabel
\end{eqnarray}

And finally, because $A^{\mu}$ is obviously a physical field, we can
naturally assume that there is some source term causing it, which we
simply call $J^{\mu}$.	This makes our final Lagrangian 
\begin{eqnarray}
\mathcal{L} = \bar \psi (i\gamma^{\mu}D_{\mu} - m)\psi -{1\over
4}F_{\mu \nu}F^{\mu \nu} - J^{\mu}A_{\mu}\nolabel
\end{eqnarray}

Comparing this to (\ref{eq:emintlag2}) we see that they are exactly the
same.  So what have we done?  We started with nothing but a Lagrangian
for a spin-$1/2$ particle, which had a global $U(1)$ symmetry.	Then,
all we did was promote the $U(1)$ symmetry to a local symmetry (we
gauged the symmetry), and then imposed what we had to impose to get a
consistent theory.  The gauge field $A_{\mu}$ was forced upon us, and
the form of the kinetic term for $A_{\mu}$ is demanded automatically by
geometric considerations we did not delve into.  

In other words, we started with nothing but a non-interacting particle,
and by specifying \it nothing \rm but $U(1)$ we have created a theory
with not only that same particle, but also electromagnetism.  The
$A_{\mu}$ field, which upon quantization will be the photon, is a
direct consequence of the $U(1)$.  

This is what we meant at the end of section \ref{sec:arbitraryj} when
we said that electromagnetism is described by $U(1)$.  We will talk
more about the weak and strong forces later, as well as the groups
that give rise to them.  

Theories of this type, where we generate forces by specifying a Lie
group, are called \bf Gauge Theories\rm, or \bf Yang-Mills Theories\rm. 

Finally, notice that (\ref{eq:emgaugefieldtrans}) has exactly the same
form as (\ref{eq:gaugetransformation}).  This is why we call $A_{\mu}$
a gauge field.	The gauge symmetry in electromagnetism is a sort of
remnant of the much deeper and more fundamental $U(1)$ structure of the
theory.  

\subsection{Quantization}

\subsubsection{Review of What Quantization Means}
\label{sec:quantreview}

In quantum mechanics (not QFT), quantization is done by taking
certain dynamical quantities and making use of the \bf Heisenberg
Uncertainty Principle\rm.  Normally we take position $\bar x$ and
momentum $\bar p$ and, according to Heisenberg, the measurement of the
particle's position will effect its momentum and vice-versa.  

To make this more precise, we promote $x$ and $p$ from merely being
variables to being Hermitian operators $\hat x$ and $\hat p$ (which can be represented by
matrices)  acting on some vector space.  Calling a
vector in this space $|\psi\rangle$, physically measurable quantities
(like position or momentum) become the eigenvalues of the operators
$\hat x$ and $\hat p$, 
\begin{eqnarray}
\hat x |\psi\rangle = x |\psi \rangle \nolabel \\
\hat p |\psi \rangle = p|\psi \rangle\nolabel
\end{eqnarray}

Heisenberg Uncertainty says that measuring $x$ will affect the value of
$p$, and vice-versa.  It is the act of measuring which enacts this
effect.  It is not an engineering problem in the sense that there is no
better measurement technique which would undo this.  It is a
fundamental fact of quantum mechanics (and therefore the universe) that
measurement of one variable affects another.  

So, if we measure $x$ (using $\hat x$) and then $p$ (using $\hat p$),
we will in general get different values for both than if we measured
$p$ and then $x$.  More mathematically, $\hat x \hat p \neq \hat p \hat
x$.  Put another way, 
\begin{eqnarray}
[\hat x, \hat p] \equiv \hat x \hat p - \hat p \hat x \neq 0 \nolabel
\end{eqnarray}
For reasons learned in an introductory quantum course, the actual
relation is 
\begin{eqnarray}
[\hat x, \hat p] = i\hbar \label{eq:cancom}
\end{eqnarray}
where $\hbar$ is Planck's Constant.  We call (\ref{eq:cancom}) the \bf
Canonical Commutation Relation\rm, and it is this structure which
allows us to determine the physical structure of the theory.  

More generally, we choose some set of operators that all commute with
each other, and then label a physical state by its eigenvectors.  For
example $\hat x$, $\hat y$ and $\hat z$ all commute with each other, so
we may label a physical state by its eigenvectors $|\psi_r\rangle =
|x,y,z\rangle$.  Or, because $\hat p_x$, $\hat p_y$, and $\hat p_z$ all
commute, we may call the state $|\psi_p\rangle = |p_x,p_y,p_z\rangle$. 
We may also include some other values like spin and angular momentum,
to have (for example) $|\psi\rangle = |x,y,z,s_z,L_z,\ldots \rangle$.  

As discussed in section \ref{sec:test}, when we make the jump to QFT,
the fields are no longer the states but the operators.	We are
therefore going to impose commutation relations on the fields, not on
the coordinates.  

Furthermore, whereas before the states were eigenvectors of the
coordinate operators, we now will expand the fields in terms of the
eigenvectors of the Hamiltonian.  

\subsubsection{Canonical Quantization of Scalar Fields}
\label{sec:scalarcanquant}

We begin with the Klein Gordon Lagrangian in equation
(\ref{eq:kleingordonlagrangian}), but we make the slight modification
of adding an arbitrary constant $\Omega$,
\begin{eqnarray}
\mathcal{L}_{KG} = -{1\over 2}\partial^{\mu}\phi \partial_{\mu}\phi -
{1\over 2}m^2 \phi^2 + \Omega\nolabel
\end{eqnarray}	
Note that $\Omega$ has absolutely no affect whatsoever on the physics.  

Quantization then comes about by defining the field momentum and
Hamiltonian (using (\ref{eq:hamiltoniandensity}) and
(\ref{eq:momentumdensity})) to get 
\begin{eqnarray}
\Pi &=& {\partial \mathcal{L} \over \partial \dot{\phi}(x)} =
\dot{\phi} \\
\mathcal{H} &=& \Pi \dot{\phi} - \mathcal{L} = {1\over 2}\Pi^2 +
{1\over 2}(\bar \nabla \phi)^2 + {1\over 2}m^2\phi^2 - \Omega
\end{eqnarray}

Now, using the canonical commutation relations (\ref{eq:cancom}) as
guides, we impose
\begin{eqnarray}
& & \;[\phi(t,\bar x),\phi(t',\bar x')]\ = 0 \nolabel\\
& & \;[\Pi(t,\bar x),\Pi(t',\bar x')] = 0 \nolabel\\
& & \;[\phi(t,\bar x),\Pi(t',\bar x')] = i\delta(t-t')\delta(\bar x -
\bar x') \label{eq:scalarcomrels}
\end{eqnarray}
(where we have set $\hbar = 1$).  

We can see more clearly what this means if we expand the solutions of
the Klein Gordon equation.  One solution is plane waves, $e^{i\bar k
\cdot \bar x \, \pm \, i \omega t}$, where 
\begin{eqnarray}
\omega = +\sqrt{\bar k^2 + m^2} \label{eq:defomega}
\end{eqnarray}
and $\bar k$ is the standard wave vector.  

So, we write the field $\phi$ as 
\begin{eqnarray}
\phi(t,\bar x) = \int {d^3 \bar k \over f(\bar k)} \big[a(\bar
k)e^{i\bar k \cdot \bar x \,- \, i \omega t} + b (\bar k)e^{i \bar k \cdot x
\, + \, i \omega t} \big]\nolabel
\end{eqnarray}
where $f(x)$ is a redundant function which we have included for later
convenience.  For now, both $a(\bar k)$ and $b(\bar k)$ are merely
arbitrary coefficients (integration constants) used to expand
$\phi(t,\bar x)$ in terms of individual solutions.  

We demand that $\phi(t,\bar x)$ be Hermitian.  This requires 
\begin{eqnarray}
\phi^{\dagger} = \phi \Rightarrow \phi^{\star} = \phi \Rightarrow
b^{\star}(\bar k) = a(-\bar k)\nolabel
\end{eqnarray}

Then, changing the sign of the integration variable $\bar k$ on the
second term in the integral allows us to use 4-vector notation, so
\begin{eqnarray}
\phi(x) = \int {d^3 \bar k \over f(\bar k)} \big[ a(\bar k)e^{ik\cdot
x} + a^{\star}(\bar k)e^{-ik\cdot x}\big]\nolabel
\end{eqnarray}
where $k\cdot x = k^{\mu} x_{\mu}$.  

Now notice that the integration measure, $d^3\bar k$, is not invariant
under Lorentz transformations (because it integrates over the spatial
part but not over the time part).  We therefore choose $f(\bar k)$ to
restore Lorentz invariance.  

We know that the measure $d^3k $ would be invariant, as would $\delta$
functions and $\Theta$ (step) functions.  So, consider the invariant
combination
\begin{eqnarray}
d^4k \delta(k^2+m^2)\Theta(k^0) \label{eq:firstinvmeasure}
\end{eqnarray}
The $\delta$ function merely requires that relativity hold ($k^2+m^2$
is simply the relativistic relation (\ref{eq:relativistichamiltonian}),
and the $\Theta$ function preserves causality.	So this is a physically
acceptable Lorentz invariant integration measure.  

Recall the general $\delta$ function identity,
\begin{eqnarray}
\int_{-\infty}^{\infty} dx \delta(g(x)) = \sum_i {1 \over \big|{dg(x)
\over dx}|_{x=x_i}\big|}\nolabel
\end{eqnarray}
where the $x_i$'s  are the zeros of the function $g(x)$.  We can do the
$k^0$ integral over measure (\ref{eq:firstinvmeasure}), and using the
fact that the zeros of $k^2+m^2 = \bar k^2 -  k^0k_0+m^2$ in terms of
$k^0$ are $k^0k_0 = \bar k^2+m^2 = \omega^2$, we get 
\begin{eqnarray}
\int d^3 \bar k dk^0\delta(k^2+m^2)\Theta(k^0) = \int {d^3\bar k \over
2\omega}\nolabel
\end{eqnarray}

So, adding a factor of $(2\pi)^3$ for later convenience, we take our
invariant measure to be 
\begin{eqnarray}
{d^3\bar k \over (2\pi)^3 2\omega}\nolabel
\end{eqnarray}

So finally, 
\begin{eqnarray}
\phi(x) = \int \widetilde{dk}\big[a(\bar k)e^{ik\cdot x} +
a^{\star}(\bar k)e^{-ik\cdot x}\big] \label{eq:scalarfield}
\end{eqnarray}
where we have defined $\widetilde{dk} \equiv {d^3\bar k \over (2\pi)^3
2\omega}$.  

The commutation relations we defined in (\ref{eq:scalarcomrels}) will
now hold provided we impose
\begin{eqnarray}
& &\;[a(\bar k),a(\bar k')] = 0 \nolabel  \\
& &\;[a^{\dagger}(\bar k),a^{\dagger} (\bar k') ] = 0 \nolabel \\
& &\;[a(\bar k),a^{\dagger}(\bar k')] = (2\pi)^3 2\omega \delta^3(\bar
k - \bar k') \label{eq:acoms}
\end{eqnarray}
(showing this is fairly tedious, but we encourage you to work it out). 
We are using $\dagger$ instead of $\star$ to emphasize that, in the
quantum theory, we are talking about Hermitian operators.  The
operators $a(\bar k)$ and $a^{\dagger}(\bar k)$ are scalars, so in this
case $a^{\star} = a^{\dagger}$.  

Furthermore, we can write the Hamiltonian $H$ in terms of
(\ref{eq:scalarfield}):
\begin{eqnarray}
H &=& \int d^3x\mathcal{H} = \int d^3x\bigg({1\over 2}\Pi^2 + {1\over
2}(\bar \nabla \phi)^2 + {1\over 2}m^2\phi^2 - \Omega \bigg) \nolabel
\\
&=& {1\over 2}\int \widetilde{dk}\widetilde{dk'} d^3x[(-i\omega a(\bar
k) e^{ik\cdot x} + i\omega a^{\star}(\bar k) e^{-ik\cdot x})(-i\omega'
a(\bar k')e^{i\bar k' \cdot x} + i\omega' a^{\star}(\bar
k')e^{-ik'\cdot x}) \nolabel \\
& & + (i\bar k a(\bar k)e^{ik\cdot x} - i\bar k  a^{\star}(\bar
k)e^{-ik\cdot x} ) \cdot (i\bar k' a(\bar k')e^{ik'\cdot x} - i\bar k'
a^{\star}(\bar k')e^{-ik'\cdot x}) \nolabel \\
& & m^2(a(\bar k)e^{ik\cdot x} + a^{\star}(\bar k)e^{-ik\cdot
x})(a(\bar k')e^{ik'\cdot x}+ a^{\star}(\bar k')e^{-ik' \cdot x})] -
\int d^3x \Omega \nolabel \\
&=& {1\over 2} \int \widetilde{dk}\widetilde{dk'}d^3x [(-\omega \omega'
a(\bar k) a(\bar k')e^{i(k+k')\cdot x}+\omega \omega' a(\bar k)
a^{\star}(\bar k') e^{i(k-k')\cdot x} \nolabel \\
& & \qquad \qquad \qquad \;\; + \omega \omega' a^{\star}(\bar k) a(\bar
k')e^{-i(k-k')\cdot} - \omega \omega' a^{\star}(\bar k)a^{\star}(\bar
k')e^{-i(k+k')\cdot x} ) \nolabel \\
& & \qquad \qquad \quad \;\; + (-\bar k\cdot \bar k' a(\bar k)a(\bar
k')e^{i(k+k')\cdot x} + \bar k \cdot \bar k' a(\bar k) a^{\star}(\bar
k') e^{i(k-k')\cdot x} \nolabel \\
& &\qquad \qquad \qquad \;\; + \bar k \cdot \bar k' a^{\star}(\bar
k)a(\bar k')e^{-i(k-k')\cdot x} - \bar k \cdot \bar k'a^{\star}(\bar k)
a^{\star}(\bar k')e^{-i(k+k')\cdot x}) \nolabel \\
& &\qquad \qquad \quad \;\; + m^2(a(\bar k)a(\bar k')e^{i(k+k')\cdot x}
+ a(\bar k)a^{\star}(\bar k')e^{i(k-k')\cdot x} \nolabel \\
& & \qquad \qquad \qquad \;\; + a^{\star}(\bar k)a(\bar
k')e^{-i(k-k')\cdot x} + a^{\star}(\bar k)a^{\star}(\bar
k')e^{-i(k+k')\cdot x} ) \nolabel \\
& & \qquad \qquad \quad \;\; - V\Omega\nolabel
\end{eqnarray}
where $V$ is the volume of the space resulting from the $\int d^3x$
integral.  Then, from the fact that $\int d^3xe^{i\bar x \cdot \bar y}
= (2\pi)^3\delta^3(\bar y )$, we have
\begin{eqnarray}
H &=& {1\over 2}(2\pi)^3 \int \widetilde{dk}\widetilde{dk'}
[\delta^3(\bar k - \bar k')(\omega \omega' + \bar k\cdot \bar k' +
m^2)(a^{\star}(\bar k)a(\bar k ')e^{-i(\omega - \omega')t} + a(\bar
k)a^{\star}(\bar k')e^{-i(\omega - \omega')t}) \nolabel \\
& & \qquad \qquad \quad \;\;\;\; + \delta^3(\bar k + \bar k')(-\omega
\omega' - \bar k \cdot \bar k' + m^2)(a(\bar k)a(\bar k')e^{-i(\omega +
\omega')t} + a^{\star}(\bar k ) a^{\star} (\bar k ') e^{i(\omega +
\omega')t}) \nolabel \\
& & \qquad \qquad \quad \;\;\;\; - V \Omega \nolabel \\
&=& {1\over 2}\int \widetilde{dk}{1\over 2\omega} [ (\omega^2+\bar k^2
+ m^2)(a^{\star}(\bar k)a(\bar k) + a(\bar k ) a^{\star}(\bar k))
\nolabel \\
& & \qquad \qquad + (-\omega^2 + \bar k^2 + m^2)(a(\bar k)a(-\bar
k)e^{-2i\omega t} + a^{\star}(\bar k) a^{\star}(-\bar k)e^{2i\omega
t})] - V\Omega\nolabel
\end{eqnarray}
and finally, using the definition of $\omega$ (equation
(\ref{eq:defomega})), this becomes 
\begin{eqnarray}
H = {1\over 2} \int \widetilde{dk}\; \omega (a^{\star}(\bar k)a(\bar k)
+ a(\bar k)a^{\star} (\bar k)) - V\Omega\nolabel
\end{eqnarray}

And now, using (\ref{eq:acoms}), we can rewrite this as (switching from
$\star$ to $\dagger$ to emphasize the Hermitian nature)
\begin{eqnarray}
H &=& {1\over 2}\int \widetilde{dk}\; \omega (a^{\dagger}(\bar k
)a(\bar k) + a(\bar k) a^{\dagger}(\bar k)) - V \Omega \nolabel \\
&=& {1\over 2}\int \widetilde{dk} \; \omega (a^{\dagger}(\bar k )a(\bar
k) + (2\pi)^3 2\omega \delta^3(\bar k - \bar k ) + a^{\dagger}(\bar k )
a(\bar k )) - V \Omega \nolabel \\
&=& \int \widetilde{dk} \; \omega a^{\dagger}(\bar k ) a (\bar k ) +
\int \widetilde{dk} \; \omega (2\pi)^3 \delta^3(0) - V\Omega \nolabel
\\
&=& \int \widetilde{dk} \omega a^{\dagger}(\bar k )a(\bar k ) + \int
{d^3 \bar k \over (2\pi)^3 2\omega} \omega (2\pi)^3 \delta^3(0) -
V\Omega \nolabel \\
&=& \int \widetilde{dk} \omega a^{\dagger}(\bar k )a(\bar k ) + {1\over
2} \delta^3(0) \int d^3\bar k - V\Omega \nolabel
\end{eqnarray}

Notice that both the second and third terms are infinite (assuming the
volume $V$ of the space we are in is infinite).  This may be troubling,
but remember that $\Omega$ is an arbitrary constant we can set to be
anything we want.  So, let's define
\begin{eqnarray}
\Omega \equiv {1\over 2V} \delta^3(0) \int d^3 \bar k\nolabel
\end{eqnarray}
leaving
\begin{eqnarray}
H = \int \widetilde{dk} \; \omega a^{\dagger} (\bar k ) a (\bar k )
\label{eq:scalarhamiltonian}
\end{eqnarray}

Remember that measurement can only detect \it changes \rm in energy,
and therefore the infinity we subtracted off does not affect the value
we will measure experimentally.  What we have done here, by subtracting
off the infinite part in a way that doesn't change the physics, is a
very primitive example of \bf Renormalization\rm.  Often, for various
reasons, measurable quantities in QFT are plagued by different types
of infinities.	However, it is possible to subtract off those
infinities in a well-defined way, leaving a finite part.  It turns out
that this finite part is the correct value seen in nature.  The reasons
for this are very deep, and we will not discuss them (or general
renormalization theory) in much depth in these notes.  For correlating
theoretical results with experiment, being able to renormalize results
correctly is vital.  However, our goal is not to understand the
subtleties of renormalization, but to understand the overall structure
of particle physics.  When you take a course on QFT you will spend a
great, great deal of time on renormalization, and a deeper
understanding of it will emerge.  

So, we have our field expansion (\ref{eq:scalarfield}) and commutation
relations (\ref{eq:acoms}).  Notice that (\ref{eq:acoms}) have the
exact form of a simple harmonic oscillator, which you learned about in
introductory quantum mechanics.  Therefore, because they have the same
structure as the harmonic oscillator, they will have the same physics. 
By doing nothing but imposing relativity, we have found that scalar
fields, which are Hermitian operators, act as raising and lowering (or
synonymously creation and annihilation) operators on the vacuum (just
like the simple harmonic oscillator).  

Comparing (\ref{eq:acoms}) with the standard harmonic oscillator
operators, it is clear that $a^{\dagger}(\bar k )$ \it creates \rm a
$\phi$ particle with momentum $\bar k$ and energy $\omega$, whereas
$a(\bar k )$ annihilates a $\phi$ particle with momentum $\bar k $ and
energy $\omega$.  A normalized state will be 
\begin{eqnarray}
|\bar k \rangle = \sqrt{2\omega} a^{\dagger} (\bar k ) |0\rangle
\label{eq:possiblescalarstates}
\end{eqnarray}
The entire spectrum of states can be studied by acting on $|0\rangle$
with creation operators, and probability amplitudes for one state to be
found in another, $\langle \bar k_f|\bar k_i\rangle$, are
straightforward to calculate (and positive semi-definite).  Naturally
this theory does not discuss any interactions between particles, and
therefore we will have to do a great deal of modification before we are
done.  But this simple exercise of merely imposing the standard
commutation relations (\ref{eq:scalarcomrels}) between the field and
its momentum, we have gained complete knowledge of the quantum
mechanical states of the theory.  

\subsubsection{The Spin-Statistics Theorem}
\label{sec:spinstat}

Notice that the states coming from (\ref{eq:possiblescalarstates}) will
include the two particle state 
\begin{eqnarray}
|\bar k ; \bar k '\rangle = 2\sqrt{\omega \omega '}\; a^{\dagger}(\bar
k ) a^{\dagger}(\bar k ')|0\rangle \label{eq:spinstat1}
\end{eqnarray}
But the commutation relations (\ref{eq:acoms}) tell us that
$a^{\dagger}(\bar k ) a^{\dagger}(\bar k') =a^{\dagger}(\bar k' )
a^{\dagger}(\bar k)$.  So, this theory also allows the state 
\begin{eqnarray}
|\bar k ' ; \bar k \rangle = 2\sqrt{\omega' \omega} a^{\dagger} (\bar
k') a^{\dagger}(\bar k)|0\rangle \label{eq:spinstat2}
\end{eqnarray}

Recall from a chemistry or modern physics course that particles with
half-integer spin obey the Pauli Exclusion Principle, whereas
particles of integer spin do not.  Our Klein Gordon scalar fields
$\phi$ are spinless ($j=0$), and therefore we would expect that they do
not obey Pauli exclusion.  The fact that our commutation relations have
allowed both states (\ref{eq:spinstat1}) and (\ref{eq:spinstat2}) is
therefore expected.  This is an indication that we quantized correctly. 

But notice that this statistical result (that the scalar fields do not
obey Pauli exclusion) is entirely a result of the commutation
relations.  Therefore, if we attempt to quantize a spin-$1/2$ field in
the same way, they will obviously not obey Pauli exclusion either.  We
must therefore quantize spin-$1/2$ differently.  

It turns out that the correct way to quantize spin-$1/2$ fields is to
use, instead of commutation relations like we used for for scalar
fields, \it anticommutation relations\rm.  If the operators of our
spin-$1/2$ fields obey
\begin{eqnarray}
\{a^{\dagger}_1,a^{\dagger}_2\} = a^{\dagger}_1a^{\dagger}_2 = 0
\Rightarrow a^{\dagger}_1 a^{\dagger}_2 =
-a^{\dagger}_2a^{\dagger}_1\nolabel
\end{eqnarray}
then if we try to act twice with the same operator, we have 
\begin{eqnarray}
a^{\dagger}_1a^{\dagger}_1|0\rangle = -a^{\dagger}_1a^{\dagger}_1
|0\rangle \Rightarrow a^{\dagger}_1a^{\dagger}_1|0\rangle = 0\nolabel
\end{eqnarray}
In other words, if we quantize with anticommutation relations, it is
not possible for two particles to occupy the same state simultaneously. 

This relationship between the spin of a particle and the statistics it
obeys (which demands that integer spin particles be quantized by
commutation relations and half-integer spin particles to be quantized
with anticommutation relations) is called the \bf Spin-Statistics
Theorem\rm.  

And, because particles obeying Pauli exclusion are said to have \bf
Bose-Einstein \rm statistics, and particles that do not obey Pauli
exclusion are said to have \bf Fermi-Dirac \rm statistics, we call
particles with integer spin \bf Bosons\rm, and particles with
half-integer spin \bf Fermions\rm.  

\subsubsection{Left-Handed and Right-Handed Fields}
\label{sec:leftright}

Recall that in the Dirac Lagrangian (\ref{eq:truediraclagrangian}), our
fundamental field was the 4-component spinor $\psi = 
\begin{pmatrix}
\psi_L \\ \psi_R
\end{pmatrix}$ where $\psi_L$ transforms under the left-handed
$(0,1/2)$ representation of the Lorentz group, and $\psi_R$ transforms
under the right-handed $(1/2,0)$ representation.  

In general, we refer to these 2-component spinors as \bf Weyl \rm
fields (usually pronounced ``vile").  So, the fermion is the spinor
combination of two Weyl fields, one being the left-handed particle, and
the other being the right-handed antiparticle.	

Also in (\ref{eq:truediraclagrangian}) was the field we defined as
$\bar \psi = \psi^{\dagger} \gamma^0 =
(\psi^{\dagger}_R,\psi^{\dagger}_L)$.  If we interpret $\bar \psi$ as
the conjugate of $\psi$ (which the form of the Dirac Lagrangian implies
we should), then we see that the right-handed field is the conjugate of
the left, and vice versa.  Or, in other words,
\begin{eqnarray}
\psi^{\dagger}_L = \psi_R \qquad \mbox{and} \qquad \psi^{\dagger}_R =
\psi_L\nolabel
\end{eqnarray}

We take advantage of the fact by writing all fields in terms of left-handed
Weyl fields.  For example, given the two left-handed Weyl fields
$\chi$ and $\xi$, we can form the 4-component spinor field $\psi = 
\begin{pmatrix}
\chi \\ \xi^{\dagger}
\end{pmatrix}$, and so $\bar \psi = (\xi, \chi^{\dagger})$.  We will
refer to such a field as a \bf Dirac Field\rm, and denote it $\psi_D$.  

On the other hand, we could define a 4-component spinor in terms of a
single left-handed Weyl field $\chi$, or $\psi = 
\begin{pmatrix}
\chi \\ \chi^{\dagger}
\end{pmatrix}$.  But now notice that $\bar \psi =
(\chi,\chi^{\dagger})$, which is equal simply to the transpose of
$\psi$.  We refer to such a field (whose conjugate is equal to its
transpose) as a \bf Majorana Field\rm, and denote it $\psi_M$.	

Recall that an antiparticle has the same mass but opposite charge and
opposite handedness of its particle.  So, working with the Dirac field
$\psi_D$, we can change the charge by merely swapping $\chi$ and $\xi$,
using the \bf Charge Conjugation \rm operator $\mathcal{C}$ defined by 
\begin{eqnarray}
\mathcal{C}\psi_D = \mathcal{C}
\begin{pmatrix}
\chi \\ \xi^{\dagger}
\end{pmatrix} = 
\begin{pmatrix}
\xi \\ \chi^{\dagger}
\end{pmatrix}\nolabel
\end{eqnarray}

Also, consider the transpose of $\bar \psi_D$ (which is just returning
the conjugate of $\psi_D$ to column form), $\bar \psi^T_D = 
\begin{pmatrix}
\xi \\ \chi^{\dagger}
\end{pmatrix}$.  Acting on this with $\mathcal{C}$ gives 
\begin{eqnarray}
\mathcal{C}\bar \psi^T_D = \mathcal{C}
\begin{pmatrix}
\xi \\ \chi^{\dagger}
\end{pmatrix} = 
\begin{pmatrix}
\chi \\ \xi^{\dagger}
\end{pmatrix} = \psi_D\nolabel
\end{eqnarray}
So, we have 
\begin{eqnarray}
\mathcal{C}\psi_D = \bar \psi^T_D \qquad and \qquad \mathcal{C}\bar
\psi^T_D = \psi_D\nolabel
\end{eqnarray}
We therefore say that $\psi_D$ and $\bar \psi_D^T$ are \bf Charge
Conjugate \rm to each other.  

However, notice that with the Majorana field,
\begin{eqnarray}
\mathcal{C}\psi_M = \mathcal{C}
\begin{pmatrix}
\chi \\ \chi^{\dagger}
\end{pmatrix} = 
\begin{pmatrix}
\chi \\ \chi^{\dagger}
\end{pmatrix} = \psi_M\nolabel
\end{eqnarray}
and
\begin{eqnarray}
\mathcal{C}\bar \psi^T_M = \bar \psi_M^T = \psi_M\nolabel
\end{eqnarray}

So in summary, Dirac fields are not equal to their charge conjugate,
while Majorana fields are.  By analogy with scalars (where the complex
conjugate of a real number is equal to itself, whereas the complex
conjugate of a complex number is not), we often refer to Majorana
fields as \bf Real\rm, and to Dirac fields as \bf Complex\rm.  

So, we can now write out the Lagrangian for Dirac and Majorana fields
in terms of their Weyl fields:
\begin{eqnarray}
\mathcal{L}_D &=& i \chi^{\dagger} \bar \sigma^{\mu} \partial_{\mu}
\chi + i \xi^{\dagger}\bar \sigma^{\mu}\partial_{\mu} \xi - m(\chi \xi
+ \chi^{\dagger}\xi^{\dagger}) \label{eq:erty1} \\
\mathcal{L}_M &=& i\chi^{\dagger} \bar \sigma^{\mu}\partial_{\mu} \chi
- {1\over 2} m(\chi \chi + \chi^{\dagger}\chi^{\dagger})
\label{eq:erty2}
\end{eqnarray}

\subsubsection{Canonical Quantization of Fermions}

We first quantize the Dirac fermion.  The general solution to the Dirac
equation is
\begin{eqnarray}
\psi_D(x) = \sum_{s=1}^2\int \widetilde{dk} [b_s(\bar k)u_s(\bar
k)e^{ik\cdot x} + d^{\dagger}_s(\bar k)v_s(\bar k)e^{-ik\cdot
x}]\nolabel
\end{eqnarray}
where $s$ =1, 2 are the two spin states, $b_s$ and $d^{\dagger}_s$ are
(respectively) the lowering operator for the particle and the raising
operator for the antiparticle.	The charge conjugate of $\psi_D$ will
have the raising operator for the particle and the lowering operator
for the antiparticle.  

The $u_s$ and $v_s$ are constant 4-component vectors which act as a
basis for all particle/antiparticle states in the spinor space (for our
purposes, they are merely present to make $\psi_D$ a 4-component
field).  

We quantize, as we said in section \ref{sec:spinstat}, using \it
anti\rm-commutation relations.	Writing only the non-zero relation,
\begin{eqnarray}
\{\psi_{\alpha}(t,\bar x),\bar \psi_{\beta}(t,\bar x)\} = \delta^3(\bar
x - \bar x')(\gamma^0)_{\alpha \beta}\nolabel
\end{eqnarray}
These imply that the only non-zero commutation relations in terms of
the operators are 
\begin{eqnarray}
\{b_s(\bar k),b^{\dagger}_{s'}(\bar k')\} = (2\pi)^3\delta^3(\bar k -
\bar k')2\omega \delta_{s s'} \nolabel\\
\{d^{\dagger}_s(\bar k),d_{s'}(\bar k')\} = (2\pi)^3\delta^3(\bar k -
\bar k')2\omega\delta_{ss'}\nolabel
\end{eqnarray}

Once again, these form the algebra of a simple harmonic oscillator, and
we can therefore find the entire spectrum of states by acting on
$|0\rangle$ with $b^{\dagger}_s$ and $d^{\dagger}_s$.  

Then, following a series of calculations nearly identical to the ones
in section \ref{sec:scalarcanquant}, we arrive at the Hamiltonian 
\begin{eqnarray}
H = \sum_{s=1}^2 \int \widetilde{dk}\; \omega [b^{\dagger}_s(\bar k)
b_s(\bar k)+ d^{\dagger}_s(\bar k)d_s(\bar k)] - \lambda
\label{eq:fermionhamiltonian}
\end{eqnarray}
where $\lambda$ is an infinite constant we can merely subtract off and
therefore ignore.  

Comparing (\ref{eq:scalarhamiltonian}) and
(\ref{eq:fermionhamiltonian}), we see that they both have essentially
the same form; $\omega$ (which is energy) to the left of the creation
operator, which is to the left of the annihilation operator.  To
understand the meaning of this, we will see how it generates energy
eigenvalues.  We will use equation (\ref{eq:scalarhamiltonian}) for
simplicity.  Consider acting with the Hamiltonian operator on some
arbitrary state $|\bar p\rangle$ with momentum $\bar p$.  Using
(\ref{eq:possiblescalarstates}), 
\begin{eqnarray}
H|\bar p\rangle &=& \int \widetilde{dk}\omega_ka^{\dagger}(\bar
k)a(\bar k)|\bar p\rangle = \int \widetilde{dk}\omega_k
a^{\dagger}(\bar k)a(\bar k)\sqrt{2\omega_p}a^{\dagger}(\bar
p)|0\rangle \nolabel \\
&=& \int \widetilde{dk}\omega_k \sqrt{2\omega_p}a^{\dagger} (\bar k)
\big((2\pi)^3 2\omega_p\delta^3(\bar k - \bar p) + a^{\dagger}(\bar
p)a(\bar k)\big)|0\rangle \nolabel \\
&=& \int \widetilde{dk} \omega_k\sqrt{2\omega_p}a^{\dagger}(\bar
k)(2\pi)^32\omega_p\delta^3(\bar k - \bar p)|0\rangle \nolabel \\
&=& \int {d^3\bar k \over (2\pi)^3 2\omega_k} \omega_k
\sqrt{2\omega_p}a^{\dagger}(\bar k)(2\pi)^32\omega_p \delta^3(\bar k -
\bar p)|0\rangle \nolabel \\
&=& \int d^3\bar k \sqrt{2\omega_p}a^{\dagger}(\bar k )\omega_p
\delta^3(\bar k - \bar p)|0\rangle \nolabel \\
&=& \omega_p \sqrt{2\omega_p}a^{\dagger}|0\rangle = \omega_p|\bar
p\rangle\nolabel
\end{eqnarray}
So, $H|\bar p\rangle = \omega_p|\bar p\rangle$, where $\omega_p = \bar
p^2 + m^2$, which is the relativistic equation for energy as in
equation (\ref{eq:defomega}).  So, the Hamiltonian operator gives the
appropriate energy eigenvalue on our physical quantum states.  

For the Dirac Hamiltonian the eigenvalue will be a linear combination
of the energies of each type of particle.  If we denote the states as
$|\bar p_b, \; s_b ; \bar p_d, \; s_d \rangle$, where the first two
elements give the state of a $b$ type particle and the second of the
$d$ type particle, we have 
\begin{eqnarray}
H|\bar p_b, \; s_b ; \bar p_d, \; s_d \rangle = \cdots =
(\omega_{p_b}+\omega_{p_d})|\bar p_b, \; s_b ; \bar p_d, \; s_d
\rangle\nolabel
\end{eqnarray}

For Majorana fields things are simpler.  We only have one type of
particle, so 
\begin{eqnarray}
\psi_M(x) = \sum_{s=1}^2 \int \widetilde{dk}\big[b_s(\bar k)u_s(\bar
k)e^{ik\cdot x} + b^{\dagger}_s(\bar k)v_s(\bar k)e^{-ik\cdot
x}\big]\nolabel
\end{eqnarray}
And quantization with anticommutation relations will give
\begin{eqnarray}
H = \sum_{s=1}^2\int \widetilde{dk}\; \omega \; b^{\dagger}_s(\bar k
)b_s(\bar k )\nolabel
\end{eqnarray}

\subsubsection{Insufficiencies of Canonical Quantization}

While the Canonical Quantization procedure we have carried out in the
past several sections has given us a tremendous amount of information
(the entire spectrum of states for bosons, Dirac fermions, and Majorana
fermions), it is still lacking quite a bit.  As we said at the
beginning of section \ref{sec:test}, we ultimately want a relativistic
quantum mechanical theory of interactions.  Canonical Quantization has
provided a relativistic quantum mechanical theory, but we aren't close
to being able to incorporate interactions into our theory.  While it is
possible to incorporate interactions, it is very difficult, and in
order to simplify we will need a new way of quantizing.  

\subsubsection{Path Integrals and Path Integral Quantization}

Perhaps the most fundamental experiment in quantum mechanics is the \bf
Double Slit \rm experiment.  In brief, what this experiment tells us is
that, when a single electron moves through a screen with two slits, and
no observation is made regarding which slit it goes through, it
actually goes through \it both \rm slits, and until a measurement is
made (for example, when it hits the observation screen behind the
double slit), it exists in a superposition of \it both \rm paths.  As a
result, the particle exhibits a wave nature, and the pattern that
emerges on the observation screen is an interference pattern---the same
as if a classical wave was passing through the double slit--all paths
in the superposition of the single electron are interfering with each
other, both destructively and constructively.  Once the electron is
observed on the observation screen, it collapses probabilistically into
one of its possible states (a particular location on the observation
screen).  

If, on the other hand, you set up some mechanism to observe \it which
\rm of the two slits the electron travels through, then the observation
has been made \it before \rm the observation screen, and you no longer
have the superposition, and therefore you no longer see any indication
of an interference pattern.  The electrons are behaving, in a sense,
classically from the double slit to the observation screen in this
case.  

The meaning of this is that a particle that has not been observed will
actually take every possible path at once.  Once an observation has
been made, there is some probability associated with each path.  Some
paths are very likely, and others are less likely (some are nearly
impossible).  But until observation, it actually exists in a
superposition of all possible states/paths.

So, to quantize, we will create a mathematical expression for a ``sum
over all possible paths".  This expression is called a \bf Path
Integral\rm, and will prove to be a much more useful way to quantize a
physical system.  

We begin this construction by considering merely the amplitude for a
particle at position $q_1$ at time $t_1$ to propagate to $q_2$ at time
$t_2$.	This amplitude will be given by 
\begin{eqnarray}
\langle q_2,t_2|q_1,t_1\rangle = \langle
q_2|e^{iH(t_2-t_1)}|q_1\rangle\nolabel
\end{eqnarray}
To evaluate this, we begin by dividing the time interval $T \equiv
t_2-t_1$ into $N+1$ equal intervals of length $\delta t = {T \over
N+1}$ each.  So, we can insert $N$ complete sets of position
eigenstates,
\begin{eqnarray}
\langle q_2,t_2|q_1,t_1\rangle = \int_{-\infty}^{\infty}
\prod_{i=1}^NdQ_i\langle q_2|e^{-iH\delta t}|Q_N\rangle \langle
Q_N|e^{-iH\delta t}|Q_{N-1}\rangle \cdots \langle Q_1|e^{-iH\delta
t}|q_1\rangle \label{eq:firstpathintegral}
\end{eqnarray}

Let's look at a single one of these amplitudes.  We know that in nearly
all physical theories, we can break the Hamiltonian up as $H = {P^2
\over 2m} + V(Q)$.  So, using the completeness of momentum eigenstates, 
\begin{eqnarray}
\langle Q_{i+1}|e^{-iH\delta t}|Q_i\rangle &=& \langle Q_{i+1} |
e^{-i\big({P^2 \over 2m}+ V(Q)\big)\delta t}|Q_i\rangle \nolabel \\
&=& \langle Q_{i+1}|e^{-i\delta t {P^2 \over 2m}}e^{-i\delta t
V(Q)}|Q_i\rangle \nolabel \\
&=& \int dP' \langle Q_{i+1}|e^{-i\delta t{P^2 \over 2m}}|P'\rangle
\langle P'|e^{-i\delta t V(Q)}|Q_i\rangle \nolabel \\
&=& \int dP' e^{-i\delta t {P'^2 \over 2m}}e^{-i\delta t V(Q_i)}\langle
Q_{i+1}|P'\rangle \langle P'|Q_i\rangle \nolabel \\
&=&  \int dP' e^{-i\delta t {P'^2 \over 2m}} e^{-i\delta t V(Q_i)}
{e^{iP' Q_{i+1}} \over \sqrt{2\pi}} {e^{-iP' Q_i} \over \sqrt{2\pi}}
\nolabel \\
&=& \int {dP' \over 2\pi} e^{iH\delta t}e^{iP'(Q_{i+1} - Q_i)} \nolabel
\\
&=& \int {dP' \over 2\pi} e^{i\big[ P' (Q_{i+1} - Q_i) - H \delta
t\big]} \nolabel \\
&=& \int {dP' \over 2\pi} e^{i\delta t\big[ P' \big({Q_{i+1} - Q_i
\over \delta t}\big) - H \big]}\nolabel
\end{eqnarray}
And taking the limit as $\delta t \rightarrow 0$, ${Q_{i+1} - Q_i \over
\delta t} \rightarrow \dot{Q}_i$.  So, 
\begin{eqnarray}
\int {dP' \over 2\pi} e^{i\delta t\big[ P' \big({Q_{i+1} - Q_i \over
\delta t}\big) - H \big]} = \int {dP' \over 2\pi} e^{i dt_{i+1}
[P'\dot{Q}_i - H]}\nolabel
\end{eqnarray}
where the subscript on $dt$ merely indicates where the infinitesimal
time interval ``ends".	So, we can plug this into
(\ref{eq:firstpathintegral}) and taking the limit as $\delta t
\rightarrow 0$, 
\begin{eqnarray}
& & \langle q_2,t_2|q_1,t_1\rangle = \int_{-\infty}^{\infty}
\prod_{i=1}^NdQ_i\langle q_2|e^{-iH\delta t}|Q_N\rangle \langle
Q_N|e^{-iH\delta t}|Q_{N-1}\rangle \cdots \langle Q_1|e^{-iH\delta
t}|q_1\rangle \nolabel \\
& & \quad = \lim_{N\rightarrow \infty}
\int_{-\infty}^{\infty}\prod_{i=1}^N dQ_i \int {dP'_i \over 2\pi} e^{i
dt_2[P'_N\dot{Q}_N - H ]} e^{idt_N[P'_{N-1}\dot{Q}_{N-1} - H ]} \cdots
e^{idt_1[P'_1 \dot{Q}_1 - H ]} \nolabel \\
& & \quad \int_{-\infty}^{\infty} \mathcal{D} p \mathcal{D}q\;
e^{-i\int_{t_1}^{t_2} dt(p\dot{q} - H)}\nolabel
\end{eqnarray}
where $\mathcal{D}p = \prod_{i=1}^{\infty} dp_i$ and $\mathcal{D}q =
\prod_{i=1}^{\infty}dq_i$.  

And if $p$ shows up quadratically (as it always does; ${p^2 \over
2m}$), then we can merely do the Gaussian integral over $p$, resulting
in an overall constant which we merely absorb back into the measure
when we normalize.  Then, recognizing that the integrand in the
exponent is $p\dot{q} - H = \mathcal{L}$, we have 
\begin{eqnarray}
\langle q_2,t_2|q_1,t_1\rangle = \int \mathcal{D}q \; e^{i
\int_{t_1}^{t_2} dt \mathcal{L}} = \int \mathcal{D}qe^{iS}
\label{eq:pathintegral}
\end{eqnarray}

Formally, the measure of (\ref{eq:pathintegral}) has an infinite number
of differentials, and therefore evaluating it would require doing an
infinite number of integrals.  This is to be expected, since the point
of the path integral is a sum over every possible path, of which there
are an infinite number.  So, because we obviously can't do an infinite
number of integrals, we will have to find a clever way of evaluating
(\ref{eq:pathintegral}).  But before doing so, we discuss what the path
integral means.  

\subsubsection{Interpretation of the Path Integral}

Equation (\ref{eq:pathintegral}) says that, given an initial and final
configuration $(q_1,t_1)$ and $(q_2,t_2)$, absolutely \it any \rm path
between them is possible.  This is the content of the $\mathcal{D}q$
part:  it is the sum over all paths.  

Then, for each of those paths, the integral assigns a \it statistical
weight \rm of $e^{iS}$ to it, where the action $S$ is calculated using
\it that \rm path (recall our comments in section
\ref{sec:hamiltonsprinciple} about $S$ being a functional, not a
function).  

So, consider an arbitrary path $q_0$, which receives statistical weight
$e^{iS[q_0]}$.	Now, consider a path $q'$ very close to $q_0$, only
varying by a small amount: $q' = q_0 + \epsilon \delta q_0$.  This will
have statistical weight $e^{iS[q_0 + \epsilon \delta q_0]} = e^{iS[q_0]
+ i\epsilon \delta q_0 {\delta S[q_0] \over \delta q}}$, where ${\delta
S \over \delta q}$ is the Euler Lagrange derivative
(\ref{eq:eulerlagrange})
\begin{eqnarray}
{\delta S \over \delta q} = {d \over dt} {\partial S \over \partial
\dot{q}} - {\partial S \over \partial q}\nolabel
\end{eqnarray}

To make our intended result more obvious, we do a Wick rotation, taking
$t \rightarrow it$, so $dt \rightarrow idt$, and $S = \int dt
\mathcal{L} \rightarrow i\int dt \mathcal{L} = iS$, and $e^{iS}
\rightarrow e^{-S}$.  Now, the path $q' = q_0 + \epsilon \delta q_0$
gets weight $e^{iS[q_0]}e^{-i \epsilon \delta q_0 {\delta S[q_0] \over
\delta q}}$.  

So, if ${\delta S \over \delta q}$ is very large, then the weight
becomes exponentially small.  In other words, the larger the variation
of the action is, the less probable that path is.  

So the \it most \rm probable path is the one for the \it smallest \rm
value of ${\delta S \over \delta t}$, or the path at which ${\delta S
\over \delta q} = 0$.  And as we discussed in
\ref{sec:hamiltonsprinciple}, this is the path of \bf Least Action\rm. 
Thus, we have recovered classical mechanics as the first order
approximation of quantum mechanics.  

So, the meaning of the path integral is that all imaginable paths are
possible for the particle to travel in moving from one configuration to
another.  However, not all paths are equally probable.	The likelihood
of a given path is given by the action exponentiated, and therefore the
most probable paths are the ones which minimize the action.  This is
the reason that, macroscopically, the world appears classical.	The
likelihood of every particle in, say, a baseball, simultaneously taking
a path noticeably far from the path of least action is negligibly
small.	

We will find that path integral quantization provides an extremely
powerful tool with which to create our relativistic quantum theory of
interactions.  

\subsubsection{Expectation Values}
\label{sec:expectationvalues}

Now that we have a way of finding $\langle q_2,t_2|q_1,t_1\rangle$, the
natural question to ask next is how do we find expectation values like
$\langle q_2,t_2|Q(t')|q_1,t_1\rangle$ or $\langle
q_2,t_2|P(t')|q_1,t_1\rangle$.	By doing a similar derivation as in the
last section, it is easy to show that 
\begin{eqnarray}
\langle q_2,t_2|Q(t')|q_1,t_1\rangle = \cdots = \int \mathcal{D}q\;
Q(t') e^{iS}\nolabel
\end{eqnarray}

We will find that evaluating integrals of this form is simplified
greatly through making use of \bf Functional Derivatives\rm.  For some
function $f(x)$, the functional derivative is defined by 
\begin{eqnarray}
{\delta \over \delta f(y)} f(x) \equiv \delta (x-y)\nolabel
\end{eqnarray}

Next, we modify our path integral by adding an \bf Auxiliary External
Source \rm function, so that 
\begin{eqnarray}
\mathcal{L} \rightarrow \mathcal{L} + f(t)Q(t) + h(t)P(t)\nolabel
\end{eqnarray}
So we now have 
\begin{eqnarray}
\langle q_2,t_2|q_1,t_1\rangle_{f,h} = \int \mathcal{D}q \; e^{\int dt
(\mathcal{L} + fQ + hP)}\nolabel
\end{eqnarray}
which allows us to write out expectation values in the simple form
\begin{eqnarray}
\langle q_2,t_2|Q(t')|q_1,t_1\rangle &=& {1\over i} {\delta \over
\delta f(t')} \langle q_2,t_2|q_1,t_1\rangle_{f,h} \bigg|_{f,h =0} =
\int \mathcal{D}q Q(t')e^{iS + i \int dt(fQ+hP)} \bigg|_{f,h = 0}
\nolabel \\
&=& \int \mathcal{D}q Q(t')e^{iS}\nolabel
\end{eqnarray}
or
\begin{eqnarray}
\langle q_2,t_2|P(t')|q_1,t_1\rangle &=& {1\over i} {\delta \over
\delta h(t')} \langle q_2,t_2|q_1,t_1\rangle_{f,h} \bigg|_{f,h = 0} =
\int \mathcal{D}q P(t')e^{iS+i\int dt(fQ+hP)}\bigg|_{f,h=0} \nolabel \\
&=& \int \mathcal{D} q P(t')e^{iS}\nolabel
\end{eqnarray}

So, once we have $\langle q_2,t_2|q_1,t_1\rangle$, we can find any
expectation value we want simply by taking successive functional
derivatives.  

\subsubsection{Path Integrals with Fields}

Because we can build whatever state we want by acting on the vacuum,
the important quantity for us to work with will be the \bf Vacuum to
Vacuum \rm expectation value, or VEV, $\langle 0 | 0 \rangle$, and
the various expectation values we can build through functional
derivatives ($\langle 0 | \phi \phi|0 \rangle$, $\langle 0 | \psi \phi
\phi |0\rangle$, etc.).  

For simplicity let's consider a scalar boson $\phi$.  The Lagrangian is
given in equation (\ref{eq:kleingordonlagrangian}).  Using this, we can
write the path integral 
\begin{eqnarray}
\langle 0 | 0 \rangle = \int \mathcal{D} \phi e^{i\int d^4x
\big[-{1\over 2} \partial^{\mu}\phi \partial_{\mu} \phi - {1\over 2}
m^2\phi^2 \big]} \equiv \int \mathcal{D}\phi e^{i\int d^4x
\mathcal{L}_0}\nolabel
\end{eqnarray}

We will eventually want to find expectation values, so we introduce the
auxiliary field $J$, creating
\begin{eqnarray}
\langle 0 | 0 \rangle_J = \int \mathcal{D}\phi e^{i\int d^4x
(\mathcal{L}_0 + J \phi)} \label{eq:vev1}
\end{eqnarray}
So, for example, $\langle 0 | \phi | 0 \rangle = {1\over i} {\delta
\over \delta J} \langle 0 | 0 \rangle_J \big|_{J=0}$.  

Of course, we still have a path integral with an infinite number of
integrals to evaluate.	But, we are finally able to discuss how we can
do the evaluation.  

We define $Z_0(J) \equiv \langle 0 | 0 \rangle_J$.  Then, making use of
the Fourier Transform of $\phi$, 
\begin{eqnarray}
\widetilde{\phi}(k) = \int d^4x \; e^{-ikx}\phi(x) \qquad \phi(x) =
\int {d^4k \over (2\pi)^4} e^{ikx} \widetilde{\phi}(k)\nolabel
\end{eqnarray}
we begin with the $\mathcal{L}_0$ part:
\begin{eqnarray}
S_0 &=& \int d^4x \mathcal{L}_0 = \int d^4x \bigg(-{1\over 2}
\partial^{\mu}\phi \partial_{\mu}\phi - {1\over 2}m^2\phi^2 \bigg)
\nolabel \\
&=& \int d^4x \bigg[-{1\over 2} \partial^{\mu}\bigg(\int {d^4k \over
(2\pi)^4} e^{ik\cdot x}\widetilde{\phi}(k)\bigg)
\partial_{\mu}\bigg(\int {d^4k' \over (2\pi)^4} e^{ik'\cdot x}
\widetilde{\phi}(k')\bigg) \nolabel \\
& & \qquad \quad \;\; -{1\over 2}m^2\bigg(\int {d^4k \over (2\pi)^4}
e^{ik\cdot x}\widetilde{\phi}(k)\bigg)\bigg(\int {d^4k' \over (2\pi)^4}
e^{ik'\cdot x} \widetilde{\phi}(k')\bigg) \nolabel \\
&=& \int d^4x \bigg[{1\over 2} \int {d^4kd^4k' \over (2\pi)^8}
e^{ik\cdot x}e^{ik'\cdot x} \widetilde{\phi}(k) \widetilde{\phi}(k')
(k^{\mu}k'_{\mu} - m^2)\bigg] \nolabel \\
&=& {1\over 2} \int {d^4k d^4k' \over (2\pi)^8} \widetilde{\phi}(k)
\widetilde{\phi}(k') (k^{\mu}k'_{\mu} - m^2)\int d^4xe^{i(k+k')\cdot x}
\nolabel \\
&=& {1\over 2}\int {d^4k d^4k' \over (2\pi)^8}\widetilde{\phi}(k)
\widetilde{\phi}(k') (k^{\mu}k'_{\mu} - m^2) (2\pi)^4 \delta^4(k+k')
\nolabel \\
&=& -{1\over 2} \int {d^4k \over (2\pi)^4} \widetilde{\phi}(k)
(k^2+m^2)\widetilde{\phi}(-k) \nolabel
\end{eqnarray}
Then, transforming the auxiliary field part,
\begin{eqnarray}
\int d^4x J(x) \phi(x) &=& \int d^4x\bigg(\int {d^4k \over (2\pi)^4}
e^{ik\cdot x}\widetilde{J}(k)\bigg)\bigg(\int {d^4k' \over (2\pi)^4}
e^{ik' \cdot x} \widetilde{\phi}(k')\bigg) \nolabel \\
&=& \int {d^4k d^4k' \over (2\pi)^8} \widetilde{J}(k)
\widetilde{\phi}(k') \int d^4x e^{i(k+k')\cdot x} \nolabel \\
&=& \int {d^4kd^4k' \over (2\pi)^8} \widetilde{J}(k)
\widetilde{\phi}(k') (2\pi)^4 \delta^4(k+k') \nolabel \\
&=& \int {d^4k \over (2\pi)^4} \widetilde{J}(k)
\widetilde{\phi}(-k)\nolabel
\end{eqnarray}
And because the integral is over all $k^{\mu}$, we can rewrite this as 
\begin{eqnarray}
\int {d^4k \over (2\pi)^4} \widetilde{J}(k) \widetilde{\phi}(-k) =
{1\over 2} \int {d^4k \over (2\pi)^4} \big(\widetilde{J}(k)
\widetilde{\phi}(-k) +
\widetilde{J}(-k)\widetilde{\phi}(k)\big)\nolabel
\end{eqnarray}
(we did this to get the factor of $1/2$ out front in order to have the
same coefficient as the $\mathcal{L}_0$ part from above).  

So, 
\begin{eqnarray}
S = {1\over 2} \int {d^4k \over (2\pi)^4} \bigg[-\widetilde{\phi}(k)
(k^2 + m^2)\widetilde{\phi}(-k)+\widetilde{J}(k)\widetilde{\phi}(-k) +
\widetilde{J}(-k)\widetilde{\phi}(k)\bigg] \nolabel
\end{eqnarray}

Now, we make a change of variables,
\begin{eqnarray}
\widetilde{\chi}(k) \equiv \widetilde{\phi}(k) - {\widetilde{J}(k)
\over k^2 + m^2}\nolabel
\end{eqnarray}
(Note that this leaves the measure of the path integral unchanged:
$\mathcal{D}\phi \rightarrow \mathcal{D}\chi$.)  

Plugging this in, we have, 
\begin{eqnarray}
S = {1\over 2} \int {d^4k \over (2\pi)^4} \bigg[ &-&
\bigg(\widetilde{\chi}(k) + {\widetilde{J}(k) \over
k^2+m^2}\bigg)(k^2+m^2)\bigg(\widetilde{\chi}(-k) + {\widetilde{J}(-k)
\over k^2+m^2}\bigg) \nolabel \\
&+& \widetilde{J}(k)\bigg(\widetilde{\chi}(-k) + {\widetilde{J}(-k)
\over k^2+m^2}\bigg) + \widetilde{J}(-k) \bigg(\widetilde{\chi}(k) +
{\widetilde{J}(k) \over k^2+m^2} \bigg) \nolabel \\
&=& {1\over 2} \int {d^4k \over (2\pi)^4}
\bigg[-\widetilde{\chi}(k)(k^2 +
m^2)\widetilde{\chi}(-k)+{\widetilde{J}(k)\widetilde{J}(-k) \over
k^2+m^2}\bigg]	\nolabel
\end{eqnarray}
(The point of all of this is that, in this form, we have all of the
$\phi$, or equivalently $\chi$, dependence in the first term, with no
$\phi$ or $\chi$ dependence on the second term.)  

Finally, our path integral (\ref{eq:vev1}) is
\begin{eqnarray}
\langle 0 | 0 \rangle_J = \int \mathcal{D}\chi e^{{i\over 2} \int {d^4k
\over  (2\pi)^4}\big[-\widetilde{\chi}(k)(k^2+m^2)\widetilde{\chi}(-k)
+ {\widetilde{J}(k) \widetilde{J}(-k) \over k^2+m^2 } \big]}\nolabel
\end{eqnarray}

Now, using some clever physical reasoning, we can see how to evaluate
the infinite number of integrals in this expression.  Notice that if we
set $J=0$, we have a free theory in which no interactions take place. 
This means that if we start with nothing (the vacuum), the probability
of having nothing later is 100\%.  Or,
\begin{eqnarray}
\langle 0 | 0 \rangle_J \big|_{J=0} = 1 =   \int \mathcal{D}\chi
e^{{i\over 2} \int {d^4k \over (2\pi)^4}
\big[-\widetilde{\chi}(k)(k^2+m^2)\widetilde{\chi}(-k)\big]}\nolabel
\end{eqnarray}
And if that part is 1, then we have 
\begin{eqnarray}
\langle 0 | 0 \rangle_J = \int \mathcal{D}\chi e^{{i \over 2} \int
{d^4k \over (2\pi)^4}{\widetilde{J}(k) \widetilde{J}(-k) \over
k^2+m^2}}\nolabel
\end{eqnarray}

And remarkably, the integrand \it has no $\chi$ dependence\rm! 
Therefore, the infinite number of integrals over all possible paths
becomes nothing more than a constant we can absorb into the
normalization, leaving 
\begin{eqnarray}
\langle 0 | 0 \rangle_J =e^{{i \over 2} \int {d^4k \over
(2\pi)^4}{\widetilde{J}(k) \widetilde{J}(-k) \over k^2+m^2}}\nolabel
\end{eqnarray}
We can Fourier Transform back to coordinate space to get 
\begin{eqnarray}
Z_0(J) = \langle 0 | 0 \rangle_J = e^{{i \over 2} \int d^4x d^4x' J(x)
\Delta(x-x')J(x')} \label{eq:z0boson}
\end{eqnarray}
where 
\begin{eqnarray}
\Delta(x-x') \equiv \int {d^4k \over (2\pi)^4} {e^{i k\cdot (x-x')}
\over k^2+m^2}\nolabel
\end{eqnarray}
is called the \bf Feynman Propagator \rm for the scalar field.	

We can then find expectation values by operating on this with ${1\over
i}{\delta \over \delta J}$ as described in section
\ref{sec:expectationvalues}.  

We can repeat everything we have just done for fermions, and while it
is a great deal more complicated (and tedious), it is in essence the
same calculation.  We begin by adding the auxiliary function $\bar \eta
\psi + \bar \psi \eta$, to get expectation values of $\bar \psi$ and
$\psi$ by using ${1\over i} {\delta \over \delta \eta}$ and ${1\over i}
{\delta \over \delta \bar \eta}$, respectively.  

We then Fourier Transform every term in the exponent and find that we
can separate out the $\bar \psi$ and $\psi$ dependence, allowing us to
set the term which does depend on $\psi$ and $\bar \psi$ equal to 1. 
Fourier Transforming back then gives
\begin{eqnarray}
Z_0(\eta,\bar \eta) = e^{i \int d^4x d^4x' \bar \eta(x)S(x-x')\eta(x')}
\label{eq:z0fermion}
\end{eqnarray}
where 
\begin{eqnarray}
S(x-x') = \int {d^4k \over (2\pi)^4} {(-\gamma^{\mu}k_{\mu} +
m)e^{ik\cdot (x-x')} \over k^2 + m^2}\nolabel
\end{eqnarray}
is the Feynman propagator for fermion fields.  

Recall that we are calling the auxiliary fields $J$, $\eta$, and $\bar
\eta$ \bf Source Fields\rm.  Comparing the form of the Lagrangian in
equation (\ref{eq:vev1}) to (\ref{eq:emlagrangian}) reveals why.  $J$,
$\eta$, and $\bar \eta$ behave mathematically as sources, giving rise
to the field they are coupled to, in the same way that the
electromagnetic source $J^{\mu}$ gives rise to the electromagnetic
field $A^{\mu}$.  The meaning behind equations (\ref{eq:z0boson}) (and
(\ref{eq:z0fermion})) is that $J$ (or $\eta$ and $\bar \eta$) act as
sources for the fields, creating a $\phi$ (or $\psi$ and $\bar \psi$)
at spacetime point $x$, and absorbing it at point $x'$.  The terms
$\Delta(x-x')$ and $S(x-x')$ then represent the expression giving the
probability amplitude $\langle 0 | 0 \rangle$ for that particular event
to occur.  In other words, the propagator is the statistical weight of
a particle going from $x$ to $x'$.  

\subsubsection{Interacting Scalar Fields and Feynman Diagrams}
\label{sec:feynmandiag}

We can now consider how to incorporate interactions into our formalism,
allowing us to finally have our relativistic quantum theory of
interactions.  

Beginning with the free scalar Lagrangian
(\ref{eq:kleingordonlagrangian}), we can add an interaction term
$\mathcal{L}_1$.  At this point, we only have one type of particle,
$\phi$, so we can only have $\phi$'s interacting with other $\phi$'s. 
Terms proportional to $\phi$ or $\phi^2$ are either constant or linear
in the equations of motion, and therefore aren't valid candidates for
interaction terms.  So, the simplest expression we can have is 
\begin{eqnarray}
\mathcal{L}_1 = {1\over 3!} g \phi^3\nolabel
\end{eqnarray}
where ${1\over 3!}$ is a conventional normalization, and $g$ is a \bf
Coupling Constant\rm.  So our total Lagrangian is 
\begin{eqnarray}
\mathcal{L} = \mathcal{L}_0 + \mathcal{L}_1 = -{1\over 2}
\partial^{\mu} \phi \partial_{\mu} \phi - {1\over 2} m^2 \phi^2 +
{1\over 6}g \phi^3\nolabel
\end{eqnarray}
and the path integral is
\begin{eqnarray}
Z(J) &=& \langle 0 | 0 \rangle_J \nolabel \\
&=& \int \mathcal{D}\phi e^{i \int d^4x [\mathcal{L}_0 + \mathcal{L}_1
+ J\phi]} \nolabel \\
&=& \int \mathcal{D}\phi e^{i\int d^4x \mathcal{L}_1}e^{i \int d^4x
[\mathcal{L}_0 + J \phi]} \nolabel \\
&=& \int \mathcal{D}\phi e^{i\int d^4x \mathcal{L}_1} Z_0(J) = \int
\mathcal{D}\phi e^{i\int d^4x\mathcal{L}_1} \langle 0 | 0
\rangle_J\nolabel
\end{eqnarray}

But, recall that we can bring out a factor of $\phi$ from $\langle 0 |
0 \rangle_J$ using the functional derivative ${1\over i} {\delta \over
\delta J}$.  So, we can make the replacement
\begin{eqnarray}
\mathcal{L}_1(\phi) \rightarrow \mathcal{L}_1\bigg({1\over i}{\delta
\over \delta J}\bigg) \Rightarrow {1\over 6} g \phi^3  \rightarrow {g
\over 6}\bigg({1\over i}{\delta \over \delta J}\bigg)^3\nolabel
\end{eqnarray}
And notice that once this is done, there is no longer any $\phi$
dependence in $Z(J)$.  So, with the free theory, we were able to remove
the $\phi$ dependence, leading to (\ref{eq:z0boson}).  And here, we
were able to remove it from the interaction term as well.  So, once
again, the infinite number of integrals in (\ref{eq:pathintegral}) will
merely give a constant which we can absorb into the normalization.  

This leaves the result
\begin{eqnarray}
Z(J) &=& e^{{i\over 6}g \int d^4x \big({1\over i}{\delta \over \delta
J(x)}\big)^3}Z_0(J) \nolabel \\
&=& e^{-{1\over 6} g \int d^4x \big({\delta \over \delta
J(x)}\big)^3}e^{{i \over 2} \int d^4x d^4x' J(x)
\Delta(x-x')J(x')}\nolabel
\end{eqnarray}

Now, we can do two separate Taylor expansions to these two
exponentials, 
\begin{eqnarray}
Z(J) = \sum_{V=0}^{\infty} {1\over V!} \bigg[-{g\over 6} \int d^4x
\bigg({\delta \over \delta J(x)}\bigg)^3 \bigg]^V \times
\sum_{P=0}^{\infty}{1\over P!}\bigg[{i\over 2} \int
d^4yd^4zJ(y)\Delta(y-z)J(z)\bigg]^P \nolabel \\
\label{eq:doubleexponential}
\end{eqnarray}

Now, recall that a functional derivative ${1\over i} {\delta \over
\delta J}$, will remove a $J$ term.  Furthermore, after taking the
functional derivatives, we will set $J=0$ to get the physical result. 
So, for a term to survive, the $2P$ sources must all be exactly removed
by the $3V$ functional derivatives.  

So, using (\ref{eq:doubleexponential}), we can expand in orders of $g$
(the coupling constant), keeping only the terms which survive, and
after removing the sources, evaluate the integrals over the propagators
$\Delta$.  The value of the integral will then be the physical
amplitude for a particular event.  

In practice, a slightly different formalism is used to organize and
keep track of each term in this  expansion.  Note that there will be
$P$ propagators $\Delta$.  We can represent each of these terms
diagrammatically, by making each source a solid dot, each propagator a
line, and let the $g$ terms be vertices joining the lines together. 
There will be a total of $V$ vertices, each joining 3 lines (matching
the fact that we are looking at $\phi^3$ theory; there would be 4
lines at each vertex for $\phi^4$ theory, etc.).  

For example, for $V=0$ and $P=1$,
\begin{eqnarray}
Z(J) = {i\over 2} \int d^4yd^4z J(y) \Delta(y-z)J(z)\nolabel
\end{eqnarray}
We have two sources, one located at $z$ and the other located at $y$,
so we draw two dots, corresponding to those locations.	Then, the
propagator $\Delta(y-z)$ connects them together, so we draw a line
between the two dots.  The diagram should look like this:
\begin{center}
\includegraphics[scale = .2]{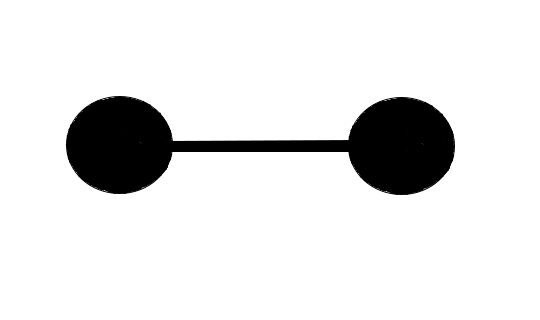}
\end{center}
Of course, once we set $J=0$, this will vanish because it contains two
sources.  

As another example, consider $V=0$ and $P=2$.  Now,
\begin{eqnarray}
Z(J) = {1\over 2!} \bigg({i\over 2}\bigg)^2 \int
d^4yd^4zd^4y'd^4z'\big(J(y)\Delta(y-z)J(z)\big)\big(J(y')
\Delta(y'-z')J(z')\big)\nolabel
\end{eqnarray}
This corresponds to four sources, located at $y,z,y'$ and $z'$, with
propagator lines connecting $y$ to $z$, and connecting $y'$ to $z'$. 
But, there are no lines connecting an unprimed source to a primed
source, so this results in two disconnected diagrams:
\begin{center}
\includegraphics[scale = .2]{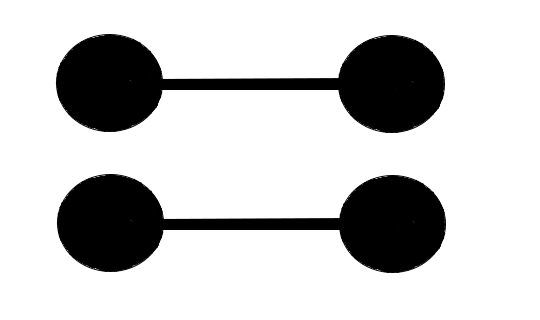}
\end{center}

As another example, consider $V=1$ and $P=2$,
\begin{eqnarray}
Z(J) &=& -{g\over 6} \int d^4x\bigg({\delta \over \delta J(x)}\bigg)^3
\nolabel \\
& & \times {1\over 2!} \bigg({i \over 2}\bigg)^2\int
d^4yd^4zd^4y'd^4z'\big(J(y)\Delta(y-z)J(z)\big)\big(J(y')
\Delta(y'-z')J(z')\big) \nolabel \\
&=& {g \over 48}\int
d^4xd^4yd^4zd^4y'd^4z'\delta(y-x)\Delta(y-z)\delta(z-x)\delta(y'-x)
\Delta(y'-z')J(z') \nolabel \\
&=& {g\over 48} \int d^4xd^4z' \Delta(x-x)\Delta(x-z') J(z')\nolabel
\end{eqnarray}
This will correspond to
\begin{center}
\includegraphics[scale = .2]{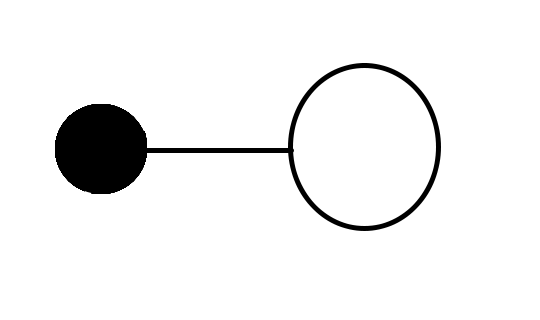}
\end{center}
where the source $J$ is located at the dot, and the vertex joining the
line to the loop is at $x$.  

You can work out the following out, and see that there are multiple
possible diagrams for $V=3,\; P=5$
\begin{center}
\includegraphics[scale = .2]{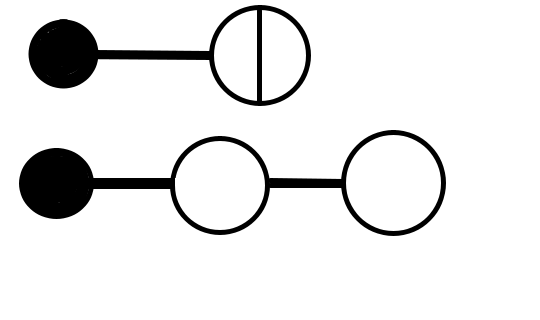}
\end{center}
\begin{center}
\includegraphics[scale = .2]{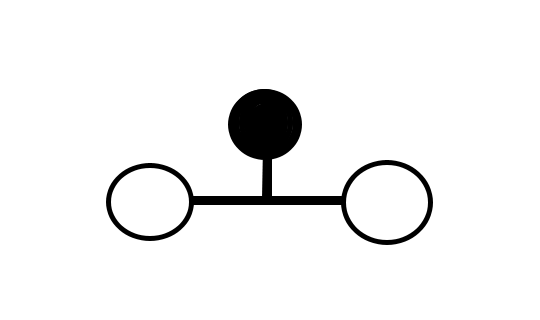}
\end{center}

And for $V=2$, $P=4$,
\begin{center}
\includegraphics[scale = .2]{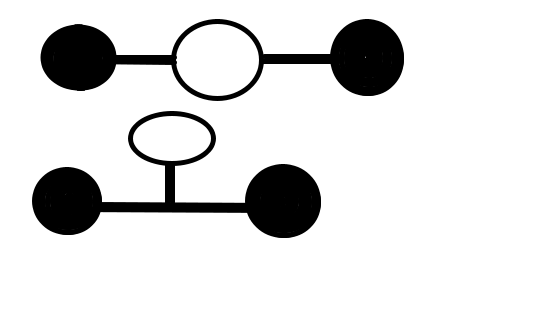}
\end{center}

And for $V=1$, $P=3$,
\begin{center}
\includegraphics[scale = .2]{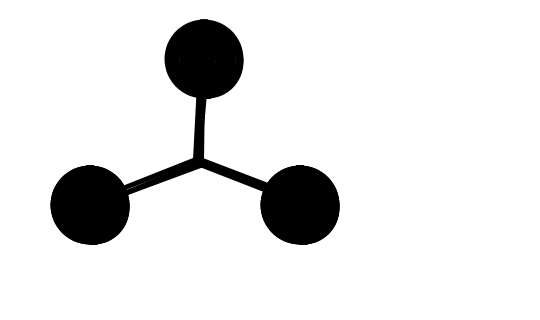}
\end{center}
and so on.  

Through a series of combinatoric and physical arguments, it can be
shown that only connected diagrams will contribute, and the ${1\over
P!}$ and ${1\over V!}$ terms will always cancel exactly.  

So, to calculate the amplitude for a particular interaction to happen
(say $N$ $\phi$'s in and $M$ $\phi$'s out), draw every connected
diagram that is topologically distinct, and has the correct number of
in and out particles.  Then, through a set of rules which you will
learn formally in a QFT course, you can reconstruct the integrals
which we started with in (\ref{eq:doubleexponential}).	

When you take a course on QFT, you will spend a tremendous amount of
time learning how to evaluate these integrals for low order (they
cannot be evaluated past about second order in most cases).  While this
is extremely important, it is not vital for the agenda of these notes,
and we therefore do not discuss how they are evaluated.  

The idea is that each diagram represents one of the possible paths the
particle can take, along with the possible interactions it can be a
part of.  Because this is a quantum mechanical theory, we know it is
actually in a superposition of all possible paths and interactions.  We
don't make a measurement or observation until the particles leave the
area in which they collide, so we have no idea about what is going on
inside the accelerator.  We know that if \it this \rm goes in and \it
this \rm comes out, we can draw a particular set of diagrams which have
the correct input and output, and the nature of the interaction terms
(which determines what types of vertices you can have) tells us what
types of interactions we can have inside the accelerator.  Evaluating
the integrals then tells us how much that particular event/diagram
contributes towards the total probability amplitude.  So, if you want
to know how likely a certain incoming/outgoing set of particles is,
write down all the diagrams, evaluate the corresponding integrals, and
add them up.  

And as we pointed out above, the classical behavior (which is more
probable) is closer to the first order approximation of the quantum
behavior.  Therefore, even though in general we can't evaluate the
integrals past about second order, the first few orders tell us to a
reasonable (in fact, exceptional in most cases) degree of accuracy what
the amplitude is.  If we want more accuracy, we can seek to evaluate
higher orders, but usually lower orders suffice for experiments at
energy levels we can currently attain.	

One of the difficulties encountered with evaluating these integrals is
that you almost always find that they yield infinite amplitudes.  Since
an amplitude (which is a probability) should be between 0 and 1,
this is obviously unacceptable.  The process of finding the infinite
parts and separating them from the finite parts of the amplitude is a
very well defined mathematical construct called \bf Renormalization\rm.
 The basic idea is that any infinite term consists of a pure infinity
and a finite part.  For example (trust us for now) the infinite sum:
\begin{eqnarray}
\sum_{n=1}^{\infty} n = \lim_{x\rightarrow 0} {1\over x^2} - {1\over
12}\nolabel
\end{eqnarray}
There is a part which is a pure infinity (the first term on the right
hand side), and a term which is finite.  While this may seem strange
and extremely unfamiliar (and a bit like hand waving), it is actually a
very rigorous and very well understood mathematical idea.  

Much of what particle physicists attempt to do is find theories (and
types of theories) that can be renormalized and theories that cannot.  For
example, the action which leads to General Relativity leads to a
quantum theory which cannot be renormalized.  Renormalization is a
fascinating and deep topic, and will be covered in great depth in any
standard QFT text or course.	Unfortunately, we will not discuss it
further.  

\subsubsection{Interacting Fermion Fields}

The analysis we performed above for scalar fields $\phi$ above is
almost identical for fermions, and we therefore won't repeat it.  The
main difference is that the interaction terms will have a field $\psi$
interacting with $\bar \psi$, and so the vertices will be slightly
different.  We won't bother with those details.  

Finally, we can have a Lagrangian with \it both \rm scalars and
fermions.  Then, naturally, you could have interaction terms where the
scalars interact with fermions.  While there are countless interaction
terms of this type, the one that will be the most interesting to us is
the \bf Yukawa \rm term,
\begin{eqnarray}
\mathcal{L}_{Yuk} = g\phi \bar \psi \psi \label{eq:yukawa}
\end{eqnarray}

If we represent $\phi$ by a dotted line, $\psi$ by a line with an arrow
in the forward time direction, and $\bar \psi$ with an arrow going
backwards in time, this interaction term will show up in a Feynman
diagram as 
\begin{center}
\includegraphics[scale = .4]{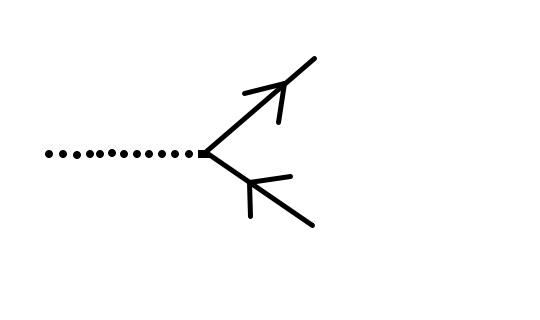}
\end{center} 

Once each diagram is drawn, there are well defined rules to write down
an integral corresponding to each diagram.  

\subsection{Final Ingredients}

The purpose of the previous section was merely to introduce the idea of
\bf Feynman Diagrams \rm as a tool to calculate amplitudes for physical
processes.  In doing so, we have met the goal set out in section
\ref{sec:test}, a relativistic quantum mechanical theory of
interactions.  We achieve such a theory by finding a Lagrangian of a
classical theory (both with and without interaction terms), and using
equation (\ref{eq:doubleexponential}) (and the analogous equation for
fermions) to write down integrals which, when evaluated, give a
contribution to a total amplitude.  It is important to remember that we
will eventually set all sources $J$ to zero, and a functional
derivative (as contained in the interaction term $\mathcal{L}_1$) will
set any term without $J$'s to zero.  So, the only non-zero terms will
be the ones where all of the $J$'s are exactly removed by the
functional derivatives.  

A large portion of understanding QFT is learning how to set these
integrals up in greater detail, and learning several methods to
evaluate them.	We will not delve into those details of \bf
Perturbative Quantum Field Theory\rm, where amplitudes are studied
order by order, here.  The goal of these notes is merely to explain
how, once given a Lagrangian, that Lagrangian can be turned into a
physically measurable quantity.  

With this done, we now set out to find the Lagrangian for the Standard
Model of Particle Physics, the theory which seems to explain our
universe (apart from gravity).	Once this Lagrangian has been
explained, we trust you have a general concept of what to do with it
from the previous sections.  

However, before we are able to explain the Standard Model Lagrangian,
there are a few final concepts we need.  They will be the subject of
this section.  Namely, we will be studying the ideas of \bf Spontaneous
Symmetry Breaking \rm and \bf Gauge Theories\rm.  In section
\ref{sec:gaugingthesymmetry}, we discussed the simple $U(1)$ gauge
theory, where we made a global $U(1)$ symmetry of the free Dirac
Lagrangian a local $U(1)$ symmetry, or a gauged symmetry, and showed
that consistency demanded the introduction of a gauge field $A^{\mu}$,
and consequently a kinetic term and a source term.  Thus we recovered
the entire electromagnetic force from nothing but $U(1)$.  Later in
this section, we generalize this to arbitrary Lie group.  Because
$U(1)$ is an Abelian group, we refer to the gauge theory of section
\ref{sec:gaugingthesymmetry} as an abelian gauge theory.  For a more
general, non-Abelian group, we refer to the theory resulting as a \bf
Non-Abelian Gauge Theory\rm.  Such theories introduce a great deal of
complexity, and we therefore consider them in detail in this section
before moving on to the Standard Model.  

However, we begin with the idea of spontaneous symmetry
breaking.  

\subsubsection{Spontaneous Symmetry Breaking}
\label{sec:spontaneoussymmetrybreaking}

Consider a complex scalar boson $\phi$ and $\phi^{\dagger}$.  The
Lagrangian will be 
\begin{eqnarray}
\mathcal{L} = -{1\over 2} \partial^{\mu}\phi^{\dagger}\partial_{\mu}
\phi - {1\over 2} m^2 \phi^{\dagger}\phi\nolabel
\end{eqnarray}
Naturally we can write this as 
\begin{eqnarray}
\mathcal{L}  = -{1\over 2}\partial^{\mu}\phi^{\dagger}\partial_{\mu}
\phi - V(\phi^{\dagger},\phi) \label{eq:morefundamental}
\end{eqnarray}
where 
\begin{eqnarray}
V(\phi^{\dagger},\phi) = {1\over 2}m^2\phi^{\dagger}\phi\nolabel
\end{eqnarray}
This Lagrangian has the $U(1)$ symmetry we discussed in
\ref{sec:gaugingthesymmetry}.  

Also, notice that we can graph $V(\phi^{\dagger},\phi)$, plotting $V$
vs. $|\phi|$,
\begin{center}
\includegraphics[scale = .5]{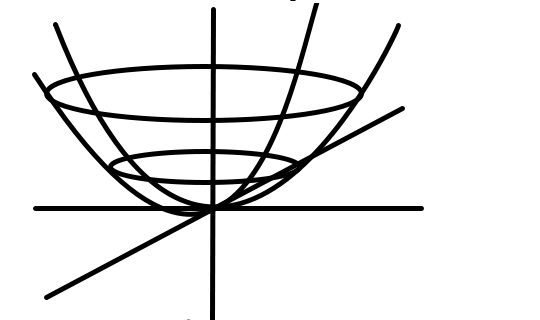}
\end{center}
We see a ``bowl" with $V_{minimum}$ at $|\phi|^2 = 0$.	The vacuum of
any theory ends up being at the lowest potential point, and therefore
the vacuum of this theory is at $\phi=0$, as we would expect.  

Now, let's change the potential.  Consider 
\begin{eqnarray}
V(\phi^{\dagger},\phi) = {1\over 2}\lambda m^2 (\phi^{\dagger} \phi -
\Phi^2)^2 \label{eq:degeneratevacuum}
\end{eqnarray}
where $\lambda$ and $\Phi$ are real constants.	Notice that the
Lagrangian will still have the global $U(1)$ symmetry from before. 
But, now if we graph $V$ vs. $|\phi |$, we get 
\begin{center}
\includegraphics[scale = .5]{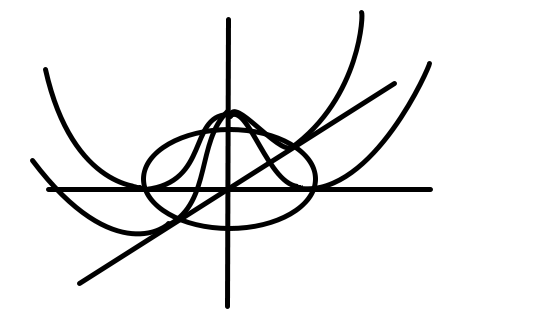}
\end{center}
where now the vacuum $V_{minimum}$ is represented by the circle at
$|\phi | = \Phi$.  In other words, there are an infinite number of
vacuums in this theory.  And because the circle drawn in the figure
above represents a rotation through field space, this degenerate vacuum
is parameterized by $e^{i\alpha}$, the global $U(1)$.  There will be a
vacuum for every value of $\alpha$, located at $|\phi| = \Phi$.  

In order to make sense of this theory, we must \it choose \rm a vacuum
by hand.  Because the theory is completely invariant under the choice
of the $U(1)$ $e^{i\alpha}$, we can choose any $\alpha$ and \it define
\rm that as our true vacuum.  So, we choose $\alpha$ to make our vacuum
at $\phi = \Phi$, or where $\phi$ is real and equal to $\Phi$.	We have
thus, in a sense, \bf Gauged Fixed \rm the symmetry in the Lagrangian,
and the $U(1)$ symmetry is no longer manifest.	

Now we need to rewrite this theory in terms of our new vacuum.	We
therefore expand around the constant vacuum value $\Phi$ to have the
new field
\begin{eqnarray}
\phi \equiv \Phi + \alpha + i \beta\nolabel
\end{eqnarray}
where $\alpha$ and $\beta$ are new real scalar fields (so
$\phi^{\dagger} = \Phi + \alpha - i\beta$).  We can now write out the
Lagrangian as
\begin{eqnarray}
\mathcal{L} &=& -{1\over 2} \partial^{\mu}[ \alpha - i
\beta]\partial_{\mu}[\alpha + i \beta] - {1\over 2}\lambda m^2[(\Phi +
\alpha - i \beta)(\Phi + \alpha + i \beta) - \Phi^2]^2 \nolabel \\
&=& \bigg[-{1\over 2} \partial^{\mu}\alpha \partial_{\mu}\alpha -
{1\over 2} 4 \lambda m^2\Phi^2\alpha^2 - {1\over 2}\partial^{\mu}\beta
\partial_{\mu}\beta \bigg] - {1\over 2}\lambda m^2\bigg[4\Phi \alpha^3
+ 4\Phi \alpha \beta^2 + \alpha^4 + \alpha^2\beta^2 + \beta^4\bigg]
\nolabel \\ \label{eq:lowenergyeff}
\end{eqnarray}
This is now a theory of a \it massive \rm real scalar field
$\alpha$ (with mass = $\sqrt{4\lambda m^2\Phi^2}$), a
\it massless \rm real scalar field $\beta$, and five different
types of interactions (one allowing three $\alpha$'s to interact, the
second allowing one $\alpha$ and two $\beta$'s, the third allowing four
$\alpha$'s, the fourth allowing two $\alpha$'s and two $\beta$'s, and
the last allowing four $\beta$'s.)  In other words, there are five
different types of vertices allowed in the Feynman diagrams for this
theory.  

Furthermore, notice that this theory has no obvious $U(1)$ symmetry. 
For this reason, writing the field in terms of fluctuations around the
vacuum we choose is called ``breaking" the symmetry.  The symmetry is
still there, but it can't be seen in this form. 

Finally, notice that breaking the symmetry has resulted in the addition
of the massless field $\beta$.	It turns out that breaking global
symmetries as we have done \it always \rm results in a massless boson. 
Such particles are called \bf Goldstone Bosons\rm.  

\subsubsection{Breaking Local Symmetries}
\label{sec:breakinglocalsymmetries}

In the previous section, we broke a global $U(1)$ symmetry.  In this
section, we will break a local $U(1)$ and see what happens.  We begin
with the Lagrangian for a complex scalar field with a gauged $U(1)$:
\begin{eqnarray}
\mathcal{L} = -{1\over 2} \big[ \big( \partial^{\mu} -
iqA^{\mu}\big)\phi^{\dagger}\big]\big[\big(\partial_{\mu}+
iqA_{\mu}\big)\phi\big] - {1\over 4}F_{\mu \nu}F^{\mu \nu} -
V(\phi^{\dagger},\phi)\nolabel
\end{eqnarray}
where we have taken the external source $J^{\mu} = 0$.	Let's once
again assume $V(\phi^{\dagger},\phi)$ has the form of equation
(\ref{eq:degeneratevacuum}), so the vacuum has the $U(1)$ degeneracy at
$|\phi| = \Phi$.  

Because our $U(1)$ is now local, we choose $\alpha(x)$ so that not only
is the vacuum real, but also so that $\phi$ is always real.  We
therefore expand
\begin{eqnarray}
\phi = \Phi + h \label{eq:firsthiggs}
\end{eqnarray}
where $h$ is a real scalar field representing fluctuations around the
vacuum we chose.  

Now,
\begin{eqnarray}
\mathcal{L} &=& -{1\over 2} \big[\big(\partial^{\mu} -
iqA^{\mu}\big)\big(\Phi + h\big)\big]\big[\big(\partial_{\mu}+
iqA_{\mu}\big)\big(\Phi + h)\big] - {1\over 4} F_{\mu \nu}F^{\mu \nu}
\nolabel \\
& & -{1\over 2}\lambda m^2\big[\big(\Phi + h\big)\big(\Phi + h \big) -
\Phi^2\big]^2 \nolabel \\
&=& \cdots \nolabel \\
&=& -{1\over 2} \partial^{\mu}h \partial_{\mu}h - {1\over 2}4 \lambda
m^2\Phi^2h^2 - {1\over 4}F^{\mu \nu}F_{\mu \nu} - {1\over 2}q^2\Phi^2
A^2 + \mathcal{L}_{interactions}\nolabel
\end{eqnarray}
where the allowed interaction terms include a vertex connecting an $h$
and two $A^{\mu}$'s, four $h$'s, and three $h$'s.  

So, before breaking, we had a complex scalar field $\phi$ and a
\it massless \rm vector field $A^{\mu}$ with two polarization
states (because it is a photon).  Now, we have a single real scalar $h$
with mass $=\sqrt{4\lambda m^2\Phi^2}$ and a field $A^{\mu}$ with mass
$=q\Phi$.  In other words, our force-carrying particle $A^{\mu}$ has
gained mass!  We started with a theory with no mass, and by  merely
breaking the symmetry, we have introduced mass into our theory.  

This mechanism for introducing mass into a theory, called the \bf Higgs
Mechanism\rm, was first discovered by Peter Higgs, and the resulting
field $h$ is called the \bf Higgs Boson\rm.  

So, whereas the consequence of global symmetry breaking is a massless
boson called a Goldstone boson, the consequence of a local symmetry
breaking is that the gauge field, which came about as a result of the
symmetry being local, acquires mass.  

\subsubsection{Non-Abelian Gauge Theory}
\label{sec:nonabeliangaugetheory}

We are now ready to generalize what we did in section
\ref{sec:gaugingthesymmetry} to an arbitrary Lie group.  

Consider a Lagrangian $\mathcal{L}$ with $N$ scalar (or spinor) fields
$\phi_i$ ($i=1,\ldots,N$) that is invariant under a continuous $SO(N)$
or $SU(N)$ symmetry, $\phi_i \rightarrow U_{ij}\phi_j$, where $U_{ij}$
is an $N\times N$ matrix of $SO(N)$ or $SU(N)$.  

In section \ref{sec:gaugingthesymmetry}, we saw that if the group is
$U(1)$, gauging it demands the introduction of the gauge field
$A^{\mu}$ to preserve the symmetry, which shows up in the covariant
derivative $D_{\mu} = \partial_{\mu} - ieA_{\mu}$.  To say
a field carried some sort of charge means that it has the corresponding
term in its covariant derivative.  We then added a kinetic term for
$A^{\mu}$ as well as an external source $J^{\mu}$.  Then, higher order
interaction terms can be included in whatever way is appropriate for
the theory.  

To generalize this, let's say for the sake of concreteness that our Lie
group is $SU(N)$.  An arbitrary element of $SU(N)$ is $e^{ig
\theta^a(x)T^a}$, where $g$ is a constant we have added for later
convenience, $\theta^a$ are the $N^2-1$ parameters of the group (cf.
section \ref{sec:su3}), and the $T^a$ are the generator matrices for
the group.  Notice that we have gauged the symmetry (in that
$\theta(x)$ is a function of spacetime).  

By definition, we know that the generators $T^a$ will obey the
commutation relations 
\begin{eqnarray}
[T^a,T^b] = if_{abc}T^c\nolabel
\end{eqnarray}
(cf equation (\ref{eq:structureconstants})), where $f_{abc}$ are the
structure constants of the group.  

When gauging the $U(1)$ in section \ref{sec:gaugingthesymmetry}, the
transformation of the gauge field was given by equation
(\ref{eq:emgaugefieldtrans}).  For the more general transformation
$\phi_i \rightarrow U_{ij}\phi_j$, the gauge field transforms according
to
\begin{eqnarray}
A^{\mu} \rightarrow U(x)A^{\mu}U^{\dagger}(x) + {i \over
g}U(x)\partial^{\mu}U^{\dagger}(x)\nolabel
\end{eqnarray}
(where we have removed the indicial notation and it is understood that
matrix multiplication is being discussed).  If $U(x)$ is an element of
$U(1)$ (so it is $e^{ig\theta(x)})$, then this transformation reduces
to
\begin{eqnarray}
A^{\mu} \rightarrow e^{ig\theta(x)}A^{\mu}e^{-ig\theta(x)} + {i \over
g}e^{ig\theta(x)}(-ig\partial^{\mu}\theta(x))e^{-ig\theta(x)} =
A^{\mu}+\partial^{\mu}\theta(x)\nolabel
\end{eqnarray}
which is exactly what we had in (\ref{eq:gaugetransformation}).  For
general $SU(N)$, however, the $U$'s are elements of a Non-Abelian
group, and the $A^{\mu}$'s are matrices of the same size.  

Generalizing, we find that a general element of the $SU(N)$ is
(changing notation slightly)
\begin{eqnarray}
U(x)  e^{-ig\Gamma^a(x)T^a}\nolabel
\end{eqnarray}
with $N^2-1$ real parameters $\Gamma^a$.  We then build the covariant
derivative in the exact same way as in equation
(\ref{eq:covariantderivative}) by adding a term proportional to the
gauge field
\begin{eqnarray}
D_{\mu} = \mathbb{I}^{N\times N} \partial_{\mu} - igA_{\mu}\nolabel
\end{eqnarray}
(Remember that each component of $A_{\mu}$ is an $N\times N$ matrix. 
They were scalars for $U(1)$ because $U(1)$ is a $1\times 1$ matrix.) 
Or, acting on the fields, the covariant derivative is
\begin{eqnarray}
(D_{\mu}\phi)_j = \partial_{\mu}\phi_j(x) - ig[A_{\mu}(x)]_{jk}
\phi_k(x) \label{eq:newcovder}
\end{eqnarray}
where $k$ is understood to be summed on the last term.	It will be
understood from now on that the normal partial derivative term (the
first term) has an $N\times N$ identity matrix multiplied by it.  

Then, just as in (\ref{eq:deffmn}), we have the field strength
\begin{eqnarray}
F_{\mu \nu}(x) \equiv {i \over g}[D_{\mu},D_{\nu}] =
\partial_{\mu}A_{\nu} - \partial_{\nu}A_{\mu} - ig[A_{\mu},A_{\nu}]
\label{eq:secondiwant}
\end{eqnarray}
where the commutator term doesn't vanish for arbitrary Lie group as it
did for Abelian $U(1)$.  

Recall from equation (\ref{eq:invarianceoffmn}) that for $U(1)$,
$F_{\mu \nu}$ is invariant under the gauge transformation
(\ref{eq:gaugetransformation}) on its own, because the commutator term
vanishes.  In general, however, the commutator term does not vanish,
and we must therefore be careful in writing down the correct kinetic
term.  It turns out that the correct choice is 
\begin{eqnarray}
\mathcal{L}_{Kin} = -{1\over 2} \Tr(F_{\mu \nu}F^{\mu \nu})
\label{eq:fullkinterm}
\end{eqnarray}
It may not be obvious, but this form is actually a consequence of
(\ref{eq:killing}).  There is algebraic machinery working under the
surface of this that, while extremely interesting, is unfortunately
beyond the scope of what we are doing.	We will discuss all of these
ideas in much greater depth later in this series.  

So, starting with a non-interacting Lagrangian that is invariant under
the global $SU(N)$, we can gauge the $SU(N)$ to create a theory with a
gauge field (or synonymously a ``force carrying'' field) $A^{\mu}$, which
is an $N\times N$ matrix.  So, every Lie group gives rise to a
particular gauge field (which is a force carrying particle, like the
photon), and therefore a particular force.  

For this reason, we discuss forces in terms of Lie groups, or
synonymously \bf Gauge Groups\rm.  Each group defines a force.	As we
said at the very end of section \ref{sec:arbitraryj}, $U(1)$ represents
the electromagnetic force (as we have seen in section
\ref{sec:gaugingthesymmetry}, while $SU(2)$ describes the weak force,
and $SU(3)$ describes the strong color force.  

\subsubsection{Representations of Gauge Groups}

As we discussed in section \ref{sec:lieintro}, given a set of structure
constants $f_{abc}$, which define the Lie algebra of some Lie group, we
can form a representation of that group, which we denote $R$.  So, $R$
will be a set of $D(R)\times D(R)$ matrices, where $D$ is the dimension
of the representation $R$.  We then call the generators of the group
(in the representation $R$) $T^a_R$, and they naturally obey
$[T^a_R,T^b_R] = if_{abc}T^c_R$.  

One representation which exists for any of the groups we have
considered is the representation of $SO(N)$ or $SU(N)$ consisting of
$N\times N$ matrices.  We denote this the \bf Fundamental
Representation \rm (also called \bf Defining Representation \rm in some
books).  Clearly, the fundamental representations of $SO(2)$, $SO(3)$,
$SU(2)$, and $SU(3)$ are the $2\times 2$, $3\times 3$, $2\times 2$, and
$3\times 3$ matrix representations, respectively.  We will denote the
fundamental representation for a given group by writing the number in
bold.  So, the fundamental representation of $SU(2)$ will be denoted
\bf 2\rm, and the generators for $SU(2)$ in the fundamental
representation will be denoted $T^a_{\bf 2\rm}$.  Obviously, the
fundamental representation of $SU(3)$ will be \bf 3 \rm with generators
$T^a_{\bf 3\rm}$.  

Furthermore, let's say we have some arbitrary representation generated
by $T^a_R$, obeying $[T^a_R,T^b_R] = if_{abc}T^c_R$.  We can take the
complex conjugate of the commutation relations to get $[T^{\star
a}_R,T^{\star b}_R] = -if_{abc}T^{\star c}_R$.	So, notice that if we
define the \it new \rm set of generators $T'^a_R \equiv -T^{\star
a}_R$, then the $T'^a_R$ will obey the correct commutation relations,
and will therefore form a representation of the group as well.	If it
turns out that $T'^a_R	= -(T^a_R)^{\star} = T^a_R$, or if there is
some unitary similarity transformation $T^a_R \rightarrow U^{-1}T^a_RU$
such that $T'^a_R = -(T^a_R)^{\star} = T^a_R$, then we call the
representation \bf Real\rm, and the complex conjugate of the $T^a_R$'s
is the same representation.  However, if no such transformation exists,
then we have a \it new \rm representation, called the \bf Complex
Conjugate \rm representation to $R$, or the \bf Anti-{\boldmath$R$} \rm
representation, which we denote $\bar R$.  

For example, there is the fundamental representation of $SU(3)$,
denoted \bf 3\rm, generated by $T^a_{\bf 3\rm}$, and then there is the
anti-fundamental representation $\bf \bar 3\rm$, generated by $T^a_{\bf
\bar 3\rm}$.  

The representations of a group which will be important to us are the
fundamental, anti-fundamental, and adjoint.  

\subsubsection{Symmetry Breaking Revisited}

As we said in section \ref{sec:nonabeliangaugetheory}, given a field
transforming in a particular representation $R$, the gauge fields
$A^{\mu}$ will be $D(R)\times D(R)$ matrices.  

Once we know what representation we are working in, and therefore know
the generators $T^a_R$, it turns out that it is always possible to
write the gauge fields in terms of the generators.  Recall in sections
\ref{sec:generatorssection} and \ref{sec:rootspace}, we encouraged you
to think of the generators as basis vectors which span the parameter
space for the group.  Because the gauge fields live in the $N\times N$
space as well, we can write them in terms of the generators.  That is,
instead of the gauge fields being $N\times N$ matrices on their own, we
will use the $N\times N$ matrix generators as basis vectors, and then
the gauge fields can be written as scalar coefficients of each
generator:
\begin{eqnarray}
A^{\mu} = A^{\mu}_aT^a_R \label{eq:thirdiwant}
\end{eqnarray}
where $a$ is understood to be summed, and each $A^{\mu}_a$ is now a
scalar function rather than a $D(R)\times D(R)$ matrix (the advantage
of this is that we can continue to think of the gauge fields as scalars
with an extra index, rather than as matrices).	As a note, we haven't
done anything particularly profound here.  We are merely writing each
component of the $D(R)\times D(R)$ matrix $A^{\mu}$ in terms of the
$D(R)\times D(R)$ generators, allowing us to work with a \it scalar \rm
field $A^{\mu}_a$ rather than the matrix field $A^{\mu}$.  We now
actually view each $A^{\mu}_a$ as a separate field.  So, if a group has
$N$ generators, we say there are $N$ gauge fields associated with it,
each one having 4 spacetime components $\mu$.  

In matrix components, we will have 
\begin{eqnarray}
(A^{\mu})_{ij} = (A^{\mu}_aT^a_R)_{ij}\nolabel
\end{eqnarray}

Then, the covariant derivative in (\ref{eq:newcovder}) will be
\begin{eqnarray}
(D_{\mu}\phi)_j = \partial_{\mu}\phi_j(x) -ig[A_{\mu
a}(x)T^a_R]_{jk}\phi_k(x) \label{eq:anothercovder}
\end{eqnarray}

We may assume that the field strength $F^{\mu\nu}$ can also be
expressed in terms of the generators, so that we have 
\begin{eqnarray}
F^{\mu \nu} = F^{\mu \nu}_aT^a \label{eq:fmnintermsofgen}
\end{eqnarray}
or
\begin{eqnarray}
(F^{\mu \nu})_{ij} = (F^{\mu \nu}_aT^a)_{ij}
\end{eqnarray}

Now, using (\ref{eq:killing}) (and taking $\kappa = 1/2$ by
convention), we can write (\ref{eq:fullkinterm}) in terms of the new
basis:
\begin{eqnarray}
\mathcal{L}_{Kin} = -{1\over 2}\Tr(F_{\mu \nu}F^{\mu \nu}) &=& -{1 \over
2}\Tr(F_a^{\mu \nu}T^aF_{\mu \nu b}T^b) \nolabel \\
&=& -{1\over 2}F^{\mu \nu}_aF_{\mu \nu b}Tr(T^aT^b) \nolabel \\
&=& -{1\over 2}F^{\mu \nu}_a F_{\mu \nu b}\kappa \delta^{ab} \nolabel
\\
&=& -{1\over 2}F^{\mu \nu}_a F^a_{\mu \nu}\kappa \nolabel \\
&=& -{1\over 4}F^{\mu \nu}_a F^a_{\mu \nu}  \label{eq:xcvzxcv}
\end{eqnarray}
(we have raised the index $a$ on the second field strength term in the
last two lines simply to explicitly imply the summation over it.  The
fact that it is raised doesn't change its value in this case; it is
merely notational).  

Furthermore, we can use (\ref{eq:killing}) to invert
(\ref{eq:fmnintermsofgen}):
\begin{eqnarray}
F^{\mu \nu} = F^{\mu \nu}_aT^a &\Rightarrow & F^{\mu \nu} T^b = F^{\mu
\nu}_a T^aT^b \nolabel \\
&\Rightarrow& \Tr(F^{\mu \nu}T^b) = F^{\mu \nu}_a \Tr(T^aT^b) \nolabel \\
&\Rightarrow& \Tr(F^{\mu \nu}T^b) = F^{\mu \nu}_a\kappa \delta^{ab}
\nolabel \\
&\Rightarrow& \Tr(F^{\mu \nu}T^b) = {1\over 2}F^{\mu \nu}_b \nolabel \\
&\Rightarrow& F^{\mu \nu}_a = 2 \Tr(F^{\mu \nu}T^a) \label{eq:firstiwant}
\end{eqnarray}

In sections \ref{sec:spontaneoussymmetrybreaking} and
\ref{sec:breakinglocalsymmetries}, we broke the $U(1)$ symmetry, which
only had one generator.  However, if we break larger groups we may only
break part of it.  For example, we will see that $SU(3)$ has an $SU(2)$
subgroup.  It is actually possible to break only the $SU(2)$ part of
the $SU(3)$.  So, three of the $SU(3)$ generators are broken (the three
corresponding to the $SU(2)$ subgroup/subalgebra), and the other five
are unbroken.  Because we are now writing our gauge fields using the
generators as a basis, this means that three of the gauge fields are
broken, while five of the gauge fields are not.  

Finally, recall from section \ref{sec:breakinglocalsymmetries} that
breaking a local symmetry results in a gauge field gaining mass.  We
seek now to elucidate the relationship between breaking a symmetry and
a field gaining mass.  First, we can summarize as follows: \it Gauge
fields corresponding to broken generators get mass, while those
corresponding to unbroken generators do not.  The unbroken generators
form a new gauge group that is smaller than the original group that was
broken\rm.  

In \ref{sec:breakinglocalsymmetries}, we saw that breaking a symmetry
gave the gauge field mass.  Now, we see that giving a gauge field mass
will break the symmetry.  

To make this clearer, we begin with a very simple example, then move on
to a more complicated example.	

\subsubsection{Simple Examples of Symmetry Breaking}

Consider a theory with three real massless scalar fields $\phi_i$
($i=1,2,3$) and with Lagrangian
\begin{eqnarray}
\mathcal{L} = -{1\over 2} \partial^{\mu}\phi_i
\partial_{\mu}\phi^i\nolabel
\end{eqnarray}
which is clearly invariant under the global $SO(3)$ rotation
\begin{eqnarray}
\phi_i \rightarrow R^{ij}\phi_j\nolabel
\end{eqnarray}
where $R^{ij}$ is an element of $SO(3)$, because the Lagrangian is
merely a dot product in field space, and we know that dot products are
invariant under $SO(3)$.  

Now, let's say that one of the fields, say $\phi_1$, gains mass.  The
new Lagrangian will then be
\begin{eqnarray}
\mathcal{L} = -{1\over 2} \partial^{\mu}\phi_i \partial_{\mu}\phi^i -
{1\over 2} m^2 \phi^2_1\nolabel
\end{eqnarray}
So this Lagrangian is no longer invariant under the full $SO(3)$ group,
which mixes any two of the three fields.  Rather, it is only invariant
under rotations in field space that mix $\phi_2$ and $\phi_3$ or
$SO(2)$.  In other words, giving one field mass broke $SO(3)$ to the
smaller $SO(2)$.  

As another simple example, if we started with five massless complex
scalar fields $\phi_i$, with Lagrangian 
\begin{eqnarray}
\mathcal{L} = -{1\over 2} \partial^{\mu}\phi^{\dagger}_i
\partial_{\mu}\phi^i\nolabel
\end{eqnarray}
This will be invariant under any $SU(5)$ transformation.  

Then let's say we give two of the fields, $\phi_1$ and $\phi_2$ (equal)
mass.  The new Lagrangian will be 
\begin{eqnarray}
\mathcal{L} =
-{1\over2}\partial^{\mu}\phi_i^{\dagger}\partial_{\mu}\phi^i - {1\over
2}m(\phi^{\dagger}_1\phi_1 + \phi^{\dagger}_2 \phi_2)\nolabel
\end{eqnarray}
So now, we no longer have the full $SU(5)$ symmetry, but we do have the
special unitary transformations mixing $\phi_3$, $\phi_4$, and
$\phi_5$.  This is an $SU(3)$ subgroup.  Furthermore, we can do a
special unitary transformation mixing $\phi_1$ and $\phi_2$.  This is
an $SU(2)$ subgroup.  So, we have broken $SU(5) \rightarrow
SU(3)\otimes SU(2)$.  

Before considering a more complicated example of this, we further
discuss the connection between symmetry breaking and fields gaining
mass.  

When we introduced spontaneous symmetry breaking in section
\ref{sec:spontaneoussymmetrybreaking}, recall that we shifted the
potential minimum from $V_{minimum}$ at $\phi=0$ to $V_{minimum}$ at
$|\phi| = \Phi$.  But we were discussing this in very classical
language.  We can interpret all of this in a more ``quantum" way in
terms of VEV's.  As we said, the vacuum of a theory is defined as the
minimum potential field configuration.	For the $V_{minimum}$ at
$\phi=0$ potential, the VEV of the field $\phi$ was at 0, or 
\begin{eqnarray}
\langle 0 | \phi | 0 \rangle = 0\nolabel
\end{eqnarray}
However, for the $V_{minimum}$ at $|\phi| = \Phi$ potential, we have
\begin{eqnarray}
\langle 0 | \phi | 0 \rangle = \Phi\nolabel
\end{eqnarray}
So, in quantum mechanical language, symmetry breaking occurs when a
field, or some components of a field, take on a non-zero VEV.  

This seems to be what is happening in nature.  At higher energies,
there is some ``Master Theory" with some gauge group defining the
physics, and all of the fields involved have 0 VEV's.  At lower
energies, for whatever reason (the reason for this is not well
understood at the time of this writing), some of the fields take on
non-zero VEV's, which break the symmetry into smaller groups, giving
mass to certain fields through the Higgs Mechanism discussed in section
\ref{sec:breakinglocalsymmetries}.  We call the theory with the
unbroken gauge symmetry at higher energies the more fundamental theory
(analogous to equation (\ref{eq:morefundamental})), and the Lagrangian
which results from breaking the symmetry (analogous to
(\ref{eq:lowenergyeff})) the \bf Low Energy Effective Theory\rm.  

And this is how mass is introduced into the Standard Model.  It turns
out that if a theory is renormalizable one can prove that any lower
energy effective theory that results from breaking the original
theory's symmetry is also renormalizable, even if it doesn't appear to
be.  And, because the actions that appear to describe the universe at
the energy level we live at (and the levels attainable by current
experiment) are not renormalizable when they have mass terms, we work
with a larger theory which has no massive particles but can be
renormalized, and use the Higgs Mechanism to give various particles
mass.  So, whereas the physics we see at low energies may not appear
renormalizable, if we can find a renormalizable theory which breaks
down to our physics, we are safe.  

Now, we consider a slightly more complicated (and realistic) example of
symmetry breaking.  

\subsubsection{A More Complicated Example of Symmetry Breaking}
\label{sec:morecomplicatedkasdk}

Consider the gauge group $SU(N)$, acting on $N$ complex scalar fields
$\phi_i$ ($i=1,\ldots,N$) in the fundamental representation $\bf N\rm$. 
Recall that in section \ref{sec:breakinglocalsymmetries}, in order to
get equation (\ref{eq:firsthiggs}), we made use of the $U(1)$ symmetry
to make the vacuum, or the VEV, real.  We can now do something
similar: we make use of the $SU(N)$ to not only make the VEV real,
but also to rotate it to a single component of the field, $\phi_N$.  In
other words, we do an $SU(N)$ rotation so that
\begin{eqnarray}
\langle 0 |\phi_i|0\rangle &=& 0 \qquad \mbox{for} \qquad i=1,\ldots,N-1
\nolabel\\
\langle 0 | \phi_N| 0 \rangle &=& \Phi \nolabel
\end{eqnarray}
So, we expand $\phi_N$ around this new vacuum:
\begin{eqnarray}
\phi_i &=& \phi_i \qquad \mbox{for} \qquad i=1,\ldots,N-1 \nolabel\\
\phi_N &=& \Phi + \chi\nolabel
\end{eqnarray}
This means that, in the vacuum configuration, the fields will have the
form
\begin{eqnarray}
\begin{pmatrix}
\phi_1 \\ \phi_2 \\ \vdots \\ \phi_N
\end{pmatrix}_{vacuum} = 
\begin{pmatrix}
0 \\ 0 \\ \vdots \\ \Phi
\end{pmatrix}\nolabel
\end{eqnarray}

So, how will the action of $SU(N)$ be affected by this VEV?  If we
consider a general element of $SU(N)$ acting on this,
\begin{eqnarray}
\begin{pmatrix}
U_{11} & U_{12} & \cdots & U_{1N} \\
U_{21} & U_{22} & \cdots & U_{2N} \\
\vdots & \vdots & \ddots & \cdots \\
U_{N1} & U_{N2} & \vdots & U_{NN}
\end{pmatrix}
\begin{pmatrix}
0 \\ 0 \\ \vdots \\ \Phi
\end{pmatrix} = 
\begin{pmatrix}
U_{1N} \\ U_{2N} \\ \vdots \\ U_{NN}
\end{pmatrix}\nolabel
\end{eqnarray}
So, only elements of $SU(N)$ with non-zero elements in the last column
will be affected by this VEV.  But the other $N-1$ elements' rows and
columns are unaffected.  This means that we have an $SU(N-1)$ symmetry
left.  Or in other words, we have broken $SU(N)\rightarrow SU(N-1)$
with this VEV.  

Let's consider a specific example of this.  Consider $SU(3)$.  The
generators are written out in (\ref{eq:gellman}).  Notice that exactly
three of them have all zeros in the last column; $\lambda^1$,
$\lambda^2$, and $\lambda^3$.  We expect these three to give an
$SU(3-1)=SU(2)$ subgroup.  And looking at the upper left $2\times 2$
boxes in those three generators, we can see that they are the Pauli
matrices, the generators of $SU(2)$.  So, if we give a non-zero VEV
to the fields transforming under $SU(3)$, we see that they do indeed
break the $SU(3)$ to $SU(2)$.  The other five generators of $SU(3)$
will be affected by the VEV, and consequently the corresponding
fields will acquire mass.  

\subsection{Particle Physics}

\subsubsection{Introduction to the Standard Model}

We are finally ready to study the \bf Standard Model of Particle
Physics\rm, which (except for gravity) appears to be the theory which
explains our universe.	To state the Standard Model in the simplest
possible terms, it is

\begin{table} [h]
\centering
\begin{tabular}{|c|}
\hline
\; \\
A Yang-Mills (Gauge) Theory with Gauge Group \\ 
\; \\
$SU(3)\otimes SU(2)\otimes U(1)$ \\
\; \\
with left-handed Weyl fields fields in three copies of the
representation \\
\; \\
$(1,\bf 2 \rm, -1/2)\oplus (1,1,1)\oplus (\bf 3\rm,\bf 2\rm,1/6)\oplus
(\bf \bar 3\rm,1,-2,3)\oplus(\bf \bar 3\rm,1,1/3)$ \\
\; \\
(where the last entry specifies the value of the $U(1)$ hypercharge),
\\
\; \\
and a single copy of a complex scalar field in the representation \\
\; \\
$(1,\bf 2\rm,-1/2)$ \\
\; \\
\hline
\end{tabular} \label{smdef}
\end{table}

Admittedly, our exposition will be somewhat cursory.  This is largely
because every concept and tool we use in this section has been
discussed in detail in the previous sections.  The purpose of these
notes is to provide an introduction to the primary concepts and
mathematical tools used in Particle Physics, not to give the details of
the theory.  We will cover the main points of the Standard Model, but
there is tremendous detail we are skipping over.  A second reason the
following section is cursory is that we will not be working out every
step in detail, as we have been doing.	For nearly all calculations
being done in this section, we have worked out a similar tedious
calculation previously.  We will therefore frequently refer to previous
sections/equations.  It will be worthwhile to go back and carefully
study the parts which we refer to.  

Because this section is slightly more experimental, or at least
phenomenological, than the rest, and because the general purpose of
these notes is to develop the mathematical tools and framework of
particle physics (especially gauge theory), undue attention should not
be given to this section.  The purpose is merely to show, as briefly as
possible, where everything we have done so far lines up with
experiment.  It will be useful to read through this section, but do not
spend too much time bogged down in the details.  

Before diving into this in detail, look over the general structure of
the Standard Model on page \pageref{standardmodelsummary}.  

\subsubsection{The Gauge and Higgs Sector}

We begin our exposition with the \bf Electroweak \rm part of the
Standard Model gauge group, the $SU(2)\otimes U(1)$ part, as well as
the Higgs.  

Beginning with the Higgs, a scalar field in the $(\bf 2\rm, -1/2)$
representation of $SU(2)\otimes U(1)$, the first step is to write down
the covariant derivative as in (\ref{eq:anothercovder}).  We denote the
generators of the $\bf 2\rm$ representation of $SU(2)$ as $T^a_{\bf
2\rm} = {1\over 2} \sigma^a$ (the Pauli matrices) and the gauge fields
as $A^a_{\mu}$.  The generator of $U(1)$ is $Y=C\begin{pmatrix} 1 & 0
\\ 0 & 1 \end{pmatrix}$ where $C$ is the hypercharge ($-1/2$ in this
case), and the $U(1)$ gauge field is $B_{\mu}$.  So, the covariant
derivative is
\begin{eqnarray}
(D_{\mu} \phi)_i = \partial_{\mu}\phi_i - i[g_2A^a_{\mu}T^a_{\bf 2\rm}
+ g_1 B_{\mu}Y]_{ij}\phi_j \label{eq:asdfasdf}
\end{eqnarray}
where $g_1$ and $g_2$ are coupling constants for the $U(1)$ part and
the $SU(2)$ part, respectively.  If the reason we wrote it down this
way isn't clear, compare this expression to equation
(\ref{eq:anothercovder}), and remember that we are saying the field
carries \it two \rm charges; one for $SU(2)$ and one for $U(1)$. 
Therefore, it has two terms in its covariant derivative.  And, as
usual, $\mu$ is a spacetime index.  

Knowing that the generators of $SU(2)$ are the Pauli matrices, we can
expand the second part of the covariant derivative in matrix form,
\begin{eqnarray}
g_2A^a_{\mu}T^a_{\bf 2\rm}+g_1B_{\mu}Y &=& {g_2 \over
2}(A^1_{\mu}\sigma^1 + A^2_{\mu}\sigma^2 + A^3_{\mu}\sigma^3) -
{g_1\over 2}B_{\mu}\mathbb{I}^{2\times 2} \nolabel \\
&=& {1\over 2} 
\begin{pmatrix}
g_2A^3_{\mu} - g_1B_{\mu} & g_2(A^1_{\mu} - iA^2_{\mu}) \\
g_2(A^1_{\mu}+iA^2_{\mu}) & -g_2A^3_{\mu} - g_1B_{\mu}
\end{pmatrix}\nolabel
\end{eqnarray}

So, the full covariant derivative is
\begin{eqnarray}
(D_{\mu}\phi)_i~ \dot{=} \begin{pmatrix} D_{\mu}\phi_1 \\ D_{\mu}\phi_2
\end{pmatrix} = 
\begin{pmatrix}
\partial_{\mu}\phi_1 + {i\over 2}(g_2A^3_{\mu} - g_1B_{\mu})\phi_1 +
{ig_2\over 2}(A^1_{\mu} - i A^2_{\mu})\phi_2 \\
\partial_{\mu}\phi_2 + {ig_2 \over 2}(A^1_{\mu} + i A^2_{\mu})\phi_1 -
{i \over 2} (g_2 A^3_{\mu}+ g_1 B_{\mu})\phi_2
\end{pmatrix} \label{eq:higgscovder}
\end{eqnarray}

Now, we know that the Lagrangian will have the kinetic term and some
potential:
\begin{eqnarray}
\mathcal{L} = -{1\over 2} D_{\mu}\phi^{\dagger}_i D^{\mu}\phi_i -
V(\phi^{\dagger},\phi) \label{eq:higgslag}
\end{eqnarray}
Let's assume that the potential has a similar form as equation
(\ref{eq:degeneratevacuum}) (we add the factors of one-half here for
the sake of convention; they don't amount to anything other than a
rescaling of $\lambda$ and $\Phi$),
\begin{eqnarray}
V(\phi^{\dagger},\phi) = {1\over 4}\lambda \bigg(\phi^{\dagger}\phi -
{1\over 2} \Phi^2\bigg)^2 \label{eq:potentialupthere}
\end{eqnarray}

Clearly the minimum field configuration is not at $\phi=0$, but at
$|\phi| = {v \over \sqrt{2}}$.	So, following what we did in section
\ref{sec:morecomplicatedkasdk}, we make a global $SU(2)$ transformation
to put the entire VEV on the \it first \rm component of $\phi$, and
then make a global $U(1)$ transformation to make the field real.  So,
\begin{eqnarray}
\langle 0 |\phi|0 \rangle = {1\over \sqrt{2}}
\begin{pmatrix}
v \\ 0
\end{pmatrix}
\end{eqnarray}
and we expand $\phi$ around this new vacuum:
\begin{eqnarray}
\phi(x) = {1\over \sqrt{2}}
\begin{pmatrix}
v+ h(x) \\ 0
\end{pmatrix} \label{eq:asdjfahs}
\end{eqnarray}
Remember that we have chosen our $SU(2)$ to keep the second component
0 and our $U(1)$ to keep the first component real.  So, $h(x)$ is a
real scalar field.  

Clearly, plugging this into the covariant derivative
(\ref{eq:higgscovder}) will give the exact same expression as before,
but with $\phi_1$ replaced by ${1\over \sqrt{2}}h(x)$ and $\phi_2$
replaced by 0, plus an extra term for $v$.  When we plug this extra
term into the kinetic term in the Lagrangian (\ref{eq:higgslag}), we
get that it is
\begin{eqnarray}
-{1\over 8}v^2
\begin{pmatrix}
1 & 0
\end{pmatrix}
\begin{pmatrix}
g_2A^3_{\mu} - g_1B_{\mu} & g_2(A^1_{\mu} - iA^2_{\mu} \\
g_2(A^1_{\mu} + i A^2_{\mu} & -g_2A^3_{\mu} - g_1B_{\mu}
\end{pmatrix}
\begin{pmatrix}
g_2A^{3\mu} - g_1B^{\mu} & g_2(A^{1\mu} - iA^{2\mu}) \\
g_2(A^{1\mu} + iA^{2\mu}) & -g_2A^{3\mu} - g_1 B^{\mu}
\end{pmatrix}
\begin{pmatrix}
1 \\ 0
\end{pmatrix} \nolabel \\ \label{eq:asdfa}
\end{eqnarray}

Before multiplying this out, we employ a trick.  Define the \bf Weak
Mixing Angle \rm
\begin{eqnarray}
\theta_w \equiv \tan^{-1}\bigg({g_1 \over g_2}\bigg)\nolabel
\end{eqnarray}
and the shorthand notation
\begin{eqnarray}
s_w \equiv \sin \theta_w \qquad and \qquad c_w \equiv
\cos\theta_w\nolabel
\end{eqnarray}
And finally, we can define four new gauge fields as linear combinations
of the four we have been using:
\begin{eqnarray}
W^+_{\mu} &\equiv& {1\over \sqrt{2}}(A^1_{\mu} - i A^2_{\mu})
\label{eq:lowenef1}\\
W^-_{\mu} &\equiv& {1\over \sqrt{2}}(A^1_{\mu} + i A^2_{\mu})
\label{eq:lowenef3}\\
Z_{\mu} &\equiv & c_wA^3_{\mu} - s_wB_{\mu} \\
A_{\mu} &\equiv & s_wA^3_{\mu}+c_wB_{\mu}  \label{eq:lowenef2}
\end{eqnarray}
These can easily be inverted to give the old fields in terms of the new
fields,
\begin{eqnarray}
A^1_{\mu} &=& {1\over \sqrt{2}} (W^+_{\mu} + W^-_{\mu})
\label{eq:thisone}\\
A^2_{\mu} &=& {i \over \sqrt{2}}(W^+_{\mu} - W^-_{\mu}) \\
A^3_{\mu} &=& c_wZ_{\mu} + s_wA_{\mu} \\
B_{\mu} &=& -s_wZ_{\mu} + c_wA_{\mu} \label{eq:thisonetoo}
\end{eqnarray}

We make a few observations about these fields before moving on.  First
of all, they are merely linear combinations of the gauge fields
introduced in equation (\ref{eq:asdfasdf}).  Second, notice that the
two fields $W^{\pm}_{\mu}$ are both linear combinations of fields
corresponding to non-Cartan generators of $SU(2)$, whereas $Z_{\mu}$
and $A_{\mu}$ are both linear combinations of fields corresponding to
Cartan generators of $SU(2)$ and $U(1)$.  So, according to our
discussion in section \ref{sec:whatpoint}, we expect that $Z_{\mu}$ and
$A_{\mu}$ will interact but not change the charge, and that
$W^{\pm}_{\mu}$ will interact and change the charge.  Incidentally,
notice that $W^{\pm}_{\mu}$ has the exact form of the raising and
lowering operators defined in (\ref{eq:jpjm}).

With these fields defined, we can now rewrite (\ref{eq:asdfa}) as 
\begin{eqnarray}
-{1\over 8} g_2^2v^2 
\begin{pmatrix}
1 & 0
\end{pmatrix}
\begin{pmatrix}
{1\over c_2}Z_{\mu} & \sqrt{2}W^+_{\mu} \\
\sqrt{2}W^-_{\mu} & \star
\end{pmatrix}^2
\begin{pmatrix}
1 \\ 0
\end{pmatrix} = -M^2_wW^{+\mu}W^-_{\mu} - {1\over
2}M^2_ZZ^{\mu}Z_{\mu}\nolabel
\end{eqnarray}
(the $\star$ is there because that matrix element will always be
multiplied by 0, so we don't bother writing it), where we have
defined
\begin{eqnarray}
M_w = {g_2 v\over 2} \qquad \mbox{and} \qquad M_Z = {M_w \over c_w} = {g_2 v
\over 2c_w}\nolabel
\end{eqnarray}
So, we see that, by symmetry breaking, we have given mass to the
$W^+_{\mu}$, the $W^-_{\mu}$, and the $Z_{\mu}$ fields.  However, the
$A_{\mu}$ has not gained mass.	

These particles are the $W$ and $Z$ vector bosons, which are the force
carrying particles of the \bf Weak Force\rm.  Each of these particles
has an extremely large mass ($M_W  \approx$~80.4~GeV, and $M_Z \approx$~91.2~GeV),
which explains why they only act over a very short range
($\approx 10^{-18}$ meters).  

Also note that the $A_{\mu}$ remains massless, implying that it did not
acquire a VEV, and because it is a single generator, we see that a
single $U(1)$ remains unbroken.  This $U(1)$ and $A_{\mu}$ are the
gauge group and field of \bf Electromagnetism\rm, as discussed in
section \ref{sec:gaugingthesymmetry}.  

The idea of all of this is that at very high energies (above the
breaking of the $SU(2)\otimes U(1)$), we have only a Higgs complex
scalar field, along with four identical massless vector boson gauge
fields ($A^1_{\mu},A^2_{\mu},A^3_{\mu},B_{\mu}$), each of which behave
basically like a photon.  At low energies, however, the $SU(2)\otimes
U(1)$ symmetry of the Higgs is broken, and the low energy effective
theory consists of a linear combination of the original four fields. 
Three of those linear combinations have gained mass, and one remains
massless, retaining the photon-like properties from before symmetry
breaking.  The theory above the symmetry breaking scale is called the
\bf Electroweak Theory \rm (with four photon-like force carrying
particles), whereas below the breaking scale they become two separate
forces; the \bf Weak \rm and the \bf Electromagnetic\rm.  This is the
first and most basic example of unification we have in our universe. 
At low energies, the electromagnetic and weak forces are separate.  At
high energies, they unify into a single theory that is described by
$SU(2)\otimes U(1)$.  

We can express the new fields as simple Euler rotations of the old
fields:
\begin{eqnarray}
\begin{pmatrix}
Z_{\mu} \\ A_{\mu}
\end{pmatrix} = 
\begin{pmatrix}
A^3_{\mu} \cos\theta_w - B_{\mu} \sin\theta_w \\
A^3_{\mu} \sin\theta_w + B_{\mu} \cos\theta_w
\end{pmatrix} \Rightarrow
\begin{pmatrix}
Z_{\mu} \\ A_{\mu}
\end{pmatrix} = R(\theta_w)
\begin{pmatrix}
A^3_{\mu} \\ B_{\mu}
\end{pmatrix}\nolabel
\end{eqnarray}
So, the $Z_{\mu}$ is a massive linear combination of the $A^3_{\mu}$
and $B_{\mu}$, while the photon $A_{\mu}$ is a massless linear
combination of the two.  

We can do the same type of analysis for the $W^{\pm}_{\mu}$, where they
are both massive linear combinations of $A^1_{\mu}$ and $A^2_{\mu}$. 
The $Z_{\mu}$ and $A_{\mu}$ are both made up of a mixture of the
$SU(2)$ and $U(1)$ gauge groups, whereas the $W^{\pm}_{\mu}$ come
solely from the $SU(2)$ part.  

Before moving on to include leptons (and then hadrons), we first write
out the full Lagrangian for the effective field theory for $h(x)$ and
the gauge fields.  

We start with the complete Lagrangian term for $h(x)$.	We have written
the original field $\phi$ as in equation (\ref{eq:asdjfahs}).  So, our
potential in equation (\ref{eq:potentialupthere}) is now
\begin{eqnarray}
V(\phi^{\dagger},\phi) = {1\over 4} \lambda (\phi^{\dagger}\phi -
{1\over 2}v^2)^2 = \cdots = {1\over 4}\lambda v^2h^2 + {1\over
4}\lambda v h^3 + {1\over 16}\lambda h^4\nolabel
\end{eqnarray}
The first term on the right hand side is clearly a mass term giving the
mass of the Higgs ($=\sqrt{{\lambda \over 2}}v$), and the second two
terms are interaction vertices.  The kinetic term for the Higgs will be
the usual $-{1\over 2} \partial_{\mu}h \partial^{\mu}h$.  

Now, following loosely what we did in section
\ref{sec:gaugingthesymmetry}, we want to find kinetic terms for the
gauge fields.  We start by finding them for the original gauge fields
before symmetry breaking ($A^1_{\mu},A^2_{\mu},A^3_{\mu}$ and
$B_{\mu}$).  Using (\ref{eq:firstiwant}), (\ref{eq:secondiwant}), and
(\ref{eq:thirdiwant}), and the $SU(2)$ structure constants given in
equation (\ref{eq:su2structurecon}), we have
\begin{eqnarray}
F^1_{\mu \nu} &=& 2 \Tr(F_{\mu \nu}T^1) \nolabel \\
&=& 2 \Tr\big((\partial_{\mu}A_{\nu}-\partial_{\nu}A_{\mu} -
ig_2[A_{\mu},A_{\nu}])T^1\big) \nolabel \\
&=& 2\Tr\big((\partial_{\mu}A^a_{\nu}T^a - \partial_{\nu}A^a_{\mu}T^a -
ig_2A^a_{\mu}A^b_{\nu}[T^a,T^b])T^1\big) \nolabel \\
&=& 2\Tr(\partial_{\mu}A^a_{\nu}T^aT^1 - \partial_{\nu}A^a_{\mu}T^aT^1 -
ig_2A^a_{\mu}A^b_{\nu}if^{abd}T^cT^1) \nolabel \\
&=& \partial_{\mu}A^a_{\nu}\delta^{a1} -
\partial_{\nu}A^a_{\mu}\delta^{a1}+g_2A^a_{\mu}
A^b_{\nu}f^{abc}\delta^{c1} \nolabel \\
&=& \partial_{\mu}A^1_{\nu}-\partial_{\nu}A^1_{\mu}+
g_2A^a_{\mu}A^b_{\nu}f^{ab1} \nolabel \\
&=& \partial_{\mu}A^1_{\nu}-\partial_{\nu}A^1_{\mu}+
g_2A^a_{\mu}A^b_{\nu}\epsilon^{ab1} \nolabel \\
&=& \partial_{\mu}A^1_{\nu}-\partial_{\nu}A^1_{\mu}+
g_2(A^2_{\mu}A^3_{\nu} - A^2_{\nu}A^3_{\mu}) \nolabel
\end{eqnarray}
And similarly,
\begin{eqnarray}
F^2_{\mu \nu} &=& \partial_{\mu}A^2_{\nu} - \partial_{\nu}A^2_{\mu} +
g_2(A^3_{\mu}A^1_{\nu} - A^3_{\nu}A^1_{\mu})\nolabel \\
F^3_{\mu \nu} &=& \partial_{\mu}A^3_{\nu} - \partial_{\nu}A^3_{\mu} +
g_2(A^1_{\mu}A^2_{\nu} - A^1_{\nu}A^2_{\mu})\nolabel
\end{eqnarray}
And the gauge field corresponding to the $U(1)$ will be defined as in
(\ref{eq:qwerqwer}):
\begin{eqnarray}
B_{\mu \nu} = \partial_{\mu}B_{\nu} - \partial_{\nu}B_{\mu}\nolabel
\end{eqnarray}

So, we can now write the kinetic term for our fields according to
equation (\ref{eq:xcvzxcv}):
\begin{eqnarray}
\mathcal{L}_{Kin} = -{1\over 4}F_a^{\mu \nu}F^a_{\mu \nu} - {1\over
4}B^{\mu \nu}B_{\mu \nu}\nolabel
\end{eqnarray}
We can then use (\ref{eq:thisone}--\ref{eq:thisonetoo}) to translate
these kinetic terms into the new fields.  We will spare the extremely
tedious detail and skip right to the Lagrangian:
\begin{eqnarray}
\mathcal{L}_{eff} &=& -{1\over 4} F^{\mu \nu}F_{\mu \nu} - {1\over
4}Z^{\mu \nu}Z_{\mu \nu} - D^{\dagger
\mu}W^{-\nu}D_{\mu}W^+_{\nu}+D^{\dagger \mu}W^{-\nu}D_{\nu}W^+_{\mu}
\nolabel \\
& & + i e (F^{\mu \nu} + \cot\theta_wZ^{\mu \nu})W^+_{\mu}W^-_{\nu}
\nolabel \\
& & -{1\over 2}\bigg({e^2 \over \sin^2\theta_w}\bigg)(W^{+\mu}W^-_{\mu}
W^{+\nu}W^-_{\nu} - W^{+\mu}W^+_{\mu} W^{-\nu}W^-_{\nu}) \nolabel \\
& & - (M^2_WW^{+\mu}W^-_{\mu} + {1\over 2}M^2_ZZ^{\mu}Z_{\mu})
\bigg(1+{h\over v}\bigg)^2 \nolabel \\
&=& -{1\over 2}\partial^{\mu}h\partial_{\mu}h - {1\over 2}m^2_hh^2 -
{1\over 2}{m_h^2\over v}h^3 - {1\over 8}{m_h^2\over v^2}h^4 \nolabel
\end{eqnarray}
where we have chosen the following definitions:
\begin{eqnarray}
F_{\mu \nu} &=& \partial_{\mu}A_{\nu} - \partial_{\nu}A_{\mu} \quad 
\quad \mbox{(Electromagnetic Field Strength)}  \nolabel\\
Z_{\mu \nu} &=& \partial_{\mu}Z_{\nu} - \partial_{\nu}Z_{\mu} \quad 
\quad \mbox{(Kinetic term for $Z_{\mu}$)} \nolabel\\
D_{\mu} &=& \partial_{\mu} - ie(A_{\mu}+\cot\theta_wZ_{\mu})\nolabel
\end{eqnarray}
and the rest of the terms were defined previously in this section.  

\subsubsection{The Lepton Sector}

We now turn to the lepton sector (which is still in the $SU(2)\otimes
U(1)$ part of the Standard Model gauge group).	A \bf Lepton \rm is a
spin-$1/2$ particle that does \it not \rm interact with the $SU(3)$
color group (the strong force).  There are six \bf Flavors \rm of
leptons arranged into three \bf Families\rm, or \bf Generations\rm. 
The table on page \pageref{standardmodelsummary} explains this.  The
first generation consists of the electron ($e$) and the electron
neutrino ($\nu_e$), the second generation the muon ($\mu$) and the muon
neutrino ($\nu_{\mu}$), and the third the tau ($\tau$) and tau neutrino
($\nu_{\tau}$).  Each family behaves exactly the same way, so we will
only discuss one generation in this section ($e$ and $\nu_e$).	To
incorporate the physics of the other families, merely change the $e$ to
either a $\mu$ or $\tau$, and the $\nu_e$ to a $\nu_{\mu}$ or
$\nu_{\tau}$ in the following notes.  

What we will see is that, in a sense, the neutrinos don't really
interact with anything on their own (which is why they are incredibly
difficult to detect).  For this reason, neutrinos don't have their own
place in a representation of $SU(3)\otimes SU(2)\otimes U(1)$ (see
table on \pageref{standardmodelsummary}).  Electrons on the other hand,
do interact with other things on their own, and we therefore see them
in the $(1,1)$ representation.	

However, the neutrino \it does \rm interact with other things \it as
part of an \rm $SU(2)$ doublet \it with \rm the electron,
\begin{eqnarray}
l  = \begin{pmatrix} \nu_e \\ e \end{pmatrix} \label{eq:leptondoublet}
\end{eqnarray}
This is why it is arranged as it is on page \pageref{standardmodelsummary} with the electron
under the $(\bf 2\rm, -1/2)$ representation of $SU(2)\otimes U(1)$.  

This may seem confusing, but we hope the following will make it clear. 
We will proceed in what we believe is the clearest way to see this
(primarily following \cite{Srednicki}).  We start with 2 fields,
$\bar e$ and $l$, where $\bar e$ is a single left-handed Weyl field (see
section \ref{sec:leftright}), and $l$ is defined in
(\ref{eq:leptondoublet}).  As we have said, $l$ is in the $(\bf
2\rm,-1/2)$ representation, $\bar e$ is in the $(1,1)$ representation,
and $\nu_e$ has no representation of its own.  

So, mimicking what we did in equation (\ref{eq:asdfasdf}) in the
previous section, we can write down the covariant derivative for each
field,
\begin{eqnarray}
(D_{\mu}l)_i &=& \partial_{\mu}l_i - ig_2A^a_{\mu}(T^a)_{ij}l_j -
ig_1B_{\mu}Y_l l_i  \label{eq:mnbcvxnbmcvxbnmxcv}\\
D_{\mu}\bar e &=& \partial_{\mu}\bar e - ig_1 B_{\mu}Y_{\bar e} \bar e
\label{eq:uqyuiqweuiryqrewruyq}
\end{eqnarray}
The field $\bar e$ has no $SU(2)$ term in its covariant derivative
because the $1$ representation of $SU(2)$ is the trivial representation
- this means it doesn't carry $SU(2)$ charge.  Also, we know that
\begin{eqnarray}
Y_l = -{1\over 2} 
\begin{pmatrix}
1 & 0 \\ 0 & 1
\end{pmatrix} \label{eq:defyl}
\end{eqnarray}
and
\begin{eqnarray}
Y_{\bar e} = (1)
\begin{pmatrix}
1 & 0 \\ 0 & 1
\end{pmatrix} \label{eq:defye}
\end{eqnarray}

Following the Lagrangian for the spin-$1/2$ fields we wrote out in
equation (\ref{eq:diraclagrangian}), we can write out the kinetic term
for both (massless) fields:
\begin{eqnarray}
\mathcal{L}_{Kin} = il^{\dagger i} \bar \sigma^{\mu}(D_{\mu}l)_i + i
\bar e^{\dagger}\bar \sigma^{\mu}D_{\mu}\bar e
\label{eq:iouwertoipuwertoipuerwt}
\end{eqnarray}

At the end of section \ref{sec:scalarcanquant} and of section
\ref{sec:feynmandiag}, we briefly discussed the idea of
renormalization.  We said that certain theories can be renormalized and
others cannot.	It turns out (for reasons beyond the scope of these
notes) that while the theory we have outlined so far is renormalizable,
if we try to add mass terms for and $l$ and $\bar e$ fields, the theory
breaks down.  Therefore we cannot add a mass term.  But, we know
experimentally that electrons and neutrinos have mass, so obviously
something is wrong.  We must incorporate mass into the theory, but in a
more subtle way than merely adding a mass term.  It turns out that we
can use the Higgs mechanism as follows.  

While adding mass terms renders the theory inconsistent, we can add a
Yukawa term (cf. equation (\ref{eq:yukawa})),
\begin{eqnarray}
\mathcal{L}_{Yuk} = -y \epsilon^{ij}\phi_i l_j \bar e + \mbox{h.c.}\nolabel
\end{eqnarray}
where $y$ is another coupling constant, $\epsilon^{ij}$ is the totally
antisymmetric tensor, and h.c.~is the Hermitian Conjugate of the
first term.  

Now that we have added $\mathcal{L}_{Yuk}$ to the Lagrangian, we want
to break the symmetry exactly as we did in the previous section. 
First, we replace $\phi_1$ with ${1\over \sqrt{2}}(v+h(x))$ and
$\phi_2$ with 0, exactly as we did in equation (\ref{eq:asdjfahs}). 
So, 
\begin{eqnarray}
\mathcal{L}_{Yuk} &=& -y\epsilon^{ij}\phi_il_j\bar e + \mbox{h.c.} \nolabel \\
&=& -y(\phi_1l_2 - \phi_2l_1)\bar e + \mbox{h.c.} \nolabel \\
&=& -{1 \over \sqrt{2}}y(v+h)l_2\bar e + \mbox{h.c.} \nolabel \\
&=& -{1\over \sqrt{2}}y(v+h)e \bar e - {1\over \sqrt{2}}y(v+h)\bar e e
\nolabel \\
&=& -{1\over \sqrt{2}} y (v+h) \mathcal{\bar E} \mathcal{E}
\label{eq:adjshflakjsdf}
\end{eqnarray}
where $\mathcal{E} = \begin{pmatrix} e \\ \bar e^{\dagger}
\end{pmatrix}$ is the Dirac field for the electron ($e$ is the electron
and $\bar e^{\dagger}$ is the antielectron, or positron).  Comparing
(\ref{eq:adjshflakjsdf}) with (\ref{eq:truediraclagrangian}), we see
that it is a mass term for the electron and positron.  

Now we want a kinetic term for the neutrino.  It is believed that
neutrinos are described by Majorana fields (see section
\ref{sec:leftright}), so we begin with the field $\mathcal{N}' = 
\begin{pmatrix} \nu_e \\ \nu^{\dagger}_e \end{pmatrix}$.  Now, we
employ a trick.  Referring back to equations (\ref{eq:erty1}) and
(\ref{eq:erty2}), the kinetic term for Majorana fields has only one
term (because Majorana fields have only one Weyl spinor), whereas the
Dirac field sums over both Weyl spinors composing it.  So, instead of
working with the Majorana field $\mathcal{N}'$, we can instead work
with the Dirac field
\begin{eqnarray}
\mathcal{N} = \begin{pmatrix} \nu_e \\ 0\end{pmatrix}\nolabel
\end{eqnarray}
So, the Dirac kinetic term $i \mathcal{\bar N}
\gamma^{\mu}\partial_{\mu} \mathcal{N}$ will clearly result in the
correct kinetic term from (\ref{eq:iouwertoipuwertoipuerwt}), or
$i\nu^{\dagger}\bar \sigma^{\mu}\partial_{\mu}\nu$.  

Now, continuing with the symmetry breaking, we want to write the
covariant derivative (\ref{eq:mnbcvxnbmcvxbnmxcv}) and
(\ref{eq:uqyuiqweuiryqrewruyq}) in terms of our low energy gauge fields
(\ref{eq:lowenef1}--\ref{eq:lowenef2}).  

We said in the previous section (which echoed our discussion in section
\ref{sec:whatpoint}) that the gauge fields corresponding to Cartan
generators ($A_{\mu}$ and $Z_{\mu}$) act as force carrying particles,
but do not change the charge of the particles they interact with.  On
the other hand, the non-Cartan generators' gauge fields
($W^{\pm}_{\mu}$) are force carrying particles which \it do \rm change
the charge of the particle they interact with.	Therefore, to make
calculations simpler, we will break the covariant derivative up into
the non-Cartan part and the Cartan part.  

The non-Cartan part of the covariant derivative
(\ref{eq:mnbcvxnbmcvxbnmxcv}) is
\begin{eqnarray}
g_2(A^1_{\mu}T^1 + A^2_{\mu}T^2) &=& {1\over 2}g_2\bigg(A^1_{\mu}
\begin{pmatrix}
0 & 1 \\ 1 & 0
\end{pmatrix} + A^2_{\mu}
\begin{pmatrix}
0 & -i \\ i & 0
\end{pmatrix}\bigg) \nolabel \\
&=& {1\over 2} g_2
\begin{pmatrix}
0 & A^1_{\mu} - iA^2_{\mu} \\ 
A^1_{\mu} + i A^2_{\mu} & 0
\end{pmatrix} \nolabel \\
&=& {g_2\over \sqrt{2}}
\begin{pmatrix}
0 & W^+_{\mu} \\ 
W^-_{\mu} & 0
\end{pmatrix}\nolabel
\end{eqnarray}
and the Cartan part is
\begin{eqnarray}
g_2A^3_{\mu}T^3 + g_1B_{\mu}Y &=& {e \over s_w}(s_wA_{\mu} +
c_wZ_{\mu})T^3 + {e \over c_w}(c_wA_{\mu}-s_wZ_{\mu})Y \nolabel \\
&=& e(A_{\mu}+\cot\theta_wZ_{\mu})T^3+e(A_{\mu}-\tan\theta_wZ_{\mu})Y
\nolabel \\
&=& e(T^3+Y)A_{\mu}+e(\cot\theta_wT^3-\tan\theta_wY)Z_{\mu} \nolabel
\end{eqnarray}
We have noted before that $A_{\mu}$ is the photon, or the
electromagnetic field, and $e$ is the electromagnetic charge. 
Therefore, the linear combination $T^3+Y$ must be the generator of
electric charge.  Notice that the electromagnetic generator is in a
linear combination of the two Cartan generators of $SU(2)\otimes U(1)$. 

We know that $T^3 = {1\over 2}\sigma^3$, and $Y_l$ and $Y_{\bar e}$ are
defined in equations (\ref{eq:defyl}) and (\ref{eq:defye}), so we can
write 
\begin{eqnarray}
T^3l &=& {1\over 2}
\begin{pmatrix}
1 & 0 \\ 0 & -1
\end{pmatrix}
\begin{pmatrix}
\nu_e \\ e
\end{pmatrix} = {1\over 2}
\begin{pmatrix}
\nu_e \\ -e
\end{pmatrix}\nolabel \\
Y_ll &=& -{1\over 2}
\begin{pmatrix}
1 & 0 \\ 0 & 1
\end{pmatrix}
\begin{pmatrix}
\nu_e \\ e
\end{pmatrix} = -{1\over 2}
\begin{pmatrix}
\nu_e \\ e
\end{pmatrix}\nolabel
\end{eqnarray}
And we know that $\bar e$ carries no $T^3$ charge, so its $T^3$
eigenvalue is 0, while $Y_{\bar e}$ is $+1$.	So, summarizing all of
this, 
\begin{eqnarray}
T^3\nu_e = +{1\over 2}\nu_e \qquad T^3 e = -{1\over 2}e \qquad T^3\bar
e = 0\nolabel \\
Y\nu_e = -{1\over 2}\nu_e \qquad Ye = -{1\over 2}e \qquad Y \bar e =
+\bar e\nolabel
\end{eqnarray}
Then defining the generator of electric charge to be $Q\equiv T^3+Y$,
we have 
\begin{eqnarray}
Q\nu_e = 0 \qquad Qe = -e \qquad Q\bar e = + \bar e\nolabel
\end{eqnarray}
So the neutrino $\nu_e$ has no electric charge, the electron $e$ has
negative electric charge, and the antielectron, or positron, has plus
one electric charge---all exactly what we would expect.  

We can now take all of the terms we have discussed so far and write out
a complete Lagrangian.	However, doing so is both tedious and
unnecessary for our purposes.  

The primary idea is that electrons/positrons and neutrinos all interact
with the $SU(2)\otimes U(1)$ gauge particles, the $W^{\pm}$, $Z_{\mu}$,
and $A_{\mu}$.	The $Z_{\mu}$ and $A_{\mu}$ (the Cartan gauge
particles) interact but do not affect the charge.  On the other hand,
the $W^{\pm}$ act as $SU(2)$ raising and lowering operators (as can
easily be seen by comparing (\ref{eq:lowenef1}) and (\ref{eq:lowenef3})
to equation (\ref{eq:jpjm})).  The $SU(2)$ doublet state acted on by these raising
and lowering operators is the doublet in equation
(\ref{eq:leptondoublet}).  The $W^+$ interacts with a left-handed electron,
raising its electric charge from minus one to zero, turning it into a
neutrino.  However $W^+$ does not interact with left-handed neutrinos. 
On the other hand, $W^-$ will lower the electric charge of a neutrino,
making it an electron.	But $W^-$ will not interact with an electron.\footnote{This
does not mean that no vertex in the Feynman diagrams will include a $W^-$ and an 
electron field, but rather that if you collide an electron and a $W^-$, there will be no
interaction}

\subsubsection{The Quark Sector}

A Quark is a spin-$1/2$ particle that interacts with the $SU(3)$ color
force.	Just as with leptons, there are six flavors of quarks, arranged
in three families or generations (see the table on page
\pageref{standardmodelsummary}).  

Following very closely what we did with the leptons, we work with only
one generation.  Extending to the other generators is then trivial.  To
begin, define three fields: $q$, $\bar u$, and $\bar d$, in the
representations $(\bf 3\rm, \bf 2\rm, +1/6)$, $(\bf \bar 3\rm,1,-2/3)$,
and $(\bf \bar 3\rm, 1,+1/2)$ of $SU(3)\otimes SU(2)\otimes U(1)$.  The
field $q$ will be the $SU(2)$ doublet
\begin{eqnarray}
q = \begin{pmatrix} u \\ d \end{pmatrix} \label{eq:quarkdoublet}
\end{eqnarray}
This is exactly analogous to equation (\ref{eq:leptondoublet}).

Again, following what we did with the leptons, we can write out the
covariant derivative for all three fields:
\begin{eqnarray}
(D_{\mu}q)_{\alpha i} &=& \partial_{\mu} q_{\alpha i} -
ig_3A^a_{\mu}(T^a_2)^{\beta}_{\alpha}q_{\beta i} - ig_2
A^a_{\mu}(T^a_2)^j_iq_{\beta j} - ig_1 \bigg({1\over 6}\bigg)
B_{\mu}q_{\alpha i} \label{eq:qcovder} \\
(D_{\mu} \bar u)^{\alpha} &=& \partial_{\mu}\bar u^{\alpha} -
ig_3A^a_{\mu}(T^a_3)^{\alpha}_{\beta}\bar u^{\beta} -
ig_1\bigg(-{2\over 3}\bigg)B_{\mu}\bar u^{\alpha} \\
(D_{\mu}\bar d)^{\alpha} &=& \partial_{\mu} \bar d^{\alpha} -
ig_3A^a_{\mu}(T^{\alpha}_3)^{\alpha}_{\beta}\bar d^{\beta} -
ig_1\bigg({1\over 3}\bigg) B_{\mu} \bar d^{\alpha} 
\end{eqnarray}
where $i$ is an $SU(2)$ index and $\alpha$ is an $SU(3)$ index.  The
$SU(3)$ index is lowered for the $\bf 3\rm$ representation and raised
for the $\bf \bar 3\rm$ representation.  

Just as with leptons, we cannot write down a mass term for these
particles, but we can include a Yukawa term coupling these fields to
the Higgs:
\begin{eqnarray}
\mathcal{L}_{Yuk} = -y'\epsilon^{ij}\phi_i q_{\alpha j} \bar d^{\alpha}
- y'' \phi^{\dagger i } q_{\alpha i } \bar u^{\alpha} + \mbox{h.c.} \nolabel
\end{eqnarray}

As with the leptons, we can break the symmetry according to equation
(\ref{eq:asdjfahs}), and writing out this Yukawa term, we get
\begin{eqnarray}
\mathcal{L}_{Yuk} &=& -{1\over \sqrt{2}} y' (v+h)(d_{\alpha} \bar
d^{\alpha} + \bar d^{\dagger}_{\alpha}d^{\dagger \alpha}) - {1\over
\sqrt{2}} y'' (v+h)(u_{\alpha}\bar u^{\alpha} + \bar
u^{\dagger}_{\alpha} u^{\dagger \alpha}) \nolabel \\
&=& -{1\over \sqrt{2}} y' (v+h)\mathcal{\bar D}^{\alpha}
\mathcal{D}^{\alpha} - {1\over \sqrt{2}} y'' (v+h)\mathcal{\bar
U}^{\alpha}\mathcal{U}_{\alpha}\nolabel
\end{eqnarray}

where we have defined the Dirac fields for the up and down quarks:
\begin{eqnarray}
\mathcal{D}_{\alpha} \equiv \begin{pmatrix} d_{\alpha} \\ \bar
d^{\dagger}_{\alpha} \end{pmatrix}
\qquad 
\mathcal{U}_{\alpha} \equiv 
\begin{pmatrix}
u_{\alpha} \\ \bar u^{\dagger}_{\alpha}
\end{pmatrix}\nolabel
\end{eqnarray}

Notice that, whereas both the up and down quarks were massless before
breaking, they have now acquired masses
\begin{eqnarray}
m_d = {y' v \over \sqrt{2}} \qquad m_u = {y'' v \over \sqrt{2}}\nolabel
\end{eqnarray}

Writing out the non-Cartan and Cartan parts of the covariant
derivatives in terms of the lower energy $SU(2)\otimes U(1)$ gauge
fields, we get
\begin{eqnarray}
g_2 A^1_{\mu}T^1 + g_2A^2_{\mu}T^2 &=& {g_2 \over \sqrt{2}}
\begin{pmatrix}
0 & W^+_{\mu} \\ W^-_{\mu} & 0 
\end{pmatrix}\nolabel \\
g_2A^3_{\mu}T^3 + g_1B_{\mu}Y &=& eQA_{\mu}+{e \over s_wc_w} (T^3 -
s^2_wQ)Z_{\mu}\nolabel
\end{eqnarray}

And it is again straightforward to find the electric charge eigenvalue
for each field:
\begin{eqnarray}
Qu = + {2\over 3}u \qquad Qd = -{1\over 3}d \qquad Q\bar u = -{2\over
3} \bar u \qquad Q\bar d = +{1\over 3} \bar d\nolabel
\end{eqnarray}

Again, we can collect all of these terms and write out a complete
Lagrangian.  But, doing so is extremely tedious and unnecessary for our
purposes.  

The primary idea to take away is that the $SU(2)$ doublet
(\ref{eq:quarkdoublet}) behaves exactly as the lepton doublet in
(\ref{eq:leptondoublet}) when interacting with the ``raising" and
``lowering" gauge particles $W^{\pm}$.	This is why the $u$ and $d$ are
arranged in the $SU(2)$ doublet $q$ in (\ref{eq:quarkdoublet}), and why
$q$ carries the $SU(2)$ index $i$ in the covariant derivative
(\ref{eq:qcovder}), whereas $\bar u$ and $\bar d$ carry only the
$SU(3)$ index.	

The $SU(3)$ index runs from 1 to 3, and the 3 values are
conventionally denoted \it red, green\rm, and \it blue \rm $(r,g,b)$. 
These obviously are merely labels and have nothing to do with the
colors in the visible spectrum.  

The eight gauge fields associated with the eight $SU(3)$ generators are
called \bf Gluons\rm, and they are represented by the matrices in
(\ref{eq:gluons}).  We label each gluon as follows:
\begin{eqnarray}
g^{\beta}_{\alpha}~\dot{=} 
\begin{pmatrix}
r\bar r & r \bar g & r \bar b \\
g \bar r & g \bar g & g \bar b \\
b \bar r & b \bar g & b \bar b
\end{pmatrix}\nolabel
\end{eqnarray}
so that the upper index is the anti-color index, and denotes the column
of the matrix, and the lower index is the color index denoting the row
of the matrix.	Then, from (\ref{eq:gluons}), consider the gluon
\begin{eqnarray}
g^{\bar g}_r \propto 
\begin{pmatrix}
0 & 1 & 0 \\
0 & 0 & 0 \\
0 & 0 & 0 
\end{pmatrix}\nolabel
\end{eqnarray}
and the quarks
\begin{eqnarray}
q_r = \begin{pmatrix} 1 \\ 0 \\ 0 \end{pmatrix} \qquad 
q_g =\begin{pmatrix} 0 \\ 1 \\ 0 \end{pmatrix} \qquad  
q_b = \begin{pmatrix} 0 \\ 0 \\ 1 \end{pmatrix} \qquad \nolabel
\end{eqnarray}
It is easy to see that this gluon will interact as 
\begin{eqnarray}
g_r^{\bar g} q_r = 0 \qquad g_r^{\bar g}q_g = q_r \qquad g_r^{\bar
g}q_b = 0\nolabel
\end{eqnarray}

Or in other words, the gluon with the anti-green index will only
interact with a green quark.  There will be no interaction with the
other quarks.  Multiplying this out, and looking more closely at the
behavior of the $SU(3)$ generators and eigentstates as discussed in
sections \ref{sec:rootspace}--\ref{sec:su3}, you can work out all of
the interaction rules between quarks and gluons.  You will see that
they behave exactly according to the root space of $SU(3)$.  

\subsection{References and Further Reading}

The primary source for these notes is \cite{Srednicki}, which is an
exceptionally clear introduction to Quantum Field Theory.  We also used
a great deal of meterial from \cite{Cottingham}, \cite{Peskin},
\cite{Ryder}, and \cite{Zee}, all of which are outstanding QFT texts.
 The derivation of the Dirac equation came from \cite{Naber}, which is
written mostly above the scope of these notes, but is an excellent
survey of some of the mathematical ideas of Non-Perturbative QFT and
Gauge Theory.  

The sections on the Standard Model come almost entirely from
\cite{Srednicki} with little change, in that Srednicki's exposition
could hardly be improved upon for the scope of these notes.  

For further reading, we also recommend \cite{Bailin}, \cite{Gilmore2},
\cite{Nakahara}, \cite{Ramond}, and \cite{Ramond2}.

\section{The Standard Model --- A Summary}

\subsection{How Does All of This Relate to Real Life?}

In the fifth century B.C., a Greek named Empedocles took the ideas of
several others before him and combined them to say that matter is made
up of earth, wind, fire, and water, and that there are two forces, Love
and Strife, that govern the way they grow and act.  More
scientifically, he was saying that matter is made of smaller substances
that interact with each other through repulsion and attraction. 
Democritus, a contemporary of Empedocles, went a step further to say
that all matter is made of fundamental particles that are
indestructible.  He called these particles atoms, meaning
``indivisible"\footnote{Of course, our modern use of the word is
different.  At their discovery, it was thought that different elements
were the indivisible particles sought for, so the name atom seemed
appropriate}~\cite{Hakim}. 

The field of particle physics seeks to continue studying these same
concepts.  Are there fundamental, indivisible particles and if so, what
are they?  How do they behave?	How do they group together to form the
matter that we see?  How do they interact with each other? 

The current answer to these questions is called the Standard Model, the
theory we spent this paper developing.	We have now spent more than one
hundred pages expositing a series of mathematical tricks for various
types of ``fields".  In doing so, we talked about ``massless scalars
with $U(1)$ charge", and about things ``in a $j={1\over 2}$
representation of $SU(2)$".  But one could easily be left wondering how
exactly this relates to the things we see in nature.  We only discussed
25 particles in the previous section and in the table on page
\pageref{standardmodelsummary} (particles and antiparticles), but you
are likely aware that there are hundreds of particles in nature.  What
about those?  How does the mathematical framework detailed so far form
the building blocks for the universe? 

While we wish to reiterate that the primary purpose of these notes is
to provide the mathematical tools with which particle physics is done,
and not to outline the phenomenological details of the theory, we are
physicists still---not mathematicians.  Therefore, before concluding
this paper, we will take a brief hiatus from the mathematical rigor and
look at a qualitative summary of particle physics. 

Throughout this section, the footnotes will provide brief explanations
of the analogous mathematical ideas from above.  This
section\footnote{Nearly everything in this section is adapted from
\cite{Dunlap}, including the tables on page \pageref{firsttab}} can be read with or without paying attention to
them.  We provide them merely for those curious. 

\subsection{The Fundamental Forces}

The two forces most familiar to people are \bf Gravity \rm and \bf
Electromagnetism\rm.  Just the act of standing on the ground or sitting
in a chair makes use of both, and every ``Physics~I'' student has drawn a
free body diagram with a gravitational force going down and a normal
force (caused by the electromagnetic repulsion between the two objects)
going up.  However, these two are only half of the four fundamental
forces in our universe (that we know of). 

We can think about the third by first considering a compact nucleus
which we know to be made of protons and neutrons.  From
electromagnetism we know that the protons should repel each other
because of their like charge.  But the nuclei of atoms somehow hold together,
which is evidence for some stronger force that causes these particles to
attract.  This force, which overcomes the electromagnetic repulsions
and allows atomic nuclei to remain stable, is called the \bf Strong \rm
force.\footnote{The $SU(3)$ color force}  Just as electrically-charged
particles are subject to the electromagnetic force, some particles have
a property similar to charge, called \bf Color\rm, and are subject to
the strong force.  The field theory that describes this is called \bf
Quantum Chromodynamics \rm (QCD)\footnote{Again, the study of the
$SU(3)$ color force} and was first proposed in 1965 by Han, Nambu,
and Greenberg~\cite{Mehra}.  This theory predicts the existence of the
gluon, which is the mediator of the strong force between two matter
particles.

The fourth force is the one we have the least familiarity with.  It is
responsible for certain types of radioactive decays; for example,
permitting a proton to turn into a neutron and vice versa.  It is
called the \bf Weak \rm force.\footnote{The $SU(2)$ part that is left
over when $SU(2)\otimes U(1)$ is broken} 

In the 1960's, Sheldon Glashow, Abdus Salum, and Steven Weinburg
independently developed a gauge-invariant theory that unified the
electromagnetic and weak force~\cite{Mehra}.  At sufficiently high
energies it is observed that the difference between these two separate
forces is negligible and that they instead act together as the \bf
Electroweak \rm force.\footnote{The unbroken $SU(2)\otimes U(1)$ force}
 For processes at lower energy scales, the symmetry between the
electromagnetic and the weak force is broken and we observe two
different forces with different properties.  Similar to QCD,
electroweak theory predicts four force-carrier
particles\footnote{Corresponding to the 3 generators of $SU(2)$ and
the 1 in $U(1)$.  Two of them are Cartan and are, therefore,
uncharged, while two are non-Cartan and therefore carry charge} that
mediate the force between matter particles.  The mediating particle for
electromagnetism is the neutral photon, and those for the weak force are
the $W^+$ (with $+1$ electron charge), $W^-$ (with $-1$ electron
charge) and $Z^0$ (neutral) bosons.

The electromagnetic, weak, and strong forces forces described above
form what is called the \bf Standard Model of Particle Physics\rm.  The
Standard Model is an incomplete theory in the sense that it fails to
describe gravitation, the force that acts on matter.  Physicists
continue to work towards a theory that describes all four fundamental
forces, with String Theory currently the most promising.  The papers
later in this series will discuss these ideas.	For the rest of the
sections in this review, however, it should be assumed that we are
talking about physics under the Standard Model only\footnote{We are not
assuming Supersymmetry in this paper, though we will consider
Supersymmetry in a later paper} which, despite the shortcoming of not
explaining gravity, has tremendous experimental support. 

\subsection{Categorizing Particles}

In the last century, experimenters were surprised as they discovered
new particle after new particle.  It seemed disorganized and
overwhelming that there could be so many elementary objects. 
Eventually, however, the properties of these particles became better
understood and it was found that there really is just a small, finite
set of fundamental particles, some of which can be grouped together to
make up larger objects.  In the next two sections, we will introduce
the elementary particles and then will discuss the types of composite
particles. 

One property of the "zoo" of discovered particles that helps in our
organizing them is their intrinsic spin.\footnote{Or in other words,
which representation of $SU(2)$ they sit in}  Any particle,
elementary or composite, that is of half-integer spin\footnote{Is in
the $j={1\over 2}$ or $j={3\over 2}$ representation of $SU(2)$} is a
\bf Fermion\rm.  Those with integer spin are \bf Bosons\rm.\footnote{Is
in the $j=0$ or $j=1$ representation of $SU(2)$}  The spins govern the
statistics of a set of such particles, so fermions and bosons may also
be defined according to the statistics they obey. 

Namely, fermions obey Fermi-Dirac statistics and therefore also obey the
Pauli Exclusion Principle.  This means that no two identical fermions can be
found in the same quantum state at the same time.  Furthermore, to
accurately display this behavior it is found that the wave function of
a system with fermions must be antisymmetric; swapping any two like
fermions causes a change in sign of the overall wave function.

Bosons on the other hand obey Bose-Einstein statistics; any number of
the same type of particle can be in the same state at the same time. 
In contrast to fermions, the wavefunction of a system of bosons is
symmetric.\footnote{Cf. section \ref{sec:spinstat}}

The Venn diagrams on page \pageref{venn}, the table provided on page
\pageref{standardmodelsummary}, 
and the table below should be referenced as you read through what
follows. 

\subsection{Elementary Particles}

The elementary particles are those that are considered fundamental, or
in other words, are not composed of smaller particles.\footnote{These
are the ones that are in some representation of $SU(3)\otimes
SU(2)\otimes U(1)$ on the table on page
\pageref{standardmodelsummary}}  They can be divided into two groups:
matter particles and non-matter particles. 

The elementary matter particles all have half-integer spin (so are
fermions) and the elementary non-matter particles all have integer spin
(so are bosons). We can then observe that an equivalent grouping is
made if we divide the elementary particles instead by their intrinsic
spin, which is commonly done. Then an Òelementary matter particleÓ is
the same thing as an Òelementary fermion,Ó and similarly for the
bosons. The two terms are used interchangeably in the discussion below. 

\subsubsection{Elementary Fermions}
The elementary fermions are the building blocks of all other matter.
For example, the proton and neutron are made up of different
combinations of three elementary quarks. Electrons, which are also
elementary, cloud around the protons and neutrons, and when all three
group together in a particular way, an atom is formed. Less familiar
examples include those that are unstable, such as the muon, which decay
into something else fairly quickly. 

For every elementary (and sometimes composite) matter particle, there
is also a corresponding particle with the same mass but of different
charge and magnetic moment.\footnote{Cf. material on spin-$1/2$
particles in section \ref{sec:primquant}} Generally the name of such a
particle is the same as the corresponding ``normal" matter particle,
but with the prefix ``anti" in front of it (e.g. antiquark, antilepton,
etc.). In this paper, whenever we discuss matter
and its properties, it is implied that the antimatter
counterparts have similar properties.

Now we further divide the elementary fermions into two groups, quarks
and leptons. A convenient way to distinguish these two sets is by
whether or not they interact via the strong force: quarks may interact
via the strong force, while leptons do not.

%\newline\\
\begin{center}
\it Quarks \rm
\end{center}

Experiments involving high energy collisions of electrons and protons
led Murray Gell-Mann to suggest in 1964~\cite{Pickering} that protons
and neutrons are actually composite particles, made of three
point-like, spin-$1/2$ particles whose charges are either $-1/3$ or
$+2/3$ units of electron charge.  He called these particles \bf
Quarks\rm.  Through further experiments it has been found that there
are six flavors of quarks total, grouped into three generations with
the first generation containing the up and down quarks, the second
generation containing the more massive charm and strange quarks, and the
third generation containing the even more massive top and bottom quarks. 

As electrically charged particles are subject to the electromagnetic
force, quarks have a property similar to charge, called color, and any
colored particle is subject to the strong force. It is found that there
are three different types of colors: (defined as) red, green, and blue
(plus three more for antiquarks: antired, antigreen, and antiblue).
Quarks are grouped together to make composite particles that are
colorless (the color charges cancel out), which is why the concept of
color was only discovered after quarks themselves were found. The
addition of color to the quark model also ensures that any quarks
contained in a composite particle will not violate the exclusion
principle since each has a different color.  Again, QCD is the field
theory that describes these properties.

Another interesting feature of quarks is that they are never found
alone, but rather always inside of a composite particle. This
phenomenon is called \bf Confinement\rm.\footnote{We did not discuss
confinement in the main body of this paper, though it can be derived
from what we did discuss} It is more a property of the strong force,
which increases in strength as two colored particles are pulled away
from each other, just as would happen when the ends of a piece of
elastic are pulled apart. We can consider reaching a distance between
the two quarks where there is sufficient potential energy built up that
it can be converted to matter, creating a quark-antiquark pair. The
pair will separate and the resulting particles will recombine with the
original quarks.  As this process repeats, and more quark-antiquark pairs
are created, the end result in the whole process is a multiplication of the number of
quarks and of the number of composite particles. In the opposite
extreme, as two quarks get closer together, the strong force between
them becomes weaker until the quarks move around freely and more
independently. This is a called \bf Asymptotic Freedom.\rm

Quarks also interact with other particles via the weak force, which is
the only force that can cause a change of flavor (changing an up into a
down, for example).  When this happens, a quark either turns into a
heavier quark by absorbing a $W$ boson, or it emits a $W$ boson and
then decays to a lighter quark.  Beta decay, a common radioactive
process, is caused by this mechanism. Instead of just thinking of beta
decay as a neutron in the nucleus of an atom decaying, or splitting,
into a proton, electron, and antineutrino, we can go a step further
with our understanding of quarks subject to the weak force.  We add
that, really, it is one of the down quarks in the neutron that emits a
$W^-$ boson and then decays to the lighter up quark, keeping charge
conserved in the process.\footnote{Other conserved quantities are
momentum, energy, quark number, lepton number, and (approximately) lepton generation
number}  The neutron, which used to have one up and two down quarks,
now has one down and two up quarks, which is the composition of a
proton.  The electron and antineutrino are created from the decay of
the $W^-$ boson.

\newpage
\begin{center}
\it Leptons \rm
\end{center}

\bf Leptons \rm interact with other matter via the electromagnetic, the
weak, and gravitational forces, but not through the strong
force.\footnote{This means that leptons carry $SU(2)\otimes U(1)$ terms
in their covariant derivatives, but not $SU(3)$ terms} There are three
charged leptons, grouped, like the quarks, into three different
generations based on their masses.\footnote{This is equivalent to the
statement above that there are three copies of the Standard Model Gauge
Group - Cf. page \pageref{smdef}} The electron is the lightest of the
charged leptons, then the muon, and the tau. There are also three
neutral leptons, called neutrinos (``little neutral one"), one type for
each of the charged leptons: the electron neutrino, the muon neutrino,
and the tau neutrino. 

Some quantities in lepton events are found to be
conserved.\footnote{These conservation laws can all be derived from the
rules we discussed above, though they are typically treated separately
because they are extremely useful when talking about specific
interactions} If we define lepton number as the number of leptons
minus the number of antileptons, then lepton number is constant in all
interactions. Additionally, the lepton number \it within each generation \rm
is also approximately conserved. For example, the number of electrons and electron neutrinos
minus the number of antielectrons and electron antineutrinos is found
to be constant in most particle reactions.

An interesting exception is in neutrino oscillations, where a neutrino
changes lepton flavors as it travels. For example, we can take a
measurement and observe an electron neutrino, even though it was known
to have been created as a muon neutrino. These oscillations of flavor
only occur if neutrinos have mass (even just very small mass), so the
fact that the Standard Model currently predicts them to be massless
demonstrates that there are some parameters in the theory that need to
be adjusted.

\subsubsection{Elementary Bosons}

Throughout the development of the Standard Model it was found that some
elementary particles play a different role than the ordinary matter
particles that make up the ÒstuffÓ of the universe. Both the gauge
bosons and the Higgs boson fall into this group. 

%\newline\\
\begin{center}
\it Gauge Bosons \rm
\end{center}

In the mathematical formulation of quantum field theory, the Lagrangian
can be made invariant under a local gauge transformation by the
addition of a vector field called a gauge field.  As with the more
familiar example of an electron, the quanta of the gauge field is a
type of particle, which in this case is called a \bf Gauge Boson\rm. 
There are three types of gauge bosons described by the Standard
Model.\footnote{This is equivalent to saying that there are three gauge
groups, each with their own set of generators}  They are the photon,
which carries the electromagnetic force, the $W$ and $Z$ bosons, which
carry the weak force, and the gluons which carry the strong force. 
Each of these bosons have been experimentally detected.

Evidence for the neutral photon first came in 1905 when Einstein
proposed an explanation of the photoelectric effect, that light was
quantized into energy packets~\cite{Ford}.  Confirmation of the $W^+$,
$W^-$, and $Z^0$ bosons came in 1983 through proton-proton collisions at
the European Organization for Nuclear Research (CERN)~\cite{Ezhela}.  

The gluons were first experimentally observed in 1979 in the
electron-position collider at the German Electron Synchroton (DESY)
in Hamburg~\cite{Ezhela}.  Further experiments have demonstrated that
the gluons have eight different color states and that, because they
interact via the strong force, they have properties similar to quarks,
such as confinement.

Taking into account their possible charge or color, we find that there
are 12 gauge bosons in all, one for the electromagnetic force, three for the
weak force, and eight for the strong force. 

%\newline
\begin{center}
\it The Higgs Boson \rm
\end{center}

The Higgs boson is the only Standard Model particle that has not yet
been observed.	It is also the only elementary boson that is not a
gauge boson.  Rather, it is the carrier particle of the scalar Higgs
field from which other particles acquire mass.	The existence of the
Higgs would explain why some particles have mass and others do not. 
For example, the $W$ and $Z$ bosons are very massive, whereas the
photon is massless.  One of the main goals of the Large Hadron Collider
(LHC), located at CERN in Switzerland, is to provide evidence for
the Higgs.  It is expected to be in full operation in 2009.

\subsection{Composite Particles}

Examples of composite particles include hadrons, nuclei, atoms, and
molecules.  The latter three are well known and will not be
described here. 

Hadrons are made up of bound quarks and interact via the strong force.
They can be either fermions or bosons, depending on the number of
quarks that make them up. An odd number of bound quarks create a
spin-$1/2$ or spin-$3/2$ hadron, which is called a baryon, and an even
number of quarks create spin-0 or spin-1 hadrons, called mesons.
Experimentally, only combinations of three quarks or two quarks have
been found, so the terms baryon and meson often just refer to three or
two bound quarks, respectively.

You can understand why mesons and baryons have the spin that they do by
considering how many spin-$1/2$ quarks compose them.  A meson has two
quarks, and therefore the total spin of a meson is the sum of an even
number of half-integer spin particles, which will be integer spin.  And
because there are only two of them, it is either spin 0 or 1. 
Baryons, on the other hand, will have a linear combination of \it three
\rm particles with half-integer spin, which will of course be
half-integer: $1/2$ or $3/2$. 

The most well known examples of baryons are protons and neutrons. 
Protons are made of two up quarks and one down quark, or $|uud\rangle$, and
neutrons are made of two down and one up, or $|udd\rangle$.  The
baryons are made of ``normal" quarks only and their antimatter
counterparts are made of the corresponding antiquarks.

The mesons are made of a quark and an antiquark pair, though not
necessarily of the same generation.  Examples include the $\pi^+$ $|u\bar
d\rangle$ and $K^+$ $|u\bar s\rangle$. 

One of the reasons for the ÒzooÓ of particles discovered in the past
century is because of the numerous possible combinations of six quarks
put into a three-quark or two-quark hadron. Additionally, each of these
combinations can be in different quantum mechanical states, thereby
displaying different properties.  For example, a rho meson $\rho$ has the
same combination of quarks as a pion $\pi$, but the $\rho$ is spin-1
whereas the pion is spin-0. 

\subsection{Visualizing It All}

Finally, we provide a few tables which should help you see all of this
more clearly. 
\newline
\begin{table} [h]
\centering
\begin{tabular}{|c|c|c|c|}
\hline
\bf Interactions & \bf Acts On & \bf Strength & \bf Range \rm \\
\hline
\hline
Strong & Hadrons & 1 & $10^{-15}$ m \\
\hline
Electromagnetism & Electric Charges  & $10^{-2}$ & $\infty$ ($1/r^2$)
\\
\hline
Weak & Leptons and Hadrons & $10^{-5}$ & $10^{-18}$m \\
\hline
Gravity & Mass & $10^{-39}$ & $\infty$ ($1/r^2$) \\
\hline
\end{tabular} \label{firsttab}
\end{table}
\newline
where the relative strengths have been normalized to unity for the
strong force. 

Also, the four classes of force-carrying gauge bosons are shown
below;\footnote{The graviton is a the hypothetical carrying particle for
the gravitational force; it is not described by the Standard Model}
\newline
\begin{table} [h]
\centering
\begin{tabular}{|c|c|c|c|}
\hline
\bf Interaction & \bf Gauge Boson & \bf Spin & \bf Acts On \rm \\
\hline
\hline
Strong & Gluon & 1 & Hadrons \\
\hline
Electromagnetism & Photon & 1 & Electric Charges \\
\hline
Weak & $W^{\pm}$, $Z^-$, & 1 & Leptons and Hadrons \\
\hline
Gravity & Graviton & 2 & Mass \\
\hline
\end{tabular} \label{secondtab}
\end{table}
\newline

\section{A Look Ahead}

Now that we have completed our introduction to basic particle theory,
we can begin our uphill climb towards more fundamental concepts.  As a
preview, notice that everything we have done so far has been an
exposition of how gauge theories work. Our investigation into gauge
theories has been purely algebraic (working entirely from group theory,
as Part II demonstrates).  As gauge theory seems to be the correct
approach to understanding our universe, everything we do for the
remainder of this series  will be focused on a more fundamental
understanding of gauge theory, culminating in String Theory. 

As we just stated, we have been treating gauge theory as a purely
algebraic construct.  However String Theory, if true, must obviously be
able to reproduce the same general framework we have seen so far.  But,
String Theory is fundamentally a geometric construct.  As we will see,
String Theory will reproduce literally everything we have seen about
gauge theory, but from a geometric framework. 

This should not be entirely foreign, though.  Recall that, for
electromagnetism, the gauge group is $U(1)$.  We can ``draw" this
geometrically as a circle in the complex plane.  The Weak force is
represented by the gauge group $SU(2)$, which we have seen is
parameterized by three numbers, and therefore has three generators.  As we
discussed in these notes, we should think of these spaces as vector
spaces and the generators as basis vectors spanning the entire space. 
The same is true of $SU(3)$, though it is an eight-dimensional space. 
So, because there is a space associated with each of these groups, it
should be somewhat obvious that there is a natural geometric picture
associated with a Lie group. 

While the idea of the parameter space of a Lie group having a geometric
picture associated with it may seem straightforward, the geometry
undergirding gauge theory can be extremely complicated, and we
therefore must spend a significant amount of time investigating it. 
Therefore, the next paper in this series will be an introduction to the
geometric structure of gauge theory.  Just as we have built gauge
theory from algebra, we will in a sense start over and rebuild it using
geometry.  However, because we have already covered a great deal of
detail in the physics and mathematics of gauge theory and particle
physics in general, we will move much more quickly to avoid being
repetitive. 

When we finally get to String Theory (later in this series), we will
see that the geometric and algebraic pictures come together
beautifully, and that a thorough understanding of both will be
necessary to understand what may be the ``ultimate" theory of our
universe.

\newpage

\begin{sidewaystable}
\begin{tabular}{|c|c|c|c|c|c|c|c|c|}
\hline
 & Leptons & & & Hadrons & & & & Higgs \\
 \hline
  & (1, \bf 2\rm, --1/2) & (1, 1, 1) & & (\bf 3\rm, \bf 2\rm, 1/6) & (\bf 3\rm, 1, --2/3) & (\bf 3\rm, 1, 1/3) & & (1, \bf 2\rm, --1/2)  \\
\hline
Generation 1 & $\begin{pmatrix} \mbox{electron neutrino} \\ \mbox{electron} \end{pmatrix}$ & \mbox{electron} & & $\begin{pmatrix} \mbox{up} \\
 \mbox{down} \end{pmatrix}$  & \mbox{up} & \mbox{down} & & 1 Generation Only \\
\hline
Generation 2 & $\begin{pmatrix} \mbox{muon neutrino} \\ \mbox{muon} \end{pmatrix}$ & \mbox{muon} & & $\begin{pmatrix} \mbox{charm} \\
 \mbox{strange} \end{pmatrix}$  & \mbox{charm} & \mbox{strange} & &   \\
\hline
Generation 3 & $\begin{pmatrix} \mbox{tau neutrino} \\ \mbox{tau} \end{pmatrix}$ & \mbox{tau} & & $\begin{pmatrix} \mbox{top} \\
 \mbox{bottom} \end{pmatrix}$  & \mbox{top} & \mbox{bottom} & &  \\
\hline
\end{tabular} \label{standardmodelsummary}
\end{sidewaystable}

\newpage

\begin{center}
\begin{tabular}{c}Particles \\ \begin{tabular}{|c|}\hline \begin{tabular}{c}Fermions \\\begin{tabular}{||ccc||}\hline\hline  & Fundamental &   \\  & \begin{tabular}{|ccccc|}\hline   & Leptons (Spin-1/2) &   & Quarks (Spin-1/2) &   \\  & \begin{tabular}{|ccc|}\hline $\nu_e$ & $\nu_{\mu}$ & $\nu_{\tau}$ \\$e$ & $\mu$ & $\tau$ \\\hline \end{tabular} &   & \begin{tabular}{|ccc|}\hline $u$ & $c$ & $t$ \\$d$ & $s$ & $b$ \\\hline \end{tabular} &   \\  &   &   &   &   \\\hline \end{tabular} &   \\  & Composite &   \\  & \begin{tabular}{|ccc|}\hline   & Baryons &   \\  & \begin{tabular}{|ccccc|}\hline   & Spin-1/2 &   & Spin-3/2 &   \\  & \begin{tabular}{|c|}\hline proton = $|uud\rangle$ \\neutron = $|udd\rangle$ \\\hline \end{tabular} &   & \begin{tabular}{|c|}\hline $\Delta^{++} = |uuu\rangle$ \\$\Delta^- = |ddd\rangle$ \\\hline \end{tabular} &   \\  &   &   &   &   \\\hline \end{tabular} &   \\  &   &   \\\hline \end{tabular} &   \\  &   &   \\\hline\hline \end{tabular}\end{tabular}
 \\ \begin{tabular}{c}Bosons \\\begin{tabular}{||ccc||}\hline\hline   & Fundamental &   \\   & \begin{tabular}{|ccccc|}\hline  & Spin-0 &  & Gauge &  \\ & \begin{tabular}{|c|}\hline \\ Higgs\\  \\\hline \end{tabular} &  & \begin{tabular}{|ccccc|}\hline  & Spin-1 &  & Spin-2 &  \\ & \begin{tabular}{|c|}\hline $A^{\mu}$ $Z^{\mu}$ \\$W^{\pm}$ $g_i$ \\\hline \end{tabular} &  & \begin{tabular}{|c|}\hline Graviton \\\hline \end{tabular} &  \\ &  &  &  &  \\\hline \end{tabular} &  \\ &  &  &  &  \\\hline \end{tabular} &  \\  & Composite &   \\  & \begin{tabular}{|ccc|}\hline   & Mesons &   \\  & \begin{tabular}{|ccccc|}\hline   & Spin-0 &   & Spin-1 &   \\  & \begin{tabular}{|c|}\hline \\ $\pi^+ = |u\bar{d} \rangle$ \\ $K^+ = |u\bar s\rangle$ \\ \\ \hline \end{tabular} &   & \begin{tabular}{|c|}\hline \\ $\rho^+ = |u\bar{d} \rangle$ \\ $K^{\star+} = |u\bar s\rangle$ \\ \\ \hline \end{tabular} &   \\  &   &   &   &   \\\hline \end{tabular} &   \\  &   &   \\\hline \end{tabular} &   \\  &   &   \\\hline \hline \end{tabular}\end{tabular} \\  \\\hline \end{tabular} \end{tabular} \label{venn}
\end{center}

\end{document}